# Effects of demographic and weather parameters on COVID-19 basic reproduction number

**Igor Salom[1], Andjela Rodic[2], Ognjen Milicevic[3], Dusan Zigic[1], Magdalena Djordjevic[1], Marko Djordjevic[2,*]**

[1]Institute of Physics Belgrade, National Institute of the Republic of Serbia, University of Belgrade, Serbia

[2]Quantitative Biology Group, Faculty of Biology, University of Belgrade, Serbia

[3]Department for Medical Statistics and Informatics, School of Medicine, University of Belgrade, Serbia

*** Correspondence:**

Marko Djordjevic, e-mail: dmarko@bio.bg.ac.rs

## Abstract

It is hard to overstate the importance of a timely prediction of the COVID-19 pandemic progression. Yet, this is not possible without a comprehensive understanding of environmental factors that may affect the infection transmissibility. Studies addressing parameters that may influence COVID-19 progression relied on either the total numbers of detected cases and similar proxies (which are highly sensitive to the testing capacity, levels of introduced social distancing measures, etc.), and/or a small number of analyzed factors, including analysis of regions that display a narrow range of these parameters. We here apply a novel approach, exploiting widespread growth regimes in COVID-19 detected case counts. By applying nonlinear dynamics methods to the exponential regime, we extract basic reproductive number $R_0$ (i.e., the measure of COVID-19 inherent biological transmissibility), applying to the completely naïve population in the absence of social distancing, for 118 different countries. We then use bioinformatics methods to systematically collect data on a large number of potentially interesting demographics and weather parameters for these countries (where data was available), and seek their correlations with the rate of COVID-19 spread. While some of the already reported or assumed tendencies (e.g. negative correlation of transmissibility with temperature and humidity, significant correlation with UV, generally positive correlation with pollution levels) are also confirmed by our analysis, we report a number of both novel results and those that help settle existing disputes: the absence of dependence on wind speed and air pressure, negative correlation with precipitation; significant positive correlation with society development level (human development index) irrespective of testing policies, and percent of the urban population, but absence of correlation with population density *per se*. We find a strong positive correlation of transmissibility on alcohol consumption, and the absence of correlation on refugee numbers, contrary to some widespread beliefs. Significant tendencies with health-related factors are reported, including a detailed analysis of the blood type group showing consistent tendencies on Rh factor, and a strong positive correlation of transmissibility with cholesterol levels. Detailed comparisons of obtained results with previous findings, and limitations of our approach, are also provided.

## 1 Introduction

The ancient wisdom teaches us that "knowing your adversary" is essential in every battle - and this equally applies to the current global struggle against the COVID-19 pandemic. Understanding the parameters that influence the course of the pandemic is of paramount importance in the ongoing worldwide attempts to minimize the devastating effects of the virus which, to the present moment, has already taken a toll of more than a million lives (Dong et al., 2020), and resulted in double-digit recession among some of the major world economies (World Bank, 2020a). Of all such factors, the ecological ones (both abiotic such as meteorological factors and biotic such as demographic and health-related population properties) likely play a prominent role in determining the dynamics of disease progression (Qu et al., 2020).

However, making good estimates of the effects that general demographic, health-related, and weather conditions, have on the spread of COVID-19 infection is beset by many difficulties. First of all, these dependencies are subtle and easily overshadowed by larger-scale effects. Furthermore, as the effective rate of disease spread is an interplay of numerous biological, medical, social, and physical factors, a particular challenge is to differentiate the dominating effects of local COVID-19-related policies, which are both highly heterogeneous and time-varying, often in an inconsistent manner. And this is precisely where, in our view, much of the previous research on this subject falls short.

There are not many directly observable variables that can be used to trace the progression of the epidemics on a global scale (i.e., for a large number of diverse countries). The most obvious one - the number of detected cases - is heavily influenced both by the excessiveness of the testing (which, in turn, depends on non-uniform medical guidelines, variable availability of testing kits, etc.) and by the introduced infection suppression measures (where the latter are not only non-homogeneous but are also erratically observed (Cohen and Kupferschmidt, 2020). Nevertheless, the majority of the research aimed to establish connections of the weather and/or demographic parameters with the spread of COVID-19 seeks correlations exactly with the raw number of detected cases (Adhikari and Yin, 2020; Correa-Araneda et al., 2020; Fareed et al., 2020; Gupta et al., 2020; Iqbal et al., 2020; Li et al., 2020; Pourghasemi et al., 2020; Rashed et al., 2020; Singh and Agarwal, 2020). For the aforementioned reasons, the conclusions reached in this way are questionable. Other variables that can be directly measured, such as the number of hospitalized patients or the number of COVID-19 induced deaths (Pranata et al., 2020; Tosepu et al., 2020; Ward, 2020), again depend on many additional parameters that are difficult to take into account: level of medical care and current hospital capacity, advancements, and changing practices in treating COVID-19 patients, the prevalence of risk groups, and even on the diverging definitions of when hospitalization or death should be attributed to the COVID-19 infection. As such, these variables are certainly not suitable as proxies of the SARS-CoV-2 transmissibility *per se*.

On the other hand, as we here empirically find (and as theoretically expected (Anderson and May, 1992; Keeling and Rohani, 2011)) the initial stage of the COVID-19 epidemic (in a given country or area) is marked by a period of a nearly perfect exponential growth for a wide range of countries, which typically lasts for about two weeks (based on our analysis of the available data). One can observe widespread dynamical growth patterns for many countries, with a sharp transition between exponential, superlinear (growth faster than linear), and sublinear (growth slower than linear) regimes (see Fig. 1) – the last two representing a subexponential growth. We here concentrate on the initial exponential growth of the detected-case data (marked in red in Fig. 1), characterizing the period *before* the control measures took effect, and with a negligible fraction of the population resistant to infection. Note that dates which correspond to the exponential growth regime (included in Supplementary Table S1) are different for each country, corresponding to the different start of COVID-19 epidemic in those countries.

We use the exponential growth regime to deduce the basic reproduction number $R_0$ (Martcheva, 2015), following a simple and robust mathematical (dynamical) model presented here. $R_0$ is a straightforward and important epidemiological parameter characterizing the inherent biological transmissibility of the virus, in a completely naïve population, and the absence of social distancing measures (Bar-On et al., 2020; Eubank et al., 2020). To emphasize the absence of social distancing in the definition (and inference) of $R_0$ used here, the term $R_{0,free}$ is also used, – for simplicity, we further denote $R_0 \equiv R_{0,free}$. $R_0$ is largely independent of the implemented COVID-19 policies and thus truly reflects the characteristics of the disease itself, as it starts to spread unhampered through the given (social and meteorological) settings. Namely, the exponential period ends precisely when the effect of control measures kick in, which happens with a delay of ~10 days after their introduction (The Novel Coronavirus Pneumonia Emergency Response Epidemiology Team, 2020), corresponding to the disease latent period, and to the time between the symptom onset and the disease confirmation. Not only that very few governments had enacted any social measures before the occurrence of a substantial number of cases (Cohen and Kupferschmidt, 2020), but also the length of the incubation period makes it likely that the infection had been already circulating for some time through the community even before the first detected case (and that the effects of the measures are inescapably delayed in general). Also, the transition from the exponential to the subsequent subexponential phase of the epidemics is readily visible in the COVID data (see Fig. 1). Furthermore, $R_0$ is invariant to the particular testing guidelines, as long as these do not significantly vary over the (here relatively short) studied period. Note that in Fig. 1 cumulative number of positive cases (also known as cumulative infection incidence) is shown, which has to monotonically increase – though with a decreasing rate, once the infection starts to slow down, i.e., once the subexponential growth (sublinear and superlinear regimes) is reached.

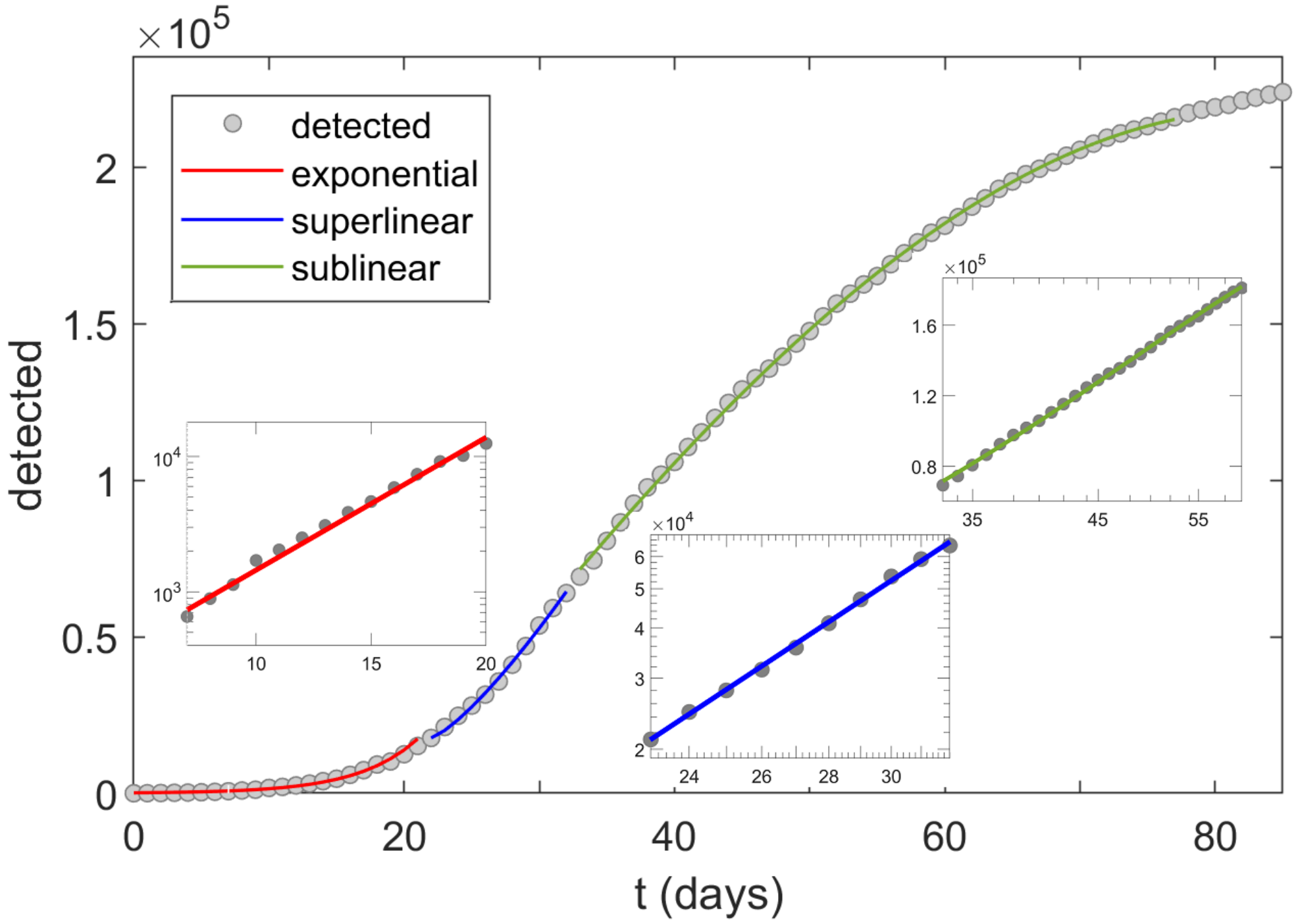

Figure 1: COVID-19 growth regimes. Transitions of the growth patterns (here shown for Italy) from exponential (red), to superlinear (blue) and sublinear (green) regime. The three insets correspond to the log-linear scale (exponential), log-log scale (superlinear), and linear-log scale (sublinear). Dots correspond to detected infections, starting from 20.02.2020. In this study, $R_{0,free}$ is extracted from the slope of the first (exponential, i.e., log-linear) inset, corresponding to dates 29.02-13.03 in the case of Italy.

In the analysis presented here, we consider 42 different weather, demographic, and health-related population factors, whose analyzed ranges correspond to their variations exhibited in 118 world countries (not all of the parameters were available for all of the countries, as discussed in section 2.4). While some authors prefer more coherent data samples to avoid confusing effects of too many different factors (Adhikari and Yin, 2020; Correa-Araneda et al., 2020; Fareed et al., 2020; Rashed et al., 2020; Singh and Agarwal, 2020; Tosepu et al., 2020), this consideration is outweighed by the fact that large ranges of the analyzed parameters serve to amplify the effects we are seeking to recognize and to more reliably determine the underlying correlations. For example, while the value of the Human Development Index (HDI, a composite index of life expectancy, education, and per capita income indicators) varies from 0.36 to 0.96 over the set of analyzed countries, this range would drop by an order of magnitude (Global Data Lab, 2020) if the states of the US were chosen as the scope of the study (other demographic parameters exhibit similar behavior). The input parameters must take values in some substantial ranges to have measurable effects on $R_0$ (i.e., small variations may lead to effects that are easily lost in statistical fluctuations).

The number of considered parameters is also significant, especially when compared to other similar studies (Adhikari and Yin, 2020; Copat et al., 2020; Fareed et al., 2020; Iqbal et al., 2020; Rashed et al., 2020; Rychter et al., 2020; Singh and Agarwal, 2020; Thangriyal et al., 2020; Tosepu et al., 2020). In a model where a large number of factors are analyzed under the same framework, consistency of the obtained results, in terms of agreement with other studies, common-sense expectations, and their self-consistency, becomes an important check of applied methodology and analysis. Furthermore, a comprehensive and robust analysis is expected to generate new findings and lead to novel hypotheses on how environmental factors influence COVID-19 spread. Overall, we expect that the understanding achieved here will contribute to the ability to understand the behavior of the pandemics in the future and, by the same token, to timely and properly take measures in an attempt to ameliorate the disease effects.

## 2 Model and parameter extraction

### *2.1 Modified SEIR model and relevant approximations*

There are various theoretical models and tools used to investigate and predict the progress of an epidemic (Keeling and Rohani, 2011; Martcheva, 2015). We here opted for the SEIR compartmental model, up to now used to predict or explain different features of COVID-19 infection dynamics (Maier and Brockmann, 2020; Maslov and Goldenfeld, 2020; Perkins and España, 2020; Tian et al., 2020; Weitz et al., 2020). The model is sufficiently simple to be applied to a wide range of countries while capturing all the features of COVID-19 progression relevant for extracting the $R_0$ values. The model assumes dividing the entire population into four (mutually exclusive) compartments with labels: (S)usceptible, (E)xposed, (I)fected, and (R)ecovered.

The dynamics of the model (which considers gradual transitions of the population from one compartment to the other) directly reflects the disease progression. Initially, a healthy individual has no developed SARS-CoV-2 virus immunity and is considered as "susceptible". Through contact with another infected individual, this person may become "exposed" - denoting that the transmission of the virus has occurred, but the newly infected person at this point has neither symptoms nor can yet transmit the disease. An exposed person becomes "infected" - in the sense of becoming contagious - on average after the so-called "latent" period which is, in the case of COVID-19, approximately 3 days. After a certain period of the disease, this person ceases to be contagious and is then considered as "recovered" (from the mathematical perspective of the model, "recovered" are all individuals who are no longer contagious, which therefore also includes deceased persons). In the present model, the recovered individuals are taken to be no longer susceptible to new infections (irrespectively of whether the COVID-19 immunity is permanent or not, it is certainly sufficiently long in the context of our analysis).

Accordingly, almost the entire population initially belongs to the susceptible class. Subsequently, parts of the population become exposed, then infected, and finally recovered. SARS-CoV-2 epidemic is characterized by a large proportion of asymptomatic cases (or cases with very mild symptoms) (Day, 2020), which leads to a large number of cases that remain undiagnosed. For this reason, only a portion of the infected will be identified (diagnosed) in the population, and we classify them as "detected". This number is important since it is the only direct observable in our model, i.e., the only number that can be directly related to the actual COVID-19 data.

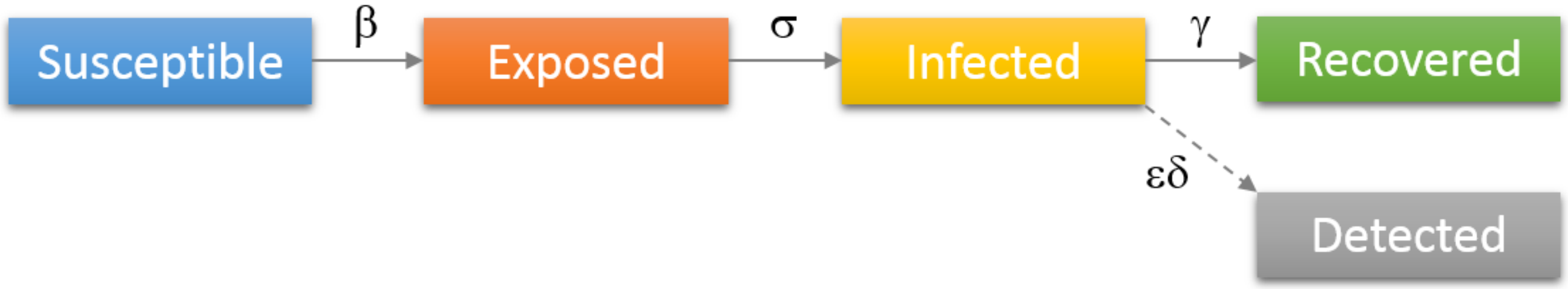

**Figure 2:** Diagrammatic representation of the SEIR model with the added class of "Detected" patients. Individuals move (denoted by solid arrows) from Susceptible to Exposed to Infected to Recovered, with the rates indicated above arrows in the figure. Some of the infected are detected (diagnosed/confirmed), indicated by the dashed arrow.

This dynamic is schematically represented in Fig. 2, and is governed by the following set of differential equations:

$$\frac{dS}{dt} = -\frac{\beta SI}{N} \quad (1.1.)$$

$$\frac{dE}{dt} = \frac{\beta SI}{N} - \sigma E \quad (1.2.)$$

$$\frac{dI}{dt} = \sigma E - \gamma I \quad (1.3.)$$

$$\frac{dR}{dt} = \gamma I \quad (1.4.)$$

$$\frac{dD}{dt} = \varepsilon\delta I \quad (1.5.)$$

In the above equations, $S$, $E$, $I$, and $R$ denote numbers of individuals belonging to, respectively, susceptible, exposed, infected, and recovered compartments, D is the cumulative number of detected cases, while $N$ is the total population. Parameter $\beta$ denotes the transmission rate, which is proportional to the probability of disease transmission in contact between a susceptible and an infectious subject. Incubation rate σ determines the rate at which exposed individuals become infected and corresponds to the inverse of the average incubation period. Recovery rate γ determines the transition rate between infected and recovered parts of the population, (i.e., 1/γ is the average period during which an individual is infectious). Finally, $\varepsilon$ and $\delta$ are detection efficiency and the detection rate. All these rate parameters are considered constant during the analyzed (brief) period. Also, note that the constants in our model do not correspond to transition probabilities per se, but rather to transition rates (with units 1/time), so that e.g. γ and $\varepsilon\delta$ do not add to one. While rates in the model can be rescaled and normalized to directly correspond to transition probabilities, our formulation (with rates rather than probabilities) is rather common (see e.g. (Keeling and Rohani, 2011)), and also has a direct intuitive interpretation, where the transition rates correspond to the inverse of the period that individuals spend in a given compartment (see e.g. the explanation for γ above).

In the first stage of the epidemic, when essentially the entire population is susceptible (i.e., $S/N \approx 1$) and no distancing measures are enforced, the average number of secondary infections, caused directly by primary infected individuals, corresponds to the basic reproduction number $R_0$. The infectious disease can spread through the population only when $R_0 > 1$ (Khajanchi et al., 2020a), and in these cases, the initial growth of the infected cases is exponential. Though $R_0$ is a characteristic of the pathogen, it also depends on environmental abiotic (e.g. local weather conditions), as well as biotic factors (e.g. prevalence of health conditions, and population mobility tightly related to the social development level).

Note that, as we seek to extract the basic reproduction number $R_0$ from the model for a wide range of countries, the social distancing effects are not included in the model presented above. That is, the introduced model serves only to explain the exponential growth phase – note that this growth regime characterizes part of the infection progression where the social distancing interventions still did not take effect, and where the fraction of resistant (non-susceptible) population is still negligible. It is only this phase which is relevant for extracting $R_0$ that is used in the subsequent analysis. $R_0$ should not be confused with the effective reproduction number $R_e$, which takes into account also the effects of social distancing interventions and the decrease in the number of susceptibles due to acquired infection resistance. $R_e$ is not considered in this work, as we are concerned with the factors that affect the inherent biological transmissibility of the virus, independently from the applied measures. That is, by considering $R_0$ rather than $R_e$, we disentangle the influence of meteorological and demographic factors on transmissibility (the goal of this study), from the effects of social distancing interventions (not analyzed here). The model can, however, be straightforwardly extended to include social distancing measures, as we did in (Djordjevic et al., 2020) - social distancing measures were also included through other frameworks (Khajanchi and Sarkar, 2020; Maier and Brockmann, 2020; Maslov and Goldenfeld, 2020; Perkins and España, 2020; Samui et al., 2020; Sarkar et al., 2020; Tian et al., 2020; Weitz et al., 2020). Such extensions are needed to explain the subexponential growth that emerges due to intervention measures (i.e., superlinear and sublinear growth regimes that are illustrated in Fig. 1 for Italy but are common for other countries as well).

### *2.2 COVID-19 growth regimes*

If we observe the number of total COVID-19 cases (e.g. in a given country) as a function of time, there is a regular pattern that we observe: the growth of the detected COVID cases is initially exponential but slows down after some time – when we say it enters the subexponential regime. The subexponential regime can be further divided into the superlinear (growing asymptotically faster than a linear function) and sublinear regime (the growth is asymptotically slower than a linear function). This typical behavior is illustrated, in the case of Italy, in Fig. 1 above. The transition to the subexponential regime occurs relatively soon, much before a significant portion of the population gains immunity, and is a consequence of the introduction of the infection suppression measures.

### *2.3 Inference of the basic reproduction number $R_0$*

In the initial exponential regime, a linear approximation to the model can be applied. Namely, in this stage, almost the entire population is susceptible to the virus, i.e., S/N≈ 1, which simplifies the equation (1.2) to:

$$\frac{dE}{dt} = -\sigma E + \beta I. \qquad (3.1.)$$

By combining expressions (1.3.) and (3.1.) one obtains:

$$\frac{d}{dt}\begin{pmatrix} E \\ I \end{pmatrix} = \begin{pmatrix} -\sigma & \beta \\ \sigma & -\gamma \end{pmatrix}\begin{pmatrix} E \\ I \end{pmatrix} = A\begin{pmatrix} E \\ I \end{pmatrix}, \qquad (3.2.)$$

where we have introduced a two-by-two matrix:

$$\mathrm{A} = \begin{pmatrix} -\sigma & \beta \\ \sigma & -\gamma \end{pmatrix} \qquad (3.3.)$$

The solution for the number of infected individuals can now be written:

$$I(t) = C_1 \cdot e^{\lambda_+ t} + C_2 \cdot e^{\lambda_- t}, \qquad (3.4.)$$

where $\lambda_+$ and $\lambda_-$ denote eigenvalues of the matrix A, i.e., the solutions of the equation:

$$\det(A - \lambda I) = 0. \qquad (3.5.)$$

The eigenvalues must satisfy:

$$\begin{vmatrix} -\sigma - \lambda & \beta \\ \sigma & -\gamma - \lambda \end{vmatrix} = 0,$$

leading to:

$$(\lambda + \sigma) \cdot (\lambda + \gamma) - \beta \cdot \sigma = 0. \qquad (3.6.)$$

The solutions of (3.6.) are:

$$\lambda_\pm = \frac{-(\gamma + \sigma) \pm \sqrt{(\gamma - \sigma)^2 + 4\beta\sigma}}{2}. \qquad (3.7.)$$

Since $\lambda_- < 0$, the second term in (3.4) can be neglected for sufficiently large t. More precisely, numerical analysis shows that this approximation is valid already after the second day, while, for the extraction of $R_0$ value we will anyhow ignore all data before the fifth day (for the analyzed countries, numbers of cases before the fifth day were generally too low, hence this early data is dominated by stochastic effects/fluctuations). Hence, $I(t)$ is proportional to $exp(\lambda_+ t)$, i.e.:

$$I(\mathrm{t}) = I(0) \cdot e^{\lambda_+ t}. \qquad (3.8.)$$

By using β from (3.7) and $R_0 = \frac{\beta}{\gamma}$ (Keeling and Rohani, 2011; Martcheva, 2015), we obtain:

$$R_0 = 1 + \frac{\lambda_+ \cdot (\gamma + \sigma) + {\lambda_+}^2}{\gamma * \sigma}. \qquad (3.9.)$$

From (3.8.) and (1.5.) we compute:

$$D(\mathrm{t}) = \varepsilon \cdot \delta \cdot I(0) \cdot \frac{(e^{\lambda_+ t} - 1)}{\lambda_+}. \qquad (3.10.)$$

By taking the logarithm, the above expression leads to:

$$\log\big(D(\mathrm{t})\big) = \log\big(\varepsilon \delta I(0)/\lambda_+\big) + \lambda_+ \cdot \mathrm{t}, \qquad (3.12.)$$

from which $\lambda_+$ can be obtained as the slope of the $\log\big(D(\mathrm{t})\big)$ function. From equation (3.9.), we thus obtain the $R_0$ value as a function of the slope of $\log\big(D(\mathrm{t})\big)$, where the latter can be efficiently inferred from the plot of the number of detected COVID-19 cases for a large set of countries.

The SEIR model and the above derivation of $R_0$ assume that the population belonging to different compartments is uniformly mixed. Possible heterogeneities may tend to increase $R_0$ values (Keeling and Rohani, 2011). However, this would not influence the results obtained below, as our $R_0$ values are consistently inferred for all analyzed countries by using the same model, methodology, and parameter set. Moreover, our $R_0$ values are in agreement with the prevailing estimates in the literature (Najafimehr et al., 2020).

### *2.4 Demographic and weather data acquisition*

For the countries for which $R_0$ was determined through the procedure above, we also collect a broad spectrum of meteorological and demographic parameter values. Overall, 118 countries were selected for our analysis, based on the relevance of the COVID-19 epidemiological data. Namely, a country was considered as relevant for the analysis if the number of detected cases on June 15th was higher than a threshold value of 1000. A few countries were then discarded from this initial set, where the case count growth was too irregular to extract any results, possibly due to inconsistent or irregular testing policies. As a source for detected cases, we used (World Bank, 2020b; Worldometer, 2020).

In the search for factors correlated with COVID-19 transmissibility, we have analyzed overall 42 parameters, 11 of which are related to weather conditions, 30 to demographics or health-related population characteristics, and one parameter quantifying a delay in the epidemic's onset. Not all of these parameters were available for all of the considered countries. In particular, data on the prevalence of blood types (Table S4 in the Supplement) was possible to find for 83 of the 118 countries, while, primarily due to scarce data on pollutant concentrations during the epidemics, almost 30 percent of entries in Table S5 in Supplement had to be left blank for this category. Nevertheless, we opted to include these parameters in our report: despite the lower number of values, some of these parameters exhibited strong and highly statistically significant correlations with $R_0$, warranting their inclusion.

Our main source of weather data was project POWER (Prediction Of Worldwide Energy Resources) of the NASA agency (NASA Langley Research Center, 2020). A dedicated Python script was written and used to acquire weather data via the provided API (Application Programming Interface). NASA project API allows a large set of weather parameters to be obtained for any given location (specified by latitude and longitude) and given date (these data are provided in the Supplementary Table S7). From this source, we gathered data on temperature (estimated at 2 meters above ground), specific humidity (estimated at 2 meters above ground), wind speed (estimated at 2 meters above ground), and precipitation (defined as the total column of precipitable water). Data on air pressure (at ground level) and UV index (international standard measurement of the strength of sunburn-producing ultraviolet radiation) were collected via similar API from World Weather Online source (World Weather Online, 2020), using the same averaging methodology. Since we needed to assign a single value to each country (for each analyzed parameter), the following method was used for averaging meteorological data. In each country, a number of largest cities[1] were selected and weather data was taken for the corresponding locations. These data was then averaged, weighted by the population of each city, followed by averaging over the period used for $R_0$ estimation (more precisely, to account for the time between disease transmission and the case confirmation, we shifted this period 12 days into the past). The applied averaging method used here can be of limited adequacy in countries spreading over multiple climate zones, but is still expected to provide reasonable single-value estimates of the weather parameters, particularly since the averaging procedure was formulated to reflect the most likely COVID-19 hotspots in a given country.

Demographic data was collected from several sources. Percentage of the urban population, refugees, net migration, social and medical insurance coverage, infant mortality, and disease (CVD, cancer, diabetes, CRD) risk was taken from the World Bank organization (World Bank, 2020b). The HDI was taken from the Our World in Data source (Our World in Data, 2020), while median age information was obtained from the CIA website. The source of most of the considered medical parameters: cholesterol, raised blood pressure, obesity, inactivity, BSG vaccination as well as data on alcohol consumption and smoking prevalence was World Health Organization (World Health Organization, 2020). Data for blood types were taken from the Wikidata web site. BUCAP parameter, representing population density in the built-up area, was taken from GHS Urban Centre Database 2015 (European Commission Global Human Settlement, 2020). The onset parameter, determining the delay (in days) of the epidemic's start, was inferred from COVID-19 counts data. We used the most recent available data for all the parameters.

[1] This number was determined for each country by the following condition: the total population of the cities taken into consideration had to surpass 10 percent of the overall population of the country.

## 3 Results

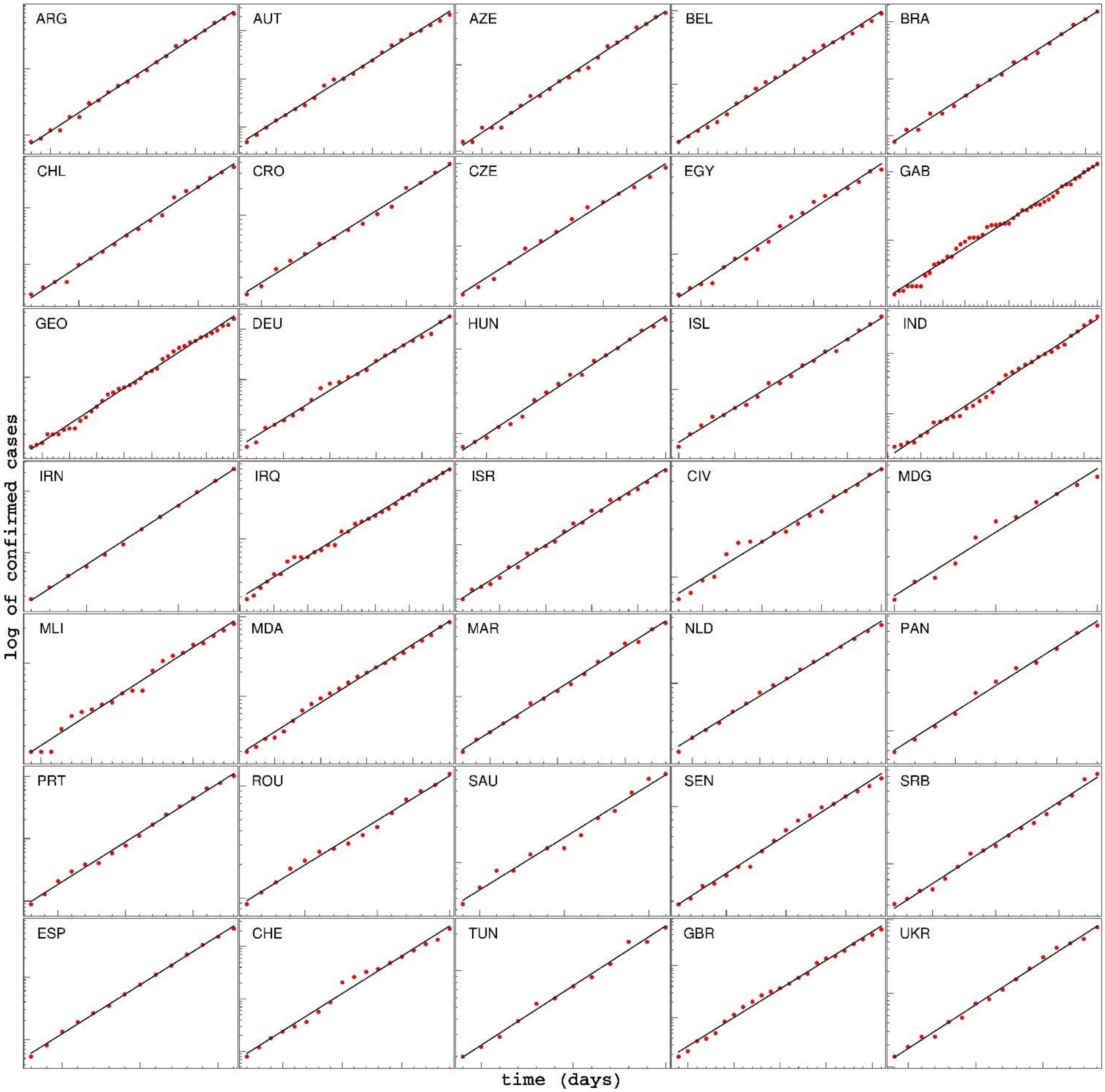

Figure 3: Time dependence of the detected cases for various countries, during the initial period of the epidemic, shown on a log-linear scale. The linear fit of log(D) shows that the spread of COVID-19 in this phase is very well approximated by exponential growth. Note that the values on axes are chosen differently for each country, in order to emphasize the exponential character of the growth. For each country, the start and end dates of the exponential regime, together with the extracted slope $\lambda_+$, are provided in the Supplementary Table S1. ARG - Argentine; AUT - Austria; AZE - Azerbaijan; BEL - Belgium; BRA - Brazil; CHL - Chile; CRO - Croatia; CZE – Czech Republic; EGY - Egypt; GAB - Gabon; GEO – Georgia; DEU - Germany; HUN - Hungary; ISL - Iceland; IND - India; IRN – Iran; IRQ - Iraq; ISR - Israel; CIV – Cote d'Ivoire; MDG - Madagascar; MLI - Mali; MDA - Moldova; MAR - Morocco; NLD - Netherlands; PAN - Panama; PRT - Portugal; ROU - Romania; SAU – Saudi Arabia; SEN – Senegal; SRB - Serbia; ESP - Spain; CHE - Switzerland; TUN – Tunis; GBR – Great Britain; UKR – Ukraine.

The $\log(D(t))$ function, for a subset of selected countries, is shown in Fig. 3. The obvious linear dependence confirms that the progression of the epidemic in this stage is almost perfectly exponential. Note that our model exactly reproduces this early exponential growth (see Eq. 3.8), happening under the assumption of a small fraction of the population being resistant, and the absence of the effect of social distancing interventions. From Fig. 3, we see that this behavior, predicted by the model for the early stage of the epidemic, is also directly supported by the data, i.e., the exponential growth in the cumulative number of confirmed cases is indeed observed for a wide range of countries. For each country, the parameter $\lambda_+$ is directly obtained as the slope of the corresponding linear fit of the $\log(D(t))$, and the basic reproduction number $R_0$ is then calculated from Equation (3.9). Here, we used the following values for the incubation rate, $\sigma = 1/3$ day$^{-1}$, and for the recovery rate $\gamma = 1/4$ day$^{-1}$, per the commonly accepted values in the literature (Bar-On et al., 2020). Note that possible variations in these two values would not significantly affect any conclusions about $R_0$ correlations, due to the mathematical properties of the relation (3.9): it is a strictly monotonous function of $\lambda_+$ and the linear term $\lambda_+(\gamma + \sigma)/(\gamma \cdot \sigma)$ dominantly determines the value of $R_0$.

Supplementary tables contain the values for 42 variables, for all countries. Correlations of each of the variables with $R_0$ are given in Table S6. Values for the Pearson correlation coefficient are further shown below, though consistent conclusions are also obtained by Kendall and Spearman correlation coefficients (which do not assume a linear relationship between variables). Correlation coefficients were calculated in the usual manner: as the correlation of the vector of parameter values with the vector of $R_0$ values, by taking into account all available data (for parameters that were available across all of the countries, both of the vectors were 118 dimensional; if values were missing for certain countries, these countries were simply ignored and lower-dimensional vectors were compared).

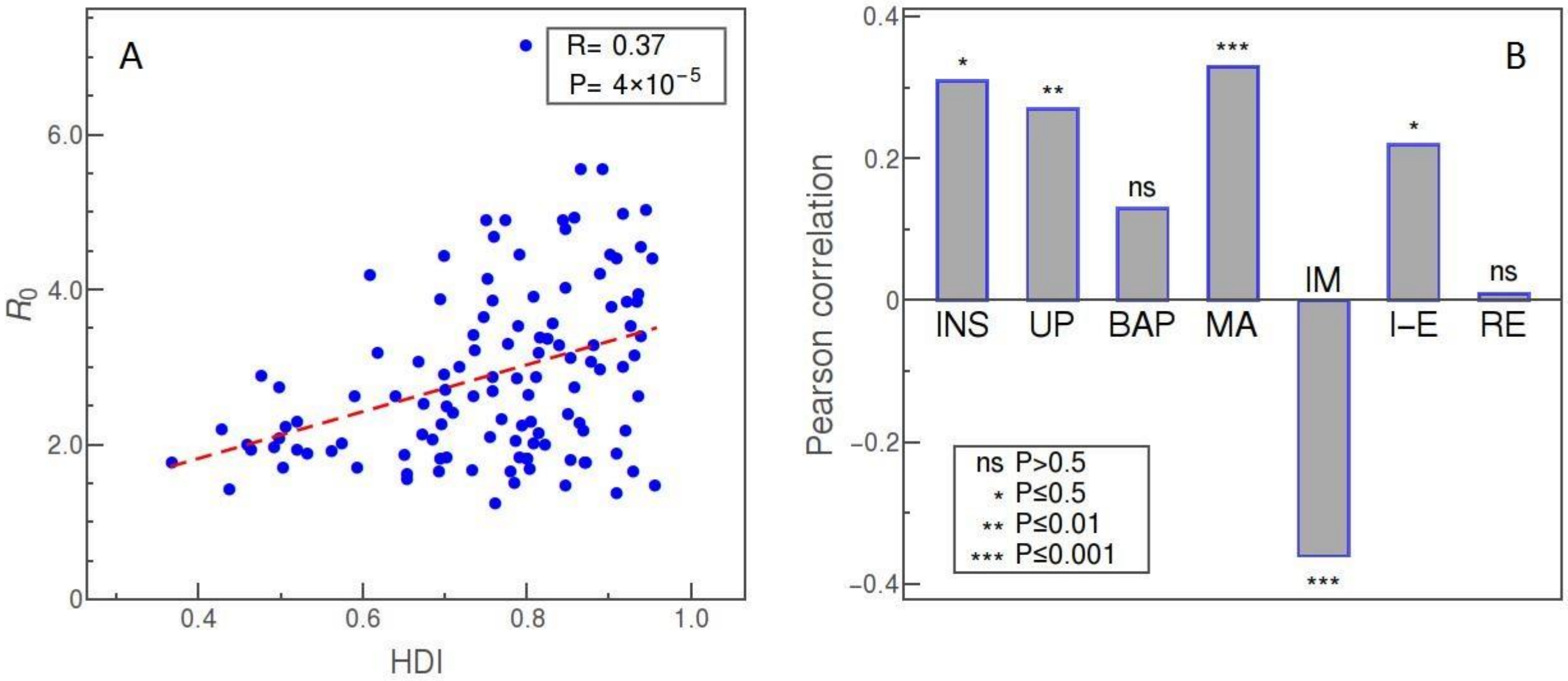

**Figure 4: A)** $R_0$ vs HDI as an example of transmissibility dependence on demographic data (Pearson correlation denoted as R). **B)** Pearson correlations of $R_0$ with (from left to right): social security and health insurance coverage (INS), percentage of urban population (UP), BUCAP measure of population density (BAP), median age (MA), infant mortality (IM), net migration (I-E), and percentage of refugee population by country or territory of asylum (RE). The statistical significance of each correlation is indicated in the legend, while "ns" stands for "no significance".

The first set of results that corresponds to, roughly speaking, general demographic data, is presented in Fig. 4. The plot in panel A shows the distribution of $R_0$ vs. HDI values for all countries, where a higher HDI score indicates the more prosperous country concerning life expectancy, education, and per capita income (Sagar and Najam, 1998). This parameter was included in the study due to a reasonable expectation that a higher level of social development also implies a higher level of population interconnectedness and mixing (stronger business and social activity, more travelers, more frequent contacts, etc.), and hence that HDI could be related to the SARS-CoV-2 transmissibility. Indeed, we note a strong, statistically highly significant correlation between the HDI and the $R_0$ value, with R = 0.37, and $p=4\cdot10^{-5}$, demonstrating that the initial expansion of COVID-19 was faster in more developed societies.

The social security and health insurance coverage (INS) "shows the percentage of population participating in programs that provide old age contributory pensions (including survivors and disability) and social security and health insurance benefits (including occupational injury benefits, paid sick leave, maternity, and other social insurance)" (World Bank, 2020b). Reflecting the percentage of the population covered by medical insurance and likely feeling more protected from the financial effects of the epidemics, this indicator shows a strong (R = 0.4) and highly significant ($p = 4\cdot10^{-4}$) positive correlation with $R_0$. The percentage of urban population (UP) and BUCAP density (BAP) are both included as measures of how concentrated is the population of the country. While the UP value simply shows what percentage of the population lives in cities, the BUCAP parameter denotes the amount of built-up area per person. Of the two, the former shows a highly significant positive correlation with the COVID-19 basic reproduction number, whereas the latter shows no correlation. Median age (MA) should be of obvious potential relevance in COVID-19 studies since it is well known that the disease more severely affects the older population (Jordan et al., 2020). Thus we wanted to investigate also if there is any connection of age with the virus transmissibility. Our results are suggestive of such a connection, since we obtained a strong positive correlation of age with $R_0$, with very high statistical confidence. Infant mortality (IM) is defined as the number of infants dying before reaching one year of age, per 1,000 live births. Lower IM rates can serve as another indicator of the prosperity of a society, and it turns out that this measure is also strongly correlated, but negatively, with R = –0.36 and $p = 8\cdot10^{-5}$ (showing again that more developed countries, i.e., those with lower IM rates, have experienced more rapid spread of the virus infection). Net migration (I-E) represents the five-year estimates of the total number of immigrants less the annual number of emigrants, including both citizens and noncitizens. This number, related to the net influx of foreigners, turns out to be positively correlated, in a statistically significant way, with $R_0$. However, according to our data, the percentage of refugees, defined as the percentage of the people in the country who are legally recognized as refugees and were granted asylum in that country, is not correlated with $R_0$ at all.

Another set of parameters corresponds to medically-related demographic parameters and is shown in the upper part of Fig. 5. The plot in panel A represents the average blood cholesterol level (in mmol/L) in the population of various countries, plotted against the value of $R_0$. The two parameters are strongly correlated, with R = 0.4, and $p=6\cdot10^{-6}$. Another demographic parameter with clear medical relevance, that has a comparatively strong and significant positive correlation with $R_0$, is the alcohol consumption per capita (ALC), as shown in panel B of Fig. 5. Our data shows that $R_0$ is also positively correlated, with high statistical significance, with the prevalence of obesity and to a somewhat smaller extent with the percentage of smokers. Here, obesity is defined as having a body-mass index over 30. A medical parameter that is strongly, but negatively, correlated with $R_0$, is a measure of prevalence and severity of COVID-19 relevant chronic diseases in the population (CD). This parameter is defined as "the percent of 30-year-old-people who would die before their 70th birthday from any of cardiovascular disease, cancer, diabetes, or chronic respiratory disease, assuming that s/he would experience current mortality rates at every age and s/he would not die from any other cause of death" (World Bank, 2020b). The percentage of people with raised blood pressure (RBP) is also negatively correlated with $R_0$, though this correlation is not as strong and as statistically significant as in the case of the CD parameter. Here, raised blood pressure is defined as systolic blood pressure over 140 or diastolic blood pressure over 90, in the population older than 18. The percentage of smokers exhibits statistically significant (though not large) positive correlation. Two medical-demographic parameters that show no correlation with $R_0$ in our data are the prevalence of insufficient physical activity among adults aged over 18 (IN) and BCG immunization coverage among 1-year-olds (BCG).

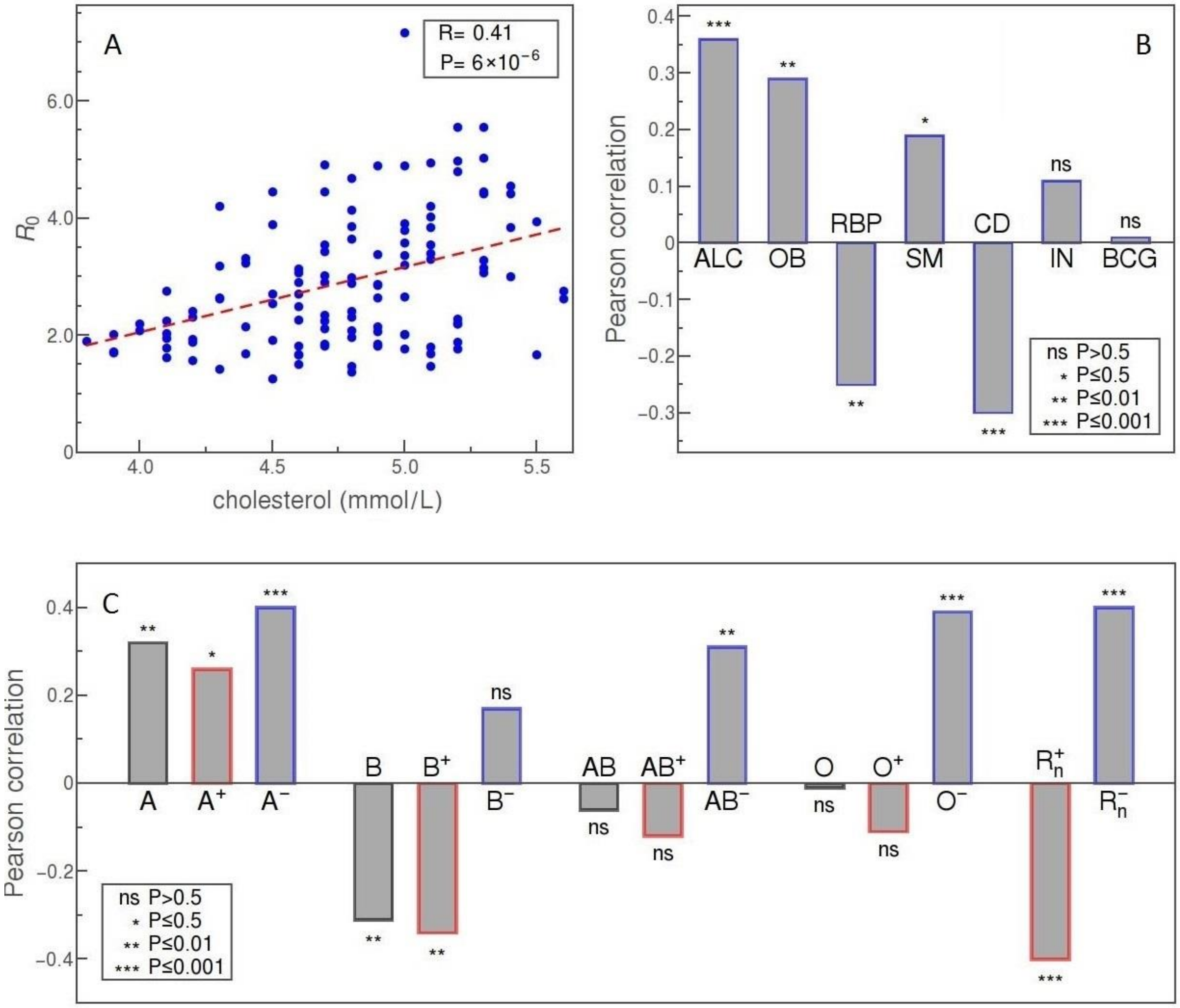

**Figure 5: A)** $R_0$ vs cholesterol level, as an example of a health-related parameter dependence. **B)** Pearson correlation of $R_0$ with (from left to right): alcohol consumption per capita (ALC); the prevalence of obesity (OB); severity of COVID-19 relevant chronic diseases in the population (CD); a percentage of people with raised blood pressure (RBP); a percentage of smokers (SM); the prevalence of insufficient physical activity among adults (IN); BCG immunization coverage among 1-year-olds (BCG) **C)** Correlation of blood types with $R_0$ in order: A, B, AB, and O (from left to right); overall value for that group, correlation only for $Rh^+$ subtype of the group, and correlation for $Rh^-$ subtype is shown. The two rightmost bars correspond to the overall correlation of $Rh^+$ and the overall correlation of the $Rh^-$ blood type with $R_0$. The convention for representing the statistical significance of each correlation is the same as in Fig. 4.

In Fig. 5C we see that blood types are, in general, strongly correlated with $R_0$. The highest positive correlation is exhibited by $A^-$ and $O^-$ types, with a Pearson correlation of 0.4 and 0.39, and a very high statistical significance of $p = 10^{-4}$ and $p = 2 \cdot 10^{-4}$, respectively. Taken as a whole, group A is still strongly and positively correlated with $R_0$, albeit with a bit lower statistical significance ($A^+$ type correlation has p-value two orders of magnitude higher than $A^-$). This is not so for group O that, overall, does not seem to be correlated to $R_0$ ($O^+$ even shows a certain negative correlation but without statistical significance). Our data reveals a highly significant positive correlation also for $AB^-$ subtype ($R = 0.31$, $p = 0.003$), while neither the $AB^+$ subtype nor overall AB group is significantly correlated with the basic reproduction number. Clear negative correlation is exhibited only by B blood group ($R = -0.31$, $p = 0.004$), mostly due to the negative correlation of its $B^+$ subtype ($R = -0.34$, $p = 0.001$), whereas $B^-$ subtype is not significantly correlated with $R_0$ in our data. If we consider the rhesus factor alone, we again observe very strong correlations with $R_0$ and with very high statistical significance: $Rh^-$ and $Rh^+$ correlate positively (R = 0.4) and negatively (R = -0.4), respectively, with very high statistical significance ($p = 2{\cdot}10^{-4}$). The tendency of $Rh^-$ and $Rh^+$ to, respectively, increase and decrease the transmissibility, is therefore consistent with the results obtained for all four individual blood-groups.

In Fig. 6, the onset represents the delay of the exponential phase and is defined, for each country, as the number of days from February 15 to the start of the exponential growth of detected cases. The motivation was to check for a possible correlation between the delay in the onset of the epidemic and the rate at which it spreads. Indeed, our data shows that such correlation exists and that it is strong and statistically significant: R = -0.48 and $p = 4{\cdot}10^{-8}$. In other words, the later the epidemic started, the lower (on average) is the basic reproduction number.

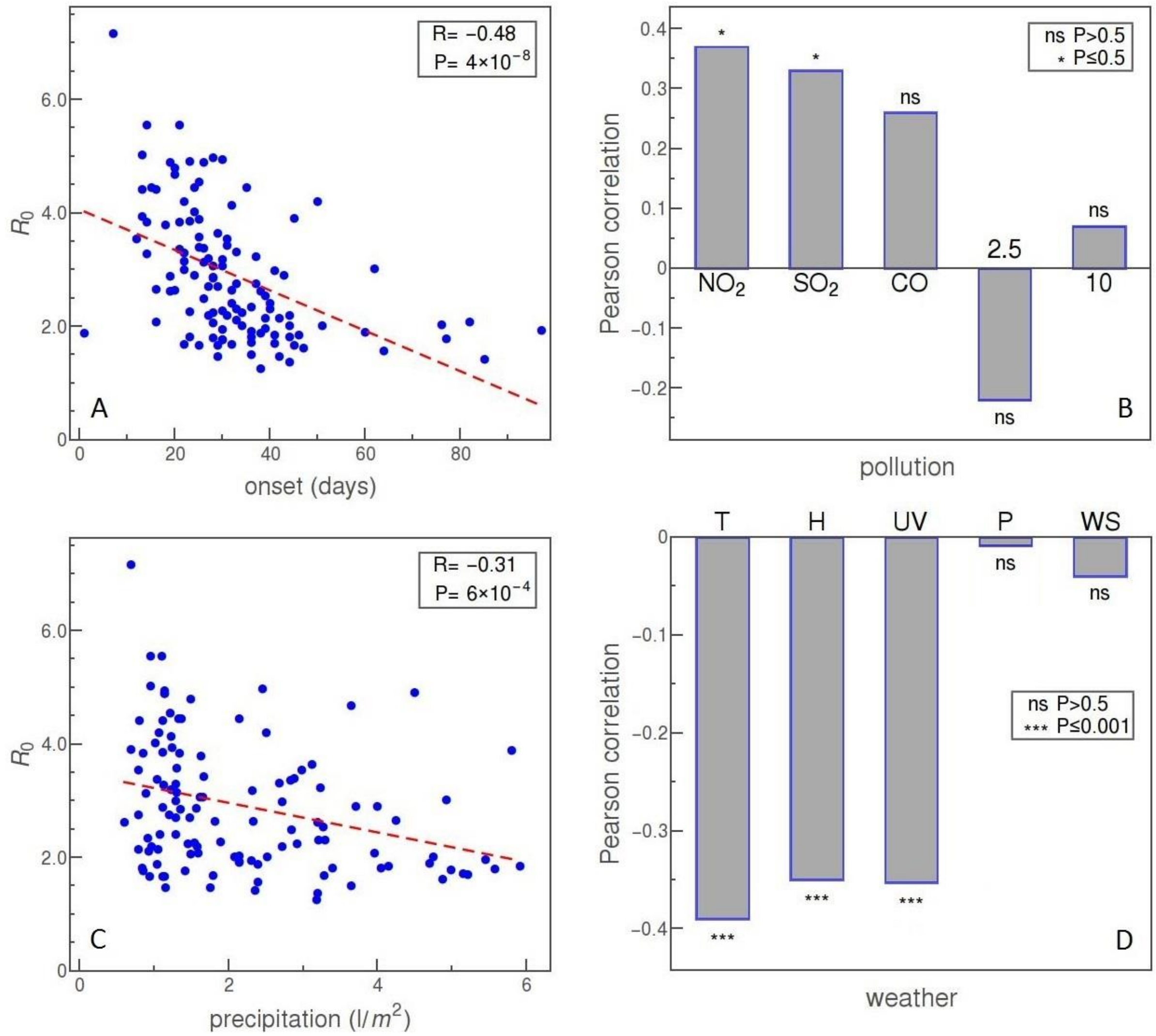

**Figure 6: A**) $R_0$ vs. the time delay of epidemic onset. **B)** Pearson correlation of $R_0$ with pollutants (from left to right): $NO_2$, $SO_2$, CO, PM2.5, and PM10 (inhalable particles with 2.5 and 10 micrometers, respectively). **C)** $R_0$ vs. precipitation. **D)** Pearson correlation of $R_0$ with (from left to right): temperature, specific humidity, UV index, air pressure, and wind speed. The convention for representing the statistical significance of each correlation is the same as before.

Panel B of Fig. 6 shows the correlation of $R_0$ with some of the commonly considered air pollutants. Our data reveal a statistically significant positive correlation of $R_0$ with $NO_2$ and $SO_2$ concentrations. Other pollutants – CO, PM2.5 (fine inhalable particles, with diameters that are generally 2.5 micrometers and smaller), and PM10 (inhalable particles, up to 10 nm in diameter) – show no statistically significant correlation with $R_0$.

Next, we consider weather factors. Panels C and D of Fig. 6 show correlations of precipitation, temperature, specific humidity, UV index, air pressure, and wind speed with the reproduction number $R_0$. Of these, precipitation, temperature, specific humidity, and UV index show a strong negative correlation, at a high level of statistical significance. Of the other two parameters, both air pressure and wind speed are not correlated at all with $R_0$ in our data.

## 4 Discussion

The present paper aimed to establish relations between the COVID-19 transmissibility and a large number of demographic and weather parameters. As a measure of COVID-19 transmissibility, we have chosen the basic reproduction number $R_0$ – a quantity that is essentially independent of the variations in both the testing policies and the introduced social measures (as discussed in the Introduction), in distinction to many studies on transmissibility that relied on the total number of detected case counts (see e.g. (Adhikari and Yin, 2020; Correa-Araneda et al., 2020; Fareed et al., 2020; Gupta et al., 2020; Iqbal et al., 2020; Li et al., 2020; Pourghasemi et al., 2020; Rashed et al., 2020; Singh and Agarwal, 2020)). We have covered a substantial number of demographic and weather parameters, and included in our analysis all world countries that were significantly affected by the COVID-19 pandemic (and had a reasonable consistency in tracking the early phase of infection progression). While a number of manuscripts have been devoted to factors that may influence COVID-19 progression, only a few used an estimate of $R_0$ or some of its proxies (Coccia, 2020; Contini and Costabile, 2020; Copiello and Grillenzoni, 2020) - these studies were however limited to China, and included a small set of meteorological variables, with conflicting results obtained for their influence on $R_0$. Therefore, a combination of *i*) using a reliable and robust measure of COVID-19 transmissibility, and *ii*) considering a large number of factors that may influence this transmissibility within the same study/framework, distinguishes our study over prior work. We, however, must be cautious when it comes to further interpretation of the obtained data. As always, we must keep in mind that "correlation does not imply causation" and that further research is necessary to identify possible confounding factors and establish which of these parameters truly affect the COVID-19 transmissibility. Due to the sheer number of studied variables, an even larger number of parameters that might be relevant but are inaccessible to study (or even impossible to quantify), as well as due to possible intricate mutual relations of the factors that may influence COVID-19 transmission, this is a highly nontrivial task. While we postpone any further analysis in this direction to future studies, we will, nevertheless, consider here the possible interpretations of the obtained correlations, assuming that they also probably indicate the existence of at least some causation. We below provide a detailed comparison between our results and previous findings. While a detailed discussion is presented, despite our best effort, we may have missed some of the relevant references due to an extremely rapidly developing field. Nevertheless, we point out a clear distinction of our work with previous studies, as outlined in this paragraph.

We will first consider the demographic variables presented in Fig. 4. The obtained correlation of the human development index (HDI) with the basic reproduction number is both strong and hardly surprising. The level of prosperity and overall development of a society is necessarily tied with the degree of population mobility and mixing, traffic intensity (in particular air traffic), business and social activity, higher local concentrations of people, and other factors that directly or indirectly increase the frequency and range of personal contacts (Gangemi et al., 2020), rendering the entire society more vulnerable to the spread of viruses. In this light, it is reasonably safe to assume that the obtained strong and highly statistically significant correlation of HDI with $R_0$ reflects a truly causal connection. However, some authors offer also a different explanation: that higher virus transmission in more developed countries is a consequence of more efficient detection of COVID-19 cases due to the better-organized health system (Gangemi et al., 2020) – but since our $R_0$ measure does not depend on detection efficiency, presented results can be taken as evidence against such hypothesis.

The interpretation is less clear for other demographic parameters, for example, the percentage of the population covered by medical and social insurance programs (INS). While there seem to be no previous studies discussing this parameter, one possibility is to attribute its strong positive correlation with $R_0$ to a hypothetical tendency of population to more easily indulge in the epidemiologically-risky behavior if they feel well-protected, both medically and financially, from the risks posed by the virus; conversely, that the population that cannot rely on professional medical care in the case of illness is likely to be more cautious not to contract the virus. The other is, of course, to see this correlation as an indirect consequence of the strong correlation of this parameter with HDI – which is also, almost certainly, the underlying explanation of the infant mortality (IM) correlation, where low mortality ratios point to a better medical system, which goes hand in hand with the overall prosperity and development of the country (Ruiz et al., 2015) (thus the negative correlation with $R_0$).

Similarly, the strong positive correlation of median age (MA) with $R_0$ might be a mere consequence of its clear relation with the overall level of development of the country (Gangemi et al., 2020), but it can be also considered in the light of the fact that clinical and epidemiological studies have unanimously shown that the elderly are at higher risk of developing a more severe clinical picture, and our result may indicate that the virus also spreads more efficiently in the elderly population. Possible explanations may include: drugs frequently prescribed to this population that increase levels of ACE2 receptors (Shahid et al., 2020), a general weakening of the immune system with age leading to a greater susceptibility to viral infections (Pawelec and Larbi, 2008), and a large number of elderly people grouped in nursing homes, where the virus can expand very quickly (Kimball et al., 2020).

The correlation of population density with $R_0$, or the lack of thereof, is more challenging to explain. Naively, one could expect that COVID-19 spreads much more rapidly in areas with a large concentration of people, but, if exists, this effect is not that easily numerically captured. As the standard population density did not show any correlation with the reproduction number $R_0$ (not shown), we explored some more subtle variants. Namely, the simplest reason why the data shows no correlation of $R_0$ with population density would be that the density, calculated in the usual way, is too averaged out: the most densely-populated country on our list, Monaco, has roughly 10.000 times more people per square kilometer than the least densely-populated Australia. However, Melbourne downtown has a similar population density as Monaco and far more people, so one would expect no a priori reason that its infection progression would be slower (and the $R_0$ rate for Australia as a whole will be dominantly determined by the fastest exponential expansion occurring anywhere on its territory). For this reason, we included the BUCAP parameter into the analysis, which takes into account only population density in built-up areas. Surprisingly, even this parameter did not exhibit any statistically significant correlation. Actually, several studies may serve as examples showing that the correlation of population density with the rate of COVID-19 expansion can be expected only under certain conditions since the frequency of contacts between people is to a large extent modulated by additional geographical, economic, and sociological factors (Berg et al., 2020; Carozzi, 2020; Pourghasemi et al., 2020; Rashed et al., 2020). Our observed absence of a correlation could be therefore expected and possibly indicates that such a correlation should be sought at the level of smaller populated areas – for example, individual cities (Yu et al., 2020). This conclusion is somewhat supported by the obtained highly significant and strong positive correlation of $R_0$ with the percentage of the population living in cities (UP) and which probably reflects the higher number of encounters between people in a more densely populated, urban environment (Li et al., 2020). It is also possible that virus spread might have a highly non-linear dependence on the population density - namely, that an outbreak in a susceptible population requires a certain threshold value of its density, while below that value population density ceases to be a significant factor influencing virus (Scheffer, 2009; Carozzi, 2020; Coro, 2020).

Another demographic parameter that exhibits a significant correlation with $R_0$ in our data is the net migration (I-E), denoting the number of immigrants less the number of emigrants. Unlike this number, which shows a positive correlation, the number of refugees (RE) seems not to be correlated at all. By definition, migrants deliberately choose to move to improve their prospects, while refugees have to move to save their lives or preserve their freedom. Migrants (e.g. in economic or academic migration), arguably tend to stay in closer contact with the country of their origin and have more financial means for that, which likely contributes to more frequent border crossings and more intensive passenger traffic (Fan et al., 2020), thereby promoting the infection spread. On the other hand, refugees are mostly stationed in refugee camps, there is less possibility of spreading the virus outside through contacts with residents, but there is a high possibility of escalation of the epidemic within camps with a high concentration of people (Hargreaves et al., 2020). We did not find any other attempt in the literature to examine this issue. In any case, our results demonstrate that refugees are certainly not a primary cause of concern in the pandemics, contrary to fears expressed in some media.

Of the medical factors, the strongest correlation of $R_0$ is established with elevated cholesterol levels, as shown in Fig. 5. Cholesterol may be associated with a viral infection and further disease development through a complex network of direct and indirect effects. In vitro studies of the role of cholesterol in virus penetration into the host body, done on several coronaviruses, indicate that its presence in the lipid rafts of the cell membrane is essential for the interaction of the virus with the ACE2 receptor, and also for the latter endocytosis of the virus (Radenkovic et al., 2020). Obesity prevalence (OB) also exhibits a highly significant, though somewhat weaker correlation with $R_0$, which might be a consequence of the common connection between obesity and cholesterol: in principle, obesity might be a relevant factor in the COVID-19 epidemic exactly due to the effects of cholesterol on SARS-CoV-2 susceptibility. Of course, other effects might be at play, e.g. the fact that the adipose tissue of obese people excessively produces pro-inflammatory cytokines (Sattar et al., 2020). In the case of obesity, a simple explanation via relation to HDI is not available, since obesity does not show a simple correlation with the society development (Haidar and Cosman, 2011). Overall, while the correlation of obesity with a more severe prognosis in COVID-19 is well established in the literature, its relation to COVID-19 transmissibility is only mentioned in (Li et al., 2020) and hitherto unexplained.

Often related to obesity is also raised blood pressure (RBP), and we have discovered that this factor is also correlated, at high statistical significance, with $R_0$. While this seems to be the first study correlating high blood pressure with the SARS-CoV-2 transmission rate, it is known that, based on clinical studies, RBP appears to be a risk factor for hospitalization and death due to COVID-19 (Ran et al., 2020a; Schiffrin et al., 2020). In this light, it might be surprising that the correlation between RBP and $R_0$ turns out to be negative. On the other hand, this result supports the existing hypothesis about the beneficial effect of ACE inhibitors and ARBs (Ran et al., 2020a; Schiffrin et al., 2020) (standardly used in the treatment of hypertension). Similarly unintuitive correlation we report in the case of chronic diseases that are known to be relevant for the COVID-19 outcome. Namely, our data show, at very high statistical significance, a strong negative correlation of $R_0$ with the risk of death from a batch of chronic diseases (cardiovascular disease, cancer, diabetes, and chronic respiratory disease), agreeing in this regard with some recent research (Chiang et al., 2020; Li et al., 2020). These diseases are identified as relevant comorbidities in the context of COVID-19, leading to a huge increase in the severity of the infection and poorer prognosis (An et al., 2020; Zheng et al., 2020) and, therefore, the discovered negative correlation comes as a surprise – particularly when contrasted to the positive correlation of obesity (where both are recognized risk factors in COVID-19 illness). One possible explanation is that the correlation may be due to potentially lower mobility of people with chronic diseases compared to the general mobility of the population. Additionally, it is possible that these people, being aware to belong to a high-risk group, behaved more cautiously even before the official introduction of social distancing measures.

According to our analysis, the prevalence of certain health-hazard habits is also significantly correlated to COVID-19 transmissibility. Chronic excessive alcohol consumption has, in general, a detrimental effect on immunity to viral and bacterial infections, which, judging by the strong positive correlation we obtained, most likely applies also to SARS-CoV-2 virus infection. This correlation contradicts the belief that alcohol can be used as a protective nostrum against COVID-19, which has spread in some countries and even led to cases of alcohol poisoning (Chick, 2020).

Regarding the impact of smoking on SARS-CoV-2 virus infection – the results are controversial (Chatkin and Godoy, 2020). The positive correlation of smoking with COVID-19 transmissibility that we obtained seems to support the reasoning that, since the SARS-CoV-2 virus enters cells by binding to angiotensin-converting enzyme 2 (ACE2) receptors and that the number of these receptors is significantly higher in the lungs of smokers, the smokers will be more affected and easily infected (Brake et al., 2020; Hoffmann et al., 2020). Accordingly, our result contradicts the hypothesis that a weakened immune response of smokers to virus infection may prove beneficial in the context of inflammation caused by intense cytokine release (Garufi et al., 2020).

Another result that addresses the association of unhealthy lifestyle with greater susceptibility to SARS-CoV-2 infection is the slight positive correlation we obtained for the prevalence of insufficient physical activity (IN) in adults, which is however not statistically significant. In this sense, in the case of COVID-19, we could not fully confirm the findings from (Jurak et al., 2020), who found that physical activity significantly reduces the risk of viral infections.

Despite the recent media interest (Gallagher, 2020), our findings neither could confirm that BCG immunization has any beneficial effect in the case of COVID-19, at least as far as reducing the risk of contracting and transmitting the disease is concerned. While it is known that the BCG vaccine provides some protection against various infectious agents, unfortunately, there is no clear evidence for such an effect against SARS-CoV-2 (O'Neill and Netea, 2020). Our analysis suggests that BCG immunization simply does not correlate with SARS-CoV-2 virus transmission.

SARS-CoV-2 target cells are typically capable of synthesizing ABH antigens and certain arguments exist, both theoretical and experimental, for a potential relation of blood groups with COVID-19 progression and transmission (Guillon et al., 2008; Dai, 2020; Gérard et al., 2020). While the results of epidemiological studies on COVID-19 patients mostly support the proposed effect of blood groups on the development of COVID-19 disease, the relationship between virus transmission and blood group prevalence and Rh phenotype has been significantly less studied. Our analysis showed strong positive correlations of virus transmission with the presence of A blood group and $Rh^-$ phenotype, as well as strong negative correlations for B blood group and $Rh^+$ phenotype, while for AB and O blood group no significant correlations were obtained (Fig. 5C). This result coincides significantly with the correlations obtained in a study conducted for 86 countries (Ansari-Lari and Saadat, 2020). However, another study focused on hospitalized patients in Turkey reported that the $Rh^+$ phenotype represents a predisposition to infection (Arac et al., 2020), contradicting our findings. Similar results regarding the Rh factor were obtained in a study (Latz et al., 2020) on hospitalized patients in the US (this study further reported no correlation of blood types with the severity of the disease). One way to reconcile these results with ours would be to speculate that the virus is more efficiently transmitted in a population with a higher proportion of $Rh^-$ phenotype because these people show a milder clinical picture compared to $Rh^+$, so their movement is not equally limited, which is why they have more ability to pass on the infection.

Our data (Fig. 6A) shows a strong negative correlation with the date of the epidemic onset. Curiously, it seems that the later the epidemic started in a given country, it is more likely that the disease expansion will be slower. Instead of interpreting this result as an indication that the virus has mutated and changed its properties over such a short period, we offer the following simpler explanation: pandemic reached first those countries that are most interconnected with the rest of the world (at the same time, those are the countries characterized by great mobility of people overall), so it is expected that also the progression of the local epidemics in these countries is more rapid. Another contributing factor could be the effect of media, which had more time to raise awareness about the risks of COVID-19 in the countries that were hit later (Khajanchi et al., 2020b).

Another segment of our interest were air pollutants, shown in Fig. 6B. Air pollution can have a detrimental effect on the human immune system and lead to the development (or to worsening) of respiratory diseases, including those caused by respiratory viral infections (Becker and Soukup, 1999; Copat et al., 2020). Several papers have already investigated air pollution in the context of COVID-19 and reported a positive correlation between the death rate due to COVID-19 and the concentration of PM2.5 in the environment (Wu et al., 2020; Yao et al., 2020c). Positive correlations were also found for the spread of the SARS-CoV-2 virus, but mainly by considering daily numbers of newly discovered cases – a method that, as we have already argued, may strongly depend on testing policies, as well as on state measures to combat the epidemic (Copat et al., 2020). It has been suggested that virus RNA can be adsorbed to airborne particles facilitating thus its spread over greater distances (Coccia, 2020; Setti et al., 2020), but these arguments were contested by examination of air samples in Wuhan (Contini and Costabile, 2020; Liu et al., 2020). The latter conclusions concur with the results of a study in which no correlation was obtained between the basic reproductive number of SARS-CoV-2 infection for 154 Chinese cities and the concentration of PM2.5 and PM10 particles, while the correlation of these factors with the death rate (CFR) was shown (Ran et al., 2020b). The statistically insignificant and relatively weak correlations we obtained for PM2.5 and PM10 pollutants also do not support the hypothesis of a potentially significant role of these particles in the transmission of this virus. In contrast, significant positive correlations were shown by our analysis for concentrations of NO2, SO2, and CO in the air (although the correlation for CO is not statistically significant), which is generally supported by the results of other studies. For example, a positive correlation of NO2 levels with the basic reproductive number of infection was obtained from data for 63 Chinese cities (Yao et al., 2020a). Also, it has been shown that the number of detected cases of COVID-19 in China is strongly positively correlated with the level of CO, while in Italy and the USA such correlation exists with NO2 (Pansini and Fornacca, 2020). The mentioned study failed to establish a clear correlation with the level of SO2. Possible mechanisms of interaction were also proposed (Daraei et al., 2020). Also, it is important to emphasize that the atmospheric concentration of NO2 strongly depends on the levels of local exhaust emissions, so its correlation with virus transmission can be interpreted by the connection with the urban environment, characterized by more intensive traffic (Goldberg et al., 2020).

Finally, we have also obtained some interesting correlations of the meteorological parameters with $R_0$, shown in Fig.s 6C and 6D. The statistically very highly significant negative correlation of the basic reproductive number of SARS-CoV-2 virus infection with both the mean temperature and humidity obtained in our research (Fig. 6D) is consistent with the results of other relevant papers, e.g. (Mecenas et al., 2020). For example, a similar correlation was obtained in a study that analyzed COVID-19 outbreak in the cities of Chile - a country that covers several climate zones, but where it is still safe to assume that social patterns of behavior and introduced epidemic control measures do not drastically differ throughout the country (Correa-Araneda et al., 2020). Effectively the same conclusion — that fewer COVID-19 cases were reported in countries with higher temperatures and humidities — was reached in a study covering over 200 countries in the world (Iqbal et al., 2020). While an established correlation between virus transmission and a certain factor is not, in general, a telltale sign of a direct causal relationship between them, in the case of temperature and humidity such connection is firmly indicated also by results of experimental research (Lowen et al., 2007; Casanova et al., 2010; Chan et al., 2011; van Doremalen et al., 2020). Nevertheless, some studies yielded different conclusions, most likely due to the method of calculating $R_0$ or due to choosing a small/uninformative sample of populations in which the number of infected cases was monitored (Guo et al., 2020; Lin et al., 2020; Yao et al., 2020b). For example, a study focused on the suburbs of New York, Queens, obtained a positive correlation between virus transmission and temperature, which seems unexpected given the prevailing observations of other studies (Adhikari and Yin, 2020). This result is most likely a consequence of analyzing data for a small area (Queens only) where the temperature varies in a relatively narrow range of values, as well as correlating the number of detected cases, which may be sensitive to variations in the testing procedure.

Another environmental agent that can destroy or inactivate viruses is UV radiation from sunlight, and the properties of a particular virus determine how long it can remain infectious when exposed to radiation. For example, epidemics of influenza have a seasonal character precisely due to the susceptibility of influenza viruses to UV radiation (Sagripanti and Lytle, 2007). Our analysis found, at very high statistical significance, a strong negative correlation between the transmission of the SARS-CoV-2 virus and the intensity of UV radiation, which is consistent with the results of other studies obtained for the cities of Brazil and the provinces of Iran (Ahmadi et al., 2020; Mendonça et al., 2020). It is worth mentioning that lower temperatures, humidity, and sunlight levels usually occur in combination and directly affect not only the virus but also the human behavior, so the observed higher transmission of the virus in such conditions can alternatively be interpreted by indirect effects of other factors that act together in cold weather, such as more time spent indoors where the virus spreads more easily, or weakening of the immune system that increases susceptibility to infections (Abdullahi et al., 2020).

While the results related to COVID-19 correlations with temperature, humidity, and UV radiation are fairly frequent in the literature, this is less so for the results on the precipitation levels. Very few other studies have examined the association of precipitation with SARS-CoV-2 transmission, with either no correlation found (Pourghasemi et al., 2020), or looking at precipitation as a surrogate for humidity and generally receiving a negative correlation with infection rate (Araujo and Naimi, 2020; Coro, 2020). Our results, however, shown in Fig. 6D, confirm natural expectations: just like humidity, the precipitation exhibits a strong negative correlation with $R_0$, only slightly lower than in the case of T, H, and UV, at a very high level of statistical significance. Such results also concur with some general conclusions about the behavior of similar viruses (Agrawal et al., 2009; Pica and Bouvier, 2012).

Our analysis did not reveal any statistically significant correlation either between the wind speed or between air pressure and SARS-CoV-2 transmissibility. In the case of wind speed, this result agrees with the findings in some other papers (Gupta et al., 2020; Oliveiros et al., 2020). A positive correlation of wind speed with COVID-19 transmissibility was obtained in a study in Chilean cities, but, as the authors themselves note, the interpretation of the effect of this factor is complicated by its observed significant interaction with temperature (Correa-Araneda et al., 2020). The role of wind in transmitting the virus to neighboring buildings is predicted by the SARS virus spread model within the Amoy Gardens residential complex in Hong Kong, but such an effect may relate to local air currents and virus transmission over relatively short distances and does not imply a correlation of mean wind speeds in the area with virus transmission (McKinney et al., 2006; Pica and Bouvier, 2012). As for the air pressure, the potential connection is hardly at all investigated in the literature. An exception is a study (Cambaza et al., 2020) reporting a positive correlation of air pressure with the number of COVID-19 cases in parts of Mozambique, but our results do not confirm such a conclusion.

## 5 Conclusion

While there is by now a significant amount of research on a crucial problem of how environmental factors affect COVID19 spread, several features set this analysis apart from the existing research. First is the applied methodology: instead of basing analysis directly on the number of detected COVID-19 cases (or some of its simple derivatives), we employ an adapted SEIR model to extract the basic reproduction number $R_0$ from the initial stage of the epidemic. By taking into account only data in the exponential growth regime, i.e., before the social measures took effect (as explained in the Methods section), we ensured that the correlations we have later identified were not confounded with the effects of local COVID-19 policies. Even more importantly, our method is also invariant to variations in COVID-19 testing practices, which, as is well known, used to vary in quite an unpredictable manner between different countries. Another important factor is the large geographical scope of our research: we collected data from 118 countries worldwide, more precisely, from all the countries that were above a certain threshold for the number of confirmed COVID-19 cases (except for several countries with clearly irregular early growth data). The third factor was the number of analyzed parameters: we calculated correlations for the selected 42 different variables (of more than a hundred that we initially considered overall) and looked for viable interpretations of the obtained results.

These results should also help in resolving some of the existing disputes in the literature. For example, our findings indicate that correlation of HDI with $R_0$ is not a consequence of the COVID-19 testing bias, as was occasionally argued. Of the opposing opinions, our data seem to support assertions that blood types are indeed related to COVID-19 transmissibility, as well as arguments that the higher prevalence of smoking does increase the virus transmissibility (though weakly). On the other hand, in the dispute about the effects of the pollution, our correlations give an edge to claims that there is no correlation between PM2.5 and PM10 particles and transmissibility (whereas we agree with the prevailing conclusions about the positive correlation of other considered pollutants). In the case of the effects of the wind, based on the obtained results we tend to side with those denying any connection. In certain cases our findings contradict popular narratives: there are no clear indications that either number of refugees or physical inactivity intensifies the spread of COVID-19. Unfortunately, our data also suggest that BCG immunization may not help in subduing the epidemic. Additionally, the obtained correlations hint to possible new alleys of research, e.g. those that would help us understand the connection between cholesterol levels and SARS-CoV-2 transmissibility.

Overall, we believe that the presented results can be a useful contribution to the ongoing attempts to better understand the first pandemic of the 21$^{st}$ century - and the better we understand it, the sooner we may hope to overcome it.

## 6 Conflict of Interest

The authors declare that the research was conducted in the absence of any commercial or financial relationships that could be construed as a potential conflict of interest.

## 7 Author Contributions

MarD and MagD conceived the research. The work was supervised by IS, MagD, and MarD. Data acquisition by OM, IS, DZ, and AR. Data analysis by DZ, OM and IS, with the help of MagD. Figures and tables made by DZ and AR with the help of MagD. A literature search by AR. Manuscript written by IS, AR, and MarD.

## 8 Acknowledgments

The authors thank Anica Brzakovic for reproducing a subset of the results presented in this manuscript.

# Supplementary tables

**Table S1.** Exponential growth parameters.

**Table S2.** Demographic factors.

**Table S3.** Medical factors.

**Table S4.** ABO and Rhesus blood group systems.

**Table S5.** Meteorological factors.

**Table S6.** Pearson correlation coefficients (R) and P values.

---

**Table S1.** Exponential growth parameters.

| Country | Start date | End date | $\lambda_+$ (1/day) |
|---|---|---|---|
| ABW | 2020-03-30 | 2020-04-12 | 0.047 |
| AFG | 2020-03-23 | 2020-04-03 | 0.173 |
| ALB | 2020-03-14 | 2020-03-26 | 0.114 |
| AND | 2020-03-19 | 2020-03-29 | 0.173 |
| ARE | 2020-03-16 | 2020-04-04 | 0.134 |
| ARG | 2020-03-07 | 2020-03-28 | 0.220 |
| ARM | 2020-03-19 | 2020-04-04 | 0.119 |
| AUS | 2020-03-11 | 2020-03-26 | 0.222 |
| AUT | 2020-02-28 | 2020-03-20 | 0.291 |
| AZE | 2020-03-13 | 2020-04-03 | 0.170 |
| BEL | 2020-03-08 | 2020-03-29 | 0.193 |
| BGD | 2020-04-05 | 2020-04-14 | 0.277 |
| BGR | 2020-03-13 | 2020-03-22 | 0.207 |
| BHR | 2020-03-15 | 2020-04-12 | 0.057 |
| BHS | 2020-03-30 | 2020-04-09 | 0.111 |
| BIH | 2020-03-22 | 2020-04-04 | 0.139 |
| BLR | 2020-03-31 | 2020-04-12 | 0.258 |
| BOL | 2020-03-31 | 2020-05-14 | 0.076 |
| BRA | 2020-03-06 | 2020-03-23 | 0.308 |
| BRB | 2020-03-22 | 2020-04-07 | 0.092 |
| BRN | 2020-03-14 | 2020-03-25 | 0.091 |
| CAF | 2020-05-02 | 2020-06-09 | 0.088 |
| CAN | 2020-02-27 | 2020-03-26 | 0.233 |
| CHE | 2020-02-28 | 2020-03-16 | 0.329 |
| CHL | 2020-03-05 | 2020-03-22 | 0.321 |
| CIV | 2020-03-25 | 2020-04-11 | 0.106 |
| COL | 2020-03-15 | 2020-03-26 | 0.240 |
| CPV | 2020-04-19 | 2020-05-15 | 0.066 |
| CRI | 2020-03-20 | 2020-03-29 | 0.131 |
| CUB | 2020-03-19 | 2020-03-31 | 0.216 |
| CYM | 2020-03-27 | 2020-04-07 | 0.191 |
| CYP | 2020-03-17 | 2020-04-02 | 0.126 |
| CZE | 2020-03-08 | 2020-03-21 | 0.278 |
| DEU | 2020-02-28 | 2020-03-21 | 0.260 |
| DJI | 2020-03-29 | 2020-04-18 | 0.185 |
| DNK | 2020-03-15 | 2020-04-09 | 0.077 |
| DOM | 2020-03-23 | 2020-03-30 | 0.210 |
| EGY | 2020-03-09 | 2020-03-27 | 0.132 |
| ESP | 2020-02-29 | 2020-03-13 | 0.359 |
| EST | 2020-03-16 | 2020-04-06 | 0.087 |
| ETH | 2020-03-16 | 2020-04-12 | 0.104 |
| FIN | 2020-03-13 | 2020-03-29 | 0.126 |
| FRA | 2020-03-01 | 2020-03-12 | 0.294 |
| GAB | 2020-04-01 | 2020-05-17 | 0.095 |
| GBR | 2020-02-29 | 2020-03-22 | 0.253 |
| GEO | 2020-03-11 | 2020-04-17 | 0.076 |
| GHA | 2020-03-27 | 2020-05-09 | 0.080 |
| GIN | 2020-04-06 | 2020-04-25 | 0.110 |
| GRC | 2020-03-16 | 2020-04-03 | 0.087 |
| GTM | 2020-03-24 | 2020-04-27 | 0.098 |
| GUY | 2020-04-02 | 2020-04-18 | 0.073 |
| HND | 2020-03-16 | 2020-04-02 | 0.207 |
| HRV | 2020-03-11 | 2020-03-25 | 0.235 |
| HTI | 2020-05-07 | 2020-06-03 | 0.117 |
| HUN | 2020-03-08 | 2020-03-25 | 0.215 |
| IDN | 2020-03-11 | 2020-03-23 | 0.256 |
| IMN | 2020-03-26 | 2020-04-03 | 0.145 |
| IND | 2020-03-06 | 2020-04-06 | 0.164 |
| IRL | 2020-03-11 | 2020-03-21 | 0.299 |
| IRN | 2020-02-22 | 2020-03-04 | 0.445 |
| IRQ | 2020-03-02 | 2020-04-01 | 0.116 |
| ISL | 2020-03-05 | 2020-03-23 | 0.163 |
| ISR | 2020-03-04 | 2020-03-26 | 0.250 |
| ITA | 2020-02-29 | 2020-03-13 | 0.214 |
| JAM | 2020-03-18 | 2020-04-06 | 0.078 |
| JOR | 2020-03-18 | 2020-03-29 | 0.164 |
| JPN | 2020-02-16 | 2020-03-11 | 0.098 |
| KAZ | 2020-03-30 | 2020-04-17 | 0.092 |
| KEN | 2020-03-24 | 2020-04-07 | 0.163 |
| KGZ | 2020-03-28 | 2020-04-12 | 0.122 |
| KWT | 2020-03-08 | 2020-05-20 | 0.079 |
| LBN | 2020-03-05 | 2020-03-17 | 0.183 |
| LTU | 2020-03-16 | 2020-03-25 | 0.323 |
| LVA | 2020-03-10 | 2020-03-21 | 0.266 |
| MAR | 2020-03-16 | 2020-03-31 | 0.198 |
| MCO | 2020-03-28 | 2020-04-11 | 0.057 |
| MDA | 2020-03-15 | 2020-04-06 | 0.170 |
| MDG | 2020-03-26 | 2020-04-05 | 0.136 |
| MDV | 2020-04-17 | 2020-05-01 | 0.194 |
| MEX | 2020-03-12 | 2020-03-21 | 0.321 |
| MKD | 2020-03-09 | 2020-03-24 | 0.255 |
| MLI | 2020-03-30 | 2020-04-19 | 0.127 |
| MLT | 2020-03-14 | 2020-03-23 | 0.198 |
| MNE | 2020-03-25 | 2020-04-07 | 0.123 |
| MOZ | 2020-05-10 | 2020-06-11 | 0.052 |
| MRT | 2020-05-22 | 2020-06-11 | 0.103 |
| MUS | 2020-03-27 | 2020-04-10 | 0.094 |
| MYS | 2020-03-02 | 2020-03-15 | 0.165 |
| NGA | 2020-04-15 | 2020-05-09 | 0.099 |
| NLD | 2020-03-08 | 2020-03-23 | 0.204 |
| NOR | 2020-03-02 | 2020-03-16 | 0.291 |
| NPL | 2020-05-01 | 2020-06-09 | 0.112 |
| NZL | 2020-03-14 | 2020-03-28 | 0.326 |
| OMN | 2020-03-20 | 2020-04-21 | 0.110 |
| PAK | 2020-03-22 | 2020-04-09 | 0.101 |
| PAN | 2020-03-17 | 2020-03-27 | 0.233 |
| PER | 2020-03-09 | 2020-03-20 | 0.321 |
| PHL | 2020-03-10 | 2020-03-31 | 0.186 |
| POL | 2020-03-07 | 2020-03-18 | 0.360 |
| PRT | 2020-03-06 | 2020-03-21 | 0.314 |
| PRY | 2020-03-12 | 2020-03-27 | 0.153 |
| ROU | 2020-03-14 | 2020-03-28 | 0.183 |
| RUS | 2020-03-12 | 2020-04-03 | 0.221 |
| SAU | 2020-03-12 | 2020-03-24 | 0.203 |
| SEN | 2020-03-14 | 2020-03-31 | 0.131 |
| SLV | 2020-03-25 | 2020-04-10 | 0.156 |
| SRB | 2020-03-14 | 2020-03-30 | 0.182 |
| SWE | 2020-03-07 | 2020-03-15 | 0.253 |
| TGO | 2020-03-22 | 2020-04-10 | 0.081 |
| TTO | 2020-03-22 | 2020-04-03 | 0.060 |
| TUN | 2020-03-17 | 2020-03-28 | 0.224 |
| TUR | 2020-03-21 | 2020-03-29 | 0.294 |
| UKR | 2020-03-18 | 2020-04-02 | 0.273 |
| URY | 2020-03-19 | 2020-03-27 | 0.136 |
| UZB | 2020-03-18 | 2020-04-17 | 0.145 |
| VEN | 2020-03-24 | 2020-04-20 | 0.033 |
| VNM | 2020-03-09 | 2020-03-30 | 0.092 |
| ZAF | 2020-03-10 | 2020-03-27 | 0.293 |

Column explanations: **Start/End date** of the exponential growth regime in the cumulative number of detected infections; $\lambda_+$ - slope of the fitted log(D(t)) line to the exponential regime growth data.

**Table S2.** Demographic factors.

| Country | HDI | INS | UP | BAP ($m^2$/person) | MA (years) | IM | I-E | RE |
|---|---|---|---|---|---|---|---|---|
| ABW | 0.91 | / | 43.4 | / | 39.9 | 10.4 | 1.0.E+03 | / |
| AFG | 0.50 | 0.5 | 25.5 | 24.8 | 19.5 | 47.9 | -3.1.E+05 | 1.9.E-01 |
| ALB | 0.79 | 41.1 | 60.3 | 78.4 | 34.3 | 7.8 | -7.0.E+04 | 4.6.E-03 |
| AND | 0.86 | / | 88.1 | / | 46.2 | 2.7 | / | / |
| ARE | 0.86 | / | 86.5 | 58.9 | 38.4 | 6.5 | 2.0.E+05 | 1.2.E-02 |
| ARG | 0.83 | 29.7 | 91.9 | 97.8 | 32.4 | 8.8 | 2.4.E+04 | 7.7.E-03 |
| ARM | 0.76 | 52.8 | 63.1 | 81.5 | 36.6 | 11.0 | -2.5.E+04 | 6.1.E-01 |
| AUS | 0.94 | / | 86.0 | 246.1 | 37.5 | 3.1 | 7.9.E+05 | 2.2.E-01 |
| AUT | 0.91 | / | 58.3 | 133.0 | 44.5 | 2.9 | 3.2.E+05 | 1.5.E+00 |
| AZE | 0.76 | 50.6 | 55.7 | 102.0 | 32.6 | 19.2 | 6.0.E+03 | 1.1.E-02 |
| BEL | 0.92 | / | 98.0 | 141.4 | 41.6 | 2.9 | 2.4.E+05 | 3.7.E-01 |
| BGD | 0.61 | 1.4 | 36.6 | 36.2 | 27.9 | 25.1 | -1.8.E+06 | 5.6.E-01 |
| BGR | 0.81 | 47.7 | 75.0 | 99.7 | 43.7 | 5.9 | -2.4.E+04 | 2.9.E-01 |
| BHR | 0.85 | / | 89.3 | 141.5 | 32.9 | 6.1 | 2.4.E+05 | 1.6.E-02 |
| BHS | 0.81 | / | 83.0 | 281.3 | 32.8 | 8.3 | 5.0.E+03 | 3.9.E-03 |
| BIH | 0.77 | 40.8 | 48.2 | 119.4 | 43.3 | 5.0 | -1.1.E+05 | 1.6.E-01 |
| BLR | 0.81 | 46.2 | 78.6 | 82.0 | 40.9 | 2.6 | 4.4.E+04 | 2.4.E-02 |
| BOL | 0.69 | 12.7 | 69.4 | 75.5 | 25.3 | 21.8 | -4.8.E+04 | 7.0.E-03 |
| BRA | 0.76 | 29.5 | 86.6 | 75.8 | 33.2 | 12.8 | 1.1.E+05 | 5.4.E-03 |
| BRB | 0.80 | / | 31.1 | 246.1 | 39.5 | 11.3 | -4.0.E+02 | 3.5.E-04 |
| BRN | 0.85 | / | 77.6 | 82.9 | 31.1 | 9.8 | 0 | / |
| CAF | 0.37 | / | 41.4 | 55.7 | 20.0 | 84.5 | -2.0.E+05 | 1.4.E-01 |
| CAN | 0.93 | / | 81.4 | 243.8 | 41.8 | 4.3 | 1.2.E+06 | 3.0.E-01 |
| CHE | 0.94 | / | 73.8 | 139.2 | 42.7 | 3.7 | 2.6.E+05 | 1.2.E+00 |
| CHL | 0.84 | 38.6 | 87.6 | 53.6 | 35.5 | 6.2 | 5.6.E+05 | 1.1.E-02 |
| CIV | 0.49 | 12.2 | 50.8 | 50.7 | 20.3 | 59.4 | -4.0.E+04 | 7.0.E-03 |
| COL | 0.75 | 10.0 | 80.8 | 25.4 | 31.2 | 12.2 | 1.0.E+06 | 6.2.E-04 |
| CPV | 0.65 | 6.0 | 65.7 | 0.0 | 26.8 | 16.7 | -6.7.E+03 | / |
| CRI | 0.79 | 15.7 | 79.3 | 113.6 | 32.6 | 7.6 | 2.1.E+04 | 9.0.E-02 |
| CUB | 0.78 | / | 77.0 | 93.1 | 42.1 | 3.7 | -7.2.E+04 | 2.5.E-03 |
| CYM | 0.89 | / | 100.0 | / | 40.5 | 5.7 | / | 5.2.E-02 |
| CYP | 0.87 | / | 66.8 | 258.9 | 37.9 | 1.9 | 2.5.E+04 | 9.2.E-01 |
| CZE | 0.89 | / | 73.8 | 147.6 | 43.3 | 2.7 | 1.1.E+05 | 2.0.E-02 |
| DEU | 0.94 | / | 77.3 | 185.0 | 47.8 | 3.1 | 2.7.E+06 | 1.3.E+00 |
| DJI | 0.48 | 8.5 | 77.8 | 3.6 | 24.9 | 49.8 | 4.5.E+03 | 1.9.E+00 |
| DNK | 0.93 | / | 87.9 | 199.5 | 42.0 | 3.6 | 7.6.E+04 | 6.3.E-01 |
| DOM | 0.74 | 5.9 | 81.1 | 64.9 | 27.9 | 24.1 | -1.5.E+05 | 1.6.E-03 |
| EGY | 0.70 | 21.3 | 42.7 | 32.3 | 24.1 | 18.1 | -1.9.E+05 | 2.5.E-01 |
| ESP | 0.89 | / | 80.3 | 79.3 | 43.9 | 2.5 | 2.0.E+05 | 4.3.E-02 |
| EST | 0.87 | / | 68.9 | 138.5 | 43.7 | 2.1 | 2.0.E+04 | 2.4.E-02 |
| ETH | 0.46 | 2.4 | 20.8 | 19.6 | 19.8 | 39.1 | 1.5.E+05 | 8.1.E-01 |
| FIN | 0.92 | / | 85.4 | 129.0 | 42.8 | 1.4 | 7.0.E+04 | 4.0.E-01 |
| FRA | 0.90 | / | 80.4 | 119.2 | 41.7 | 3.4 | 1.8.E+05 | 5.5.E-01 |
| GAB | 0.70 | 13.4 | 89.4 | 74.5 | 21.0 | 32.7 | 1.6.E+04 | 3.2.E-02 |
| GBR | 0.92 | / | 83.4 | 134.7 | 40.6 | 3.6 | 1.3.E+06 | 1.9.E-01 |
| GEO | 0.78 | 61.2 | 58.6 | 57.1 | 38.6 | 8.7 | -5.0.E+04 | 5.4.E-02 |
| GHA | 0.59 | 36.9 | 56.1 | 112.2 | 21.4 | 34.9 | -5.0.E+04 | 3.9.E-02 |
| GIN | 0.46 | 2.3 | 36.1 | 64.5 | 19.1 | 64.9 | -2.0.E+04 | 3.4.E-02 |
| GRC | 0.87 | / | 79.1 | 98.4 | 45.3 | 3.6 | -8.0.E+04 | 5.7.E-01 |
| GTM | 0.65 | 4.2 | 51.1 | 69.3 | 23.2 | 22.1 | -4.6.E+04 | 2.3.E-03 |
| GUY | 0.65 | / | 26.6 | 35.7 | 27.5 | 25.1 | -3.0.E+04 | 2.9.E-03 |
| HND | 0.62 | 2.7 | 57.1 | 72.2 | 24.4 | 15.1 | -3.4.E+04 | 2.8.E-04 |
| HRV | 0.83 | 55.3 | 56.9 | 155.7 | 43.9 | 4.0 | -4.0.E+04 | 1.9.E-02 |
| HTI | 0.50 | 0.4 | 55.3 | 52.3 | 24.1 | 49.5 | -1.8.E+05 | 7.1.E-05 |
| HUN | 0.84 | 57.1 | 71.4 | 163.5 | 43.6 | 3.6 | 3.0.E+04 | 6.2.E-02 |
| IDN | 0.69 | 8.9 | 55.3 | 52.0 | 31.1 | 21.1 | -4.9.E+05 | 4.0.E-03 |
| IMN | 0.85 | / | 52.6 | / | 44.6 | 5.6 | / | / |
| IND | 0.64 | 17.7 | 34.0 | 35.9 | 28.7 | 29.9 | -2.7.E+06 | 1.4.E-02 |
| IRL | 0.94 | / | 63.2 | 174.2 | 37.8 | 3.1 | 1.2.E+05 | 1.2.E-01 |
| IRN | 0.80 | / | 74.9 | 54.6 | 31.7 | 12.4 | -2.7.E+05 | 1.2.E+00 |
| IRQ | 0.69 | 27.3 | 70.5 | 80.9 | 21.2 | 22.5 | 3.9.E+04 | 7.2.E-01 |
| ISL | 0.94 | / | 93.8 | 88.4 | 37.1 | 1.5 | 1.9.E+03 | 1.6.E-01 |
| ISR | 0.90 | / | 92.4 | 94.6 | 30.4 | 3.0 | 5.0.E+04 | 2.1.E-01 |
| ITA | 0.88 | / | 70.4 | 169.7 | 46.5 | 2.6 | 7.4.E+05 | 3.1.E-01 |
| JAM | 0.73 | 5.0 | 55.7 | 129.6 | 29.4 | 12.4 | -5.7.E+04 | 5.1.E-04 |
| JOR | 0.74 | 26.6 | 91.0 | 75.2 | 23.5 | 13.9 | 5.1.E+04 | 2.9.E+01 |
| JPN | 0.91 | / | 91.6 | 111.0 | 48.6 | 1.8 | 3.6.E+05 | 1.5.E-03 |
| KAZ | 0.80 | 22.0 | 57.4 | 120.4 | 31.6 | 8.8 | -9.0.E+04 | 3.1.E-03 |
| KEN | 0.59 | 1.5 | 27.0 | 23.1 | 20.0 | 30.6 | -5.0.E+04 | 8.0.E-01 |
| KGZ | 0.67 | 37.9 | 36.4 | 100.7 | 27.3 | 16.9 | -2.0.E+04 | 5.2.E-03 |
| KWT | 0.80 | / | 100.0 | 96.2 | 29.7 | 6.7 | 2.0.E+05 | 1.6.E-02 |
| LBN | 0.76 | 53.7 | 88.6 | 56.0 | 33.7 | 6.4 | -1.5.E+05 | 2.1.E+01 |
| LTU | 0.86 | 45.0 | 67.7 | 132.9 | 44.5 | 3.3 | -1.6.E+05 | 6.2.E-02 |
| LVA | 0.85 | 49.1 | 68.1 | 135.0 | 44.4 | 3.3 | -7.4.E+04 | 3.5.E-02 |
| MAR | 0.67 | / | 62.5 | 39.4 | 29.1 | 19.2 | -2.6.E+05 | 1.6.E-02 |
| MCO | 0.96 | / | 100.0 | 98.4 | 55.4 | 2.6 | / | 6.4.E-02 |
| MDA | 0.70 | 42.7 | 42.6 | 164.0 | 37.7 | 13.6 | -6.9.E+03 | 1.6.E-02 |
| MDG | 0.52 | / | 37.2 | 28.6 | 20.3 | 38.2 | -7.5.E+03 | 1.6.E-04 |
| MDV | 0.72 | 4.6 | 39.8 | 15.1 | 29.5 | 7.4 | 5.7.E+04 | / |
| MEX | 0.77 | 36.2 | 80.2 | 66.4 | 29.3 | 11.0 | -3.0.E+05 | 1.3.E-02 |
| MKD | 0.76 | / | 58.0 | 107.5 | 39.0 | 8.7 | -5.0.E+03 | 2.0.E-02 |
| MLI | 0.43 | 1.3 | 42.4 | 50.5 | 16.0 | 62.0 | -2.0.E+05 | 1.4.E-01 |
| MLT | 0.88 | / | 94.6 | 153.6 | 42.3 | 6.1 | 4.5.E+03 | 1.7.E+00 |
| MNE | 0.81 | 48.3 | 66.8 | 157.9 | 39.6 | 2.3 | -2.4.E+03 | 1.2.E-01 |
| MOZ | 0.44 | 2.8 | 36.0 | 94.2 | 17.0 | 54.0 | -2.5.E+04 | 1.6.E-02 |
| MRT | 0.52 | 9.2 | 53.7 | 74.4 | 21.0 | 51.5 | 2.5.E+04 | 1.8.E+00 |
| MUS | 0.79 | 11.0 | 40.8 | 104.6 | 36.3 | 13.6 | 0 | 1.1.E-03 |
| MYS | 0.80 | 7.7 | 76.0 | 123.5 | 29.2 | 6.7 | 2.5.E+05 | 3.8.E-01 |
| NGA | 0.53 | 2.0 | 50.3 | 63.7 | 18.6 | 75.7 | -3.0.E+05 | 1.7.E-02 |
| NLD | 0.93 | / | 91.5 | 219.2 | 42.8 | 3.3 | 8.0.E+04 | 5.9.E-01 |
| NOR | 0.95 | / | 82.2 | 132.4 | 39.5 | 2.1 | 1.4.E+05 | 1.1.E+00 |
| NPL | 0.57 | 5.3 | 19.7 | 15.7 | 25.3 | 26.7 | 2.1.E+05 | 7.3.E-02 |
| NZL | 0.92 | / | 86.5 | 245.9 | 37.2 | 4.7 | 7.4.E+04 | 3.2.E-02 |
| OMN | 0.82 | / | 84.5 | 80.5 | 26.2 | 9.8 | 4.4.E+05 | 6.2.E-03 |
| PAK | 0.56 | 6.6 | 36.7 | 28.8 | 22.0 | 57.2 | -1.2.E+06 | 6.5.E-01 |
| PAN | 0.79 | 18.2 | 67.7 | 100.7 | 30.1 | 13.1 | 5.6.E+04 | 5.9.E-02 |
| PER | 0.75 | 10.9 | 77.9 | 42.2 | 29.1 | 11.1 | 5.0.E+05 | 7.8.E-03 |
| PHL | 0.70 | 9.2 | 46.9 | 29.7 | 24.1 | 22.5 | -3.4.E+05 | 5.9.E-04 |
| POL | 0.87 | 46.0 | 60.1 | 167.8 | 41.9 | 3.8 | -1.5.E+05 | 3.3.E-02 |
| PRT | 0.85 | / | 65.2 | 116.7 | 44.6 | 3.1 | -3.0.E+04 | 2.1.E-02 |
| PRY | 0.70 | 6.4 | 61.6 | 116.7 | 29.7 | 17.2 | -8.3.E+04 | 3.8.E-03 |
| ROU | 0.81 | 47.6 | 54.0 | 88.9 | 42.5 | 6.1 | -3.7.E+05 | 2.1.E-02 |
| RUS | 0.82 | 52.7 | 74.4 | 86.6 | 40.3 | 6.1 | 9.1.E+05 | 5.4.E-02 |
| SAU | 0.85 | / | 83.8 | 111.9 | 30.8 | 6.0 | 6.7.E+05 | 7.8.E-04 |
| SEN | 0.51 | 8.3 | 47.2 | 50.5 | 19.4 | 31.8 | -1.0.E+05 | 8.8.E-02 |
| SLV | 0.67 | 5.9 | 72.0 | 75.6 | 27.7 | 11.8 | -2.0.E+05 | 7.4.E-04 |
| SRB | 0.79 | 55.6 | 56.1 | 108.0 | 43.4 | 4.8 | 2.0.E+04 | 4.5.E-01 |
| SWE | 0.93 | / | 87.4 | 121.0 | 41.1 | 2.2 | 2.0.E+05 | 2.4.E+00 |
| TGO | 0.50 | 3.0 | 41.7 | 72.8 | 20.0 | 47.4 | -1.0.E+04 | 1.5.E-01 |
| TTO | 0.78 | / | 53.2 | 204.8 | 37.8 | 16.4 | -4.0.E+03 | 5.6.E-02 |
| TUN | 0.74 | / | 68.9 | 103.6 | 32.7 | 14.6 | -2.0.E+04 | 9.1.E-03 |
| TUR | 0.79 | 33.9 | 75.1 | 50.9 | 32.2 | 9.1 | 1.4.E+06 | 4.4.E+00 |
| UKR | 0.75 | 51.1 | 69.4 | 110.8 | 41.2 | 7.5 | 5.0.E+04 | 5.9.E-03 |
| URY | 0.80 | 36.1 | 95.3 | 111.9 | 35.5 | 6.4 | -1.5.E+04 | 1.1.E-02 |
| UZB | 0.71 | 20.1 | 50.5 | 90.2 | 30.1 | 19.1 | -4.4.E+04 | 4.2.E-05 |
| VEN | 0.76 | 10.2 | 88.2 | 43.1 | 30.0 | 21.4 | -3.3.E+06 | 2.4.E-01 |
| VNM | 0.69 | 15.2 | 35.9 | 53.7 | 31.9 | 16.5 | -4.0.E+05 | / |
| ZAF | 0.70 | 3.2 | 66.4 | 138.2 | 28.0 | 28.5 | 7.3.E+05 | 1.5.E-01 |

Column explanations: **HDI** - Human Development Index; **INS** - social security and health insurance coverage; **UP** - percentage of urban population; **BAP** – build-up area per person; **MA** - median age; **IM** - infant mortality; **I-E** - net migration; **RE** – percentage of refugee population by country or territory of asylum.

**Table S3.** Medical factors.

| Country | CH (mmol/L) | ALC (l/person) | OB (%) | RBP (%) | SM (%) | CD (%) | IN (%) | BCG (%) |
|---|---|---|---|---|---|---|---|---|
| ABW | 4.9 | 6.7 | 22.1 | 23.7 | 23.5 | / | 31.0 | 95 |
| AFG | 4.2 | 0.2 | 5.5 | 30.6 | 23.5 | 29.8 | 31.0 | 78 |
| ALB | 4.9 | 7.2 | 21.7 | 29.0 | 29.2 | 17.0 | 31.0 | 99 |
| AND | 5.6 | 11.0 | 25.6 | 18.7 | 33.8 | / | 38.4 | 95 |
| ARE | 5.3 | 3.9 | 31.7 | 21.1 | 18.2 | 16.8 | 41.4 | 95 |
| ARG | 5.0 | 9.7 | 28.3 | 22.6 | 21.8 | 15.8 | 41.6 | 93 |
| ARM | 4.8 | 5.5 | 20.2 | 25.5 | 26.7 | 22.3 | 22.6 | 99 |
| AUS | 5.2 | 10.5 | 29.0 | 15.2 | 16.2 | 9.1 | 30.4 | 95 |
| AUT | 5.3 | 12.0 | 20.1 | 21.0 | 29.1 | 11.4 | 30.1 | 95 |
| AZE | 4.6 | 4.4 | 19.9 | 21.0 | 19.6 | 22.2 | 31.0 | 97 |
| BEL | 5.4 | 11.1 | 22.1 | 21.0 | 25.0 | 11.4 | 35.7 | 95 |
| BGD | 4.4 | 0.0 | 3.6 | 21.0 | 39.1 | 21.6 | 27.8 | 99 |
| BGR | 5.0 | 12.7 | 25.0 | 21.0 | 38.9 | 23.6 | 38.6 | 96 |
| BHR | 5.2 | 1.1 | 29.8 | 21.0 | 25.1 | 11.3 | 31.0 | 95 |
| BHS | 5.1 | 4.8 | 31.6 | 21.0 | 10.9 | 15.5 | 43.3 | 95 |
| BIH | 4.8 | 7.1 | 17.9 | 21.0 | 38.3 | 17.8 | 25.5 | 95 |
| BLR | 5.1 | 11.4 | 24.5 | 21.0 | 26.6 | 23.7 | 14.1 | 98 |
| BOL | 4.7 | 4.4 | 20.2 | 21.0 | 23.5 | 17.2 | 31.0 | 90 |
| BRA | 4.9 | 7.4 | 22.1 | 21.0 | 16.5 | 16.6 | 47.0 | 90 |
| BRB | 4.8 | 9.7 | 23.1 | 21.0 | 8.7 | 16.2 | 42.9 | 95 |
| BRN | 5.2 | 0.5 | 14.1 | 21.0 | 15.5 | 16.6 | 27.3 | 99 |
| CAF | 4.2 | 2.4 | 7.5 | 21.0 | 23.5 | 23.1 | 14.3 | 74 |
| CAN | 5.1 | 8.9 | 29.4 | 21.0 | 17.5 | 9.8 | 28.6 | 95 |
| CHE | 5.3 | 11.5 | 19.5 | 21.0 | 25.1 | 8.6 | 23.7 | 95 |
| CHL | 5.0 | 9.1 | 28.0 | 21.0 | 44.7 | 12.4 | 26.6 | 96 |
| CIV | 4.9 | 6.7 | 22.1 | 21.0 | 23.5 | 29.1 | 31.0 | 95 |
| COL | 4.8 | 5.7 | 22.3 | 21.0 | 7.9 | 15.8 | 44.0 | 89 |
| CPV | 4.3 | 5.6 | 11.8 | 21.0 | 23.5 | 17.2 | 19.7 | 96 |
| CRI | 4.8 | 4.9 | 25.7 | 21.0 | 9.8 | 11.5 | 46.1 | 92 |
| CUB | 4.5 | 5.8 | 24.6 | 21.0 | 27.1 | 16.4 | 36.9 | 99 |
| CYM | 4.9 | 6.7 | 22.1 | 21.0 | 23.5 | / | 31.0 | 95 |
| CYP | 5.2 | 10.8 | 21.8 | 21.0 | 36.7 | 11.3 | 44.4 | 95 |
| CZE | 5.1 | 14.4 | 26.0 | 21.0 | 31.5 | 15.0 | 31.1 | 95 |
| DEU | 5.4 | 12.9 | 22.3 | 21.0 | 28.0 | 12.1 | 42.2 | 95 |
| DJI | 4.7 | 0.4 | 13.5 | 21.0 | 23.5 | 19.6 | 31.0 | 93 |
| DNK | 5.5 | 10.3 | 19.7 | 21.0 | 18.6 | 11.3 | 28.5 | 95 |
| DOM | 4.6 | 6.7 | 27.6 | 21.0 | 9.4 | 19.0 | 39.0 | 99 |
| EGY | 4.8 | 0.4 | 32.0 | 21.0 | 21.4 | 27.7 | 31.0 | 95 |
| ESP | 5.2 | 12.7 | 23.8 | 21.0 | 27.9 | 9.9 | 26.8 | 95 |
| EST | 5.2 | 9.2 | 21.2 | 21.0 | 30.5 | 17.0 | 32.0 | 92 |
| ETH | 4.2 | 2.4 | 4.5 | 21.0 | 4.6 | 18.3 | 14.9 | 85 |
| FIN | 5.3 | 10.8 | 22.2 | 21.0 | 19.7 | 10.2 | 16.6 | 95 |
| FRA | 5.3 | 12.3 | 21.6 | 21.0 | 34.6 | 10.6 | 29.3 | 95 |
| GAB | 4.8 | 8.7 | 15.0 | 21.0 | 23.5 | 14.4 | 25.3 | 87 |
| GBR | 5.4 | 11.4 | 27.8 | 21.0 | 19.2 | 10.9 | 35.9 | 95 |
| GEO | 4.7 | 8.2 | 21.7 | 21.0 | 29.7 | 24.9 | 18.0 | 97 |
| GHA | 4.1 | 2.8 | 10.9 | 21.0 | 3.7 | 20.8 | 21.8 | 98 |
| GIN | 4.1 | 1.1 | 7.7 | 21.0 | 23.5 | 22.4 | 14.5 | 72 |
| GRC | 5.0 | 10.2 | 24.9 | 21.0 | 39.1 | 12.4 | 37.7 | 95 |
| GTM | 4.4 | 2.4 | 21.2 | 21.0 | 23.5 | 14.9 | 37.1 | 88 |
| GUY | 4.3 | 6.9 | 20.2 | 21.0 | 12.2 | 30.5 | 31.0 | 99 |
| HND | 4.5 | 3.8 | 21.4 | 21.0 | 23.5 | 14.0 | 31.0 | 94 |
| HRV | 5.0 | 9.2 | 24.4 | 21.0 | 36.6 | 16.7 | 31.1 | 98 |
| HTI | 4.2 | 2.7 | 22.7 | 21.0 | 8.3 | 26.5 | 31.0 | 83 |
| HUN | 5.2 | 11.3 | 26.4 | 21.0 | 30.6 | 23.0 | 38.5 | 99 |
| IDN | 4.6 | 0.6 | 6.9 | 21.0 | 37.9 | 26.4 | 22.6 | 81 |
| IMN | 4.9 | 6.7 | 22.1 | 21.0 | 23.5 | / | 31.0 | 95 |
| IND | 4.4 | 5.5 | 3.9 | 21.0 | 27.0 | 23.3 | 34.0 | 92 |
| IRL | 5.4 | 12.9 | 25.3 | 21.0 | 23.6 | 10.3 | 32.7 | 95 |
| IRN | 5.2 | 1.0 | 25.8 | 21.0 | 14.0 | 14.8 | 33.2 | 99 |
| IRQ | 4.9 | 0.4 | 30.4 | 21.0 | 22.2 | 21.3 | 52.0 | 95 |
| ISL | 5.6 | 9.1 | 21.9 | 21.0 | 13.8 | 9.1 | 31.0 | 95 |
| ISR | 5.1 | 4.2 | 26.1 | 21.0 | 25.5 | 9.6 | 31.0 | 95 |
| ITA | 5.4 | 7.8 | 19.9 | 21.0 | 23.4 | 9.5 | 41.4 | 95 |
| JAM | 4.5 | 4.2 | 24.7 | 21.0 | 11.0 | 14.7 | 32.6 | 93 |
| JOR | 5.0 | 0.7 | 35.5 | 21.0 | 23.5 | 19.2 | 11.9 | 94 |
| JPN | 5.2 | 8.0 | 4.3 | 21.0 | 21.9 | 8.4 | 35.5 | 99 |
| KAZ | 4.9 | 4.8 | 21.0 | 21.0 | 24.4 | 26.8 | 27.5 | 95 |
| KEN | 4.4 | 2.8 | 7.1 | 21.0 | 11.8 | 13.4 | 15.4 | 95 |
| KGZ | 4.5 | 6.3 | 16.6 | 21.0 | 27.9 | 24.9 | 13.9 | 97 |
| KWT | 5.2 | 0.0 | 37.9 | 21.0 | 22.1 | 17.4 | 67.0 | 99 |
| LBN | 4.9 | 1.7 | 32.0 | 21.0 | 42.6 | 17.9 | 36.4 | 95 |
| LTU | 5.1 | 13.2 | 26.3 | 21.0 | 27.1 | 20.7 | 26.5 | 96 |
| LVA | 5.2 | 12.8 | 23.6 | 21.0 | 36.7 | 21.9 | 29.5 | 95 |
| MAR | 4.7 | 0.7 | 26.1 | 21.0 | 14.7 | 12.4 | 26.2 | 99 |
| MCO | 4.9 | 6.7 | 22.1 | 21.0 | 23.5 | / | 31.0 | 89 |
| MDA | 4.7 | 11.4 | 18.9 | 21.0 | 25.3 | 24.9 | 11.5 | 96 |
| MDG | 4.3 | 2.0 | 5.3 | 21.0 | 28.9 | 22.9 | 17.2 | 70 |
| MDV | 4.9 | 2.2 | 8.6 | 21.0 | 23.5 | 13.4 | 30.3 | 99 |
| MEX | 5.1 | 5.0 | 28.9 | 21.0 | 13.9 | 15.7 | 28.9 | 96 |
| MKD | 4.9 | 6.2 | 22.4 | 21.0 | 23.5 | 20.3 | 31.0 | 97 |
| MLI | 4.1 | 1.3 | 8.6 | 21.0 | 12.0 | 24.6 | 40.4 | 83 |
| MLT | 5.3 | 8.0 | 28.9 | 21.0 | 25.1 | 10.8 | 41.7 | 95 |
| MNE | 5.0 | 11.5 | 23.3 | 21.0 | 23.5 | 20.6 | 31.0 | 83 |
| MOZ | 4.4 | 2.3 | 7.2 | 21.0 | 14.4 | 18.4 | 5.6 | 95 |
| MRT | 4.3 | 0.0 | 12.7 | 21.0 | 23.5 | 18.1 | 41.3 | 90 |
| MUS | 4.9 | 4.3 | 10.8 | 21.0 | 26.9 | 22.6 | 29.8 | 99 |
| MYS | 5.1 | 0.9 | 15.6 | 21.0 | 21.8 | 17.2 | 38.8 | 98 |
| NGA | 4.0 | 10.8 | 8.9 | 21.0 | 4.8 | 22.5 | 27.1 | 53 |
| NLD | 5.3 | 9.6 | 20.4 | 21.0 | 23.4 | 11.2 | 27.2 | 95 |
| NOR | 5.4 | 7.4 | 23.1 | 21.0 | 13.0 | 9.2 | 31.7 | 95 |
| NPL | 4.2 | 2.1 | 4.1 | 21.0 | 31.9 | 21.8 | 13.4 | 96 |
| NZL | 5.2 | 10.6 | 30.8 | 21.0 | 14.8 | 10.1 | 42.4 | 95 |
| OMN | 5.1 | 0.8 | 27.0 | 21.0 | 9.6 | 17.8 | 32.9 | 99 |
| PAK | 4.5 | 0.3 | 8.6 | 21.0 | 20.0 | 24.7 | 33.7 | 86 |
| PAN | 4.8 | 8.0 | 22.7 | 21.0 | 6.9 | 13.0 | 31.0 | 99 |
| PER | 4.7 | 6.4 | 19.7 | 21.0 | 9.6 | 12.6 | 31.0 | 81 |
| PHL | 4.9 | 6.9 | 6.4 | 21.0 | 24.3 | 26.8 | 39.7 | 75 |
| POL | 5.2 | 11.7 | 23.1 | 21.0 | 26.0 | 18.7 | 32.5 | 92 |
| PRT | 5.2 | 12.0 | 20.8 | 21.0 | 27.9 | 11.1 | 43.4 | 95 |
| PRY | 4.7 | 7.6 | 20.3 | 21.0 | 12.8 | 17.5 | 37.4 | 91 |
| ROU | 4.9 | 11.7 | 22.5 | 21.0 | 25.5 | 21.4 | 35.4 | 96 |
| RUS | 5.0 | 11.2 | 23.1 | 21.0 | 28.3 | 25.4 | 17.1 | 95 |
| SAU | 4.7 | 0.2 | 35.4 | 21.0 | 16.6 | 16.4 | 53.1 | 98 |
| SEN | 4.2 | 0.8 | 8.8 | 21.0 | 9.1 | 18.1 | 23.1 | 83 |
| SLV | 4.6 | 3.9 | 24.6 | 21.0 | 12.7 | 14.0 | 31.0 | 81 |
| SRB | 5.0 | 8.7 | 21.5 | 21.0 | 40.6 | 19.1 | 39.5 | 98 |
| SWE | 5.0 | 8.9 | 20.6 | 21.0 | 28.8 | 9.1 | 23.1 | 26 |
| TGO | 4.0 | 2.5 | 8.4 | 21.0 | 7.6 | 23.6 | 9.8 | 83 |
| TTO | 4.7 | 6.7 | 18.6 | 21.0 | 23.5 | 21.3 | 38.2 | 95 |
| TUN | 4.8 | 2.1 | 26.9 | 21.0 | 26.0 | 16.1 | 30.4 | 92 |
| TUR | 4.8 | 2.0 | 32.1 | 21.0 | 29.3 | 16.1 | 30.6 | 96 |
| UKR | 4.9 | 8.3 | 24.1 | 21.0 | 25.5 | 24.7 | 19.6 | 90 |
| URY | 4.8 | 6.9 | 27.9 | 21.0 | 21.8 | 16.7 | 22.4 | 98 |
| UZB | 4.3 | 2.6 | 16.6 | 21.0 | 12.3 | 24.5 | 19.1 | 96 |
| VEN | 4.7 | 4.1 | 25.6 | 21.0 | 23.5 | 18.1 | 31.4 | 92 |
| VNM | 4.7 | 8.7 | 2.1 | 21.0 | 23.5 | 17.1 | 25.4 | 95 |
| ZAF | 4.6 | 9.5 | 28.3 | 21.0 | 31.4 | 26.2 | 38.2 | 70 |

Column explanations: **CH** - cholesterol level; **ALC** - alcohol consumption per capita; **OB** - prevalence of obesity; **RBP** - percentage of people with raised blood pressure; **SM** – percentage of smokers; **CD** - severity of COVID-19 relevant chronic diseases in the population; **IN** - prevalence of insufficient physical activity among adults; **BCG** - BCG immunization coverage among 1-year-olds.

**Table S4.** ABO and Rhesus blood group systems.

| Country | A | A+ | A- | B | B+ | B- | AB | AB+ | AB- | O | O+ | O- | Rh+ | Rh- |
|---|---|---|---|---|---|---|---|---|---|---|---|---|---|---|
| ABW | / | / | / | / | / | / | / | / | / | / | / | / | / | / |
| AFG | / | / | / | / | / | / | / | / | / | / | / | / | / | / |
| ALB | 0.37 | 0.31 | 0.06 | 0.17 | 0.15 | 0.03 | 0.06 | 0.05 | 0.01 | 0.40 | 0.34 | 0.06 | 0.85 | 0.15 |
| AND | / | / | / | / | / | / | / | / | / | / | / | / | / | / |
| ARE | 0.24 | 0.22 | 0.02 | 0.23 | 0.21 | 0.02 | 0.05 | 0.04 | 0.00 | 0.48 | 0.44 | 0.04 | 0.91 | 0.09 |
| ARG | 0.35 | 0.34 | 0.00 | 0.09 | 0.09 | 0.00 | 0.03 | 0.03 | 0.00 | 0.54 | 0.45 | 0.08 | 0.91 | 0.09 |
| ARM | 0.50 | 0.46 | 0.04 | 0.13 | 0.12 | 0.01 | 0.06 | 0.06 | 0.00 | 0.31 | 0.29 | 0.02 | 0.93 | 0.07 |
| AUS | 0.38 | 0.31 | 0.07 | 0.10 | 0.08 | 0.02 | 0.03 | 0.02 | 0.01 | 0.49 | 0.40 | 0.09 | 0.81 | 0.19 |
| AUT | 0.44 | 0.37 | 0.07 | 0.14 | 0.12 | 0.02 | 0.06 | 0.05 | 0.01 | 0.36 | 0.30 | 0.06 | 0.84 | 0.16 |
| AZE | 0.33 | 0.30 | 0.03 | 0.24 | 0.21 | 0.02 | 0.10 | 0.09 | 0.01 | 0.33 | 0.30 | 0.03 | 0.90 | 0.10 |
| BEL | 0.44 | 0.37 | 0.07 | 0.10 | 0.09 | 0.01 | 0.04 | 0.03 | 0.01 | 0.42 | 0.36 | 0.06 | 0.85 | 0.15 |
| BGD | 0.22 | 0.21 | 0.01 | 0.36 | 0.35 | 0.01 | 0.09 | 0.09 | 0.01 | 0.33 | 0.32 | 0.01 | 0.97 | 0.04 |
| BGR | 0.44 | 0.37 | 0.07 | 0.15 | 0.13 | 0.02 | 0.08 | 0.07 | 0.01 | 0.33 | 0.28 | 0.05 | 0.85 | 0.15 |
| BHR | 0.21 | 0.19 | 0.01 | 0.24 | 0.23 | 0.01 | 0.04 | 0.04 | 0.00 | 0.52 | 0.48 | 0.03 | 0.94 | 0.06 |
| BHS | / | / | / | / | / | / | / | / | / | / | / | / | / | / |
| BIH | 0.43 | 0.36 | 0.07 | 0.14 | 0.12 | 0.02 | 0.07 | 0.06 | 0.01 | 0.36 | 0.31 | 0.05 | 0.85 | 0.15 |
| BLR | / | / | / | / | / | / | / | / | / | / | / | / | / | / |
| BOL | 0.32 | 0.29 | 0.03 | 0.11 | 0.10 | 0.01 | 0.01 | 0.01 | 0.00 | 0.56 | 0.52 | 0.04 | 0.92 | 0.08 |
| BRA | 0.42 | 0.34 | 0.08 | 0.10 | 0.08 | 0.02 | 0.03 | 0.03 | 0.01 | 0.45 | 0.36 | 0.09 | 0.81 | 0.20 |
| BRB | / | / | / | / | / | / | / | / | / | / | / | / | / | / |
| BRN | / | / | / | / | / | / | / | / | / | / | / | / | / | / |
| CAF | / | / | / | / | / | / | / | / | / | / | / | / | / | / |
| CAN | 0.42 | 0.36 | 0.06 | 0.09 | 0.08 | 0.01 | 0.03 | 0.03 | 0.01 | 0.46 | 0.39 | 0.07 | 0.85 | 0.15 |
| CHE | 0.45 | 0.38 | 0.07 | 0.09 | 0.08 | 0.01 | 0.05 | 0.04 | 0.01 | 0.41 | 0.35 | 0.06 | 0.85 | 0.15 |
| CHL | 0.29 | 0.27 | 0.02 | 0.09 | 0.09 | 0.00 | 0.02 | 0.02 | 0.00 | 0.60 | 0.57 | 0.03 | 0.94 | 0.06 |
| CIV | 0.24 | 0.23 | 0.01 | 0.24 | 0.23 | 0.01 | 0.05 | 0.04 | 0.00 | 0.49 | 0.47 | 0.02 | 0.96 | 0.04 |
| COL | 0.29 | 0.26 | 0.03 | 0.08 | 0.07 | 0.01 | 0.02 | 0.01 | 0.00 | 0.61 | 0.56 | 0.05 | 0.91 | 0.09 |
| CPV | / | / | / | / | / | / | / | / | / | / | / | / | / | / |
| CRI | 0.30 | 0.29 | 0.02 | 0.13 | 0.12 | 0.01 | 0.03 | 0.03 | 0.00 | 0.53 | 0.50 | 0.03 | 0.94 | 0.06 |
| CUB | 0.36 | 0.34 | 0.03 | 0.11 | 0.10 | 0.01 | 0.03 | 0.03 | 0.00 | 0.49 | 0.46 | 0.04 | 0.92 | 0.08 |
| CYM | / | / | / | / | / | / | / | / | / | / | / | / | / | / |
| CYP | 0.44 | 0.40 | 0.03 | 0.12 | 0.11 | 0.01 | 0.05 | 0.05 | 0.00 | 0.39 | 0.35 | 0.04 | 0.91 | 0.09 |
| CZE | 0.42 | 0.36 | 0.06 | 0.18 | 0.15 | 0.03 | 0.08 | 0.07 | 0.01 | 0.32 | 0.27 | 0.05 | 0.85 | 0.15 |
| DEU | 0.43 | 0.37 | 0.06 | 0.11 | 0.09 | 0.02 | 0.05 | 0.04 | 0.01 | 0.41 | 0.35 | 0.06 | 0.85 | 0.15 |
| DJI | / | / | / | / | / | / | / | / | / | / | / | / | / | / |
| DNK | 0.44 | 0.37 | 0.07 | 0.10 | 0.08 | 0.02 | 0.05 | 0.04 | 0.01 | 0.41 | 0.35 | 0.06 | 0.84 | 0.16 |
| DOM | 0.29 | 0.26 | 0.02 | 0.18 | 0.17 | 0.01 | 0.03 | 0.03 | 0.00 | 0.50 | 0.46 | 0.04 | 0.93 | 0.07 |
| EGY | 0.26 | 0.24 | 0.02 | 0.13 | 0.12 | 0.01 | 0.04 | 0.04 | 0.00 | 0.57 | 0.52 | 0.05 | 0.92 | 0.08 |
| ESP | 0.43 | 0.36 | 0.07 | 0.10 | 0.08 | 0.02 | 0.03 | 0.03 | 0.01 | 0.44 | 0.35 | 0.09 | 0.82 | 0.19 |
| EST | 0.35 | 0.31 | 0.05 | 0.24 | 0.21 | 0.03 | 0.07 | 0.06 | 0.01 | 0.34 | 0.30 | 0.04 | 0.87 | 0.13 |
| ETH | 0.30 | 0.28 | 0.02 | 0.22 | 0.21 | 0.01 | 0.06 | 0.05 | 0.01 | 0.42 | 0.39 | 0.03 | 0.93 | 0.07 |
| FIN | 0.41 | 0.35 | 0.06 | 0.18 | 0.16 | 0.02 | 0.08 | 0.07 | 0.01 | 0.33 | 0.28 | 0.05 | 0.86 | 0.14 |
| FRA | 0.44 | 0.37 | 0.07 | 0.10 | 0.09 | 0.01 | 0.04 | 0.03 | 0.01 | 0.42 | 0.36 | 0.06 | 0.85 | 0.15 |
| GAB | / | / | / | / | / | / | / | / | / | / | / | / | / | / |
| GBR | 0.39 | 0.32 | 0.07 | 0.10 | 0.08 | 0.02 | 0.04 | 0.03 | 0.01 | 0.47 | 0.38 | 0.09 | 0.81 | 0.19 |
| GEO | / | / | / | / | / | / | / | / | / | / | / | / | / | / |
| GHA | 0.19 | 0.18 | 0.01 | 0.20 | 0.18 | 0.01 | 0.03 | 0.03 | 0.00 | 0.58 | 0.54 | 0.05 | 0.93 | 0.07 |
| GIN | 0.23 | 0.22 | 0.01 | 0.24 | 0.23 | 0.01 | 0.05 | 0.05 | 0.00 | 0.49 | 0.47 | 0.02 | 0.96 | 0.04 |
| GRC | 0.38 | 0.32 | 0.06 | 0.13 | 0.11 | 0.02 | 0.05 | 0.04 | 0.01 | 0.44 | 0.38 | 0.07 | 0.85 | 0.15 |
| GTM | / | / | / | / | / | / | / | / | / | / | / | / | / | / |
| GUY | / | / | / | / | / | / | / | / | / | / | / | / | / | / |
| HND | 0.29 | 0.27 | 0.02 | 0.08 | 0.08 | 0.01 | 0.03 | 0.03 | 0.00 | 0.60 | 0.58 | 0.03 | 0.95 | 0.05 |
| HRV | 0.42 | 0.36 | 0.06 | 0.18 | 0.15 | 0.03 | 0.06 | 0.05 | 0.01 | 0.34 | 0.29 | 0.05 | 0.85 | 0.15 |
| HTI | / | / | / | / | / | / | / | / | / | / | / | / | / | / |
| HUN | 0.40 | 0.33 | 0.07 | 0.19 | 0.16 | 0.03 | 0.09 | 0.08 | 0.01 | 0.32 | 0.27 | 0.05 | 0.84 | 0.16 |
| IDN | 0.26 | 0.26 | 0.00 | 0.29 | 0.29 | 0.00 | 0.08 | 0.08 | 0.00 | 0.37 | 0.37 | 0.00 | 1.00 | 0.01 |
| IMN | / | / | / | / | / | / | / | / | / | / | / | / | / | / |
| IND | 0.21 | 0.21 | 0.01 | 0.40 | 0.38 | 0.02 | 0.09 | 0.09 | 0.00 | 0.29 | 0.28 | 0.01 | 0.96 | 0.04 |
| IRL | 0.31 | 0.26 | 0.05 | 0.11 | 0.09 | 0.02 | 0.03 | 0.02 | 0.01 | 0.55 | 0.47 | 0.08 | 0.84 | 0.16 |
| IRN | 0.30 | 0.27 | 0.03 | 0.25 | 0.22 | 0.03 | 0.08 | 0.07 | 0.01 | 0.38 | 0.34 | 0.04 | 0.90 | 0.10 |
| IRQ | 0.28 | 0.25 | 0.03 | 0.28 | 0.26 | 0.03 | 0.08 | 0.07 | 0.01 | 0.36 | 0.32 | 0.04 | 0.90 | 0.10 |
| ISL | 0.32 | 0.27 | 0.05 | 0.11 | 0.09 | 0.02 | 0.02 | 0.02 | 0.00 | 0.55 | 0.47 | 0.08 | 0.85 | 0.15 |
| ISR | 0.38 | 0.34 | 0.04 | 0.19 | 0.17 | 0.02 | 0.08 | 0.07 | 0.01 | 0.35 | 0.32 | 0.03 | 0.90 | 0.10 |
| ITA | 0.42 | 0.36 | 0.06 | 0.09 | 0.08 | 0.02 | 0.03 | 0.03 | 0.01 | 0.46 | 0.39 | 0.07 | 0.85 | 0.15 |
| JAM | 0.25 | 0.23 | 0.02 | 0.21 | 0.20 | 0.01 | 0.04 | 0.03 | 0.01 | 0.51 | 0.47 | 0.04 | 0.93 | 0.07 |
| JOR | / | / | / | / | / | / | / | / | / | / | / | / | / | / |
| JPN | 0.40 | 0.40 | 0.00 | 0.20 | 0.20 | 0.00 | 0.10 | 0.10 | 0.00 | 0.30 | 0.30 | 0.00 | 1.00 | 0.01 |
| KAZ | 0.32 | 0.30 | 0.02 | 0.26 | 0.24 | 0.02 | 0.09 | 0.08 | 0.01 | 0.33 | 0.31 | 0.02 | 0.93 | 0.07 |
| KEN | 0.26 | 0.25 | 0.01 | 0.22 | 0.21 | 0.01 | 0.04 | 0.04 | 0.00 | 0.47 | 0.46 | 0.02 | 0.96 | 0.04 |
| KGZ | / | / | / | / | / | / | / | / | / | / | / | / | / | / |
| KWT | / | / | / | / | / | / | / | / | / | / | / | / | / | / |
| LBN | 0.39 | 0.32 | 0.07 | 0.11 | 0.10 | 0.02 | 0.04 | 0.03 | 0.01 | 0.46 | 0.38 | 0.08 | 0.83 | 0.17 |
| LTU | 0.39 | 0.33 | 0.06 | 0.13 | 0.11 | 0.02 | 0.05 | 0.04 | 0.01 | 0.43 | 0.36 | 0.07 | 0.84 | 0.16 |
| LVA | 0.37 | 0.31 | 0.06 | 0.20 | 0.17 | 0.03 | 0.07 | 0.06 | 0.01 | 0.36 | 0.31 | 0.05 | 0.85 | 0.15 |
| MAR | 0.33 | 0.30 | 0.03 | 0.16 | 0.14 | 0.02 | 0.05 | 0.04 | 0.00 | 0.47 | 0.42 | 0.05 | 0.91 | 0.10 |
| MCO | / | / | / | / | / | / | / | / | / | / | / | / | / | / |
| MDA | 0.38 | 0.32 | 0.06 | 0.21 | 0.18 | 0.03 | 0.08 | 0.07 | 0.01 | 0.34 | 0.29 | 0.05 | 0.85 | 0.15 |
| MDG | / | / | / | / | / | / | / | / | / | / | / | / | / | / |
| MDV | / | / | / | / | / | / | / | / | / | / | / | / | / | / |
| MEX | 0.27 | 0.26 | 0.01 | 0.09 | 0.09 | 0.00 | 0.02 | 0.02 | 0.00 | 0.62 | 0.59 | 0.03 | 0.96 | 0.04 |
| MKD | 0.40 | 0.34 | 0.06 | 0.18 | 0.15 | 0.03 | 0.07 | 0.06 | 0.01 | 0.35 | 0.30 | 0.05 | 0.85 | 0.15 |
| MLI | / | / | / | / | / | / | / | / | / | / | / | / | / | / |
| MLT | 0.46 | 0.41 | 0.05 | 0.08 | 0.07 | 0.01 | 0.04 | 0.03 | 0.01 | 0.43 | 0.38 | 0.05 | 0.89 | 0.11 |
| MNE | / | / | / | / | / | / | / | / | / | / | / | / | / | / |
| MOZ | / | / | / | / | / | / | / | / | / | / | / | / | / | / |
| MRT | 0.28 | 0.27 | 0.02 | 0.19 | 0.17 | 0.01 | 0.04 | 0.04 | 0.00 | 0.49 | 0.46 | 0.03 | 0.94 | 0.06 |
| MUS | 0.27 | 0.26 | 0.01 | 0.26 | 0.25 | 0.01 | 0.07 | 0.07 | 0.00 | 0.40 | 0.38 | 0.02 | 0.96 | 0.04 |
| MYS | 0.31 | 0.30 | 0.00 | 0.28 | 0.27 | 0.00 | 0.08 | 0.07 | 0.00 | 0.34 | 0.34 | 0.00 | 1.00 | 0.01 |
| NGA | 0.23 | 0.22 | 0.01 | 0.21 | 0.21 | 0.01 | 0.03 | 0.03 | 0.00 | 0.53 | 0.51 | 0.02 | 0.97 | 0.03 |
| NLD | 0.42 | 0.35 | 0.07 | 0.08 | 0.07 | 0.01 | 0.03 | 0.03 | 0.01 | 0.47 | 0.40 | 0.08 | 0.84 | 0.16 |
| NOR | 0.49 | 0.42 | 0.07 | 0.08 | 0.07 | 0.01 | 0.04 | 0.03 | 0.01 | 0.39 | 0.33 | 0.06 | 0.85 | 0.15 |
| NPL | 0.29 | 0.28 | 0.00 | 0.27 | 0.27 | 0.00 | 0.09 | 0.09 | 0.00 | 0.36 | 0.35 | 0.00 | 0.99 | 0.01 |
| NZL | 0.38 | 0.32 | 0.06 | 0.11 | 0.09 | 0.02 | 0.04 | 0.03 | 0.01 | 0.47 | 0.38 | 0.09 | 0.82 | 0.18 |
| OMN | / | / | / | / | / | / | / | / | / | / | / | / | / | / |
| PAK | 0.23 | 0.22 | 0.02 | 0.38 | 0.34 | 0.04 | 0.10 | 0.10 | 0.00 | 0.29 | 0.27 | 0.02 | 0.92 | 0.08 |
| PAN | / | / | / | / | / | / | / | / | / | / | / | / | / | / |
| PER | 0.19 | 0.18 | 0.01 | 0.08 | 0.08 | 0.00 | 0.02 | 0.02 | 0.00 | 0.71 | 0.70 | 0.01 | 0.98 | 0.02 |
| PHL | 0.23 | 0.23 | 0.00 | 0.25 | 0.25 | 0.00 | 0.06 | 0.06 | 0.00 | 0.46 | 0.46 | 0.00 | 1.00 | 0.00 |
| POL | 0.38 | 0.32 | 0.06 | 0.17 | 0.15 | 0.02 | 0.08 | 0.07 | 0.01 | 0.37 | 0.31 | 0.06 | 0.85 | 0.15 |
| PRT | 0.47 | 0.40 | 0.07 | 0.08 | 0.07 | 0.01 | 0.03 | 0.03 | 0.01 | 0.42 | 0.36 | 0.06 | 0.86 | 0.15 |
| PRY | / | / | / | / | / | / | / | / | / | / | / | / | / | / |
| ROU | 0.43 | 0.37 | 0.06 | 0.16 | 0.14 | 0.02 | 0.08 | 0.07 | 0.01 | 0.33 | 0.28 | 0.05 | 0.86 | 0.14 |
| RUS | 0.36 | 0.30 | 0.05 | 0.23 | 0.20 | 0.04 | 0.08 | 0.07 | 0.01 | 0.33 | 0.28 | 0.05 | 0.85 | 0.15 |
| SAU | 0.26 | 0.24 | 0.02 | 0.18 | 0.17 | 0.01 | 0.04 | 0.04 | 0.00 | 0.52 | 0.48 | 0.04 | 0.93 | 0.07 |
| SEN | / | / | / | / | / | / | / | / | / | / | / | / | / | / |
| SLV | 0.24 | 0.23 | 0.01 | 0.12 | 0.11 | 0.01 | 0.01 | 0.01 | 0.00 | 0.63 | 0.62 | 0.01 | 0.97 | 0.03 |
| SRB | 0.42 | 0.35 | 0.07 | 0.15 | 0.13 | 0.02 | 0.05 | 0.04 | 0.01 | 0.38 | 0.32 | 0.06 | 0.84 | 0.16 |
| SWE | 0.44 | 0.37 | 0.07 | 0.12 | 0.10 | 0.02 | 0.06 | 0.05 | 0.01 | 0.38 | 0.32 | 0.06 | 0.84 | 0.16 |
| TGO | / | / | / | / | / | / | / | / | / | / | / | / | / | / |
| TTO | / | / | / | / | / | / | / | / | / | / | / | / | / | / |
| TUN | / | / | / | / | / | / | / | / | / | / | / | / | / | / |
| TUR | 0.43 | 0.38 | 0.05 | 0.16 | 0.14 | 0.02 | 0.08 | 0.07 | 0.01 | 0.34 | 0.30 | 0.04 | 0.89 | 0.11 |
| UKR | 0.40 | 0.34 | 0.06 | 0.17 | 0.15 | 0.02 | 0.06 | 0.05 | 0.01 | 0.37 | 0.32 | 0.05 | 0.86 | 0.14 |
| URY | / | / | / | / | / | / | / | / | / | / | / | / | / | / |
| UZB | / | / | / | / | / | / | / | / | / | / | / | / | / | / |
| VEN | 0.30 | 0.28 | 0.02 | 0.06 | 0.06 | 0.00 | 0.02 | 0.02 | 0.00 | 0.62 | 0.58 | 0.04 | 0.94 | 0.06 |
| VNM | 0.22 | 0.22 | 0.00 | 0.31 | 0.31 | 0.00 | 0.05 | 0.05 | 0.00 | 0.42 | 0.42 | 0.00 | 0.99 | 0.01 |
| ZAF | 0.37 | 0.32 | 0.05 | 0.14 | 0.12 | 0.02 | 0.04 | 0.03 | 0.01 | 0.45 | 0.39 | 0.06 | 0.86 | 0.14 |

**Table S5.** Meteorological factors.

| Country | ONS (days) | $NO_2$ | $SO_2$ | CO | 2.5 | 10 | PC ($l/m^2$) | T (°C) | H | UV | P (millibars) | WS (km/h) |
|---|---|---|---|---|---|---|---|---|---|---|---|---|
| ABW | 44 | / | / | / | / | / | 3.2 | 26.8 | 2.0.E-02 | 6.3 | 1014.9 | 7.3 |
| AFG | 37 | / | / | / | 115.0 | / | 0.8 | 2.9 | 1.7.E-03 | 3.4 | 1020.8 | 1.8 |
| ALB | 28 | / | / | / | / | / | 1.5 | 10.6 | 1.0.E-02 | 3.8 | 1014.8 | 2.3 |
| AND | 33 | / | / | / | / | / | 1.2 | 5.3 | 6.4.E-03 | 2.1 | 1023.5 | 1.2 |
| ARE | 30 | / | / | / | 90.2 | / | 1.9 | 23.0 | 1.0.E-02 | 6.4 | 1014.7 | 3.1 |
| ARG | 21 | 7.2 | 4.8 | 2.8 | 39.5 | 18.2 | 2.8 | 23.7 | 1.0.E-02 | 6.9 | 1016.8 | 2.2 |
| ARM | 33 | / | / | / | / | / | 0.9 | 5.9 | 5.9.E-04 | 3.4 | 1019.8 | 1.6 |
| AUS | 25 | 5.7 | 1.9 | 2.0 | 16.6 | 14.6 | 2.9 | 20.5 | 1.1.E-02 | 5.4 | 1019.9 | 2.4 |
| AUT | 13 | 7.7 | 1.0 | 0.1 | 26.9 | 8.8 | 1.1 | 5.1 | 2.7.E-03 | 2.8 | 1015.8 | 3.4 |
| AZE | 27 | / | / | / | / | / | 1.5 | 8.4 | 8.2.E-03 | 3.2 | 1022.7 | 4.0 |
| BEL | 22 | 10.3 | 0.9 | 0.1 | 30.7 | 12.3 | 1.3 | 6.3 | 5.0.E-03 | 2.1 | 1009.8 | 4.5 |
| BGD | 50 | / | / | / | 162.3 | / | 2.5 | 28.4 | 1.0.E-02 | 8.3 | 1010.6 | 1.5 |
| BGR | 27 | 5.5 | / | 0.1 | / | 16.7 | 1.2 | 4.5 | 5.0.E-03 | 2.5 | 1013.2 | 2.1 |
| BHR | 29 | / | / | / | 110.6 | / | 1.8 | 21.3 | 1.0.E-02 | 5.7 | 1014.9 | 4.3 |
| BHS | 44 | / | / | / | / | / | 2.5 | 24.3 | 1.2.E-02 | 5.7 | 1022.1 | 5.1 |
| BIH | 36 | 5.8 | 14.2 | / | 77.2 | 29.4 | 0.9 | 5.3 | 2.6.E-03 | 3.1 | 1024.3 | 1.0 |
| BLR | 45 | / | / | / | / | / | 0.7 | 1.4 | 0.0.E+00 | 2.2 | 1029.2 | 2.9 |
| BOL | 45 | / | / | / | / | / | 1.1 | 8.2 | 9.1.E-03 | 3.3 | 1020.6 | 1.9 |
| BRA | 20 | 9.0 | 1.0 | 3.1 | 34.6 | 15.6 | 3.6 | 21.5 | 1.2.E-02 | 6.3 | 1012.4 | 2.5 |
| BRB | 36 | / | / | / | / | / | 3.4 | 26.3 | 2.0.E-02 | 6.7 | 1015.8 | 5.9 |
| BRN | 28 | / | / | / | / | / | 5.6 | 26.3 | 2.0.E-02 | 5.8 | 1016.6 | 1.4 |
| CAF | 77 | / | / | / | / | / | 5.0 | 27.1 | 2.0.E-02 | 6.7 | 1012.6 | 1.4 |
| CAN | 12 | 8.2 | 0.4 | 2.5 | 27.2 | / | 0.8 | -1.5 | 3.4.E-04 | 1.7 | 1020.8 | 4.8 |
| CHE | 13 | / | / | / | / | / | 1.0 | 2.8 | 2.3.E-03 | 2.2 | 1020.9 | 2.4 |
| CHL | 19 | 15.5 | 11.0 | 6.2 | 66.0 | 61.4 | 1.1 | 21.8 | 7.2.E-03 | 6.3 | 1016.7 | 2.1 |
| CIV | 39 | / | / | / | 56.8 | / | 5.4 | 27.6 | 2.0.E-02 | 6.4 | 1011.7 | 2.7 |
| COL | 29 | / | / | / | / | / | 3.1 | 19.4 | 1.0.E-02 | 4.0 | 1015.8 | 0.8 |
| CPV | 64 | / | / | / | / | / | 2.4 | 23.1 | 1.2.E-02 | 6.0 | 1015.7 | 6.3 |
| CRI | 34 | / | / | / | 23.6 | / | 2.9 | 25.3 | 1.2.E-02 | 5.8 | 1015.2 | 0.7 |
| CUB | 33 | / | / | / | / | / | 2.7 | 25.0 | 1.1.E-02 | 6.4 | 1022.2 | 6.2 |
| CYM | 41 | / | / | / | / | / | 2.7 | 26.5 | 1.9.E-02 | 6.8 | 1019.3 | 5.4 |
| CYP | 31 | 8.5 | 1.5 | 2.8 | 46.0 | 26.6 | 1.6 | 14.7 | 8.2.E-03 | 4.4 | 1014.9 | 3.8 |
| CZE | 22 | 7.7 | 3.5 | / | 25.7 | 9.0 | 1.1 | 3.5 | 7.1.E-04 | 2.1 | 1005.9 | 3.7 |
| DEU | 13 | 10.8 | 1.2 | 0.1 | 31.7 | 12.3 | 1.2 | 5.0 | 3.6.E-03 | 2.2 | 1009.0 | 4.8 |
| DJI | 43 | / | / | / | / | / | 4.0 | 27.3 | 2.0.E-02 | 6.5 | 1012.6 | 3.9 |
| DNK | 29 | 5.8 | 0.4 | 2.0 | 36.0 | 19.9 | 0.9 | 4.3 | 2.7.E-03 | 2.4 | 1019.0 | 4.7 |
| DOM | 37 | / | / | / | / | / | 3.2 | 25.4 | 1.8.E-02 | 6.9 | 1018.1 | 4.0 |
| EGY | 23 | / | / | / | / | / | 1.5 | 16.3 | 1.0.E-02 | 5.8 | 1015.9 | 2.7 |
| ESP | 14 | 19.7 | 2.7 | 0.1 | 44.9 | 21.2 | 1.0 | 8.9 | 4.3.E-03 | 4.1 | 1027.0 | 2.0 |
| EST | 30 | 3.9 | 0.3 | / | 15.3 | 10.3 | 0.8 | 1.5 | 0.0.E+00 | 1.5 | 1016.4 | 5.6 |
| ETH | 30 | / | / | / | 71.5 | / | 2.3 | 20.8 | 9.9.E-03 | 5.4 | 1015.8 | 2.1 |
| FIN | 27 | 6.6 | 0.3 | / | 23.1 | 10.1 | 1.0 | 1.2 | 0.0.E+00 | 1.3 | 1007.4 | 2.7 |
| FRA | 15 | 12.9 | 0.4 | 0.1 | 29.5 | 14.8 | 1.4 | 6.9 | 6.7.E-03 | 2.4 | 1019.9 | 5.4 |
| GAB | 46 | / | / | / | / | / | 5.9 | 27.7 | 2.0.E-02 | 6.3 | 1012.3 | 2.2 |
| GBR | 14 | 13.0 | 2.7 | 5.0 | 27.8 | 12.3 | 1.3 | 6.3 | 6.5.E-03 | 2.6 | 1009.3 | 5.7 |
| GEO | 25 | 12.1 | 2.2 | 3.5 | 67.6 | 37.1 | 1.1 | 7.2 | 1.6.E-03 | 3.3 | 1019.6 | 1.7 |
| GHA | 41 | / | / | / | 61.5 | / | 5.2 | 27.7 | 2.0.E-02 | 6.6 | 1011.8 | 2.2 |
| GIN | 51 | / | / | / | 69.4 | / | 4.7 | 29.4 | 2.0.E-02 | 6.8 | 1012.3 | 3.1 |
| GRC | 30 | 12.8 | 5.0 | 0.1 | 45.8 | 18.1 | 1.4 | 13.2 | 8.9.E-03 | 4.2 | 1018.3 | 4.2 |
| GTM | 38 | / | / | / | 81.8 | / | 2.4 | 22.2 | 1.0.E-02 | 5.9 | 1016.1 | 1.3 |
| GUY | 47 | / | / | / | / | / | 4.9 | 27.0 | 2.0.E-02 | 6.2 | 1013.9 | 2.1 |
| HND | 30 | / | / | / | / | / | 2.3 | 24.2 | 1.0.E-02 | 5.9 | 1018.1 | 2.7 |
| HRV | 25 | 11.7 | 1.1 | 0.1 | / | 12.7 | 1.3 | 6.7 | 5.3.E-03 | 3.1 | 1014.9 | 1.3 |
| HTI | 82 | / | / | / | / | / | 4.0 | 27.3 | 1.7.E-02 | 6.5 | 1016.3 | 1.6 |
| HUN | 22 | 12.3 | 2.9 | 5.0 | 46.4 | 15.2 | 1.3 | 6.4 | 4.4.E-03 | 2.9 | 1012.7 | 3.3 |
| IDN | 25 | / | / | / | 85.6 | 22.8 | 5.8 | 26.4 | 2.0.E-02 | 6.8 | 1010.8 | 2.2 |
| IMN | 40 | / | / | / | / | / | 1.1 | 6.9 | 4.4.E-03 | 2.6 | 1021.1 | 6.7 |
| IND | 20 | 11.7 | 4.8 | 6.0 | 110.2 | 67.0 | 2.3 | 24.6 | 1.1.E-02 | 6.9 | 1013.8 | 2.3 |
| IRL | 25 | 9.0 | 1.2 | / | 23.0 | 8.4 | 1.2 | 5.4 | 3.6.E-03 | 1.8 | 997.2 | 5.9 |
| IRN | 7 | 34.7 | 16.1 | 23.8 | 74.5 | 44.7 | 0.7 | -1.7 | 0.0.E+00 | 2.3 | 1025.5 | 2.1 |
| IRQ | 16 | / | / | / | 107.2 | / | 1.6 | 16.7 | 7.1.E-03 | 5.3 | 1016.3 | 1.9 |
| ISL | 19 | / | / | / | / | / | 0.6 | -2.8 | 0.0.E+00 | 1.6 | 992.8 | 4.6 |
| ISR | 18 | 16.2 | 1.6 | 4.4 | 66.0 | 46.6 | 1.6 | 15.3 | 1.0.E-02 | 4.8 | 1017.1 | 3.0 |
| ITA | 14 | 24.5 | 3.0 | 3.3 | 74.5 | 34.9 | 1.1 | 8.9 | 6.2.E-03 | 3.5 | 1022.6 | 2.4 |
| JAM | 32 | / | / | / | / | / | 3.3 | 26.6 | 2.0.E-02 | 6.8 | 1018.4 | 4.9 |
| JOR | 32 | 12.3 | 3.8 | 24.6 | / | 54.5 | 1.8 | 16.3 | 1.0.E-02 | 4.3 | 1014.7 | 2.8 |
| JPN | 1 | 18.0 | 2.1 | 4.3 | 41.9 | 12.2 | 1.0 | 6.0 | 3.2.E-03 | 3.0 | 1023.1 | 2.3 |
| KAZ | 44 | / | / | / | 66.6 | / | 0.8 | 4.6 | 1.1.E-03 | 2.7 | 1023.6 | 1.8 |
| KEN | 38 | / | / | / | / | / | 3.2 | 22.3 | 1.2.E-02 | 5.4 | 1015.0 | 3.0 |
| KGZ | 42 | / | / | / | 59.3 | / | 0.8 | 4.2 | 1.3.E-03 | 3.1 | 1023.2 | 1.7 |
| KWT | 22 | / | / | / | / | / | 1.8 | 22.6 | 1.0.E-02 | 6.5 | 1013.0 | 3.5 |
| LBN | 19 | / | / | / | / | / | 1.1 | 10.5 | 1.0.E-02 | 4.5 | 1018.3 | 2.3 |
| LTU | 30 | / | / | / | / | / | 1.1 | 3.4 | 1.0.E-03 | 2.0 | 1007.8 | 2.0 |
| LVA | 24 | / | / | / | / | / | 1.0 | 2.4 | 0.0.E+00 | 1.7 | 1004.8 | 3.0 |
| MAR | 30 | / | / | / | / | / | 1.6 | 16.1 | 1.0.E-02 | 4.6 | 1021.8 | 3.2 |
| MCO | 42 | / | / | / | / | / | 1.1 | 8.1 | 6.7.E-03 | 3.7 | 1021.5 | 3.0 |
| MDA | 29 | / | / | / | / | / | 1.3 | 7.0 | 4.8.E-03 | 3.1 | 1020.9 | 3.2 |
| MDG | 40 | / | / | / | / | / | 3.3 | 19.9 | 1.5.E-02 | 5.0 | 1016.2 | 1.5 |
| MDV | 62 | / | / | / | / | / | 4.9 | 28.9 | 2.0.E-02 | 6.5 | 1012.1 | 2.3 |
| MEX | 26 | 19.7 | 5.7 | 6.8 | 72.8 | 46.8 | 1.1 | 16.3 | 1.0.E-02 | 5.6 | 1020.7 | 1.1 |
| MKD | 23 | 7.2 | 1.3 | 5.9 | 64.4 | 30.1 | 1.1 | 4.7 | 3.8.E-03 | 2.8 | 1015.8 | 2.1 |
| MLI | 44 | / | / | / | 136.8 | / | 2.7 | 30.9 | 1.0.E-02 | 8.5 | 1011.4 | 2.4 |
| MLT | 28 | / | / | / | / | / | 1.7 | 14.8 | 1.0.E-02 | 3.9 | 1016.0 | 5.8 |
| MNE | 39 | / | / | / | / | / | 1.0 | 8.0 | 4.3.E-03 | 3.5 | 1022.1 | 3.1 |
| MOZ | 85 | / | / | / | / | / | 2.4 | 22.5 | 1.0.E-02 | 6.0 | 1022.8 | 2.9 |
| MRT | 97 | / | / | / | / | / | 2.1 | 28.8 | 1.0.E-02 | 6.7 | 1012.7 | 4.1 |
| MUS | 41 | / | / | / | / | / | 4.1 | 26.9 | 1.9.E-02 | 6.3 | 1013.1 | 5.1 |
| MYS | 16 | / | / | / | / | / | 4.2 | 25.6 | 2.0.E-02 | 7.9 | 1012.2 | 1.7 |
| NGA | 60 | / | / | / | / | / | 4.7 | 27.8 | 1.8.E-02 | 7.8 | 1011.5 | 1.8 |
| NLD | 22 | / | / | / | / | / | 1.3 | 6.4 | 6.3.E-03 | 2.1 | 1004.4 | 5.6 |
| NOR | 16 | 12.0 | / | 0.1 | 26.5 | 17.7 | 0.8 | -0.3 | 0.0.E+00 | 1.5 | 998.1 | 0.5 |
| NPL | 76 | / | / | / | 60.5 | 20.5 | 2.1 | 19.0 | 1.0.E-02 | 5.8 | 1012.5 | 1.4 |
| NZL | 28 | 2.3 | 1.0 | / | 22.4 | 9.6 | 2.5 | 20.0 | 1.0.E-02 | 5.1 | 1018.9 | 4.5 |
| OMN | 34 | / | / | / | / | / | 2.1 | 23.8 | 1.0.E-02 | 6.8 | 1014.8 | 2.6 |
| PAK | 36 | / | / | / | 114.2 | / | 2.1 | 21.7 | 1.0.E-02 | 6.4 | 1013.9 | 2.4 |
| PAN | 31 | / | / | / | / | / | 3.0 | 27.0 | 2.0.E-02 | 7.0 | 1014.1 | 6.8 |
| PER | 23 | 5.1 | 7.9 | 9.8 | 57.3 | 29.8 | 4.5 | 22.5 | 1.8.E-02 | 6.4 | 1016.7 | 2.7 |
| PHL | 24 | / | / | / | / | / | 3.7 | 24.7 | 2.0.E-02 | 7.3 | 1014.4 | 1.3 |
| POL | 21 | 11.1 | 4.5 | 5.0 | 58.5 | 19.1 | 1.1 | 3.3 | 1.7.E-03 | 2.2 | 1005.4 | 3.8 |
| PRT | 20 | 8.0 | 0.3 | / | 43.3 | 21.6 | 1.5 | 13.8 | 1.0.E-02 | 3.9 | 1026.4 | 4.3 |
| PRY | 26 | / | / | / | / | / | 2.8 | 30.1 | 1.0.E-02 | 8.7 | 1013.3 | 2.1 |
| ROU | 28 | 14.2 | 3.1 | / | / | 33.3 | 1.6 | 9.6 | 7.6.E-03 | 3.7 | 1016.6 | 2.4 |
| RUS | 26 | 13.5 | 1.0 | 2.8 | 48.9 | 23.2 | 1.0 | 0.8 | 0.0.E+00 | 1.6 | 1015.6 | 4.3 |
| SAU | 26 | 8.5 | 1.4 | 15.1 | / | 29.5 | 0.9 | 18.9 | 0.0.E+00 | 5.8 | 1018.5 | 2.7 |
| SEN | 28 | / | / | / | / | / | 1.4 | 24.0 | 1.0.E-02 | 6.7 | 1012.6 | 4.2 |
| SLV | 39 | / | / | / | 64.5 | / | 3.3 | 28.1 | 1.2.E-02 | 7.1 | 1016.1 | 2.4 |
| SRB | 28 | 11.2 | 4.8 | 7.3 | 72.1 | 27.8 | 1.3 | 8.2 | 4.7.E-03 | 3.1 | 1019.2 | 2.5 |
| SWE | 21 | 10.2 | / | / | 19.7 | 15.8 | 0.9 | 1.2 | 0.0.E+00 | 1.3 | 997.2 | 2.6 |
| TGO | 36 | / | / | / | / | / | 5.1 | 28.4 | 2.0.E-02 | 7.3 | 1011.2 | 3.1 |
| TTO | 36 | / | / | / | / | / | 3.6 | 26.5 | 2.0.E-02 | 6.9 | 1015.1 | 5.1 |
| TUN | 31 | / | / | / | / | / | 1.7 | 13.8 | 1.0.E-02 | 4.5 | 1019.9 | 3.8 |
| TUR | 35 | 20.3 | 7.4 | 20.7 | 70.9 | 35.3 | 1.3 | 9.4 | 7.8.E-03 | 3.4 | 1021.8 | 2.8 |
| UKR | 32 | / | / | / | 44.3 | 19.1 | 1.2 | 6.3 | 4.9.E-03 | 3.0 | 1019.7 | 3.4 |
| URY | 33 | / | / | / | / | / | 3.2 | 22.9 | 1.0.E-02 | 6.4 | 1016.3 | 3.5 |
| UZB | 32 | / | / | / | 102.5 | / | 1.3 | 11.0 | 4.8.E-03 | 4.7 | 1020.7 | 1.6 |
| VEN | 38 | / | / | / | / | / | 3.2 | 25.4 | 1.1.E-02 | 6.2 | 1014.4 | 1.9 |
| VNM | 23 | / | / | / | 65.2 | / | 4.0 | 30.0 | 1.5.E-02 | 8.0 | 1013.1 | 3.5 |
| ZAF | 24 | 8.8 | 3.0 | 6.3 | 65.4 | 32.1 | 2.1 | 19.5 | 1.0.E-02 | 6.2 | 1019.2 | 1.8 |

Column explanations: **ONS** - epidemic onset; **$NO_2$** - ambient level of $NO_2$; **$SO_2$** - ambient level of $SO_2$; **CO** - ambient level of CO; **2.5** - ambient level of inhalable particles with 2.5 micrometers; **10** - ambient level of inhalable particles with 10 micrometers; **PC** - level of precipitation; **T** - temperature; **H** - specific humidity; **UV** - UV index; **P** - air pressure; **WS** - wind speed. Values for all air pollutants correspond to US EPA Air Quality Index.

**Table S6.** Pearson correlation coefficients (R) and P values.

| Factor | R | P | Factor | R | P |
|---|---|---|---|---|---|
| **Demographic** | | | **Medical** | | |
| HDI | 0.37 | 4.E-05 | CH | 0.41 | 5.E-06 |
| INS | 0.40 | 4.E-04 | ALC | 0.36 | 7.E-05 |
| UP | 0.27 | 3.E-03 | OB | 0.29 | 2.E-03 |
| BAP | 0.13 | ns | RBP | -0.25 | 7.E-03 |
| MA | 0.33 | 3.E-04 | SM | 0.19 | 4.E-02 |
| IM | -0.36 | 8.E-05 | CD | -0.30 | 1.E-03 |
| I-E | 0.22 | 2.E-02 | IN | 0.11 | ns |
| RE | 0.15 | ns | BCG | 0.01 | ns |
| | | | | | |
| **Meteorological** | | | **Blood groups** | | |
| ONS | -0.48 | 4.E-08 | A | 0.32 | 3.E-03 |
| NO2 | 0.37 | 1.E-02 | A+ | 0.26 | 2.E-02 |
| SO2 | 0.33 | 4.E-02 | A- | 0.40 | 2.E-04 |
| CO | 0.26 | 1.E-01 | B | -0.31 | 4.E-03 |
| 2.5 | -0.22 | ns | B+ | -0.34 | 2.E-03 |
| 10 | 0.07 | ns | B- | 0.17 | ns |
| PC | -0.31 | 6.E-04 | AB | -0.06 | ns |
| T | -0.39 | 2.E-05 | AB+ | -0.12 | ns |
| H | -0.35 | 1.E-04 | AB- | 0.31 | 4.E-03 |
| UV | -0.35 | 4.E-05 | O | -0.01 | ns |
| P | -0.01 | ns | O+ | -0.11 | ns |
| WS | -0.04 | ns | O- | 0.39 | 3.E-04 |
| | | | Rh+ | -0.40 | 2.E-04 |
| | | | Rh- | 0.40 | 2.E-04 |

| Country | Date | Precipitation | Temperature | Specific Humidity | UV Index | Air Pressure | Wind Speed |
|---|---|---|---|---|---|---|---|
| Afghanistan | 3/11/2020 | 0.84 | -1.09 | 3.91 | 2.00 | 1016.00 | 2.53 |
| Afghanistan | 3/12/2020 | 0.76 | -1.52 | 3.46 | 2.00 | 1021.00 | 2.84 |
| Afghanistan | 3/13/2020 | 0.52 | -0.28 | 3.14 | 3.00 | 1022.00 | 1.79 |
| Afghanistan | 3/14/2020 | 0.34 | -0.46 | 2.83 | 4.00 | 1024.00 | 2.83 |
| Afghanistan | 3/15/2020 | 0.40 | 2.43 | 3.03 | 4.00 | 1022.00 | 1.52 |
| Afghanistan | 3/16/2020 | 0.89 | 4.95 | 4.34 | 3.00 | 1022.00 | 1.09 |
| Afghanistan | 3/17/2020 | 1.12 | 4.36 | 5.31 | 4.00 | 1023.00 | 0.60 |
| Afghanistan | 3/18/2020 | 0.88 | 5.49 | 4.98 | 4.00 | 1023.00 | 2.11 |
| Afghanistan | 3/19/2020 | 0.80 | 6.68 | 4.48 | 4.00 | 1020.00 | 1.38 |
| Afghanistan | 3/20/2020 | 1.27 | 5.62 | 5.88 | 4.00 | 1016.00 | 1.38 |
| Afghanistan | 3/21/2020 | 0.96 | 4.17 | 4.91 | 3.00 | 1020.00 | 1.59 |
| Afghanistan | 3/22/2020 | 0.65 | 4.79 | 3.97 | 4.00 | 1021.00 | 1.39 |
| Albania | 3/2/2020 | 1.48 | 12.01 | 7.85 | 5.00 | 1014.00 | 4.22 |
| Albania | 3/3/2020 | 1.96 | 11.88 | 7.80 | 4.00 | 1010.00 | 4.08 |
| Albania | 3/4/2020 | 1.43 | 8.44 | 5.96 | 2.00 | 1010.00 | 1.28 |
| Albania | 3/5/2020 | 1.08 | 8.40 | 5.12 | 4.00 | 1011.00 | 1.97 |
| Albania | 3/6/2020 | 1.70 | 9.58 | 6.61 | 4.00 | 1011.00 | 4.68 |
| Albania | 3/7/2020 | 1.68 | 10.38 | 7.07 | 3.00 | 1009.00 | 1.66 |
| Albania | 3/8/2020 | 1.53 | 9.84 | 6.66 | 3.00 | 1016.00 | 2.29 |
| Albania | 3/9/2020 | 1.22 | 8.21 | 5.81 | 3.00 | 1015.00 | 1.13 |
| Albania | 3/10/2020 | 1.04 | 8.19 | 5.12 | 3.00 | 1018.00 | 2.27 |
| Albania | 3/11/2020 | 1.36 | 9.62 | 5.41 | 4.00 | 1023.00 | 1.49 |
| Albania | 3/12/2020 | 1.33 | 13.06 | 6.62 | 5.00 | 1023.00 | 1.42 |
| Albania | 3/13/2020 | 1.58 | 14.70 | 6.98 | 5.00 | 1017.00 | 1.05 |
| Albania | 3/14/2020 | 1.92 | 13.07 | 7.46 | 5.00 | 1015.00 | 1.90 |
| Andorra | 3/7/2020 | 0.78 | 0.83 | 4.17 | 1.00 | 1025.00 | 2.67 |
| Andorra | 3/8/2020 | 0.84 | 4.61 | 4.67 | 1.00 | 1025.00 | 0.73 |
| Andorra | 3/9/2020 | 0.99 | 1.88 | 4.56 | 2.00 | 1025.00 | 2.45 |
| Andorra | 3/10/2020 | 1.48 | 4.29 | 5.05 | 2.00 | 1025.00 | 1.19 |
| Andorra | 3/11/2020 | 1.22 | 8.49 | 5.73 | 4.00 | 1024.00 | 0.72 |
| Andorra | 3/12/2020 | 1.66 | 10.98 | 6.29 | 4.00 | 1022.00 | 0.69 |
| Andorra | 3/13/2020 | 1.57 | 4.55 | 5.48 | 2.00 | 1021.00 | 0.74 |
| Andorra | 3/14/2020 | 0.74 | 4.19 | 4.56 | 2.00 | 1021.00 | 0.83 |
| Andorra | 3/15/2020 | 0.98 | 6.82 | 5.55 | 2.00 | 1020.00 | 1.40 |
| Andorra | 3/16/2020 | 1.45 | 4.72 | 5.70 | 1.00 | 1022.00 | 1.62 |
| Andorra | 3/17/2020 | 1.55 | 7.12 | 6.21 | 2.00 | 1029.00 | 0.55 |
| Argentina | 2/24/2020 | 2.67 | 23.28 | 10.01 | 7.00 | 1013.00 | 2.89 |
| Argentina | 2/25/2020 | 1.97 | 21.31 | 7.44 | 6.00 | 1012.00 | 2.18 |
| Argentina | 2/26/2020 | 2.18 | 21.14 | 8.53 | 6.00 | 1016.00 | 1.80 |
| Argentina | 2/27/2020 | 2.22 | 22.60 | 8.26 | 7.00 | 1015.00 | 1.86 |
| Argentina | 2/28/2020 | 2.71 | 24.61 | 9.48 | 7.00 | 1017.00 | 1.47 |
| Argentina | 2/29/2020 | 2.84 | 25.83 | 11.00 | 7.00 | 1021.00 | 1.83 |
| Argentina | 3/1/2020 | 2.70 | 27.11 | 10.79 | 8.00 | 1020.00 | 1.77 |
| Argentina | 3/2/2020 | 2.73 | 27.08 | 11.39 | 8.00 | 1020.00 | 2.22 |
| Argentina | 3/3/2020 | 2.21 | 26.03 | 10.64 | 8.00 | 1018.00 | 1.98 |
| Argentina | 3/4/2020 | 2.20 | 26.22 | 10.81 | 8.00 | 1018.00 | 1.66 |
| Argentina | 3/5/2020 | 2.08 | 25.16 | 10.73 | 7.00 | 1020.00 | 2.29 |
| Argentina | 3/6/2020 | 2.14 | 24.92 | 10.29 | 7.00 | 1018.00 | 2.16 |
| Argentina | 3/7/2020 | 2.98 | 26.32 | 10.40 | 7.00 | 1013.00 | 2.29 |
| Argentina | 3/8/2020 | 4.00 | 26.77 | 11.46 | 8.00 | 1012.00 | 2.96 |
| Argentina | 3/9/2020 | 4.03 | 23.53 | 11.76 | 7.00 | 1019.00 | 3.01 |
| Argentina | 3/10/2020 | 3.65 | 22.49 | 10.26 | 5.00 | 1020.00 | 3.72 |
| Argentina | 3/11/2020 | 4.47 | 21.07 | 13.87 | 6.00 | 1016.00 | 2.75 |
| Argentina | 3/12/2020 | 3.01 | 23.54 | 14.25 | 7.00 | 1012.00 | 1.16 |
| Argentina | 3/13/2020 | 3.35 | 25.24 | 14.60 | 7.00 | 1014.00 | 2.51 |
| Argentina | 3/14/2020 | 4.16 | 20.68 | 13.12 | 7.00 | 1017.00 | 2.64 |
| Argentina | 3/15/2020 | 2.12 | 17.67 | 9.45 | 5.00 | 1019.00 | 2.24 |
| Argentina | 3/16/2020 | 1.91 | 18.45 | 9.89 | 6.00 | 1019.00 | 1.69 |
| Armenia | 3/7/2020 | 0.95 | 7.24 | 3.64 | 3.00 | 1025.00 | 1.64 |
| Armenia | 3/8/2020 | 1.04 | 7.40 | 4.24 | 4.00 | 1027.00 | 0.58 |
| Armenia | 3/9/2020 | 0.48 | 7.15 | 3.04 | 4.00 | 1028.00 | 1.84 |
| Armenia | 3/10/2020 | 0.45 | 7.26 | 2.26 | 4.00 | 1025.00 | 1.88 |
| Armenia | 3/11/2020 | 0.91 | 7.91 | 3.83 | 4.00 | 1023.00 | 1.57 |
| Armenia | 3/12/2020 | 0.94 | 8.47 | 4.14 | 4.00 | 1022.00 | 2.13 |
| Armenia | 3/13/2020 | 1.37 | 7.87 | 4.63 | 4.00 | 1021.00 | 1.19 |
| Armenia | 3/14/2020 | 1.32 | 7.03 | 5.07 | 4.00 | 1016.00 | 2.05 |
| Armenia | 3/15/2020 | 1.16 | 7.27 | 4.57 | 3.00 | 1015.00 | 1.85 |
| Armenia | 3/16/2020 | 1.22 | 4.87 | 4.51 | 4.00 | 1015.00 | 2.05 |
| Armenia | 3/17/2020 | 1.14 | 5.01 | 4.57 | 2.00 | 1018.00 | 2.15 |
| Armenia | 3/18/2020 | 1.18 | 2.33 | 4.21 | 2.00 | 1018.00 | 3.11 |
| Armenia | 3/19/2020 | 0.76 | 3.01 | 3.67 | 2.00 | 1014.00 | 0.89 |
| Armenia | 3/20/2020 | 0.79 | 3.70 | 3.41 | 3.00 | 1015.00 | 1.14 |
| Armenia | 3/21/2020 | 0.82 | 4.25 | 3.62 | 4.00 | 1017.00 | 1.35 |
| Armenia | 3/22/2020 | 0.72 | 3.85 | 3.50 | 3.00 | 1019.00 | 1.20 |
| Armenia | 3/23/2020 | 0.54 | 5.81 | 3.11 | 4.00 | 1019.00 | 1.08 |
| Aruba | 3/18/2020 | 3.15 | 26.77 | 17.08 | 6.00 | 1016.00 | 6.59 |
| Aruba | 3/19/2020 | 2.91 | 27.12 | 16.99 | 6.00 | 1016.00 | 7.50 |
| Aruba | 3/20/2020 | 3.26 | 26.95 | 17.25 | 6.00 | 1017.00 | 7.21 |
| Aruba | 3/21/2020 | 4.04 | 26.74 | 16.90 | 6.00 | 1017.00 | 6.87 |
| Aruba | 3/22/2020 | 4.10 | 26.79 | 17.03 | 7.00 | 1016.00 | 7.01 |
| Aruba | 3/23/2020 | 3.82 | 27.01 | 17.54 | 6.00 | 1017.00 | 7.55 |
| Aruba | 3/24/2020 | 3.31 | 26.77 | 17.16 | 7.00 | 1016.00 | 7.32 |
| Aruba | 3/25/2020 | 3.00 | 26.83 | 17.03 | 6.00 | 1014.00 | 6.58 |
| Aruba | 3/26/2020 | 3.22 | 26.80 | 17.08 | 7.00 | 1013.00 | 6.90 |
| Aruba | 3/27/2020 | 2.89 | 26.62 | 16.75 | 6.00 | 1012.00 | 6.77 |
| Aruba | 3/28/2020 | 2.89 | 26.63 | 17.16 | 7.00 | 1013.00 | 7.38 |
| Aruba | 3/29/2020 | 2.81 | 27.07 | 17.31 | 6.00 | 1015.00 | 8.45 |
| Aruba | 3/30/2020 | 2.95 | 26.90 | 17.00 | 6.00 | 1014.00 | 8.16 |
| Aruba | 3/31/2020 | 2.42 | 26.79 | 15.56 | 6.00 | 1012.00 | 7.50 |
| Australia | 2/28/2020 | 2.06 | 19.92 | 11.26 | 5.00 | 1017.00 | 2.47 |
| Australia | 2/29/2020 | 2.51 | 21.32 | 11.85 | 5.00 | 1018.00 | 1.77 |
| Australia | 3/1/2020 | 3.32 | 22.90 | 15.00 | 7.00 | 1015.00 | 1.93 |
| Australia | 3/2/2020 | 2.70 | 23.09 | 14.16 | 7.00 | 1017.00 | 2.36 |
| Australia | 3/3/2020 | 3.97 | 20.38 | 12.10 | 5.00 | 1021.00 | 2.16 |
| Australia | 3/4/2020 | 4.42 | 21.50 | 13.81 | 5.00 | 1020.00 | 2.62 |
| Australia | 3/5/2020 | 4.83 | 22.01 | 15.47 | 5.00 | 1015.00 | 3.41 |
| Australia | 3/6/2020 | 3.53 | 22.13 | 14.60 | 5.00 | 1015.00 | 2.65 |
| Australia | 3/7/2020 | 2.89 | 20.22 | 12.10 | 5.00 | 1021.00 | 3.00 |
| Australia | 3/8/2020 | 2.82 | 19.10 | 10.92 | 5.00 | 1021.00 | 2.33 |
| Australia | 3/9/2020 | 2.66 | 18.73 | 10.82 | 4.00 | 1022.00 | 2.25 |
| Australia | 3/10/2020 | 2.11 | 19.27 | 9.94 | 5.00 | 1022.00 | 1.65 |
| Australia | 3/11/2020 | 1.94 | 20.07 | 10.41 | 6.00 | 1025.00 | 1.74 |
| Australia | 3/12/2020 | 1.85 | 20.29 | 10.57 | 6.00 | 1026.00 | 1.92 |
| Australia | 3/13/2020 | 1.70 | 20.02 | 10.63 | 6.00 | 1023.00 | 1.71 |
| Australia | 3/14/2020 | 2.86 | 17.38 | 10.30 | 5.00 | 1020.00 | 3.96 |
| Austria | 2/16/2020 | 1.05 | 5.29 | 4.65 | 3.00 | 1028.00 | 3.71 |
| Austria | 2/17/2020 | 1.79 | 6.98 | 5.40 | 3.00 | 1022.00 | 1.17 |
| Austria | 2/18/2020 | 0.99 | 5.94 | 4.50 | 2.00 | 1027.00 | 3.79 |
| Austria | 2/19/2020 | 0.89 | 3.31 | 4.00 | 3.00 | 1027.00 | 3.09 |
| Austria | 2/20/2020 | 0.82 | 3.25 | 3.54 | 2.00 | 1026.00 | 2.83 |
| Austria | 2/21/2020 | 0.78 | 2.92 | 3.50 | 3.00 | 1031.00 | 4.54 |
| Austria | 2/22/2020 | 1.11 | 5.14 | 3.71 | 3.00 | 1031.00 | 2.17 |
| Austria | 2/23/2020 | 2.00 | 10.03 | 5.56 | 3.00 | 1025.00 | 5.49 |
| Austria | 2/24/2020 | 1.18 | 6.41 | 4.15 | 3.00 | 1024.00 | 4.58 |
| Austria | 2/25/2020 | 1.29 | 8.85 | 5.89 | 4.00 | 1014.00 | 2.64 |
| Austria | 2/26/2020 | 0.94 | 4.32 | 4.37 | 2.00 | 999.00 | 3.00 |
| Austria | 2/27/2020 | 0.72 | 2.81 | 3.41 | 3.00 | 998.00 | 3.95 |
| Austria | 2/28/2020 | 0.72 | 1.89 | 3.42 | 2.00 | 999.00 | 5.16 |
| Austria | 2/29/2020 | 1.28 | 3.48 | 3.74 | 3.00 | 1020.00 | 3.05 |
| Austria | 3/1/2020 | 1.55 | 6.46 | 5.37 | 3.00 | 1008.00 | 2.62 |
| Austria | 3/2/2020 | 1.14 | 7.99 | 5.04 | 2.00 | 999.00 | 3.60 |
| Austria | 3/3/2020 | 1.41 | 6.35 | 5.09 | 2.00 | 999.00 | 3.79 |
| Austria | 3/4/2020 | 0.81 | 3.71 | 3.59 | 3.00 | 1014.00 | 3.09 |
| Austria | 3/5/2020 | 0.92 | 4.10 | 3.71 | 3.00 | 1014.00 | 3.88 |
| Austria | 3/6/2020 | 1.42 | 4.76 | 4.79 | 3.00 | 999.00 | 2.31 |
| Austria | 3/7/2020 | 0.95 | 4.07 | 4.10 | 3.00 | 1021.00 | 4.02 |
| Austria | 3/8/2020 | 0.65 | 3.40 | 3.40 | 3.00 | 1023.00 | 2.20 |
| Azerbaijan | 3/1/2020 | 1.69 | 7.73 | 5.50 | 3.00 | 1021.00 | 5.76 |
| Azerbaijan | 3/2/2020 | 1.80 | 6.57 | 5.25 | 3.00 | 1023.00 | 7.86 |
| Azerbaijan | 3/3/2020 | 1.56 | 7.90 | 5.40 | 2.00 | 1025.00 | 2.40 |
| Azerbaijan | 3/4/2020 | 1.37 | 8.49 | 6.00 | 4.00 | 1025.00 | 3.06 |
| Azerbaijan | 3/5/2020 | 1.15 | 9.01 | 6.39 | 4.00 | 1021.00 | 3.91 |
| Azerbaijan | 3/6/2020 | 1.03 | 9.41 | 5.98 | 4.00 | 1021.00 | 2.66 |
| Azerbaijan | 3/7/2020 | 0.83 | 8.54 | 5.29 | 3.00 | 1027.00 | 4.37 |
| Azerbaijan | 3/8/2020 | 0.46 | 6.71 | 4.23 | 3.00 | 1029.00 | 3.33 |
| Azerbaijan | 3/9/2020 | 0.43 | 7.32 | 4.82 | 3.00 | 1029.00 | 3.52 |
| Azerbaijan | 3/10/2020 | 0.57 | 8.23 | 5.86 | 4.00 | 1024.00 | 4.69 |
| Azerbaijan | 3/11/2020 | 1.09 | 9.86 | 6.13 | 4.00 | 1019.00 | 3.72 |
| Azerbaijan | 3/12/2020 | 1.60 | 10.44 | 6.21 | 4.00 | 1018.00 | 1.98 |
| Azerbaijan | 3/13/2020 | 1.94 | 10.28 | 6.21 | 4.00 | 1023.00 | 3.24 |
| Azerbaijan | 3/14/2020 | 2.28 | 11.33 | 6.88 | 4.00 | 1015.00 | 3.32 |
| Azerbaijan | 3/15/2020 | 1.99 | 10.84 | 6.34 | 4.00 | 1016.00 | 4.18 |
| Azerbaijan | 3/16/2020 | 2.08 | 8.08 | 5.51 | 3.00 | 1025.00 | 7.42 |
| Azerbaijan | 3/17/2020 | 1.74 | 5.74 | 4.47 | 2.00 | 1030.00 | 7.53 |
| Azerbaijan | 3/18/2020 | 2.13 | 6.22 | 4.99 | 2.00 | 1028.00 | 5.45 |
| Azerbaijan | 3/19/2020 | 1.76 | 7.30 | 5.53 | 2.00 | 1019.00 | 2.87 |
| Azerbaijan | 3/20/2020 | 1.53 | 7.68 | 5.31 | 2.00 | 1018.00 | 2.35 |
| Azerbaijan | 3/21/2020 | 1.63 | 8.03 | 5.30 | 2.00 | 1022.00 | 2.40 |
| Azerbaijan | 3/22/2020 | 1.80 | 8.45 | 5.77 | 4.00 | 1021.00 | 1.96 |
| Bahamas | 3/18/2020 | 2.75 | 23.92 | 14.41 | 6.00 | 1024.00 | 5.13 |
| Bahamas | 3/19/2020 | 2.51 | 24.06 | 14.23 | 5.00 | 1026.00 | 7.08 |
| Bahamas | 3/20/2020 | 2.44 | 24.17 | 14.35 | 6.00 | 1026.00 | 6.69 |
| Bahamas | 3/21/2020 | 2.26 | 24.28 | 13.81 | 6.00 | 1025.00 | 5.32 |
| Bahamas | 3/22/2020 | 2.47 | 24.35 | 13.79 | 6.00 | 1023.00 | 5.32 |
| Bahamas | 3/23/2020 | 2.57 | 24.38 | 14.10 | 5.00 | 1024.00 | 4.85 |
| Bahamas | 3/24/2020 | 2.52 | 24.25 | 14.21 | 6.00 | 1022.00 | 3.82 |
| Bahamas | 3/25/2020 | 2.72 | 24.86 | 15.06 | 6.00 | 1018.00 | 3.45 |
| Bahamas | 3/26/2020 | 2.96 | 25.11 | 15.93 | 5.00 | 1017.00 | 3.37 |
| Bahamas | 3/27/2020 | 2.04 | 23.75 | 14.14 | 6.00 | 1018.00 | 5.13 |
| Bahamas | 3/28/2020 | 2.41 | 24.11 | 14.60 | 6.00 | 1020.00 | 5.55 |
| Bahrain | 3/3/2020 | 0.92 | 19.27 | 8.68 | 5.00 | 1020.00 | 7.32 |
| Bahrain | 3/4/2020 | 0.94 | 19.16 | 8.63 | 5.00 | 1019.00 | 6.49 |
| Bahrain | 3/5/2020 | 1.00 | 19.44 | 10.13 | 5.00 | 1017.00 | 5.12 |
| Bahrain | 3/6/2020 | 0.79 | 19.72 | 9.42 | 6.00 | 1017.00 | 2.33 |
| Bahrain | 3/7/2020 | 1.18 | 20.19 | 10.34 | 6.00 | 1015.00 | 4.06 |
| Bahrain | 3/8/2020 | 1.58 | 20.50 | 9.98 | 6.00 | 1016.00 | 2.41 |
| Bahrain | 3/9/2020 | 1.49 | 20.01 | 10.52 | 5.00 | 1016.00 | 5.42 |
| Bahrain | 3/10/2020 | 0.97 | 20.22 | 10.27 | 6.00 | 1018.00 | 3.00 |
| Bahrain | 3/11/2020 | 0.78 | 20.46 | 9.85 | 6.00 | 1019.00 | 1.47 |
| Bahrain | 3/12/2020 | 0.85 | 20.57 | 9.67 | 6.00 | 1018.00 | 2.17 |
| Bahrain | 3/13/2020 | 1.21 | 21.36 | 10.34 | 6.00 | 1018.00 | 3.61 |
| Bahrain | 3/14/2020 | 2.53 | 22.11 | 12.16 | 6.00 | 1015.00 | 3.27 |
| Bahrain | 3/15/2020 | 2.34 | 22.62 | 11.44 | 6.00 | 1011.00 | 4.55 |
| Bahrain | 3/16/2020 | 2.31 | 21.94 | 13.99 | 6.00 | 1014.00 | 1.68 |
| Bahrain | 3/17/2020 | 1.90 | 22.47 | 13.40 | 6.00 | 1016.00 | 2.61 |
| Bahrain | 3/18/2020 | 2.76 | 23.11 | 13.63 | 6.00 | 1013.00 | 5.79 |
| Bahrain | 3/19/2020 | 2.61 | 22.67 | 14.50 | 5.00 | 1013.00 | 3.74 |
| Bahrain | 3/20/2020 | 2.15 | 22.33 | 12.37 | 6.00 | 1013.00 | 3.06 |
| Bahrain | 3/21/2020 | 3.52 | 23.23 | 13.93 | 6.00 | 1009.00 | 4.31 |
| Bahrain | 3/22/2020 | 2.06 | 22.09 | 12.38 | 6.00 | 1012.00 | 6.02 |
| Bahrain | 3/23/2020 | 1.16 | 20.03 | 8.32 | 5.00 | 1015.00 | 6.52 |
| Bahrain | 3/24/2020 | 1.30 | 20.49 | 8.67 | 6.00 | 1016.00 | 3.84 |
| Bahrain | 3/25/2020 | 1.66 | 21.36 | 8.50 | 5.00 | 1014.00 | 2.23 |
| Bahrain | 3/26/2020 | 2.11 | 21.50 | 10.99 | 6.00 | 1014.00 | 4.52 |
| Bahrain | 3/27/2020 | 2.21 | 22.66 | 12.80 | 6.00 | 1014.00 | 6.54 |
| Bahrain | 3/28/2020 | 3.21 | 22.50 | 13.35 | 6.00 | 1011.00 | 6.90 |
| Bahrain | 3/29/2020 | 2.60 | 23.08 | 13.25 | 5.00 | 1009.00 | 6.34 |
| Bahrain | 3/30/2020 | 1.34 | 21.91 | 10.59 | 6.00 | 1014.00 | 6.36 |
| Bahrain | 3/31/2020 | 1.42 | 22.09 | 10.43 | 6.00 | 1017.00 | 3.88 |
| Bangladesh | 3/24/2020 | 2.65 | 26.09 | 10.84 | 8.00 | 1013.00 | 0.62 |
| Bangladesh | 3/25/2020 | 2.51 | 26.74 | 11.14 | 8.00 | 1011.00 | 0.81 |
| Bangladesh | 3/26/2020 | 2.71 | 27.92 | 11.28 | 8.00 | 1011.00 | 1.05 |
| Bangladesh | 3/27/2020 | 2.80 | 28.66 | 11.57 | 9.00 | 1011.00 | 1.54 |
| Bangladesh | 3/28/2020 | 3.39 | 29.35 | 12.55 | 8.00 | 1010.00 | 1.83 |
| Bangladesh | 3/29/2020 | 2.29 | 29.23 | 11.63 | 8.00 | 1010.00 | 2.06 |
| Bangladesh | 3/30/2020 | 1.67 | 28.82 | 9.05 | 8.00 | 1010.00 | 1.81 |
| Bangladesh | 3/31/2020 | 1.62 | 28.55 | 8.68 | 9.00 | 1011.00 | 1.32 |
| Bangladesh | 4/1/2020 | 2.06 | 29.20 | 9.35 | 9.00 | 1010.00 | 1.48 |
| Bangladesh | 4/2/2020 | 3.33 | 29.59 | 14.52 | 8.00 | 1009.00 | 2.37 |
| Barbados | 3/10/2020 | 3.09 | 26.50 | 15.68 | 7.00 | 1018.00 | 6.68 |
| Barbados | 3/11/2020 | 3.09 | 26.43 | 15.61 | 7.00 | 1015.00 | 5.99 |
| Barbados | 3/12/2020 | 3.28 | 26.18 | 16.26 | 6.00 | 1013.00 | 6.36 |
| Barbados | 3/13/2020 | 3.53 | 26.30 | 16.32 | 6.00 | 1015.00 | 6.17 |
| Barbados | 3/14/2020 | 4.01 | 26.12 | 16.98 | 6.00 | 1015.00 | 5.59 |
| Barbados | 3/15/2020 | 3.77 | 25.89 | 16.54 | 7.00 | 1015.00 | 4.40 |
| Barbados | 3/16/2020 | 3.25 | 26.09 | 16.07 | 7.00 | 1015.00 | 3.57 |
| Barbados | 3/17/2020 | 3.04 | 26.19 | 16.38 | 6.00 | 1015.00 | 4.54 |
| Barbados | 3/18/2020 | 2.85 | 26.50 | 16.50 | 7.00 | 1017.00 | 5.40 |
| Barbados | 3/19/2020 | 3.07 | 26.47 | 16.86 | 7.00 | 1017.00 | 6.39 |
| Barbados | 3/20/2020 | 3.31 | 26.35 | 16.75 | 7.00 | 1016.00 | 6.79 |
| Barbados | 3/21/2020 | 3.68 | 26.49 | 16.80 | 6.00 | 1016.00 | 6.13 |
| Barbados | 3/22/2020 | 3.77 | 26.47 | 16.83 | 7.00 | 1017.00 | 6.91 |
| Barbados | 3/23/2020 | 3.62 | 26.60 | 16.09 | 7.00 | 1018.00 | 6.69 |
| Barbados | 3/24/2020 | 3.80 | 26.38 | 16.16 | 7.00 | 1018.00 | 6.94 |
| Barbados | 3/25/2020 | 3.28 | 26.33 | 15.87 | 7.00 | 1015.00 | 6.90 |
| Barbados | 3/26/2020 | 3.29 | 26.11 | 15.82 | 7.00 | 1014.00 | 5.68 |
| Belarus | 3/19/2020 | 1.42 | 6.11 | 4.98 | 2.00 | 1021.00 | 4.14 |
| Belarus | 3/20/2020 | 0.63 | 2.68 | 3.63 | 3.00 | 1023.00 | 2.34 |
| Belarus | 3/21/2020 | 0.65 | 0.15 | 3.23 | 2.00 | 1024.00 | 4.34 |
| Belarus | 3/22/2020 | 0.45 | -2.16 | 2.29 | 1.00 | 1037.00 | 5.89 |
| Belarus | 3/23/2020 | 0.36 | -3.15 | 2.20 | 2.00 | 1040.00 | 2.94 |
| Belarus | 3/24/2020 | 0.36 | -1.13 | 2.29 | 2.00 | 1041.00 | 1.52 |
| Belarus | 3/25/2020 | 0.49 | 1.14 | 2.56 | 2.00 | 1040.00 | 1.11 |
| Belarus | 3/26/2020 | 0.44 | 2.64 | 2.89 | 3.00 | 1038.00 | 1.01 |
| Belarus | 3/27/2020 | 0.80 | 3.28 | 3.83 | 3.00 | 1035.00 | 1.78 |
| Belarus | 3/28/2020 | 1.01 | 6.30 | 4.83 | 4.00 | 1026.00 | 1.52 |
| Belarus | 3/29/2020 | 1.21 | 5.53 | 4.68 | 3.00 | 1013.00 | 3.26 |
| Belarus | 3/30/2020 | 0.68 | -1.96 | 2.45 | 1.00 | 1019.00 | 5.17 |
| Belarus | 3/31/2020 | 0.40 | -1.74 | 2.19 | 1.00 | 1023.00 | 2.89 |
| Belgium | 2/25/2020 | 1.01 | 6.67 | 5.23 | 2.00 | 999.00 | 5.39 |
| Belgium | 2/26/2020 | 0.92 | 3.22 | 4.24 | 1.00 | 999.00 | 5.24 |
| Belgium | 2/27/2020 | 1.00 | 2.13 | 4.17 | 1.00 | 998.00 | 4.15 |
| Belgium | 2/28/2020 | 1.18 | 3.52 | 4.45 | 1.00 | 1017.00 | 5.51 |
| Belgium | 2/29/2020 | 1.16 | 7.65 | 5.48 | 2.00 | 999.00 | 7.58 |
| Belgium | 3/1/2020 | 1.06 | 5.75 | 4.91 | 2.00 | 995.00 | 5.42 |
| Belgium | 3/2/2020 | 1.25 | 4.22 | 4.93 | 2.00 | 996.00 | 3.73 |
| Belgium | 3/3/2020 | 0.96 | 3.53 | 4.51 | 2.00 | 999.00 | 3.39 |
| Belgium | 3/4/2020 | 1.10 | 4.59 | 4.72 | 2.00 | 1011.00 | 3.23 |
| Belgium | 3/5/2020 | 1.70 | 5.36 | 5.44 | 2.00 | 999.00 | 3.24 |
| Belgium | 3/6/2020 | 1.25 | 4.93 | 4.99 | 2.00 | 999.00 | 4.15 |
| Belgium | 3/7/2020 | 1.21 | 5.57 | 4.51 | 2.00 | 1019.00 | 3.40 |
| Belgium | 3/8/2020 | 1.36 | 7.23 | 5.84 | 2.00 | 1017.00 | 5.44 |
| Belgium | 3/9/2020 | 1.23 | 6.89 | 5.18 | 2.00 | 1015.00 | 4.34 |
| Belgium | 3/10/2020 | 2.58 | 10.13 | 7.30 | 4.00 | 1013.00 | 7.13 |
| Belgium | 3/11/2020 | 2.08 | 12.22 | 7.87 | 3.00 | 1013.00 | 5.98 |
| Belgium | 3/12/2020 | 1.11 | 8.42 | 5.13 | 2.00 | 1014.00 | 6.66 |
| Belgium | 3/13/2020 | 1.00 | 6.49 | 4.82 | 3.00 | 1021.00 | 4.39 |
| Belgium | 3/14/2020 | 1.35 | 6.67 | 5.12 | 2.00 | 1021.00 | 3.68 |
| Belgium | 3/15/2020 | 1.34 | 8.91 | 6.03 | 2.00 | 1017.00 | 4.29 |
| Belgium | 3/16/2020 | 1.59 | 8.39 | 5.88 | 3.00 | 1025.00 | 1.18 |
| Belgium | 3/17/2020 | 0.96 | 6.74 | 4.52 | 2.00 | 1030.00 | 2.31 |
| Bolivia | 3/19/2020 | 1.44 | 9.38 | 9.27 | 4.00 | 1018.00 | 1.99 |
| Bolivia | 3/20/2020 | 1.42 | 9.02 | 9.20 | 2.00 | 1017.00 | 2.25 |
| Bolivia | 3/21/2020 | 1.59 | 7.90 | 9.00 | 2.00 | 1017.00 | 2.24 |
| Bolivia | 3/22/2020 | 1.44 | 9.02 | 8.42 | 3.00 | 1018.00 | 2.55 |
| Bolivia | 3/23/2020 | 1.24 | 9.03 | 8.37 | 3.00 | 1021.00 | 2.56 |
| Bolivia | 3/24/2020 | 1.56 | 8.19 | 8.77 | 4.00 | 1020.00 | 1.59 |
| Bolivia | 3/25/2020 | 1.13 | 9.91 | 8.55 | 4.00 | 1017.00 | 1.96 |
| Bolivia | 3/26/2020 | 1.20 | 9.80 | 8.85 | 3.00 | 1015.00 | 2.33 |
| Bolivia | 3/27/2020 | 1.57 | 7.50 | 9.06 | 3.00 | 1016.00 | 2.21 |
| Bolivia | 3/28/2020 | 1.28 | 9.02 | 8.65 | 4.00 | 1016.00 | 2.43 |
| Bolivia | 3/29/2020 | 1.41 | 9.27 | 9.25 | 4.00 | 1019.00 | 3.00 |
| Bolivia | 3/30/2020 | 1.25 | 8.79 | 8.75 | 4.00 | 1021.00 | 2.75 |
| Bolivia | 3/31/2020 | 1.41 | 8.69 | 9.02 | 4.00 | 1021.00 | 2.79 |
| Bolivia | 4/1/2020 | 1.14 | 8.83 | 8.27 | 4.00 | 1021.00 | 2.22 |
| Bolivia | 4/2/2020 | 1.03 | 8.96 | 8.11 | 4.00 | 1023.00 | 2.53 |
| Bolivia | 4/3/2020 | 1.34 | 8.42 | 7.97 | 4.00 | 1023.00 | 2.24 |
| Bolivia | 4/4/2020 | 1.46 | 8.89 | 8.47 | 3.00 | 1019.00 | 1.71 |
| Bolivia | 4/5/2020 | 1.47 | 7.96 | 7.55 | 3.00 | 1019.00 | 1.72 |
| Bolivia | 4/6/2020 | 1.33 | 9.19 | 8.02 | 3.00 | 1020.00 | 2.17 |
| Bolivia | 4/7/2020 | 1.59 | 5.81 | 7.84 | 3.00 | 1022.00 | 1.56 |
| Bolivia | 4/8/2020 | 1.04 | 8.78 | 7.54 | 3.00 | 1022.00 | 0.95 |
| Bolivia | 4/9/2020 | 0.73 | 8.59 | 6.40 | 3.00 | 1021.00 | 1.09 |
| Bolivia | 4/10/2020 | 0.56 | 7.69 | 5.31 | 4.00 | 1021.00 | 1.18 |
| Bolivia | 4/11/2020 | 0.48 | 7.06 | 4.73 | 3.00 | 1020.00 | 1.35 |
| Bolivia | 4/12/2020 | 0.37 | 6.56 | 4.38 | 4.00 | 1020.00 | 1.39 |
| Bolivia | 4/13/2020 | 0.34 | 6.77 | 4.19 | 4.00 | 1019.00 | 1.69 |
| Bolivia | 4/14/2020 | 0.46 | 6.65 | 4.27 | 4.00 | 1019.00 | 1.14 |
| Bolivia | 4/15/2020 | 0.72 | 6.36 | 5.25 | 3.00 | 1020.00 | 1.79 |
| Bolivia | 4/16/2020 | 0.95 | 6.30 | 6.43 | 3.00 | 1023.00 | 1.87 |
| Bolivia | 4/17/2020 | 1.12 | 5.94 | 6.80 | 2.00 | 1026.00 | 2.22 |
| Bolivia | 4/18/2020 | 1.12 | 6.33 | 6.91 | 2.00 | 1025.00 | 1.94 |
| Bolivia | 4/19/2020 | 1.06 | 6.64 | 6.61 | 3.00 | 1024.00 | 1.69 |
| Bolivia | 4/20/2020 | 1.01 | 7.30 | 6.58 | 3.00 | 1023.00 | 1.74 |
| Bolivia | 4/21/2020 | 1.12 | 7.46 | 7.20 | 2.00 | 1023.00 | 1.43 |
| Bolivia | 4/22/2020 | 0.91 | 7.97 | 7.40 | 3.00 | 1022.00 | 1.45 |
| Bolivia | 4/23/2020 | 1.01 | 8.70 | 7.83 | 4.00 | 1020.00 | 1.51 |
| Bolivia | 4/24/2020 | 0.95 | 9.15 | 7.78 | 4.00 | 1018.00 | 1.59 |
| Bolivia | 4/25/2020 | 0.96 | 9.05 | 7.40 | 4.00 | 1019.00 | 1.63 |
| Bolivia | 4/26/2020 | 1.19 | 8.35 | 8.29 | 3.00 | 1022.00 | 1.93 |
| Bolivia | 4/27/2020 | 1.18 | 8.10 | 7.87 | 3.00 | 1024.00 | 2.32 |
| Bolivia | 4/28/2020 | 1.13 | 9.65 | 8.29 | 4.00 | 1022.00 | 1.50 |
| Bolivia | 4/29/2020 | 1.22 | 8.91 | 8.81 | 3.00 | 1021.00 | 2.36 |
| Bolivia | 4/30/2020 | 1.08 | 9.36 | 8.63 | 4.00 | 1023.00 | 2.15 |
| Bolivia | 5/1/2020 | 1.05 | 9.17 | 8.22 | 4.00 | 1024.00 | 2.23 |
| Bolivia | 5/2/2020 | 1.24 | 8.11 | 8.42 | 3.00 | 1024.00 | 1.50 |
| Bosnia and Herzegovina | 3/10/2020 | 0.82 | 2.88 | 4.08 | 1.60 | 1019.70 | 1.04 |
| Bosnia and Herzegovina | 3/11/2020 | 1.45 | 6.73 | 5.15 | 2.30 | 1022.70 | 0.47 |
| Bosnia and Herzegovina | 3/12/2020 | 1.14 | 9.58 | 5.90 | 4.30 | 1022.70 | 0.90 |
| Bosnia and Herzegovina | 3/13/2020 | 1.67 | 8.82 | 5.87 | 4.00 | 1019.30 | 0.91 |
| Bosnia and Herzegovina | 3/14/2020 | 1.21 | 3.68 | 4.26 | 2.00 | 1020.30 | 1.31 |
| Bosnia and Herzegovina | 3/15/2020 | 0.70 | 1.06 | 3.30 | 1.60 | 1026.30 | 1.66 |
| Bosnia and Herzegovina | 3/16/2020 | 0.43 | 4.32 | 3.60 | 3.70 | 1028.70 | 0.71 |
| Bosnia and Herzegovina | 3/17/2020 | 0.65 | 6.37 | 3.97 | 4.00 | 1030.00 | 0.69 |
| Bosnia and Herzegovina | 3/18/2020 | 1.00 | 6.32 | 4.27 | 4.00 | 1030.00 | 1.11 |
| Bosnia and Herzegovina | 3/19/2020 | 0.63 | 8.17 | 3.86 | 4.00 | 1027.70 | 0.58 |
| Bosnia and Herzegovina | 3/20/2020 | 0.86 | 9.70 | 4.65 | 4.00 | 1024.70 | 0.58 |
| Bosnia and Herzegovina | 3/21/2020 | 0.93 | 8.23 | 4.99 | 4.30 | 1019.00 | 0.71 |
| Bosnia and Herzegovina | 3/22/2020 | 0.91 | 1.72 | 3.76 | 2.00 | 1023.90 | 1.65 |
| Bosnia and Herzegovina | 3/23/2020 | 0.38 | -4.01 | 2.01 | 1.00 | 1025.60 | 2.15 |
| Brazil | 2/23/2020 | 3.10 | 20.31 | 12.58 | 5.34 | 1019.00 | 2.68 |
| Brazil | 2/24/2020 | 4.35 | 22.02 | 14.51 | 6.00 | 1017.34 | 1.90 |
| Brazil | 2/25/2020 | 4.99 | 24.24 | 17.06 | 7.00 | 1014.34 | 1.57 |
| Brazil | 2/26/2020 | 5.26 | 24.82 | 17.94 | 6.34 | 1010.34 | 2.98 |
| Brazil | 2/27/2020 | 4.11 | 20.68 | 13.91 | 5.00 | 1016.00 | 3.34 |
| Brazil | 2/28/2020 | 3.60 | 20.36 | 13.59 | 5.00 | 1018.00 | 2.85 |
| Brazil | 2/29/2020 | 4.26 | 21.36 | 14.92 | 5.34 | 1018.00 | 3.19 |
| Brazil | 3/1/2020 | 4.06 | 20.94 | 14.58 | 5.00 | 1017.66 | 4.04 |
| Brazil | 3/2/2020 | 4.30 | 20.87 | 14.66 | 5.00 | 1017.34 | 3.52 |
| Brazil | 3/3/2020 | 4.02 | 21.41 | 14.50 | 5.34 | 1015.66 | 1.62 |
| Brazil | 3/4/2020 | 3.60 | 21.26 | 14.17 | 5.34 | 1015.66 | 1.89 |
| Brazil | 3/5/2020 | 2.76 | 20.54 | 13.12 | 6.00 | 1017.66 | 2.36 |
| Brazil | 3/6/2020 | 2.19 | 19.95 | 12.46 | 6.00 | 1017.66 | 2.26 |
| Brazil | 3/7/2020 | 2.40 | 20.60 | 12.87 | 6.00 | 1016.00 | 2.08 |
| Brazil | 3/8/2020 | 3.31 | 21.10 | 13.76 | 6.34 | 1016.00 | 1.97 |
| Brazil | 3/9/2020 | 3.51 | 21.94 | 14.66 | 6.66 | 1018.00 | 2.46 |
| Brazil | 3/10/2020 | 3.04 | 22.28 | 14.73 | 6.66 | 1018.00 | 2.08 |
| Brazil | 3/11/2020 | 2.82 | 22.35 | 14.67 | 6.66 | 1017.00 | 2.30 |
| Brunei | 3/2/2020 | 6.24 | 26.42 | 19.78 | 6.00 | 1011.00 | 1.86 |
| Brunei | 3/3/2020 | 6.73 | 26.43 | 20.07 | 6.00 | 1011.00 | 1.82 |
| Brunei | 3/4/2020 | 6.68 | 25.57 | 19.26 | 6.00 | 1012.00 | 1.48 |
| Brunei | 3/5/2020 | 6.13 | 25.27 | 18.42 | 6.00 | 1013.00 | 1.11 |
| Brunei | 3/6/2020 | 5.23 | 25.95 | 17.95 | 6.00 | 1014.00 | 1.16 |
| Brunei | 3/7/2020 | 5.04 | 25.97 | 18.26 | 7.00 | 1014.00 | 1.37 |
| Brunei | 3/8/2020 | 4.27 | 26.28 | 18.00 | 6.00 | 1013.00 | 1.13 |
| Brunei | 3/9/2020 | 4.94 | 26.10 | 18.09 | 7.00 | 1011.00 | 1.06 |
| Brunei | 3/10/2020 | 5.80 | 26.27 | 18.30 | 6.00 | 1012.00 | 1.31 |
| Brunei | 3/11/2020 | 5.47 | 26.55 | 18.68 | 6.00 | 1012.00 | 1.58 |
| Brunei | 3/12/2020 | 5.35 | 27.58 | 20.17 | 6.00 | 1013.00 | 2.01 |
| Brunei | 3/13/2020 | 4.99 | 27.59 | 20.02 | 7.00 | 1013.00 | 1.08 |
| Bulgaria | 3/1/2020 | 1.02 | 2.96 | 3.54 | 2.00 | 1020.00 | 2.04 |
| Bulgaria | 3/2/2020 | 1.03 | 7.08 | 5.24 | 4.00 | 1015.00 | 3.09 |
| Bulgaria | 3/3/2020 | 1.22 | 9.97 | 5.95 | 4.00 | 1010.00 | 3.18 |
| Bulgaria | 3/4/2020 | 1.48 | 7.47 | 6.06 | 4.00 | 1007.00 | 2.19 |
| Bulgaria | 3/5/2020 | 0.98 | 2.13 | 4.02 | 1.00 | 1010.00 | 2.08 |
| Bulgaria | 3/6/2020 | 0.92 | 2.40 | 3.69 | 3.00 | 1009.00 | 1.90 |
| Bulgaria | 3/7/2020 | 1.39 | 5.93 | 5.49 | 3.00 | 1012.00 | 1.56 |
| Bulgaria | 3/8/2020 | 1.43 | 4.94 | 5.32 | 2.00 | 1016.00 | 1.13 |
| Bulgaria | 3/9/2020 | 1.46 | 1.28 | 4.23 | 1.00 | 1016.00 | 1.44 |
| Bulgaria | 3/10/2020 | 1.33 | 0.66 | 4.10 | 1.00 | 1017.00 | 1.89 |
| Canada | 2/15/2020 | 0.56 | -7.96 | 1.83 | 2.00 | 1035.00 | 6.04 |
| Canada | 2/16/2020 | 0.55 | -2.92 | 2.74 | 1.00 | 1020.00 | 4.38 |
| Canada | 2/17/2020 | 0.53 | -6.19 | 1.91 | 1.00 | 1028.00 | 2.65 |
| Canada | 2/18/2020 | 1.00 | -0.22 | 3.47 | 1.00 | 1024.00 | 6.80 |
| Canada | 2/19/2020 | 0.32 | -5.32 | 1.91 | 2.00 | 1032.00 | 5.41 |
| Canada | 2/20/2020 | 0.25 | -9.26 | 1.45 | 1.00 | 1035.00 | 3.84 |
| Canada | 2/21/2020 | 0.36 | -7.38 | 1.74 | 2.00 | 1035.00 | 5.79 |
| Canada | 2/22/2020 | 0.54 | -2.74 | 2.72 | 2.00 | 1023.00 | 5.64 |
| Canada | 2/23/2020 | 0.69 | 0.81 | 3.47 | 2.00 | 1020.00 | 4.71 |
| Canada | 2/24/2020 | 0.73 | 1.75 | 3.82 | 2.00 | 1015.00 | 3.03 |
| Canada | 2/25/2020 | 1.19 | 0.90 | 3.61 | 2.00 | 1016.00 | 3.66 |
| Canada | 2/26/2020 | 1.21 | -1.34 | 3.15 | 1.00 | 1016.00 | 5.07 |
| Canada | 2/27/2020 | 0.58 | -5.53 | 1.95 | 1.00 | 999.00 | 9.95 |
| Canada | 2/28/2020 | 0.43 | -6.43 | 1.85 | 1.00 | 1010.00 | 7.92 |
| Canada | 2/29/2020 | 0.39 | -8.88 | 1.46 | 1.00 | 1019.00 | 4.69 |
| Canada | 3/1/2020 | 0.85 | -3.69 | 2.46 | 2.00 | 1021.00 | 3.35 |
| Canada | 3/2/2020 | 1.41 | 2.44 | 4.43 | 1.00 | 1012.00 | 4.08 |
| Canada | 3/3/2020 | 1.18 | 0.51 | 3.94 | 2.00 | 999.00 | 2.45 |
| Canada | 3/4/2020 | 0.65 | 0.48 | 3.37 | 1.00 | 1015.00 | 5.19 |
| Canada | 3/5/2020 | 0.61 | -0.25 | 2.90 | 2.00 | 1021.00 | 2.44 |
| Canada | 3/6/2020 | 0.76 | -0.77 | 3.11 | 1.00 | 1026.00 | 5.44 |
| Canada | 3/7/2020 | 0.46 | -2.76 | 2.27 | 2.00 | 1033.00 | 3.29 |
| Canada | 3/8/2020 | 0.90 | 3.32 | 3.70 | 3.00 | 1029.00 | 4.80 |
| Canada | 3/9/2020 | 1.68 | 8.29 | 5.86 | 3.00 | 1022.00 | 5.89 |
| Canada | 3/10/2020 | 1.46 | 4.56 | 4.80 | 2.00 | 1023.00 | 6.24 |
| Canada | 3/11/2020 | 0.80 | -0.69 | 2.93 | 2.00 | 1024.00 | 2.52 |
| Canada | 3/12/2020 | 1.29 | 2.63 | 4.09 | 2.00 | 1018.00 | 2.67 |
| Canada | 3/13/2020 | 0.92 | 4.09 | 3.97 | 2.00 | 999.00 | 7.49 |
| Canada | 3/14/2020 | 0.63 | -0.25 | 2.81 | 2.00 | 1034.00 | 4.37 |
| Cape Verde | 4/7/2020 | 1.48 | 23.39 | 14.85 | 6.00 | 1015.00 | 6.14 |
| Cape Verde | 4/8/2020 | 1.62 | 23.58 | 15.04 | 6.00 | 1014.00 | 6.38 |
| Cape Verde | 4/9/2020 | 1.32 | 23.66 | 14.79 | 7.00 | 1013.00 | 6.45 |
| Cape Verde | 4/10/2020 | 1.53 | 23.52 | 15.44 | 6.00 | 1013.00 | 5.25 |
| Cape Verde | 4/11/2020 | 1.75 | 23.47 | 15.41 | 6.00 | 1014.00 | 5.36 |
| Cape Verde | 4/12/2020 | 1.83 | 23.35 | 14.74 | 6.00 | 1014.00 | 5.80 |
| Cape Verde | 4/13/2020 | 2.22 | 23.42 | 14.19 | 5.00 | 1014.00 | 5.37 |
| Cape Verde | 4/14/2020 | 2.21 | 23.14 | 13.69 | 6.00 | 1014.00 | 6.23 |
| Cape Verde | 4/15/2020 | 2.49 | 22.95 | 13.48 | 6.00 | 1016.00 | 5.67 |
| Cape Verde | 4/16/2020 | 2.80 | 22.90 | 13.25 | 6.00 | 1017.00 | 5.69 |
| Cape Verde | 4/17/2020 | 2.88 | 22.81 | 13.52 | 6.00 | 1017.00 | 6.78 |
| Cape Verde | 4/18/2020 | 1.72 | 22.87 | 12.79 | 6.00 | 1017.00 | 7.06 |
| Cape Verde | 4/19/2020 | 1.68 | 22.73 | 12.79 | 6.00 | 1017.00 | 6.71 |
| Cape Verde | 4/20/2020 | 2.46 | 22.64 | 12.60 | 6.00 | 1018.00 | 6.50 |
| Cape Verde | 4/21/2020 | 2.65 | 22.94 | 13.24 | 6.00 | 1017.00 | 6.74 |
| Cape Verde | 4/22/2020 | 2.71 | 22.82 | 13.55 | 6.00 | 1014.00 | 7.03 |
| Cape Verde | 4/23/2020 | 2.43 | 22.39 | 13.14 | 6.00 | 1015.00 | 6.66 |
| Cape Verde | 4/24/2020 | 2.25 | 22.48 | 12.79 | 6.00 | 1016.00 | 6.33 |
| Cape Verde | 4/25/2020 | 2.47 | 22.61 | 12.95 | 6.00 | 1017.00 | 6.05 |
| Cape Verde | 4/26/2020 | 3.00 | 22.68 | 13.23 | 6.00 | 1016.00 | 5.88 |
| Cape Verde | 4/27/2020 | 3.33 | 22.95 | 13.49 | 6.00 | 1017.00 | 6.22 |
| Cape Verde | 4/28/2020 | 3.72 | 23.20 | 13.92 | 6.00 | 1017.00 | 7.01 |
| Cape Verde | 4/29/2020 | 3.39 | 23.47 | 14.39 | 6.00 | 1017.00 | 7.91 |
| Cape Verde | 4/30/2020 | 2.91 | 23.28 | 14.31 | 6.00 | 1017.00 | 7.31 |
| Cape Verde | 5/1/2020 | 2.84 | 23.34 | 14.40 | 6.00 | 1016.00 | 6.20 |
| Cape Verde | 5/2/2020 | 2.64 | 23.48 | 14.83 | 6.00 | 1016.00 | 4.85 |
| Cape Verde | 5/3/2020 | 2.45 | 23.45 | 14.79 | 6.00 | 1015.00 | 5.81 |
| Cayman Islands | 3/15/2020 | 2.97 | 26.44 | 16.61 | 7.00 | 1020.00 | 6.58 |
| Cayman Islands | 3/16/2020 | 2.95 | 26.30 | 16.26 | 7.00 | 1019.00 | 6.68 |
| Cayman Islands | 3/17/2020 | 3.06 | 26.30 | 16.20 | 7.00 | 1019.00 | 6.08 |
| Cayman Islands | 3/18/2020 | 3.01 | 26.55 | 16.42 | 7.00 | 1020.00 | 6.45 |
| Cayman Islands | 3/19/2020 | 2.96 | 26.72 | 16.60 | 7.00 | 1020.00 | 6.74 |
| Cayman Islands | 3/20/2020 | 3.26 | 26.64 | 16.98 | 7.00 | 1021.00 | 5.90 |
| Cayman Islands | 3/21/2020 | 2.63 | 26.59 | 16.56 | 6.00 | 1020.00 | 5.90 |
| Cayman Islands | 3/22/2020 | 2.74 | 26.65 | 16.40 | 7.00 | 1019.00 | 5.72 |
| Cayman Islands | 3/23/2020 | 2.44 | 26.54 | 16.01 | 7.00 | 1021.00 | 4.64 |
| Cayman Islands | 3/24/2020 | 2.00 | 26.53 | 15.22 | 6.00 | 1019.00 | 3.75 |
| Cayman Islands | 3/25/2020 | 2.53 | 26.20 | 15.71 | 7.00 | 1017.00 | 2.90 |
| Cayman Islands | 3/26/2020 | 2.05 | 26.66 | 14.73 | 7.00 | 1017.00 | 2.96 |
| Central African Republic | 4/20/2020 | 4.73 | 26.75 | 16.61 | 7.00 | 1013.00 | 1.49 |
| Central African Republic | 4/21/2020 | 4.93 | 26.20 | 17.28 | 7.00 | 1013.00 | 1.05 |
| Central African Republic | 4/22/2020 | 5.08 | 25.70 | 18.42 | 7.00 | 1011.00 | 1.42 |
| Central African Republic | 4/23/2020 | 4.80 | 25.70 | 18.51 | 6.00 | 1011.00 | 1.42 |
| Central African Republic | 4/24/2020 | 4.13 | 26.38 | 18.01 | 7.00 | 1011.00 | 1.19 |
| Central African Republic | 4/25/2020 | 4.40 | 26.62 | 18.63 | 8.00 | 1009.00 | 0.99 |
| Central African Republic | 4/26/2020 | 4.80 | 26.08 | 18.41 | 7.00 | 1011.00 | 1.30 |
| Central African Republic | 4/27/2020 | 4.94 | 25.30 | 18.17 | 5.00 | 1013.00 | 1.27 |
| Central African Republic | 4/28/2020 | 4.92 | 25.63 | 17.85 | 7.00 | 1013.00 | 1.26 |
| Central African Republic | 4/29/2020 | 4.47 | 26.25 | 17.70 | 6.00 | 1013.00 | 1.24 |
| Central African Republic | 4/30/2020 | 4.75 | 26.14 | 18.32 | 8.00 | 1013.00 | 1.23 |
| Central African Republic | 5/1/2020 | 4.65 | 26.50 | 18.33 | 7.00 | 1013.00 | 1.25 |
| Central African Republic | 5/2/2020 | 4.53 | 26.60 | 17.96 | 7.00 | 1013.00 | 1.37 |
| Central African Republic | 5/3/2020 | 4.52 | 26.91 | 17.55 | 8.00 | 1012.00 | 1.41 |
| Central African Republic | 5/4/2020 | 4.83 | 27.05 | 17.88 | 8.00 | 1011.00 | 0.91 |
| Central African Republic | 5/5/2020 | 5.57 | 27.15 | 18.58 | 8.00 | 1012.00 | 0.77 |
| Central African Republic | 5/6/2020 | 5.61 | 26.51 | 19.01 | 6.00 | 1011.00 | 0.72 |
| Central African Republic | 5/7/2020 | 5.26 | 25.47 | 17.56 | 6.00 | 1011.00 | 1.21 |
| Central African Republic | 5/8/2020 | 5.23 | 26.82 | 16.97 | 6.00 | 1013.00 | 0.72 |
| Central African Republic | 5/9/2020 | 5.60 | 25.73 | 18.25 | 5.00 | 1013.00 | 0.87 |
| Central African Republic | 5/10/2020 | 5.14 | 25.68 | 17.48 | 6.00 | 1012.00 | 1.72 |
| Central African Republic | 5/11/2020 | 5.22 | 25.77 | 17.34 | 6.00 | 1014.00 | 1.13 |
| Central African Republic | 5/12/2020 | 5.29 | 26.45 | 17.77 | 7.00 | 1014.00 | 1.04 |
| Central African Republic | 5/13/2020 | 5.32 | 25.75 | 17.36 | 7.00 | 1014.00 | 1.07 |
| Central African Republic | 5/14/2020 | 4.93 | 26.64 | 17.67 | 7.00 | 1013.00 | 1.05 |
| Central African Republic | 5/15/2020 | 4.81 | 26.85 | 18.21 | 6.00 | 1013.00 | 1.46 |
| Central African Republic | 5/16/2020 | 4.93 | 26.28 | 18.13 | 8.00 | 1012.00 | 1.49 |
| Central African Republic | 5/17/2020 | 4.80 | 26.61 | 17.38 | 8.00 | 1012.00 | 1.34 |
| Central African Republic | 5/18/2020 | 5.25 | 25.99 | 17.66 | 7.00 | 1013.00 | 1.36 |
| Central African Republic | 5/19/2020 | 5.22 | 26.08 | 18.02 | 8.00 | 1014.00 | 1.29 |
| Central African Republic | 5/20/2020 | 5.18 | 25.89 | 18.16 | 7.00 | 1013.00 | 1.05 |
| Central African Republic | 5/21/2020 | 5.31 | 24.63 | 18.07 | 5.00 | 1015.00 | 0.93 |
| Central African Republic | 5/22/2020 | 5.36 | 24.95 | 16.79 | 6.00 | 1014.00 | 0.77 |
| Central African Republic | 5/23/2020 | 5.45 | 25.57 | 17.66 | 6.00 | 1014.00 | 0.85 |
| Central African Republic | 5/24/2020 | 5.59 | 25.24 | 18.50 | 6.00 | 1014.00 | 1.34 |
| Central African Republic | 5/25/2020 | 5.11 | 25.60 | 18.14 | 6.00 | 1013.00 | 1.36 |
| Central African Republic | 5/26/2020 | 5.01 | 25.53 | 17.53 | 7.00 | 1013.00 | 0.77 |
| Central African Republic | 5/27/2020 | 4.91 | 26.16 | 17.97 | 6.00 | 1013.00 | 0.87 |
| Central African Republic | 5/28/2020 | 5.14 | 25.70 | 18.57 | 7.00 | 1012.00 | 0.98 |
| Chile | 2/22/2020 | 0.84 | 20.72 | 5.16 | 6.00 | 1019.00 | 1.92 |
| Chile | 2/23/2020 | 0.73 | 20.14 | 4.37 | 6.00 | 1016.00 | 1.82 |
| Chile | 2/24/2020 | 1.10 | 18.53 | 4.68 | 6.00 | 1017.00 | 2.04 |
| Chile | 2/25/2020 | 1.03 | 20.14 | 4.52 | 6.00 | 1018.00 | 1.98 |
| Chile | 2/26/2020 | 0.79 | 22.01 | 3.77 | 7.00 | 1018.00 | 2.15 |
| Chile | 2/27/2020 | 0.90 | 21.30 | 4.21 | 6.00 | 1017.00 | 2.01 |
| Chile | 2/28/2020 | 1.11 | 21.90 | 5.03 | 6.00 | 1018.00 | 2.26 |
| Chile | 2/29/2020 | 1.54 | 22.87 | 6.50 | 7.00 | 1018.00 | 2.20 |
| Chile | 3/1/2020 | 1.45 | 21.65 | 6.76 | 6.00 | 1017.00 | 2.03 |
| Chile | 3/2/2020 | 1.10 | 21.38 | 6.55 | 6.00 | 1017.00 | 2.17 |
| Chile | 3/3/2020 | 1.03 | 21.81 | 6.25 | 6.00 | 1017.00 | 2.24 |
| Chile | 3/4/2020 | 1.19 | 22.91 | 6.30 | 6.00 | 1016.00 | 2.27 |
| Chile | 3/5/2020 | 1.24 | 25.27 | 5.25 | 7.00 | 1015.00 | 2.17 |
| Chile | 3/6/2020 | 1.22 | 23.03 | 5.65 | 7.00 | 1015.00 | 2.07 |
| Chile | 3/7/2020 | 1.56 | 21.91 | 5.64 | 6.00 | 1014.00 | 2.18 |
| Chile | 3/8/2020 | 1.38 | 22.01 | 5.97 | 6.00 | 1016.00 | 1.96 |
| Chile | 3/9/2020 | 1.07 | 22.57 | 5.02 | 6.00 | 1017.00 | 2.04 |
| Chile | 3/10/2020 | 1.10 | 21.85 | 5.20 | 7.00 | 1016.00 | 1.67 |
| Colombia | 3/3/2020 | 3.33 | 20.32 | 14.19 | 4.00 | 1015.00 | 0.64 |
| Colombia | 3/4/2020 | 2.83 | 19.97 | 13.38 | 4.00 | 1016.00 | 1.02 |
| Colombia | 3/5/2020 | 1.49 | 18.54 | 10.21 | 4.00 | 1016.00 | 1.50 |
| Colombia | 3/6/2020 | 2.04 | 18.82 | 10.59 | 4.00 | 1017.00 | 0.98 |
| Colombia | 3/7/2020 | 3.02 | 19.81 | 13.08 | 4.00 | 1015.00 | 0.90 |
| Colombia | 3/8/2020 | 3.75 | 19.39 | 14.44 | 4.00 | 1015.00 | 0.78 |
| Colombia | 3/9/2020 | 3.63 | 19.34 | 14.10 | 4.00 | 1017.00 | 0.73 |
| Colombia | 3/10/2020 | 3.33 | 19.88 | 13.66 | 4.00 | 1017.00 | 0.92 |
| Colombia | 3/11/2020 | 3.34 | 19.47 | 13.85 | 4.00 | 1015.00 | 0.70 |
| Colombia | 3/12/2020 | 3.43 | 19.70 | 14.05 | 4.00 | 1014.00 | 0.65 |
| Colombia | 3/13/2020 | 3.57 | 19.30 | 14.32 | 4.00 | 1015.00 | 0.59 |
| Colombia | 3/14/2020 | 3.67 | 17.93 | 14.01 | 4.00 | 1017.00 | 0.51 |
| Costa Rica | 3/8/2020 | 2.43 | 25.24 | 13.31 | 6.00 | 1014.00 | 0.88 |
| Costa Rica | 3/9/2020 | 2.15 | 25.53 | 13.49 | 6.00 | 1015.00 | 0.57 |
| Costa Rica | 3/10/2020 | 2.38 | 25.35 | 13.83 | 6.00 | 1016.00 | 0.44 |
| Costa Rica | 3/11/2020 | 3.16 | 25.35 | 14.71 | 6.00 | 1015.00 | 0.71 |
| Costa Rica | 3/12/2020 | 3.42 | 24.84 | 14.73 | 6.00 | 1013.00 | 0.88 |
| Costa Rica | 3/13/2020 | 3.45 | 25.06 | 15.20 | 5.00 | 1015.00 | 0.74 |
| Costa Rica | 3/14/2020 | 3.35 | 25.34 | 15.04 | 5.00 | 1017.00 | 0.71 |
| Costa Rica | 3/15/2020 | 3.04 | 25.32 | 14.64 | 6.00 | 1016.00 | 0.52 |
| Costa Rica | 3/16/2020 | 2.72 | 25.16 | 13.90 | 6.00 | 1015.00 | 1.02 |
| Costa Rica | 3/17/2020 | 3.05 | 25.43 | 14.33 | 6.00 | 1016.00 | 0.99 |
| Croatia | 2/28/2020 | 0.68 | 4.39 | 3.86 | 2.00 | 1023.00 | 1.99 |
| Croatia | 2/29/2020 | 1.20 | 3.33 | 4.32 | 3.00 | 1023.00 | 1.84 |
| Croatia | 3/1/2020 | 1.82 | 8.08 | 6.46 | 4.00 | 1010.00 | 1.60 |
| Croatia | 3/2/2020 | 1.45 | 9.73 | 6.61 | 3.00 | 999.00 | 1.84 |
| Croatia | 3/3/2020 | 1.40 | 10.06 | 6.45 | 3.00 | 999.00 | 1.54 |
| Croatia | 3/4/2020 | 1.01 | 5.57 | 4.28 | 2.00 | 1014.00 | 1.42 |
| Croatia | 3/5/2020 | 1.03 | 5.35 | 4.14 | 3.00 | 1014.00 | 0.85 |
| Croatia | 3/6/2020 | 1.49 | 5.93 | 5.49 | 2.00 | 999.00 | 0.84 |
| Croatia | 3/7/2020 | 1.11 | 5.25 | 4.76 | 3.00 | 1020.00 | 0.94 |
| Croatia | 3/8/2020 | 0.73 | 3.64 | 3.68 | 2.00 | 1022.00 | 1.23 |
| Croatia | 3/9/2020 | 1.10 | 4.76 | 3.98 | 3.00 | 1019.00 | 0.79 |
| Croatia | 3/10/2020 | 1.40 | 7.03 | 5.03 | 4.00 | 1018.00 | 0.88 |
| Croatia | 3/11/2020 | 1.77 | 9.99 | 6.20 | 4.00 | 1021.00 | 1.13 |
| Croatia | 3/12/2020 | 1.45 | 10.00 | 6.31 | 4.00 | 1021.00 | 1.66 |
| Croatia | 3/13/2020 | 1.83 | 7.61 | 6.24 | 4.00 | 1021.00 | 0.94 |
| Cuba | 3/7/2020 | 1.65 | 22.85 | 11.09 | 6.00 | 1023.00 | 7.71 |
| Cuba | 3/8/2020 | 2.15 | 23.01 | 10.78 | 5.00 | 1025.00 | 8.49 |
| Cuba | 3/9/2020 | 2.38 | 24.22 | 11.53 | 6.00 | 1026.00 | 7.99 |
| Cuba | 3/10/2020 | 2.86 | 24.42 | 12.85 | 6.00 | 1024.00 | 6.31 |
| Cuba | 3/11/2020 | 2.90 | 24.57 | 13.01 | 6.00 | 1021.00 | 5.08 |
| Cuba | 3/12/2020 | 2.99 | 24.80 | 13.81 | 6.00 | 1017.00 | 5.09 |
| Cuba | 3/13/2020 | 2.83 | 25.44 | 14.28 | 7.00 | 1020.00 | 5.56 |
| Cuba | 3/14/2020 | 3.01 | 26.13 | 15.03 | 7.00 | 1022.00 | 5.75 |
| Cuba | 3/15/2020 | 2.84 | 26.03 | 14.30 | 6.00 | 1022.00 | 5.33 |
| Cuba | 3/16/2020 | 2.69 | 25.44 | 13.73 | 7.00 | 1022.00 | 5.49 |
| Cuba | 3/17/2020 | 2.99 | 25.74 | 14.12 | 7.00 | 1022.00 | 5.38 |
| Cuba | 3/18/2020 | 2.86 | 25.77 | 14.55 | 7.00 | 1022.00 | 5.89 |
| Cuba | 3/19/2020 | 2.66 | 26.12 | 14.22 | 7.00 | 1023.00 | 6.51 |
| Cyprus | 3/5/2020 | 1.50 | 16.24 | 9.25 | 5.00 | 1013.00 | 3.29 |
| Cyprus | 3/6/2020 | 1.63 | 14.72 | 8.19 | 5.00 | 1015.00 | 3.26 |
| Cyprus | 3/7/2020 | 1.34 | 15.47 | 8.58 | 5.00 | 1021.00 | 3.66 |
| Cyprus | 3/8/2020 | 1.31 | 15.93 | 8.40 | 5.00 | 1022.00 | 1.61 |
| Cyprus | 3/9/2020 | 1.45 | 16.37 | 9.55 | 4.00 | 1019.00 | 4.89 |
| Cyprus | 3/10/2020 | 1.18 | 15.16 | 8.07 | 5.00 | 1018.00 | 3.73 |
| Cyprus | 3/11/2020 | 1.33 | 15.56 | 7.98 | 5.00 | 1018.00 | 2.52 |
| Cyprus | 3/12/2020 | 2.60 | 17.74 | 9.07 | 4.00 | 1013.00 | 7.32 |
| Cyprus | 3/13/2020 | 2.71 | 15.94 | 10.11 | 5.00 | 1005.00 | 6.58 |
| Cyprus | 3/14/2020 | 2.32 | 16.72 | 9.84 | 4.00 | 1014.00 | 3.55 |
| Cyprus | 3/15/2020 | 2.01 | 16.41 | 9.59 | 5.00 | 1016.00 | 2.65 |
| Cyprus | 3/16/2020 | 1.68 | 15.40 | 8.98 | 5.00 | 1014.00 | 2.73 |
| Cyprus | 3/17/2020 | 1.63 | 12.49 | 7.10 | 2.00 | 1012.00 | 4.59 |
| Cyprus | 3/18/2020 | 0.83 | 11.96 | 4.90 | 4.00 | 1012.00 | 4.68 |
| Cyprus | 3/19/2020 | 1.00 | 10.55 | 4.88 | 4.00 | 1013.00 | 2.78 |
| Cyprus | 3/20/2020 | 1.08 | 11.17 | 4.91 | 4.00 | 1013.00 | 2.81 |
| Cyprus | 3/21/2020 | 1.13 | 12.57 | 6.06 | 3.00 | 1015.00 | 4.64 |
| Czech Republic | 2/25/2020 | 1.45 | 6.62 | 5.68 | 2.00 | 999.00 | 4.48 |
| Czech Republic | 2/26/2020 | 0.83 | 2.69 | 3.98 | 2.00 | 999.00 | 4.70 |
| Czech Republic | 2/27/2020 | 0.75 | 1.52 | 3.60 | 1.00 | 999.00 | 3.66 |
| Czech Republic | 2/28/2020 | 0.71 | 1.05 | 3.52 | 1.00 | 998.00 | 4.53 |
| Czech Republic | 2/29/2020 | 1.50 | 3.23 | 4.30 | 2.00 | 1017.00 | 4.59 |
| Czech Republic | 3/1/2020 | 1.05 | 5.04 | 4.89 | 2.00 | 999.00 | 3.84 |
| Czech Republic | 3/2/2020 | 0.94 | 5.13 | 4.41 | 3.00 | 998.00 | 3.50 |
| Czech Republic | 3/3/2020 | 1.34 | 3.50 | 4.74 | 3.00 | 999.00 | 2.98 |
| Czech Republic | 3/4/2020 | 0.79 | 1.76 | 3.72 | 2.00 | 1013.00 | 2.53 |
| Czech Republic | 3/5/2020 | 0.97 | 3.64 | 3.87 | 3.00 | 999.00 | 3.52 |
| Czech Republic | 3/6/2020 | 1.36 | 5.35 | 4.84 | 2.00 | 999.00 | 4.29 |
| Czech Republic | 3/7/2020 | 0.97 | 2.61 | 4.00 | 1.00 | 1022.00 | 4.08 |
| Czech Republic | 3/8/2020 | 0.93 | 3.15 | 3.90 | 3.00 | 1023.00 | 2.46 |
| Czech Republic | 3/9/2020 | 1.23 | 3.35 | 4.47 | 2.00 | 1018.00 | 2.06 |
| Côte d'Ivoire | 3/13/2020 | 5.26 | 28.59 | 20.43 | 6.00 | 1012.00 | 2.78 |
| Côte d'Ivoire | 3/14/2020 | 5.75 | 28.12 | 20.36 | 7.00 | 1012.00 | 3.03 |
| Côte d'Ivoire | 3/15/2020 | 5.63 | 27.70 | 19.38 | 6.00 | 1012.00 | 2.50 |
| Côte d'Ivoire | 3/16/2020 | 4.84 | 28.31 | 19.55 | 6.00 | 1011.00 | 2.69 |
| Côte d'Ivoire | 3/17/2020 | 5.39 | 28.31 | 20.13 | 7.00 | 1011.00 | 3.19 |
| Côte d'Ivoire | 3/18/2020 | 5.49 | 28.47 | 19.64 | 7.00 | 1012.00 | 3.43 |
| Côte d'Ivoire | 3/19/2020 | 5.35 | 27.94 | 19.34 | 6.00 | 1012.00 | 3.73 |
| Côte d'Ivoire | 3/20/2020 | 4.52 | 28.23 | 18.09 | 6.00 | 1012.00 | 2.73 |
| Côte d'Ivoire | 3/21/2020 | 5.26 | 27.93 | 19.23 | 7.00 | 1011.00 | 1.86 |
| Côte d'Ivoire | 3/22/2020 | 5.57 | 27.75 | 19.79 | 6.00 | 1011.00 | 2.71 |
| Côte d'Ivoire | 3/23/2020 | 6.04 | 26.65 | 19.05 | 7.00 | 1013.00 | 3.40 |
| Côte d'Ivoire | 3/24/2020 | 6.02 | 26.18 | 18.33 | 6.00 | 1013.00 | 3.33 |
| Côte d'Ivoire | 3/25/2020 | 5.83 | 26.53 | 17.92 | 6.00 | 1012.00 | 3.23 |
| Côte d'Ivoire | 3/26/2020 | 5.40 | 27.11 | 18.68 | 6.00 | 1011.00 | 2.06 |
| Côte d'Ivoire | 3/27/2020 | 5.45 | 26.99 | 19.29 | 6.00 | 1011.00 | 1.85 |
| Côte d'Ivoire | 3/28/2020 | 5.37 | 27.15 | 19.59 | 7.00 | 1012.00 | 1.95 |
| Côte d'Ivoire | 3/29/2020 | 5.30 | 27.24 | 19.75 | 7.00 | 1012.00 | 2.03 |
| Côte d'Ivoire | 3/30/2020 | 5.59 | 27.12 | 19.82 | 6.00 | 1011.00 | 2.10 |
| Denmark | 3/3/2020 | 1.21 | 3.71 | 4.79 | 3.00 | 999.00 | 2.51 |
| Denmark | 3/4/2020 | 0.97 | 4.23 | 4.67 | 2.00 | 1005.00 | 4.34 |
| Denmark | 3/5/2020 | 0.87 | 3.11 | 4.43 | 1.00 | 1005.00 | 3.06 |
| Denmark | 3/6/2020 | 1.05 | 3.76 | 4.50 | 3.00 | 999.00 | 5.95 |
| Denmark | 3/7/2020 | 1.03 | 3.77 | 4.48 | 2.00 | 1014.00 | 3.32 |
| Denmark | 3/8/2020 | 1.51 | 5.19 | 5.08 | 3.00 | 1014.00 | 6.08 |
| Denmark | 3/9/2020 | 1.33 | 5.94 | 5.46 | 3.00 | 1008.00 | 3.26 |
| Denmark | 3/10/2020 | 1.39 | 5.36 | 5.35 | 3.00 | 999.00 | 6.37 |
| Denmark | 3/11/2020 | 0.88 | 6.15 | 5.06 | 2.00 | 999.00 | 7.28 |
| Denmark | 3/12/2020 | 1.04 | 5.63 | 4.83 | 3.00 | 999.00 | 10.12 |
| Denmark | 3/13/2020 | 0.85 | 3.90 | 4.31 | 2.00 | 999.00 | 7.10 |
| Denmark | 3/14/2020 | 0.39 | 2.22 | 2.95 | 2.00 | 1024.00 | 3.39 |
| Denmark | 3/15/2020 | 1.34 | 5.39 | 4.70 | 3.00 | 1021.00 | 7.29 |
| Denmark | 3/16/2020 | 1.16 | 6.94 | 5.48 | 2.00 | 1022.00 | 3.89 |
| Denmark | 3/17/2020 | 1.50 | 5.43 | 5.08 | 3.00 | 1023.00 | 5.15 |
| Denmark | 3/18/2020 | 1.92 | 6.82 | 5.53 | 2.00 | 1022.00 | 6.00 |
| Denmark | 3/19/2020 | 0.87 | 5.43 | 4.65 | 2.00 | 1025.00 | 3.55 |
| Denmark | 3/20/2020 | 0.61 | 4.47 | 4.19 | 3.00 | 1033.00 | 4.89 |
| Denmark | 3/21/2020 | 0.46 | 2.62 | 3.33 | 2.00 | 1037.00 | 3.95 |
| Denmark | 3/22/2020 | 0.44 | 1.24 | 3.00 | 2.00 | 1042.00 | 3.69 |
| Denmark | 3/23/2020 | 0.46 | 1.54 | 2.82 | 2.00 | 1042.00 | 3.84 |
| Denmark | 3/24/2020 | 0.54 | 2.93 | 3.22 | 2.00 | 1038.00 | 4.85 |
| Denmark | 3/25/2020 | 0.51 | 3.55 | 3.43 | 1.00 | 1035.00 | 4.17 |
| Denmark | 3/26/2020 | 0.46 | 3.75 | 3.71 | 3.00 | 1033.00 | 2.84 |
| Denmark | 3/27/2020 | 0.71 | 4.57 | 4.24 | 3.00 | 1031.00 | 2.42 |
| Denmark | 3/28/2020 | 0.87 | 4.64 | 4.36 | 3.00 | 1025.00 | 3.51 |
| Djibouti | 3/17/2020 | 4.82 | 25.81 | 17.37 | 6.00 | 1012.00 | 3.58 |
| Djibouti | 3/18/2020 | 4.76 | 26.29 | 17.69 | 6.00 | 1013.00 | 4.91 |
| Djibouti | 3/19/2020 | 4.87 | 26.55 | 17.57 | 6.00 | 1014.00 | 4.01 |
| Djibouti | 3/20/2020 | 4.32 | 26.92 | 17.33 | 7.00 | 1012.00 | 2.66 |
| Djibouti | 3/21/2020 | 5.39 | 27.41 | 18.00 | 7.00 | 1012.00 | 2.41 |
| Djibouti | 3/22/2020 | 5.38 | 27.32 | 18.27 | 6.00 | 1012.00 | 2.63 |
| Djibouti | 3/23/2020 | 4.69 | 28.35 | 17.49 | 7.00 | 1014.00 | 3.00 |
| Djibouti | 3/24/2020 | 4.59 | 28.28 | 17.06 | 7.00 | 1011.00 | 2.97 |
| Djibouti | 3/25/2020 | 4.55 | 27.38 | 18.11 | 7.00 | 1010.00 | 3.39 |
| Djibouti | 3/26/2020 | 3.43 | 27.08 | 18.13 | 7.00 | 1011.00 | 4.03 |
| Djibouti | 3/27/2020 | 3.41 | 26.85 | 17.54 | 6.00 | 1014.00 | 4.90 |
| Djibouti | 3/28/2020 | 3.15 | 26.84 | 17.73 | 6.00 | 1014.00 | 5.13 |
| Djibouti | 3/29/2020 | 3.37 | 27.26 | 17.98 | 6.00 | 1012.00 | 4.14 |
| Djibouti | 3/30/2020 | 3.51 | 27.80 | 18.20 | 7.00 | 1012.00 | 2.60 |
| Djibouti | 3/31/2020 | 3.77 | 27.70 | 18.45 | 7.00 | 1013.00 | 2.82 |
| Djibouti | 4/1/2020 | 3.87 | 27.74 | 18.72 | 7.00 | 1013.00 | 4.30 |
| Djibouti | 4/2/2020 | 3.68 | 27.60 | 18.75 | 6.00 | 1012.00 | 4.26 |
| Djibouti | 4/3/2020 | 3.44 | 27.36 | 18.08 | 6.00 | 1012.00 | 4.31 |
| Djibouti | 4/4/2020 | 2.90 | 27.22 | 18.03 | 7.00 | 1013.00 | 5.01 |
| Djibouti | 4/5/2020 | 2.84 | 27.30 | 18.08 | 6.00 | 1014.00 | 5.51 |
| Djibouti | 4/6/2020 | 3.15 | 27.37 | 17.97 | 6.00 | 1014.00 | 4.94 |
| Dominican Republic | 3/11/2020 | 3.84 | 25.06 | 15.42 | 7.00 | 1018.00 | 2.66 |
| Dominican Republic | 3/12/2020 | 4.23 | 25.09 | 15.68 | 7.00 | 1015.00 | 2.27 |
| Dominican Republic | 3/13/2020 | 3.79 | 25.87 | 15.60 | 7.00 | 1017.00 | 2.30 |
| Dominican Republic | 3/14/2020 | 3.16 | 25.74 | 15.66 | 7.00 | 1019.00 | 4.76 |
| Dominican Republic | 3/15/2020 | 2.36 | 25.34 | 13.93 | 6.00 | 1019.00 | 5.00 |
| Dominican Republic | 3/16/2020 | 2.41 | 25.03 | 13.47 | 7.00 | 1018.00 | 4.99 |
| Dominican Republic | 3/17/2020 | 3.06 | 25.53 | 15.10 | 7.00 | 1019.00 | 4.79 |
| Dominican Republic | 3/18/2020 | 3.00 | 25.78 | 15.03 | 7.00 | 1020.00 | 5.14 |
| Egypt | 2/26/2020 | 1.39 | 14.78 | 6.39 | 6.00 | 1021.00 | 1.68 |
| Egypt | 2/27/2020 | 1.57 | 17.37 | 5.73 | 6.00 | 1020.00 | 1.54 |
| Egypt | 2/28/2020 | 1.50 | 18.45 | 6.08 | 7.00 | 1017.00 | 2.09 |
| Egypt | 2/29/2020 | 1.31 | 14.69 | 6.98 | 5.00 | 1018.00 | 3.18 |
| Egypt | 3/1/2020 | 1.02 | 13.88 | 5.80 | 5.00 | 1021.00 | 2.51 |
| Egypt | 3/2/2020 | 0.99 | 14.09 | 5.90 | 6.00 | 1022.00 | 1.59 |
| Egypt | 3/3/2020 | 0.95 | 14.58 | 6.04 | 6.00 | 1021.00 | 2.83 |
| Egypt | 3/4/2020 | 0.82 | 17.00 | 5.26 | 7.00 | 1016.00 | 1.97 |
| Egypt | 3/5/2020 | 1.42 | 18.71 | 5.02 | 7.00 | 1011.00 | 2.88 |
| Egypt | 3/6/2020 | 1.77 | 15.55 | 6.71 | 5.00 | 1017.00 | 3.30 |
| Egypt | 3/7/2020 | 1.44 | 16.11 | 6.72 | 6.00 | 1020.00 | 1.65 |
| Egypt | 3/8/2020 | 0.96 | 17.32 | 6.49 | 6.00 | 1020.00 | 2.90 |
| Egypt | 3/9/2020 | 1.03 | 18.73 | 5.58 | 7.00 | 1015.00 | 3.65 |
| Egypt | 3/10/2020 | 1.13 | 15.21 | 6.49 | 6.00 | 1017.00 | 2.15 |
| Egypt | 3/11/2020 | 2.16 | 17.24 | 7.56 | 6.00 | 1015.00 | 3.92 |
| Egypt | 3/12/2020 | 3.11 | 15.59 | 9.96 | 4.00 | 999.00 | 3.50 |
| Egypt | 3/13/2020 | 2.62 | 15.44 | 9.16 | 4.00 | 999.00 | 6.34 |
| Egypt | 3/14/2020 | 2.40 | 17.32 | 9.18 | 5.00 | 1016.00 | 2.81 |
| Egypt | 3/15/2020 | 1.58 | 18.41 | 7.64 | 6.00 | 1017.00 | 1.34 |
| El Salvador | 3/13/2020 | 3.45 | 27.43 | 13.63 | 7.00 | 1016.00 | 2.12 |
| El Salvador | 3/14/2020 | 3.44 | 27.91 | 13.80 | 7.00 | 1017.00 | 2.14 |
| El Salvador | 3/15/2020 | 3.18 | 28.27 | 13.45 | 8.00 | 1017.00 | 2.48 |
| El Salvador | 3/16/2020 | 2.84 | 28.02 | 12.84 | 7.00 | 1017.00 | 2.16 |
| El Salvador | 3/17/2020 | 3.15 | 27.78 | 13.19 | 7.00 | 1017.00 | 2.56 |
| El Salvador | 3/18/2020 | 3.16 | 28.03 | 13.65 | 7.00 | 1016.00 | 2.50 |
| El Salvador | 3/19/2020 | 3.01 | 28.23 | 13.32 | 7.00 | 1018.00 | 2.52 |
| El Salvador | 3/20/2020 | 3.37 | 27.84 | 14.08 | 7.00 | 1018.00 | 2.03 |
| El Salvador | 3/21/2020 | 3.37 | 28.61 | 13.29 | 7.00 | 1018.00 | 1.87 |
| El Salvador | 3/22/2020 | 3.27 | 28.85 | 12.61 | 6.00 | 1016.00 | 2.79 |
| El Salvador | 3/23/2020 | 2.92 | 28.88 | 12.36 | 7.00 | 1018.00 | 2.51 |
| El Salvador | 3/24/2020 | 3.32 | 28.16 | 13.62 | 7.00 | 1017.00 | 2.42 |
| El Salvador | 3/25/2020 | 3.15 | 27.63 | 14.28 | 7.00 | 1015.00 | 2.29 |
| El Salvador | 3/26/2020 | 3.40 | 28.13 | 14.46 | 8.00 | 1014.00 | 2.45 |
| El Salvador | 3/27/2020 | 3.22 | 27.96 | 15.28 | 7.00 | 1013.00 | 2.68 |
| El Salvador | 3/28/2020 | 3.42 | 27.68 | 15.53 | 7.00 | 1013.00 | 2.47 |
| El Salvador | 3/29/2020 | 3.81 | 27.70 | 15.10 | 7.00 | 1014.00 | 2.19 |
| Estonia | 3/4/2020 | 1.36 | 1.47 | 4.08 | 1.00 | 1010.00 | 7.22 |
| Estonia | 3/5/2020 | 1.08 | 2.10 | 4.31 | 1.00 | 1004.00 | 4.30 |
| Estonia | 3/6/2020 | 0.77 | 2.12 | 3.94 | 1.00 | 1010.00 | 3.42 |
| Estonia | 3/7/2020 | 0.97 | 0.84 | 3.80 | 2.00 | 1012.00 | 2.38 |
| Estonia | 3/8/2020 | 0.92 | 0.94 | 3.75 | 1.00 | 1015.00 | 2.64 |
| Estonia | 3/9/2020 | 1.36 | 2.24 | 4.05 | 1.00 | 1013.00 | 5.89 |
| Estonia | 3/10/2020 | 1.22 | 2.33 | 4.33 | 1.00 | 999.00 | 5.72 |
| Estonia | 3/11/2020 | 1.25 | 2.84 | 4.55 | 1.00 | 998.00 | 6.76 |
| Estonia | 3/12/2020 | 1.02 | 3.60 | 4.68 | 1.00 | 988.00 | 6.93 |
| Estonia | 3/13/2020 | 0.74 | 1.92 | 3.78 | 1.00 | 999.00 | 7.20 |
| Estonia | 3/14/2020 | 0.36 | -1.71 | 2.38 | 2.00 | 1018.00 | 5.87 |
| Estonia | 3/15/2020 | 0.57 | 0.54 | 3.04 | 2.00 | 1020.00 | 7.41 |
| Estonia | 3/16/2020 | 1.15 | 1.37 | 3.86 | 1.00 | 1017.00 | 6.76 |
| Estonia | 3/17/2020 | 0.78 | 1.06 | 3.73 | 2.00 | 1020.00 | 5.31 |
| Estonia | 3/18/2020 | 1.27 | 3.82 | 4.80 | 2.00 | 1013.00 | 7.10 |
| Estonia | 3/19/2020 | 0.69 | 3.33 | 4.26 | 1.00 | 1014.00 | 6.50 |
| Estonia | 3/20/2020 | 0.69 | 1.33 | 3.57 | 2.00 | 1025.00 | 3.76 |
| Estonia | 3/21/2020 | 0.45 | -1.23 | 2.49 | 2.00 | 1035.00 | 3.11 |
| Estonia | 3/22/2020 | 0.32 | -0.84 | 2.21 | 2.00 | 1041.00 | 2.08 |
| Estonia | 3/23/2020 | 0.58 | -0.27 | 2.75 | 2.00 | 1042.00 | 4.62 |
| Estonia | 3/24/2020 | 0.56 | 1.68 | 3.53 | 1.00 | 1037.00 | 7.93 |
| Estonia | 3/25/2020 | 0.54 | 3.42 | 3.83 | 2.00 | 1030.00 | 9.37 |
| Ethiopia | 3/4/2020 | 1.96 | 21.73 | 8.68 | 5.74 | 1017.52 | 2.22 |
| Ethiopia | 3/5/2020 | 2.24 | 21.54 | 9.66 | 5.50 | 1017.68 | 2.04 |
| Ethiopia | 3/6/2020 | 2.40 | 21.94 | 9.30 | 5.49 | 1016.65 | 1.74 |
| Ethiopia | 3/7/2020 | 2.29 | 21.96 | 9.11 | 5.16 | 1015.81 | 2.16 |
| Ethiopia | 3/8/2020 | 2.12 | 21.66 | 8.85 | 6.04 | 1015.41 | 2.71 |
| Ethiopia | 3/9/2020 | 2.02 | 22.11 | 8.53 | 6.03 | 1014.81 | 2.31 |
| Ethiopia | 3/10/2020 | 2.31 | 21.98 | 9.74 | 5.95 | 1014.37 | 2.34 |
| Ethiopia | 3/11/2020 | 2.64 | 20.29 | 10.46 | 5.35 | 1014.72 | 2.59 |
| Ethiopia | 3/12/2020 | 2.43 | 20.84 | 10.26 | 5.40 | 1015.28 | 2.10 |
| Ethiopia | 3/13/2020 | 2.11 | 22.22 | 9.24 | 6.13 | 1014.96 | 1.86 |
| Ethiopia | 3/14/2020 | 2.50 | 21.25 | 10.19 | 5.05 | 1016.59 | 2.20 |
| Ethiopia | 3/15/2020 | 2.50 | 21.47 | 10.93 | 4.80 | 1014.98 | 2.08 |
| Ethiopia | 3/16/2020 | 2.86 | 20.75 | 11.86 | 4.71 | 1014.92 | 2.42 |
| Ethiopia | 3/17/2020 | 2.84 | 20.15 | 11.98 | 4.81 | 1015.61 | 2.12 |
| Ethiopia | 3/18/2020 | 2.44 | 21.04 | 10.89 | 4.93 | 1017.03 | 1.79 |
| Ethiopia | 3/19/2020 | 2.24 | 21.11 | 10.46 | 5.41 | 1016.72 | 2.37 |
| Ethiopia | 3/20/2020 | 2.53 | 20.93 | 11.33 | 5.24 | 1016.45 | 1.93 |
| Ethiopia | 3/21/2020 | 2.97 | 19.71 | 12.01 | 5.28 | 1015.68 | 1.72 |
| Ethiopia | 3/22/2020 | 2.95 | 19.09 | 11.91 | 4.77 | 1016.48 | 1.46 |
| Ethiopia | 3/23/2020 | 2.83 | 18.31 | 11.76 | 4.72 | 1017.63 | 1.46 |
| Ethiopia | 3/24/2020 | 2.25 | 20.33 | 10.81 | 5.11 | 1014.79 | 1.11 |
| Ethiopia | 3/25/2020 | 2.15 | 19.88 | 10.49 | 4.87 | 1013.35 | 1.69 |
| Ethiopia | 3/26/2020 | 1.74 | 19.76 | 9.49 | 4.98 | 1014.40 | 1.94 |
| Ethiopia | 3/27/2020 | 1.74 | 20.01 | 8.76 | 5.70 | 1016.18 | 2.20 |
| Ethiopia | 3/28/2020 | 1.70 | 20.43 | 9.21 | 6.16 | 1016.74 | 1.99 |
| Ethiopia | 3/29/2020 | 1.87 | 21.25 | 9.47 | 6.21 | 1015.75 | 2.23 |
| Ethiopia | 3/30/2020 | 1.83 | 21.24 | 10.01 | 6.08 | 1015.08 | 2.31 |
| Ethiopia | 3/31/2020 | 2.12 | 20.74 | 10.99 | 5.66 | 1015.62 | 2.63 |
| Finland | 3/1/2020 | 1.06 | 1.37 | 4.13 | 1.00 | 997.00 | 3.18 |
| Finland | 3/2/2020 | 0.99 | 2.25 | 4.43 | 1.00 | 999.00 | 2.48 |
| Finland | 3/3/2020 | 1.14 | 1.14 | 4.01 | 1.00 | 1011.00 | 2.29 |
| Finland | 3/4/2020 | 1.24 | 1.22 | 4.02 | 1.00 | 1012.00 | 3.83 |
| Finland | 3/5/2020 | 1.05 | 1.99 | 4.32 | 1.00 | 1004.00 | 2.06 |
| Finland | 3/6/2020 | 0.74 | 1.09 | 3.81 | 1.00 | 1010.00 | 1.96 |
| Finland | 3/7/2020 | 0.86 | 0.85 | 3.90 | 2.00 | 1012.00 | 1.26 |
| Finland | 3/8/2020 | 0.96 | 0.81 | 3.74 | 2.00 | 1014.00 | 1.54 |
| Finland | 3/9/2020 | 1.49 | 2.43 | 4.28 | 1.00 | 1012.00 | 2.83 |
| Finland | 3/10/2020 | 1.26 | 2.31 | 4.39 | 1.00 | 999.00 | 2.63 |
| Finland | 3/11/2020 | 1.21 | 1.76 | 4.29 | 1.00 | 997.00 | 2.67 |
| Finland | 3/12/2020 | 1.02 | 2.53 | 4.46 | 1.00 | 986.00 | 2.65 |
| Finland | 3/13/2020 | 0.67 | 0.45 | 3.41 | 1.00 | 999.00 | 3.25 |
| Finland | 3/14/2020 | 0.33 | -2.68 | 2.27 | 2.00 | 1018.00 | 3.05 |
| Finland | 3/15/2020 | 0.65 | 0.29 | 3.10 | 2.00 | 1019.00 | 3.97 |
| Finland | 3/16/2020 | 0.99 | 2.07 | 3.92 | 1.00 | 1017.00 | 3.24 |
| Finland | 3/17/2020 | 0.79 | 0.98 | 3.67 | 2.00 | 1020.00 | 2.88 |
| France | 2/18/2020 | 1.12 | 6.40 | 4.90 | 3.00 | 1028.00 | 5.22 |
| France | 2/19/2020 | 1.29 | 6.96 | 5.31 | 2.00 | 1026.00 | 4.42 |
| France | 2/20/2020 | 1.07 | 7.42 | 5.72 | 2.00 | 1026.00 | 5.43 |
| France | 2/21/2020 | 1.13 | 4.70 | 4.32 | 2.00 | 1031.00 | 3.66 |
| France | 2/22/2020 | 1.75 | 7.44 | 5.57 | 3.00 | 1029.00 | 5.15 |
| France | 2/23/2020 | 2.42 | 10.90 | 7.32 | 3.00 | 1029.00 | 6.38 |
| France | 2/24/2020 | 1.75 | 10.51 | 6.91 | 3.00 | 1027.00 | 5.92 |
| France | 2/25/2020 | 1.25 | 7.88 | 5.52 | 2.00 | 1012.00 | 5.62 |
| France | 2/26/2020 | 0.87 | 4.14 | 4.19 | 1.00 | 1014.00 | 5.48 |
| France | 2/27/2020 | 1.12 | 3.90 | 4.58 | 3.00 | 999.00 | 5.75 |
| France | 2/28/2020 | 1.43 | 4.78 | 5.12 | 3.00 | 1019.00 | 4.68 |
| France | 2/29/2020 | 1.18 | 8.21 | 5.62 | 2.00 | 999.00 | 7.08 |
| Gabon | 3/20/2020 | 6.04 | 27.34 | 19.19 | 6.00 | 1012.00 | 3.10 |
| Gabon | 3/21/2020 | 5.84 | 27.92 | 18.07 | 7.00 | 1011.00 | 1.40 |
| Gabon | 3/22/2020 | 6.08 | 28.32 | 19.44 | 7.00 | 1012.00 | 1.44 |
| Gabon | 3/23/2020 | 6.16 | 27.45 | 19.52 | 7.00 | 1012.00 | 3.02 |
| Gabon | 3/24/2020 | 5.69 | 27.54 | 17.65 | 7.00 | 1012.00 | 1.92 |
| Gabon | 3/25/2020 | 5.68 | 27.37 | 18.81 | 6.00 | 1012.00 | 2.97 |
| Gabon | 3/26/2020 | 5.78 | 27.02 | 18.10 | 6.00 | 1012.00 | 2.20 |
| Gabon | 3/27/2020 | 5.48 | 28.21 | 18.14 | 7.00 | 1012.00 | 1.55 |
| Gabon | 3/28/2020 | 5.66 | 28.10 | 20.47 | 7.00 | 1013.00 | 2.78 |
| Gabon | 3/29/2020 | 5.51 | 27.94 | 20.80 | 6.00 | 1012.00 | 2.70 |
| Gabon | 3/30/2020 | 5.37 | 28.14 | 19.14 | 6.00 | 1010.00 | 2.48 |
| Gabon | 3/31/2020 | 5.43 | 28.38 | 19.49 | 6.00 | 1011.00 | 2.36 |
| Gabon | 4/1/2020 | 6.47 | 27.19 | 20.08 | 6.00 | 1012.00 | 3.05 |
| Gabon | 4/2/2020 | 5.99 | 27.86 | 19.82 | 6.00 | 1011.00 | 2.43 |
| Gabon | 4/3/2020 | 6.42 | 27.91 | 20.94 | 6.00 | 1011.00 | 2.48 |
| Gabon | 4/4/2020 | 6.07 | 27.79 | 19.87 | 6.00 | 1014.00 | 2.57 |
| Gabon | 4/5/2020 | 5.83 | 28.76 | 19.65 | 7.00 | 1012.00 | 1.90 |
| Gabon | 4/6/2020 | 5.85 | 28.24 | 20.07 | 7.00 | 1013.00 | 2.22 |
| Gabon | 4/7/2020 | 5.58 | 28.16 | 19.74 | 6.00 | 1012.00 | 1.66 |
| Gabon | 4/8/2020 | 6.38 | 27.65 | 20.91 | 6.00 | 1011.00 | 2.69 |
| Gabon | 4/9/2020 | 5.96 | 27.35 | 18.73 | 6.00 | 1012.00 | 1.69 |
| Gabon | 4/10/2020 | 5.95 | 27.91 | 19.38 | 7.00 | 1012.00 | 2.62 |
| Gabon | 4/11/2020 | 6.21 | 26.97 | 19.39 | 6.00 | 1013.00 | 2.97 |
| Gabon | 4/12/2020 | 5.92 | 27.93 | 18.48 | 6.00 | 1013.00 | 1.82 |
| Gabon | 4/13/2020 | 6.43 | 27.36 | 19.51 | 6.00 | 1014.00 | 2.25 |
| Gabon | 4/14/2020 | 6.17 | 28.27 | 19.53 | 6.00 | 1012.00 | 1.77 |
| Gabon | 4/15/2020 | 6.43 | 28.29 | 20.41 | 6.00 | 1011.00 | 2.10 |
| Gabon | 4/16/2020 | 7.02 | 27.58 | 21.17 | 6.00 | 1011.00 | 2.45 |
| Gabon | 4/17/2020 | 5.97 | 27.53 | 19.95 | 6.00 | 1011.00 | 2.78 |
| Gabon | 4/18/2020 | 6.00 | 27.58 | 19.76 | 6.00 | 1012.00 | 2.50 |
| Gabon | 4/19/2020 | 5.75 | 27.77 | 18.98 | 6.00 | 1014.00 | 2.90 |
| Gabon | 4/20/2020 | 5.66 | 26.55 | 18.52 | 5.00 | 1015.00 | 2.73 |
| Gabon | 4/21/2020 | 5.34 | 27.10 | 17.87 | 7.00 | 1012.00 | 1.60 |
| Gabon | 4/22/2020 | 5.67 | 27.47 | 18.38 | 6.00 | 1013.00 | 1.04 |
| Gabon | 4/23/2020 | 5.68 | 27.92 | 19.41 | 6.00 | 1012.00 | 2.35 |
| Gabon | 4/24/2020 | 5.72 | 27.77 | 19.65 | 6.00 | 1012.00 | 1.98 |
| Gabon | 4/25/2020 | 5.75 | 27.66 | 20.30 | 6.00 | 1011.00 | 2.96 |
| Gabon | 4/26/2020 | 5.87 | 27.16 | 19.35 | 6.00 | 1013.00 | 2.21 |
| Gabon | 4/27/2020 | 5.70 | 27.57 | 18.53 | 7.00 | 1013.00 | 1.73 |
| Gabon | 4/28/2020 | 5.96 | 27.65 | 19.22 | 6.00 | 1014.00 | 1.79 |
| Gabon | 4/29/2020 | 5.95 | 27.99 | 20.16 | 6.00 | 1013.00 | 2.24 |
| Gabon | 4/30/2020 | 6.13 | 27.80 | 20.73 | 6.00 | 1014.00 | 2.80 |
| Gabon | 5/1/2020 | 6.07 | 27.43 | 20.24 | 6.00 | 1014.00 | 2.49 |
| Gabon | 5/2/2020 | 5.81 | 27.60 | 19.83 | 6.00 | 1014.00 | 2.10 |
| Gabon | 5/3/2020 | 5.93 | 27.80 | 20.05 | 6.00 | 1012.00 | 2.14 |
| Gabon | 5/4/2020 | 6.11 | 27.54 | 19.60 | 7.00 | 1012.00 | 1.49 |
| Gabon | 5/5/2020 | 6.10 | 27.54 | 19.36 | 7.00 | 1013.00 | 1.72 |
| Georgia | 2/28/2020 | 0.83 | 7.78 | 3.59 | 3.00 | 1016.00 | 2.70 |
| Georgia | 2/29/2020 | 1.19 | 5.43 | 4.31 | 3.00 | 1016.00 | 1.81 |
| Georgia | 3/1/2020 | 1.08 | 4.77 | 4.02 | 2.00 | 1020.00 | 2.96 |
| Georgia | 3/2/2020 | 1.17 | 2.83 | 3.91 | 3.00 | 1022.00 | 1.51 |
| Georgia | 3/3/2020 | 0.74 | 6.01 | 3.24 | 2.00 | 1024.00 | 1.11 |
| Georgia | 3/4/2020 | 0.67 | 8.94 | 3.16 | 4.00 | 1025.00 | 0.66 |
| Georgia | 3/5/2020 | 0.59 | 10.19 | 3.14 | 4.00 | 1021.00 | 0.66 |
| Georgia | 3/6/2020 | 0.90 | 8.63 | 4.01 | 4.00 | 1019.00 | 1.21 |
| Georgia | 3/7/2020 | 1.14 | 7.31 | 4.06 | 4.00 | 1024.00 | 1.63 |
| Georgia | 3/8/2020 | 1.13 | 9.15 | 4.29 | 4.00 | 1025.00 | 0.70 |
| Georgia | 3/9/2020 | 0.80 | 8.84 | 4.28 | 4.00 | 1026.00 | 1.11 |
| Georgia | 3/10/2020 | 0.71 | 9.06 | 3.51 | 4.00 | 1022.00 | 1.09 |
| Georgia | 3/11/2020 | 1.17 | 12.23 | 4.21 | 5.00 | 1019.00 | 2.21 |
| Georgia | 3/12/2020 | 1.28 | 12.38 | 5.12 | 5.00 | 1020.00 | 2.88 |
| Georgia | 3/13/2020 | 1.55 | 11.47 | 5.43 | 3.00 | 1021.00 | 1.71 |
| Georgia | 3/14/2020 | 1.85 | 9.63 | 6.06 | 4.00 | 1015.00 | 1.47 |
| Georgia | 3/15/2020 | 1.41 | 10.22 | 4.87 | 3.00 | 1014.00 | 2.27 |
| Georgia | 3/16/2020 | 1.33 | 4.94 | 4.32 | 3.00 | 1021.00 | 2.50 |
| Georgia | 3/17/2020 | 1.18 | 0.91 | 3.51 | 1.00 | 1024.00 | 2.14 |
| Georgia | 3/18/2020 | 1.30 | -0.28 | 3.56 | 1.00 | 1024.00 | 2.37 |
| Georgia | 3/19/2020 | 1.02 | 3.87 | 3.95 | 2.00 | 1017.00 | 0.65 |
| Georgia | 3/20/2020 | 1.06 | 5.51 | 4.16 | 4.00 | 1016.00 | 0.62 |
| Georgia | 3/21/2020 | 1.09 | 4.94 | 4.46 | 3.00 | 1020.00 | 1.20 |
| Georgia | 3/22/2020 | 0.98 | 4.87 | 4.32 | 2.00 | 1020.00 | 1.41 |
| Georgia | 3/23/2020 | 0.83 | 7.98 | 4.03 | 4.00 | 1019.00 | 1.08 |
| Georgia | 3/24/2020 | 1.26 | 7.30 | 4.89 | 4.00 | 1019.00 | 1.31 |
| Georgia | 3/25/2020 | 1.45 | 8.01 | 5.06 | 3.00 | 1022.00 | 1.51 |
| Georgia | 3/26/2020 | 1.03 | 7.06 | 3.82 | 4.00 | 1024.00 | 1.84 |
| Georgia | 3/27/2020 | 0.88 | 7.00 | 3.26 | 4.00 | 1024.00 | 1.40 |
| Georgia | 3/28/2020 | 1.33 | 8.28 | 4.10 | 4.00 | 1020.00 | 1.68 |
| Georgia | 3/29/2020 | 1.51 | 4.78 | 4.60 | 4.00 | 1012.00 | 1.16 |
| Georgia | 3/30/2020 | 1.31 | 5.65 | 4.68 | 3.00 | 1007.00 | 3.14 |
| Georgia | 3/31/2020 | 1.11 | 7.90 | 4.37 | 3.00 | 1014.00 | 3.84 |
| Georgia | 4/1/2020 | 1.33 | 8.88 | 4.83 | 4.00 | 1015.00 | 1.33 |
| Georgia | 4/2/2020 | 1.14 | 6.63 | 4.48 | 2.00 | 1014.00 | 3.02 |
| Georgia | 4/3/2020 | 1.17 | 7.98 | 4.45 | 4.00 | 1015.00 | 0.92 |
| Georgia | 4/4/2020 | 1.48 | 7.29 | 5.46 | 2.00 | 1024.00 | 1.37 |
| Georgia | 4/5/2020 | 1.33 | 10.57 | 5.06 | 3.00 | 1025.00 | 0.63 |
| Germany | 2/16/2020 | 2.41 | 10.38 | 6.40 | 3.81 | 1011.08 | 8.31 |
| Germany | 2/17/2020 | 1.08 | 7.50 | 5.21 | 2.39 | 1013.06 | 6.31 |
| Germany | 2/18/2020 | 1.09 | 5.62 | 4.85 | 2.39 | 1018.67 | 6.08 |
| Germany | 2/19/2020 | 1.04 | 4.34 | 4.56 | 2.00 | 1019.05 | 5.30 |
| Germany | 2/20/2020 | 1.30 | 4.76 | 4.90 | 1.80 | 1019.42 | 5.39 |
| Germany | 2/21/2020 | 1.22 | 4.91 | 4.54 | 2.00 | 1023.25 | 5.81 |
| Germany | 2/22/2020 | 1.66 | 6.62 | 5.20 | 2.00 | 1021.45 | 7.32 |
| Germany | 2/23/2020 | 1.61 | 6.45 | 5.61 | 2.41 | 1017.52 | 6.33 |
| Germany | 2/24/2020 | 1.35 | 4.88 | 5.02 | 2.22 | 1016.84 | 5.02 |
| Germany | 2/25/2020 | 0.96 | 6.33 | 5.14 | 2.00 | 998.42 | 5.74 |
| Germany | 2/26/2020 | 0.93 | 2.49 | 4.29 | 1.59 | 998.61 | 4.02 |
| Germany | 2/27/2020 | 0.87 | 2.03 | 4.04 | 1.00 | 999.00 | 3.01 |
| Germany | 2/28/2020 | 0.90 | 2.54 | 3.97 | 1.19 | 1014.86 | 4.28 |
| Germany | 2/29/2020 | 1.32 | 5.73 | 5.31 | 2.81 | 999.00 | 5.81 |
| Germany | 3/1/2020 | 1.05 | 5.24 | 4.77 | 2.61 | 995.63 | 4.87 |
| Germany | 3/2/2020 | 1.15 | 4.85 | 4.90 | 2.59 | 993.60 | 2.91 |
| Germany | 3/3/2020 | 1.07 | 3.67 | 4.56 | 2.39 | 999.00 | 2.90 |
| Germany | 3/4/2020 | 0.97 | 3.29 | 4.37 | 2.39 | 1010.03 | 2.82 |
| Germany | 3/5/2020 | 1.22 | 3.92 | 4.46 | 2.39 | 999.00 | 3.40 |
| Germany | 3/6/2020 | 1.41 | 3.83 | 4.81 | 1.81 | 997.82 | 3.58 |
| Germany | 3/7/2020 | 1.05 | 3.89 | 4.38 | 2.00 | 1010.16 | 3.58 |
| Germany | 3/8/2020 | 1.39 | 5.80 | 5.09 | 2.19 | 1018.01 | 4.25 |
| Germany | 3/9/2020 | 1.29 | 6.36 | 5.32 | 2.00 | 1013.84 | 3.12 |
| Ghana | 3/15/2020 | 4.62 | 28.57 | 18.43 | 7.00 | 1011.00 | 2.48 |
| Ghana | 3/16/2020 | 4.19 | 28.59 | 18.91 | 7.00 | 1011.00 | 2.43 |
| Ghana | 3/17/2020 | 5.03 | 28.46 | 19.61 | 7.00 | 1011.00 | 2.51 |
| Ghana | 3/18/2020 | 5.18 | 28.90 | 18.95 | 7.00 | 1012.00 | 2.81 |
| Ghana | 3/19/2020 | 5.14 | 28.06 | 19.03 | 7.00 | 1011.00 | 3.33 |
| Ghana | 3/20/2020 | 4.06 | 27.97 | 16.88 | 6.00 | 1011.00 | 2.51 |
| Ghana | 3/21/2020 | 4.77 | 27.66 | 18.96 | 6.00 | 1011.00 | 1.95 |
| Ghana | 3/22/2020 | 5.35 | 27.56 | 19.16 | 6.00 | 1011.00 | 2.29 |
| Ghana | 3/23/2020 | 5.71 | 26.68 | 18.99 | 6.00 | 1012.00 | 2.81 |
| Ghana | 3/24/2020 | 5.82 | 26.10 | 17.96 | 6.00 | 1013.00 | 2.54 |
| Ghana | 3/25/2020 | 5.59 | 26.40 | 17.74 | 6.00 | 1012.00 | 2.15 |
| Ghana | 3/26/2020 | 5.47 | 27.54 | 18.09 | 6.00 | 1012.00 | 1.91 |
| Ghana | 3/27/2020 | 5.36 | 27.73 | 18.44 | 7.00 | 1011.00 | 1.46 |
| Ghana | 3/28/2020 | 5.00 | 27.20 | 18.79 | 7.00 | 1012.00 | 1.87 |
| Ghana | 3/29/2020 | 4.85 | 27.40 | 18.76 | 7.00 | 1012.00 | 2.07 |
| Ghana | 3/30/2020 | 5.24 | 27.48 | 18.33 | 7.00 | 1010.00 | 2.23 |
| Ghana | 3/31/2020 | 5.29 | 27.27 | 18.28 | 6.00 | 1010.00 | 1.78 |
| Ghana | 4/1/2020 | 5.93 | 26.50 | 19.03 | 6.00 | 1011.00 | 1.89 |
| Ghana | 4/2/2020 | 5.11 | 27.32 | 18.13 | 7.00 | 1011.00 | 2.16 |
| Ghana | 4/3/2020 | 5.42 | 27.32 | 18.83 | 7.00 | 1011.00 | 2.26 |
| Ghana | 4/4/2020 | 5.29 | 26.60 | 19.17 | 7.00 | 1013.00 | 2.47 |
| Ghana | 4/5/2020 | 4.82 | 28.53 | 17.66 | 7.00 | 1013.00 | 1.43 |
| Ghana | 4/6/2020 | 5.32 | 28.07 | 19.15 | 7.00 | 1013.00 | 2.68 |
| Ghana | 4/7/2020 | 5.23 | 27.73 | 19.11 | 7.00 | 1012.00 | 2.10 |
| Ghana | 4/8/2020 | 5.41 | 28.08 | 18.53 | 6.00 | 1011.00 | 2.37 |
| Ghana | 4/9/2020 | 5.48 | 27.78 | 18.71 | 6.00 | 1011.00 | 1.87 |
| Ghana | 4/10/2020 | 5.37 | 28.42 | 18.59 | 7.00 | 1012.00 | 1.23 |
| Ghana | 4/11/2020 | 5.67 | 27.59 | 19.13 | 6.00 | 1013.00 | 1.29 |
| Ghana | 4/12/2020 | 4.96 | 28.11 | 18.67 | 6.00 | 1013.00 | 1.73 |
| Ghana | 4/13/2020 | 4.80 | 28.25 | 18.96 | 7.00 | 1012.00 | 2.31 |
| Ghana | 4/14/2020 | 5.25 | 28.33 | 19.40 | 6.00 | 1012.00 | 2.29 |
| Ghana | 4/15/2020 | 5.12 | 28.87 | 19.24 | 7.00 | 1011.00 | 2.06 |
| Ghana | 4/16/2020 | 5.79 | 29.18 | 19.79 | 8.00 | 1011.00 | 2.17 |
| Ghana | 4/17/2020 | 5.39 | 28.22 | 19.63 | 6.00 | 1011.00 | 2.92 |
| Ghana | 4/18/2020 | 5.08 | 28.15 | 19.13 | 6.00 | 1012.00 | 2.76 |
| Ghana | 4/19/2020 | 5.42 | 26.98 | 19.25 | 7.00 | 1014.00 | 2.63 |
| Ghana | 4/20/2020 | 5.29 | 27.95 | 18.02 | 6.00 | 1014.00 | 2.22 |
| Ghana | 4/21/2020 | 4.43 | 27.74 | 17.59 | 8.00 | 1013.00 | 1.87 |
| Ghana | 4/22/2020 | 5.32 | 27.37 | 19.19 | 7.00 | 1012.00 | 2.16 |
| Ghana | 4/23/2020 | 5.43 | 27.04 | 19.13 | 6.00 | 1011.00 | 1.94 |
| Ghana | 4/24/2020 | 5.15 | 27.80 | 18.53 | 7.00 | 1012.00 | 2.19 |
| Ghana | 4/25/2020 | 5.22 | 27.27 | 18.51 | 7.00 | 1012.00 | 2.75 |
| Ghana | 4/26/2020 | 5.69 | 26.21 | 18.23 | 6.00 | 1012.00 | 2.01 |
| Ghana | 4/27/2020 | 5.32 | 26.91 | 18.07 | 7.00 | 1013.00 | 1.41 |
| Greece | 3/4/2020 | 1.72 | 14.73 | 8.44 | 5.00 | 1010.00 | 5.71 |
| Greece | 3/5/2020 | 1.42 | 13.39 | 6.16 | 4.00 | 1009.00 | 2.09 |
| Greece | 3/6/2020 | 1.47 | 13.15 | 6.95 | 4.00 | 1012.00 | 3.40 |
| Greece | 3/7/2020 | 2.17 | 14.01 | 8.80 | 5.00 | 1013.00 | 4.90 |
| Greece | 3/8/2020 | 2.14 | 14.92 | 9.80 | 4.00 | 1015.00 | 3.64 |
| Greece | 3/9/2020 | 2.10 | 12.34 | 7.70 | 3.00 | 1013.00 | 3.23 |
| Greece | 3/10/2020 | 1.34 | 12.88 | 6.54 | 4.00 | 1015.00 | 3.93 |
| Greece | 3/11/2020 | 1.50 | 13.40 | 6.82 | 4.00 | 1020.00 | 2.55 |
| Greece | 3/12/2020 | 1.16 | 13.67 | 7.06 | 5.00 | 1020.00 | 4.89 |
| Greece | 3/13/2020 | 1.05 | 15.34 | 7.95 | 5.00 | 1016.00 | 4.27 |
| Greece | 3/14/2020 | 1.53 | 15.95 | 8.00 | 5.00 | 1014.00 | 1.45 |
| Greece | 3/15/2020 | 1.98 | 14.07 | 8.54 | 5.00 | 1019.00 | 4.29 |
| Greece | 3/16/2020 | 1.17 | 9.83 | 5.59 | 2.00 | 1030.00 | 9.40 |
| Greece | 3/17/2020 | 0.51 | 10.04 | 4.92 | 4.00 | 1030.00 | 6.56 |
| Greece | 3/18/2020 | 1.00 | 11.74 | 5.53 | 4.00 | 1026.00 | 4.11 |
| Greece | 3/19/2020 | 0.64 | 11.00 | 4.96 | 3.00 | 1027.00 | 5.48 |
| Greece | 3/20/2020 | 1.16 | 12.96 | 5.80 | 5.00 | 1024.00 | 3.95 |
| Greece | 3/21/2020 | 1.24 | 14.13 | 7.05 | 5.00 | 1020.00 | 2.40 |
| Greece | 3/22/2020 | 1.65 | 13.95 | 8.26 | 4.00 | 1015.00 | 3.24 |
| Guatemala | 3/12/2020 | 2.10 | 19.74 | 10.23 | 5.00 | 1015.00 | 1.30 |
| Guatemala | 3/13/2020 | 2.24 | 20.98 | 11.53 | 5.00 | 1017.00 | 1.00 |
| Guatemala | 3/14/2020 | 2.29 | 21.06 | 12.25 | 5.00 | 1018.00 | 1.07 |
| Guatemala | 3/15/2020 | 2.12 | 20.86 | 11.64 | 6.00 | 1018.00 | 1.91 |
| Guatemala | 3/16/2020 | 1.80 | 20.04 | 10.66 | 6.00 | 1018.00 | 1.86 |
| Guatemala | 3/17/2020 | 1.92 | 20.09 | 10.62 | 6.00 | 1018.00 | 1.42 |
| Guatemala | 3/18/2020 | 2.10 | 20.86 | 11.63 | 6.00 | 1018.00 | 1.27 |
| Guatemala | 3/19/2020 | 2.17 | 21.53 | 11.82 | 5.00 | 1019.00 | 1.33 |
| Guatemala | 3/20/2020 | 2.24 | 21.43 | 12.11 | 5.00 | 1019.00 | 1.06 |
| Guatemala | 3/21/2020 | 1.98 | 21.12 | 11.47 | 5.00 | 1019.00 | 2.01 |
| Guatemala | 3/22/2020 | 2.00 | 20.51 | 10.89 | 5.00 | 1018.00 | 2.32 |
| Guatemala | 3/23/2020 | 1.87 | 20.98 | 10.92 | 6.00 | 1019.00 | 1.83 |
| Guatemala | 3/24/2020 | 2.05 | 21.37 | 11.23 | 6.00 | 1018.00 | 1.16 |
| Guatemala | 3/25/2020 | 1.92 | 22.04 | 11.54 | 7.00 | 1016.00 | 1.22 |
| Guatemala | 3/26/2020 | 2.08 | 22.16 | 11.70 | 7.00 | 1015.00 | 1.33 |
| Guatemala | 3/27/2020 | 2.07 | 22.68 | 12.14 | 7.00 | 1014.00 | 1.53 |
| Guatemala | 3/28/2020 | 2.18 | 22.96 | 12.79 | 6.00 | 1014.00 | 1.16 |
| Guatemala | 3/29/2020 | 2.22 | 22.45 | 12.63 | 6.00 | 1015.00 | 1.15 |
| Guatemala | 3/30/2020 | 2.53 | 21.42 | 12.47 | 5.00 | 1016.00 | 1.17 |
| Guatemala | 3/31/2020 | 2.44 | 22.29 | 12.62 | 5.00 | 1014.00 | 1.06 |
| Guatemala | 4/1/2020 | 2.96 | 22.41 | 13.06 | 6.00 | 1014.00 | 1.04 |
| Guatemala | 4/2/2020 | 3.23 | 22.11 | 13.40 | 6.00 | 1015.00 | 0.97 |
| Guatemala | 4/3/2020 | 3.01 | 22.54 | 12.92 | 5.00 | 1014.00 | 1.05 |
| Guatemala | 4/4/2020 | 2.85 | 23.06 | 12.80 | 6.00 | 1013.00 | 0.71 |
| Guatemala | 4/5/2020 | 2.68 | 23.48 | 12.76 | 6.00 | 1014.00 | 0.94 |
| Guatemala | 4/6/2020 | 2.54 | 23.32 | 13.18 | 7.00 | 1016.00 | 1.32 |
| Guatemala | 4/7/2020 | 2.04 | 23.50 | 11.55 | 7.00 | 1017.00 | 1.50 |
| Guatemala | 4/8/2020 | 2.10 | 23.65 | 11.27 | 7.00 | 1016.00 | 1.53 |
| Guatemala | 4/9/2020 | 2.41 | 24.65 | 11.83 | 7.00 | 1015.00 | 1.26 |
| Guatemala | 4/10/2020 | 2.83 | 24.32 | 12.77 | 7.00 | 1014.00 | 0.92 |
| Guatemala | 4/11/2020 | 2.98 | 23.66 | 13.59 | 5.00 | 1014.00 | 1.29 |
| Guatemala | 4/12/2020 | 2.86 | 23.01 | 14.55 | 5.00 | 1015.00 | 1.88 |
| Guatemala | 4/13/2020 | 2.57 | 24.26 | 13.23 | 6.00 | 1016.00 | 1.58 |
| Guatemala | 4/14/2020 | 3.17 | 23.67 | 13.66 | 6.00 | 1016.00 | 1.09 |
| Guatemala | 4/15/2020 | 3.10 | 23.32 | 13.89 | 6.00 | 1015.00 | 0.84 |
| Guinea | 3/25/2020 | 4.40 | 27.70 | 15.49 | 6.00 | 1013.00 | 2.61 |
| Guinea | 3/26/2020 | 5.03 | 28.28 | 16.90 | 7.00 | 1011.00 | 3.26 |
| Guinea | 3/27/2020 | 4.86 | 29.02 | 15.97 | 7.00 | 1012.00 | 2.79 |
| Guinea | 3/28/2020 | 4.89 | 29.20 | 16.30 | 7.00 | 1012.00 | 2.67 |
| Guinea | 3/29/2020 | 5.00 | 30.36 | 16.00 | 6.00 | 1013.00 | 2.85 |
| Guinea | 3/30/2020 | 4.76 | 29.87 | 15.22 | 7.00 | 1011.00 | 3.52 |
| Guinea | 3/31/2020 | 4.55 | 30.12 | 15.54 | 6.00 | 1011.00 | 3.51 |
| Guinea | 4/1/2020 | 4.44 | 29.72 | 16.34 | 7.00 | 1012.00 | 3.33 |
| Guinea | 4/2/2020 | 5.10 | 29.20 | 18.14 | 7.00 | 1013.00 | 3.51 |
| Guinea | 4/3/2020 | 5.10 | 29.31 | 18.12 | 7.00 | 1012.00 | 3.91 |
| Guinea | 4/4/2020 | 4.82 | 29.56 | 17.97 | 7.00 | 1012.00 | 3.13 |
| Guinea | 4/5/2020 | 5.05 | 30.01 | 18.53 | 7.00 | 1013.00 | 3.06 |
| Guinea | 4/6/2020 | 4.54 | 29.99 | 17.52 | 7.00 | 1013.00 | 3.38 |
| Guinea | 4/7/2020 | 4.73 | 29.63 | 18.10 | 7.00 | 1013.00 | 3.29 |
| Guinea | 4/8/2020 | 4.46 | 29.86 | 17.56 | 7.00 | 1012.00 | 3.27 |
| Guinea | 4/9/2020 | 3.86 | 29.32 | 17.21 | 7.00 | 1011.00 | 3.50 |
| Guinea | 4/10/2020 | 4.16 | 29.21 | 17.84 | 7.00 | 1012.00 | 3.31 |
| Guinea | 4/11/2020 | 4.99 | 29.33 | 18.46 | 7.00 | 1013.00 | 2.99 |
| Guinea | 4/12/2020 | 5.32 | 29.25 | 17.93 | 6.00 | 1013.00 | 2.74 |
| Guinea | 4/13/2020 | 4.90 | 29.25 | 17.10 | 6.00 | 1013.00 | 2.31 |
| Guyana | 3/21/2020 | 4.76 | 26.40 | 16.59 | 6.00 | 1014.00 | 2.12 |
| Guyana | 3/22/2020 | 4.24 | 27.05 | 16.04 | 7.00 | 1015.00 | 2.22 |
| Guyana | 3/23/2020 | 4.40 | 27.35 | 17.07 | 6.00 | 1016.00 | 1.98 |
| Guyana | 3/24/2020 | 4.28 | 26.99 | 16.37 | 6.00 | 1016.00 | 2.10 |
| Guyana | 3/25/2020 | 4.38 | 26.73 | 16.37 | 6.00 | 1014.00 | 2.09 |
| Guyana | 3/26/2020 | 4.28 | 27.05 | 15.38 | 7.00 | 1013.00 | 1.94 |
| Guyana | 3/27/2020 | 4.74 | 27.01 | 16.16 | 6.00 | 1012.00 | 1.69 |
| Guyana | 3/28/2020 | 5.80 | 26.91 | 18.20 | 7.00 | 1012.00 | 2.18 |
| Guyana | 3/29/2020 | 5.91 | 26.69 | 18.59 | 5.00 | 1014.00 | 2.66 |
| Guyana | 3/30/2020 | 5.58 | 26.24 | 17.83 | 6.00 | 1014.00 | 2.51 |
| Guyana | 3/31/2020 | 4.84 | 26.05 | 16.32 | 6.00 | 1013.00 | 2.13 |
| Guyana | 4/1/2020 | 4.71 | 26.40 | 16.22 | 6.00 | 1013.00 | 1.80 |
| Guyana | 4/2/2020 | 5.24 | 27.10 | 17.18 | 7.00 | 1015.00 | 1.85 |
| Guyana | 4/3/2020 | 5.65 | 27.32 | 18.33 | 7.00 | 1014.00 | 2.07 |
| Guyana | 4/4/2020 | 4.75 | 27.63 | 17.86 | 6.00 | 1013.00 | 2.09 |
| Guyana | 4/5/2020 | 4.56 | 27.99 | 18.16 | 6.00 | 1013.00 | 2.43 |
| Guyana | 4/6/2020 | 4.72 | 28.07 | 18.01 | 6.00 | 1015.00 | 2.39 |
| Haiti | 4/25/2020 | 3.36 | 29.01 | 14.41 | 6.00 | 1018.00 | 1.01 |
| Haiti | 4/26/2020 | 3.28 | 28.85 | 14.45 | 6.00 | 1018.00 | 1.13 |
| Haiti | 4/27/2020 | 3.30 | 28.61 | 14.43 | 7.00 | 1017.00 | 1.94 |
| Haiti | 4/28/2020 | 3.63 | 28.18 | 14.26 | 6.00 | 1017.00 | 1.54 |
| Haiti | 4/29/2020 | 3.38 | 28.35 | 13.67 | 6.00 | 1016.00 | 1.44 |
| Haiti | 4/30/2020 | 3.91 | 28.06 | 14.18 | 6.00 | 1017.00 | 1.60 |
| Haiti | 5/1/2020 | 3.98 | 28.31 | 13.89 | 6.00 | 1017.00 | 1.88 |
| Haiti | 5/2/2020 | 4.49 | 27.95 | 14.81 | 6.00 | 1018.00 | 1.05 |
| Haiti | 5/3/2020 | 4.74 | 27.44 | 14.42 | 6.00 | 1016.00 | 1.61 |
| Haiti | 5/4/2020 | 4.60 | 28.04 | 14.70 | 6.00 | 1015.00 | 1.60 |
| Haiti | 5/5/2020 | 4.83 | 28.21 | 15.26 | 6.00 | 1015.00 | 1.47 |
| Haiti | 5/6/2020 | 4.66 | 28.30 | 14.69 | 6.00 | 1015.00 | 1.68 |
| Haiti | 5/7/2020 | 4.29 | 28.94 | 15.13 | 7.00 | 1016.00 | 1.52 |
| Haiti | 5/8/2020 | 3.69 | 29.17 | 14.64 | 7.00 | 1017.00 | 2.14 |
| Haiti | 5/9/2020 | 3.20 | 28.82 | 14.27 | 7.00 | 1017.00 | 2.76 |
| Haiti | 5/10/2020 | 2.99 | 28.93 | 13.92 | 8.00 | 1017.00 | 2.89 |
| Haiti | 5/11/2020 | 2.99 | 28.42 | 13.82 | 7.00 | 1016.00 | 2.16 |
| Haiti | 5/12/2020 | 4.34 | 28.10 | 15.17 | 6.00 | 1016.00 | 0.93 |
| Haiti | 5/13/2020 | 4.92 | 27.78 | 15.16 | 6.00 | 1016.00 | 1.55 |
| Haiti | 5/14/2020 | 4.78 | 26.93 | 15.69 | 6.00 | 1016.00 | 1.86 |
| Haiti | 5/15/2020 | 3.86 | 27.14 | 15.16 | 7.00 | 1016.00 | 2.42 |
| Haiti | 5/16/2020 | 3.24 | 27.51 | 13.92 | 6.00 | 1015.00 | 2.35 |
| Haiti | 5/17/2020 | 3.35 | 28.41 | 14.54 | 7.00 | 1016.00 | 1.70 |
| Haiti | 5/18/2020 | 3.56 | 28.52 | 15.30 | 7.00 | 1016.00 | 1.57 |
| Haiti | 5/19/2020 | 3.79 | 28.21 | 14.78 | 7.00 | 1016.00 | 1.35 |
| Haiti | 5/20/2020 | 4.72 | 27.54 | 15.83 | 6.00 | 1016.00 | 1.08 |
| Haiti | 5/21/2020 | 4.81 | 25.96 | 14.85 | 6.00 | 1016.00 | 1.89 |
| Haiti | 5/22/2020 | 4.34 | 27.72 | 15.47 | 8.00 | 1016.00 | 1.84 |
| Honduras | 3/4/2020 | 2.01 | 24.60 | 10.30 | 7.00 | 1016.00 | 1.64 |
| Honduras | 3/5/2020 | 2.20 | 25.65 | 11.19 | 7.00 | 1016.00 | 1.70 |
| Honduras | 3/6/2020 | 2.18 | 23.99 | 12.56 | 6.00 | 1018.00 | 3.72 |
| Honduras | 3/7/2020 | 1.99 | 21.59 | 11.61 | 5.00 | 1018.00 | 4.22 |
| Honduras | 3/8/2020 | 2.65 | 21.94 | 12.37 | 5.00 | 1018.00 | 3.51 |
| Honduras | 3/9/2020 | 2.81 | 22.85 | 12.90 | 5.00 | 1019.00 | 3.78 |
| Honduras | 3/10/2020 | 2.72 | 23.29 | 12.62 | 5.00 | 1020.00 | 3.64 |
| Honduras | 3/11/2020 | 2.66 | 23.09 | 12.26 | 5.00 | 1018.00 | 3.38 |
| Honduras | 3/12/2020 | 2.64 | 23.64 | 11.10 | 5.00 | 1015.00 | 2.47 |
| Honduras | 3/13/2020 | 2.35 | 24.54 | 10.76 | 6.00 | 1017.00 | 1.60 |
| Honduras | 3/14/2020 | 2.20 | 24.70 | 10.72 | 6.00 | 1019.00 | 2.03 |
| Honduras | 3/15/2020 | 2.19 | 25.08 | 11.24 | 6.00 | 1019.00 | 2.47 |
| Honduras | 3/16/2020 | 2.13 | 24.60 | 11.16 | 6.00 | 1018.00 | 2.70 |
| Honduras | 3/17/2020 | 2.19 | 24.66 | 10.79 | 6.00 | 1018.00 | 2.50 |
| Honduras | 3/18/2020 | 2.29 | 25.25 | 11.43 | 6.00 | 1018.00 | 2.43 |
| Honduras | 3/19/2020 | 1.95 | 24.84 | 10.32 | 6.00 | 1019.00 | 2.23 |
| Honduras | 3/20/2020 | 2.15 | 25.29 | 10.46 | 7.00 | 1020.00 | 2.00 |
| Honduras | 3/21/2020 | 2.34 | 25.25 | 10.93 | 7.00 | 1019.00 | 2.70 |
| Hungary | 2/25/2020 | 1.32 | 7.84 | 5.24 | 2.00 | 1017.00 | 4.03 |
| Hungary | 2/26/2020 | 1.32 | 6.10 | 5.31 | 2.00 | 999.00 | 2.98 |
| Hungary | 2/27/2020 | 0.72 | 3.22 | 3.57 | 3.00 | 1007.00 | 4.07 |
| Hungary | 2/28/2020 | 0.91 | 3.03 | 3.77 | 3.00 | 1020.00 | 5.29 |
| Hungary | 2/29/2020 | 1.07 | 3.17 | 3.45 | 3.00 | 1020.00 | 3.82 |
| Hungary | 3/1/2020 | 1.73 | 7.81 | 5.43 | 3.00 | 1009.00 | 3.77 |
| Hungary | 3/2/2020 | 1.42 | 9.55 | 5.84 | 3.00 | 999.00 | 4.42 |
| Hungary | 3/3/2020 | 1.89 | 9.91 | 6.78 | 3.00 | 999.00 | 3.01 |
| Hungary | 3/4/2020 | 1.33 | 4.59 | 4.54 | 2.00 | 1012.00 | 3.69 |
| Hungary | 3/5/2020 | 0.67 | 3.88 | 3.51 | 3.00 | 1013.00 | 1.81 |
| Hungary | 3/6/2020 | 1.63 | 4.14 | 4.88 | 2.00 | 1004.00 | 2.76 |
| Hungary | 3/7/2020 | 1.12 | 6.00 | 4.69 | 2.00 | 1017.00 | 3.96 |
| Hungary | 3/8/2020 | 0.93 | 4.82 | 4.01 | 3.00 | 1020.00 | 2.66 |
| Hungary | 3/9/2020 | 0.65 | 5.22 | 3.33 | 4.00 | 1018.00 | 0.93 |
| Hungary | 3/10/2020 | 1.04 | 5.93 | 4.17 | 3.00 | 1016.00 | 2.38 |
| Hungary | 3/11/2020 | 2.19 | 9.41 | 6.52 | 4.00 | 1019.00 | 3.50 |
| Hungary | 3/12/2020 | 1.75 | 11.90 | 6.67 | 4.00 | 1019.00 | 3.02 |
| Hungary | 3/13/2020 | 1.50 | 8.14 | 5.41 | 4.00 | 1020.00 | 3.44 |
| Iceland | 2/22/2020 | 0.58 | -2.89 | 2.79 | 2.00 | 991.00 | 3.18 |
| Iceland | 2/23/2020 | 0.62 | -0.89 | 3.16 | 2.00 | 998.00 | 4.98 |
| Iceland | 2/24/2020 | 0.74 | -0.36 | 3.49 | 1.00 | 999.00 | 3.33 |
| Iceland | 2/25/2020 | 0.59 | -3.76 | 2.63 | 1.00 | 992.00 | 3.51 |
| Iceland | 2/26/2020 | 0.46 | -4.47 | 2.31 | 2.00 | 996.00 | 5.61 |
| Iceland | 2/27/2020 | 0.60 | -4.56 | 2.46 | 2.00 | 996.00 | 6.99 |
| Iceland | 2/28/2020 | 0.71 | -1.25 | 3.30 | 1.00 | 988.00 | 3.25 |
| Iceland | 2/29/2020 | 0.68 | -2.25 | 3.10 | 2.00 | 983.00 | 3.16 |
| Iceland | 3/1/2020 | 0.75 | 0.14 | 3.43 | 2.00 | 983.00 | 6.54 |
| Iceland | 3/2/2020 | 0.66 | 0.26 | 3.45 | 1.00 | 993.00 | 7.40 |
| Iceland | 3/3/2020 | 0.62 | -2.58 | 2.86 | 2.00 | 994.00 | 3.19 |
| Iceland | 3/4/2020 | 0.59 | -2.56 | 2.86 | 2.00 | 984.00 | 2.12 |
| Iceland | 3/5/2020 | 0.54 | -4.03 | 2.68 | 1.00 | 999.00 | 4.53 |
| Iceland | 3/6/2020 | 0.38 | -5.44 | 2.30 | 2.00 | 999.00 | 3.70 |
| Iceland | 3/7/2020 | 0.44 | -4.72 | 2.44 | 2.00 | 996.00 | 4.61 |
| Iceland | 3/8/2020 | 0.38 | -5.09 | 2.30 | 1.00 | 997.00 | 3.58 |
| Iceland | 3/9/2020 | 0.57 | -4.96 | 2.45 | 2.00 | 998.00 | 5.10 |
| Iceland | 3/10/2020 | 0.69 | -2.46 | 2.91 | 1.00 | 986.00 | 6.05 |
| Iceland | 3/11/2020 | 0.75 | -1.31 | 3.11 | 1.00 | 992.00 | 7.39 |
| India | 2/23/2020 | 2.27 | 23.53 | 9.56 | 6.84 | 1017.62 | 2.08 |
| India | 2/24/2020 | 2.14 | 23.59 | 9.64 | 7.18 | 1014.73 | 2.29 |
| India | 2/25/2020 | 2.05 | 23.43 | 9.45 | 6.59 | 1014.31 | 2.39 |
| India | 2/26/2020 | 1.97 | 23.42 | 9.29 | 7.00 | 1013.97 | 2.21 |
| India | 2/27/2020 | 1.96 | 23.27 | 9.08 | 7.24 | 1015.35 | 2.11 |
| India | 2/28/2020 | 2.17 | 23.87 | 9.53 | 6.92 | 1013.65 | 2.34 |
| India | 2/29/2020 | 2.42 | 23.97 | 10.28 | 6.83 | 1012.67 | 2.32 |
| India | 3/1/2020 | 2.47 | 24.24 | 10.75 | 6.43 | 1013.09 | 2.09 |
| India | 3/2/2020 | 2.42 | 24.40 | 10.52 | 7.06 | 1013.99 | 1.80 |
| India | 3/3/2020 | 2.23 | 24.47 | 10.64 | 7.00 | 1013.20 | 1.97 |
| India | 3/4/2020 | 2.42 | 24.42 | 11.04 | 6.92 | 1013.11 | 2.04 |
| India | 3/5/2020 | 2.53 | 23.87 | 11.90 | 6.36 | 1012.22 | 2.41 |
| India | 3/6/2020 | 2.66 | 23.07 | 12.09 | 5.78 | 1012.21 | 3.06 |
| India | 3/7/2020 | 2.50 | 22.79 | 11.05 | 5.83 | 1015.14 | 2.23 |
| India | 3/8/2020 | 2.27 | 23.47 | 10.55 | 6.59 | 1015.08 | 2.09 |
| India | 3/9/2020 | 2.01 | 23.26 | 10.17 | 6.61 | 1013.16 | 2.50 |
| India | 3/10/2020 | 2.15 | 23.66 | 10.22 | 6.91 | 1013.00 | 2.43 |
| India | 3/11/2020 | 2.43 | 23.83 | 11.08 | 6.44 | 1012.82 | 2.58 |
| India | 3/12/2020 | 2.32 | 24.06 | 11.11 | 6.39 | 1014.62 | 2.38 |
| India | 3/13/2020 | 1.94 | 23.95 | 9.58 | 6.91 | 1015.24 | 2.34 |
| India | 3/14/2020 | 1.84 | 23.76 | 9.08 | 6.80 | 1014.99 | 2.72 |
| India | 3/15/2020 | 2.09 | 23.97 | 9.20 | 6.96 | 1015.10 | 2.57 |
| India | 3/16/2020 | 2.44 | 25.22 | 9.97 | 6.74 | 1014.72 | 2.27 |
| India | 3/17/2020 | 2.22 | 25.98 | 9.78 | 7.60 | 1015.44 | 2.28 |
| India | 3/18/2020 | 2.16 | 26.23 | 9.74 | 7.60 | 1014.75 | 2.25 |
| India | 3/19/2020 | 2.01 | 26.07 | 10.15 | 7.50 | 1013.11 | 2.11 |
| India | 3/20/2020 | 2.49 | 25.93 | 11.02 | 6.96 | 1013.01 | 2.23 |
| India | 3/21/2020 | 2.55 | 26.42 | 10.64 | 7.57 | 1012.81 | 1.95 |
| India | 3/22/2020 | 2.68 | 26.47 | 11.35 | 7.64 | 1013.62 | 2.14 |
| India | 3/23/2020 | 2.89 | 26.96 | 11.36 | 7.34 | 1012.54 | 2.12 |
| India | 3/24/2020 | 3.00 | 27.42 | 11.46 | 7.60 | 1011.43 | 2.03 |
| India | 3/25/2020 | 2.98 | 27.70 | 11.69 | 7.77 | 1011.09 | 2.16 |
| Indonesia | 2/28/2020 | 5.98 | 26.29 | 19.52 | 6.00 | 1010.00 | 1.54 |
| Indonesia | 2/29/2020 | 6.15 | 26.33 | 19.67 | 6.00 | 1009.00 | 2.00 |
| Indonesia | 3/1/2020 | 6.05 | 26.30 | 20.02 | 6.00 | 1010.00 | 2.49 |
| Indonesia | 3/2/2020 | 5.27 | 26.71 | 19.20 | 7.00 | 1010.00 | 1.88 |
| Indonesia | 3/3/2020 | 5.64 | 26.89 | 19.38 | 8.00 | 1010.00 | 1.62 |
| Indonesia | 3/4/2020 | 6.15 | 26.43 | 19.58 | 7.00 | 1012.00 | 2.18 |
| Indonesia | 3/5/2020 | 5.81 | 25.84 | 18.96 | 6.00 | 1012.00 | 3.31 |
| Indonesia | 3/6/2020 | 5.18 | 26.49 | 18.73 | 7.00 | 1012.00 | 2.36 |
| Indonesia | 3/7/2020 | 5.82 | 26.29 | 18.92 | 6.00 | 1012.00 | 2.03 |
| Indonesia | 3/8/2020 | 6.05 | 25.94 | 18.75 | 6.00 | 1012.00 | 1.55 |
| Indonesia | 3/9/2020 | 5.99 | 26.16 | 19.21 | 7.00 | 1011.00 | 3.34 |
| Indonesia | 3/10/2020 | 5.60 | 26.71 | 19.44 | 8.00 | 1010.00 | 2.32 |
| Indonesia | 3/11/2020 | 5.71 | 26.60 | 19.62 | 8.00 | 1010.00 | 1.88 |
| Iran | 2/10/2020 | 0.47 | -2.94 | 2.50 | 1.00 | 1020.00 | 1.62 |
| Iran | 2/11/2020 | 0.28 | -8.34 | 1.71 | 1.00 | 1021.00 | 2.80 |
| Iran | 2/12/2020 | 0.24 | -10.17 | 1.37 | 2.00 | 1022.00 | 2.72 |
| Iran | 2/13/2020 | 0.68 | -2.41 | 2.81 | 3.00 | 1025.00 | 2.46 |
| Iran | 2/14/2020 | 0.45 | -2.68 | 2.68 | 2.00 | 1033.00 | 3.37 |
| Iran | 2/15/2020 | 0.40 | -4.99 | 2.02 | 2.00 | 1035.00 | 2.27 |
| Iran | 2/16/2020 | 0.75 | 0.02 | 2.59 | 2.00 | 1031.00 | 1.31 |
| Iran | 2/17/2020 | 1.05 | 0.63 | 3.81 | 2.00 | 1029.00 | 1.21 |
| Iran | 2/18/2020 | 0.75 | 2.18 | 4.19 | 3.00 | 1030.00 | 1.09 |
| Iran | 2/19/2020 | 1.24 | 2.99 | 4.40 | 3.00 | 1026.00 | 1.92 |
| Iran | 2/20/2020 | 1.28 | 2.05 | 4.88 | 3.00 | 1016.00 | 2.37 |
| Iran | 2/21/2020 | 0.65 | 2.70 | 4.05 | 3.00 | 1018.00 | 2.23 |
| Iraq | 2/19/2020 | 2.37 | 16.36 | 8.23 | 5.00 | 1016.00 | 1.84 |
| Iraq | 2/20/2020 | 1.94 | 14.07 | 7.17 | 5.00 | 1016.00 | 2.31 |
| Iraq | 2/21/2020 | 1.09 | 14.16 | 5.21 | 5.00 | 1018.00 | 1.21 |
| Iraq | 2/22/2020 | 1.33 | 15.57 | 4.77 | 5.00 | 1017.00 | 2.33 |
| Iraq | 2/23/2020 | 1.17 | 14.81 | 5.36 | 5.00 | 1019.00 | 1.42 |
| Iraq | 2/24/2020 | 2.19 | 16.67 | 6.05 | 5.00 | 1018.00 | 1.49 |
| Iraq | 2/25/2020 | 3.03 | 15.81 | 9.64 | 5.00 | 1011.00 | 0.93 |
| Iraq | 2/26/2020 | 1.93 | 14.07 | 7.01 | 3.00 | 1019.00 | 2.48 |
| Iraq | 2/27/2020 | 1.09 | 14.30 | 5.89 | 5.00 | 1022.00 | 1.12 |
| Iraq | 2/28/2020 | 1.07 | 15.30 | 5.35 | 5.00 | 1019.00 | 1.17 |
| Iraq | 2/29/2020 | 1.42 | 16.15 | 5.33 | 6.00 | 1014.00 | 1.63 |
| Iraq | 3/1/2020 | 1.41 | 14.21 | 5.35 | 5.00 | 1013.00 | 2.69 |
| Iraq | 3/2/2020 | 1.17 | 13.37 | 5.17 | 4.00 | 1020.00 | 2.08 |
| Iraq | 3/3/2020 | 1.02 | 13.43 | 4.68 | 5.00 | 1023.00 | 1.75 |
| Iraq | 3/4/2020 | 0.93 | 14.49 | 4.91 | 5.00 | 1020.00 | 2.09 |
| Iraq | 3/5/2020 | 0.78 | 16.07 | 4.80 | 6.00 | 1018.00 | 1.45 |
| Iraq | 3/6/2020 | 1.29 | 18.35 | 4.48 | 6.00 | 1015.00 | 2.74 |
| Iraq | 3/7/2020 | 1.83 | 17.33 | 6.16 | 4.00 | 1012.00 | 2.61 |
| Iraq | 3/8/2020 | 1.83 | 17.46 | 7.09 | 6.00 | 1017.00 | 1.89 |
| Iraq | 3/9/2020 | 1.20 | 19.50 | 5.58 | 6.00 | 1017.00 | 1.42 |
| Iraq | 3/10/2020 | 0.84 | 20.42 | 4.35 | 7.00 | 1017.00 | 0.96 |
| Iraq | 3/11/2020 | 1.12 | 19.99 | 4.36 | 7.00 | 1018.00 | 1.43 |
| Iraq | 3/12/2020 | 1.68 | 19.66 | 4.64 | 7.00 | 1017.00 | 1.37 |
| Iraq | 3/13/2020 | 2.20 | 21.28 | 5.22 | 7.00 | 1014.00 | 3.16 |
| Iraq | 3/14/2020 | 2.46 | 18.06 | 7.85 | 4.00 | 1010.00 | 3.22 |
| Iraq | 3/15/2020 | 2.42 | 17.83 | 8.46 | 5.00 | 1015.00 | 0.74 |
| Iraq | 3/16/2020 | 2.01 | 18.84 | 7.55 | 5.00 | 1016.00 | 0.86 |
| Iraq | 3/17/2020 | 1.88 | 19.73 | 6.72 | 6.00 | 1015.00 | 2.03 |
| Iraq | 3/18/2020 | 1.97 | 18.06 | 7.53 | 5.00 | 1011.00 | 3.99 |
| Iraq | 3/19/2020 | 1.10 | 15.35 | 5.37 | 5.00 | 1015.00 | 1.86 |
| Iraq | 3/20/2020 | 1.47 | 16.27 | 4.85 | 6.00 | 1014.00 | 1.98 |
| Ireland | 2/28/2020 | 1.80 | 7.80 | 6.44 | 2.00 | 998.00 | 6.29 |
| Ireland | 2/29/2020 | 1.01 | 4.35 | 4.65 | 1.00 | 979.00 | 10.10 |
| Ireland | 3/1/2020 | 0.81 | 4.11 | 4.65 | 1.00 | 984.00 | 7.52 |
| Ireland | 3/2/2020 | 0.85 | 3.61 | 4.54 | 1.00 | 994.00 | 6.25 |
| Ireland | 3/3/2020 | 0.83 | 4.74 | 4.70 | 2.00 | 999.00 | 5.15 |
| Ireland | 3/4/2020 | 1.08 | 4.13 | 4.49 | 1.00 | 1009.00 | 1.45 |
| Ireland | 3/5/2020 | 0.92 | 4.04 | 4.48 | 3.00 | 999.00 | 2.43 |
| Ireland | 3/6/2020 | 1.07 | 4.43 | 4.82 | 1.00 | 999.00 | 4.12 |
| Ireland | 3/7/2020 | 1.93 | 8.36 | 6.53 | 3.00 | 998.00 | 7.64 |
| Ireland | 3/8/2020 | 1.16 | 6.79 | 5.60 | 2.00 | 999.00 | 6.98 |
| Ireland | 3/9/2020 | 1.90 | 7.14 | 6.03 | 3.00 | 1011.00 | 6.97 |
| Isle of Man | 3/14/2020 | 1.60 | 7.48 | 6.05 | 2.00 | 999.00 | 8.40 |
| Isle of Man | 3/15/2020 | 1.02 | 7.46 | 5.50 | 2.00 | 999.00 | 8.35 |
| Isle of Man | 3/16/2020 | 1.35 | 7.01 | 5.24 | 3.00 | 1018.00 | 7.72 |
| Isle of Man | 3/17/2020 | 2.06 | 8.32 | 6.38 | 2.00 | 1021.00 | 7.59 |
| Isle of Man | 3/18/2020 | 1.23 | 6.68 | 4.95 | 2.00 | 1026.00 | 4.65 |
| Isle of Man | 3/19/2020 | 0.87 | 6.43 | 4.66 | 3.00 | 1031.00 | 3.84 |
| Isle of Man | 3/20/2020 | 0.63 | 6.04 | 4.54 | 3.00 | 1034.00 | 6.15 |
| Isle of Man | 3/21/2020 | 0.49 | 6.14 | 4.47 | 3.00 | 1033.00 | 8.24 |
| Isle of Man | 3/22/2020 | 0.47 | 6.23 | 4.52 | 3.00 | 1029.00 | 5.59 |
| Israel | 2/21/2020 | 1.56 | 14.16 | 7.63 | 5.00 | 1018.00 | 3.96 |
| Israel | 2/22/2020 | 1.51 | 14.16 | 7.69 | 4.00 | 1020.00 | 4.01 |
| Israel | 2/23/2020 | 1.73 | 13.78 | 7.53 | 3.00 | 1020.00 | 2.24 |
| Israel | 2/24/2020 | 1.66 | 13.79 | 8.20 | 5.00 | 1017.00 | 3.00 |
| Israel | 2/25/2020 | 1.71 | 14.84 | 8.50 | 5.00 | 1018.00 | 4.28 |
| Israel | 2/26/2020 | 1.35 | 14.61 | 8.34 | 5.00 | 1021.00 | 2.57 |
| Israel | 2/27/2020 | 1.42 | 15.02 | 8.04 | 5.00 | 1021.00 | 1.92 |
| Israel | 2/28/2020 | 1.56 | 16.20 | 7.75 | 6.00 | 1018.00 | 1.34 |
| Israel | 2/29/2020 | 1.53 | 14.51 | 8.21 | 4.00 | 1015.00 | 4.86 |
| Israel | 3/1/2020 | 1.38 | 14.50 | 7.60 | 4.00 | 1020.00 | 4.57 |
| Israel | 3/2/2020 | 1.30 | 14.13 | 7.20 | 4.00 | 1022.00 | 2.28 |
| Israel | 3/3/2020 | 1.17 | 14.09 | 7.57 | 5.00 | 1022.00 | 2.57 |
| Israel | 3/4/2020 | 0.97 | 16.78 | 7.81 | 6.00 | 1017.00 | 2.15 |
| Israel | 3/5/2020 | 1.09 | 16.53 | 8.03 | 6.00 | 1012.00 | 2.41 |
| Israel | 3/6/2020 | 1.92 | 14.31 | 8.41 | 5.00 | 1015.00 | 3.08 |
| Israel | 3/7/2020 | 1.57 | 15.19 | 8.65 | 4.00 | 1021.00 | 2.75 |
| Israel | 3/8/2020 | 1.45 | 16.22 | 8.94 | 6.00 | 1021.00 | 2.56 |
| Israel | 3/9/2020 | 1.29 | 17.58 | 8.83 | 6.00 | 1017.00 | 2.67 |
| Israel | 3/10/2020 | 1.43 | 15.21 | 8.77 | 5.00 | 1017.00 | 2.52 |
| Israel | 3/11/2020 | 1.89 | 16.22 | 9.06 | 5.00 | 1016.00 | 2.27 |
| Israel | 3/12/2020 | 2.95 | 18.96 | 10.57 | 5.00 | 1011.00 | 3.32 |
| Israel | 3/13/2020 | 2.45 | 15.44 | 9.00 | 4.00 | 999.00 | 4.83 |
| Israel | 3/14/2020 | 2.48 | 15.89 | 10.19 | 4.00 | 1015.00 | 3.83 |
| Italy | 2/17/2020 | 1.38 | 9.25 | 6.17 | 3.51 | 1028.51 | 1.28 |
| Italy | 2/18/2020 | 1.01 | 8.79 | 5.45 | 3.00 | 1026.49 | 2.02 |
| Italy | 2/19/2020 | 1.10 | 8.46 | 5.65 | 3.00 | 1024.95 | 2.02 |
| Italy | 2/20/2020 | 0.87 | 7.91 | 4.43 | 3.49 | 1025.49 | 2.40 |
| Italy | 2/21/2020 | 0.96 | 8.32 | 4.46 | 4.00 | 1027.97 | 1.52 |
| Italy | 2/22/2020 | 1.13 | 8.95 | 4.38 | 4.00 | 1031.51 | 2.15 |
| Italy | 2/23/2020 | 1.60 | 10.47 | 6.08 | 4.00 | 1031.51 | 1.86 |
| Italy | 2/24/2020 | 1.32 | 11.25 | 6.93 | 4.49 | 1025.03 | 1.83 |
| Italy | 2/25/2020 | 1.16 | 10.00 | 7.01 | 3.49 | 1018.54 | 1.68 |
| Italy | 2/26/2020 | 0.86 | 8.72 | 5.62 | 3.51 | 1004.64 | 4.57 |
| Italy | 2/27/2020 | 0.77 | 6.78 | 4.10 | 3.00 | 1014.51 | 3.57 |
| Italy | 2/28/2020 | 0.84 | 8.85 | 4.80 | 3.49 | 1021.51 | 3.41 |
| Italy | 2/29/2020 | 1.22 | 8.05 | 5.08 | 3.51 | 1021.51 | 2.43 |
| Italy | 3/1/2020 | 1.57 | 8.91 | 6.94 | 3.03 | 1013.56 | 2.96 |
| Jamaica | 3/6/2020 | 3.19 | 26.87 | 16.29 | 7.00 | 1018.00 | 4.45 |
| Jamaica | 3/7/2020 | 3.44 | 26.71 | 15.43 | 7.00 | 1016.00 | 1.78 |
| Jamaica | 3/8/2020 | 3.97 | 26.12 | 15.85 | 6.00 | 1018.00 | 8.35 |
| Jamaica | 3/9/2020 | 4.17 | 25.97 | 15.71 | 6.00 | 1020.00 | 7.54 |
| Jamaica | 3/10/2020 | 3.94 | 26.54 | 15.90 | 6.00 | 1019.00 | 5.85 |
| Jamaica | 3/11/2020 | 3.44 | 26.23 | 15.37 | 7.00 | 1017.00 | 3.71 |
| Jamaica | 3/12/2020 | 3.21 | 26.12 | 14.91 | 6.00 | 1014.00 | 2.79 |
| Jamaica | 3/13/2020 | 3.26 | 26.59 | 15.67 | 7.00 | 1017.00 | 3.86 |
| Jamaica | 3/14/2020 | 3.04 | 26.90 | 16.52 | 7.00 | 1019.00 | 5.75 |
| Jamaica | 3/15/2020 | 2.78 | 26.91 | 15.78 | 7.00 | 1019.00 | 5.32 |
| Jamaica | 3/16/2020 | 2.68 | 26.56 | 15.03 | 7.00 | 1018.00 | 5.36 |
| Jamaica | 3/17/2020 | 2.98 | 26.78 | 15.21 | 7.00 | 1018.00 | 5.21 |
| Jamaica | 3/18/2020 | 2.79 | 26.75 | 15.53 | 7.00 | 1019.00 | 5.61 |
| Jamaica | 3/19/2020 | 3.22 | 26.92 | 16.02 | 7.00 | 1020.00 | 5.66 |
| Jamaica | 3/20/2020 | 3.88 | 26.78 | 16.04 | 7.00 | 1020.00 | 5.35 |
| Jamaica | 3/21/2020 | 3.29 | 26.56 | 15.96 | 7.00 | 1020.00 | 4.71 |
| Jamaica | 3/22/2020 | 3.06 | 26.41 | 15.39 | 7.00 | 1019.00 | 4.78 |
| Jamaica | 3/23/2020 | 2.57 | 26.82 | 15.31 | 6.00 | 1020.00 | 4.62 |
| Jamaica | 3/24/2020 | 3.27 | 26.77 | 16.13 | 7.00 | 1019.00 | 4.70 |
| Jamaica | 3/25/2020 | 3.42 | 26.93 | 16.11 | 7.00 | 1017.00 | 3.29 |
| Japan | 2/4/2020 | 0.69 | 5.35 | 3.95 | 3.00 | 1022.00 | 1.72 |
| Japan | 2/5/2020 | 0.61 | 4.41 | 3.75 | 3.00 | 1018.00 | 3.18 |
| Japan | 2/6/2020 | 0.36 | -0.58 | 2.32 | 2.00 | 1029.00 | 4.31 |
| Japan | 2/7/2020 | 0.52 | 1.19 | 2.71 | 2.00 | 1030.00 | 1.23 |
| Japan | 2/8/2020 | 0.77 | 3.55 | 3.56 | 2.00 | 1020.00 | 2.33 |
| Japan | 2/9/2020 | 0.45 | 0.51 | 2.66 | 2.00 | 1018.00 | 3.72 |
| Japan | 2/10/2020 | 0.68 | 2.22 | 3.12 | 3.00 | 1017.00 | 1.67 |
| Japan | 2/11/2020 | 0.58 | 3.58 | 3.19 | 3.00 | 1030.00 | 2.37 |
| Japan | 2/12/2020 | 1.12 | 6.53 | 4.66 | 4.00 | 1030.00 | 2.07 |
| Japan | 2/13/2020 | 1.99 | 12.23 | 8.05 | 3.00 | 1021.00 | 2.38 |
| Japan | 2/14/2020 | 1.65 | 9.91 | 6.57 | 4.00 | 1020.00 | 1.01 |
| Japan | 2/15/2020 | 1.12 | 9.14 | 6.03 | 4.00 | 1022.00 | 1.46 |
| Japan | 2/16/2020 | 3.08 | 9.42 | 7.42 | 4.00 | 1019.00 | 1.42 |
| Japan | 2/17/2020 | 1.27 | 9.62 | 5.54 | 3.00 | 1005.00 | 2.73 |
| Japan | 2/18/2020 | 0.52 | 4.07 | 3.25 | 2.00 | 1020.00 | 3.65 |
| Japan | 2/19/2020 | 0.69 | 5.24 | 3.48 | 3.00 | 1028.00 | 1.55 |
| Japan | 2/20/2020 | 1.08 | 6.57 | 4.39 | 3.00 | 1032.00 | 1.20 |
| Japan | 2/21/2020 | 0.98 | 7.48 | 4.81 | 3.00 | 1036.00 | 1.80 |
| Japan | 2/22/2020 | 1.67 | 10.56 | 6.77 | 3.00 | 1032.00 | 3.87 |
| Japan | 2/23/2020 | 0.60 | 6.25 | 3.81 | 4.00 | 1022.00 | 3.59 |
| Japan | 2/24/2020 | 0.85 | 7.30 | 4.05 | 4.00 | 1024.00 | 1.34 |
| Japan | 2/25/2020 | 1.68 | 9.30 | 5.48 | 3.00 | 1021.00 | 1.48 |
| Japan | 2/26/2020 | 1.59 | 7.29 | 5.43 | 2.00 | 1015.00 | 3.32 |
| Japan | 2/27/2020 | 0.64 | 3.84 | 3.49 | 2.00 | 1022.00 | 3.46 |
| Japan | 2/28/2020 | 0.72 | 4.52 | 3.39 | 3.00 | 1025.00 | 1.47 |
| Jordan | 3/6/2020 | 1.63 | 13.99 | 6.76 | 4.00 | 1014.00 | 2.89 |
| Jordan | 3/7/2020 | 1.64 | 13.98 | 7.67 | 3.00 | 1021.00 | 2.38 |
| Jordan | 3/8/2020 | 1.34 | 17.11 | 7.28 | 5.00 | 1021.00 | 2.28 |
| Jordan | 3/9/2020 | 1.17 | 18.84 | 6.82 | 5.00 | 1018.00 | 3.02 |
| Jordan | 3/10/2020 | 1.52 | 16.27 | 8.14 | 5.00 | 1016.00 | 1.33 |
| Jordan | 3/11/2020 | 1.90 | 17.14 | 8.13 | 6.00 | 1015.00 | 1.64 |
| Jordan | 3/12/2020 | 2.87 | 18.79 | 9.36 | 4.00 | 1012.00 | 4.77 |
| Jordan | 3/13/2020 | 2.31 | 15.58 | 8.03 | 4.00 | 999.00 | 5.36 |
| Jordan | 3/14/2020 | 2.32 | 14.96 | 8.88 | 3.00 | 1015.00 | 3.98 |
| Jordan | 3/15/2020 | 1.86 | 16.82 | 8.53 | 4.00 | 1017.00 | 0.76 |
| Jordan | 3/16/2020 | 1.61 | 17.39 | 8.32 | 5.00 | 1015.00 | 1.71 |
| Jordan | 3/17/2020 | 1.59 | 14.31 | 7.86 | 4.00 | 1013.00 | 3.20 |
| Kazakhstan | 3/18/2020 | 1.33 | 6.52 | 5.50 | 4.00 | 1027.00 | 1.71 |
| Kazakhstan | 3/19/2020 | 1.08 | 6.46 | 4.56 | 2.00 | 1025.00 | 1.66 |
| Kazakhstan | 3/20/2020 | 0.84 | 5.79 | 4.25 | 2.00 | 1020.00 | 1.45 |
| Kazakhstan | 3/21/2020 | 1.11 | 8.26 | 4.86 | 4.00 | 1019.00 | 2.48 |
| Kazakhstan | 3/22/2020 | 1.03 | 4.83 | 4.71 | 2.00 | 1025.00 | 1.59 |
| Kazakhstan | 3/23/2020 | 0.74 | 1.09 | 3.48 | 2.00 | 1025.00 | 1.73 |
| Kazakhstan | 3/24/2020 | 0.39 | -1.05 | 2.32 | 2.00 | 1024.00 | 1.73 |
| Kazakhstan | 3/25/2020 | 0.36 | -3.67 | 1.88 | 2.00 | 1027.00 | 1.91 |
| Kazakhstan | 3/26/2020 | 0.58 | 0.11 | 2.64 | 2.00 | 1027.00 | 1.40 |
| Kazakhstan | 3/27/2020 | 0.64 | 3.51 | 3.32 | 3.00 | 1026.00 | 1.67 |
| Kazakhstan | 3/28/2020 | 0.82 | 3.52 | 3.40 | 3.00 | 1023.00 | 2.48 |
| Kazakhstan | 3/29/2020 | 0.34 | 1.61 | 2.44 | 1.00 | 1026.00 | 2.00 |
| Kazakhstan | 3/30/2020 | 0.36 | 4.28 | 2.53 | 3.00 | 1025.00 | 1.57 |
| Kazakhstan | 3/31/2020 | 0.84 | 8.69 | 3.88 | 4.00 | 1018.00 | 1.37 |
| Kazakhstan | 4/1/2020 | 1.39 | 6.39 | 4.99 | 3.00 | 1017.00 | 2.31 |
| Kazakhstan | 4/2/2020 | 1.36 | 5.34 | 5.25 | 2.00 | 1024.00 | 1.87 |
| Kazakhstan | 4/3/2020 | 1.03 | 6.45 | 4.82 | 3.00 | 1024.00 | 1.01 |
| Kazakhstan | 4/4/2020 | 0.62 | 8.39 | 3.79 | 3.00 | 1024.00 | 1.82 |
| Kazakhstan | 4/5/2020 | 0.99 | 10.73 | 4.53 | 4.00 | 1022.00 | 1.65 |
| Kenya | 3/12/2020 | 3.82 | 22.41 | 15.16 | 5.21 | 1014.38 | 3.51 |
| Kenya | 3/13/2020 | 3.77 | 22.81 | 15.38 | 5.41 | 1014.38 | 3.50 |
| Kenya | 3/14/2020 | 3.47 | 22.81 | 15.38 | 5.21 | 1015.38 | 3.01 |
| Kenya | 3/15/2020 | 2.45 | 23.10 | 13.89 | 5.41 | 1015.18 | 2.82 |
| Kenya | 3/16/2020 | 2.79 | 23.26 | 14.79 | 5.41 | 1013.18 | 2.78 |
| Kenya | 3/17/2020 | 3.19 | 22.44 | 15.02 | 5.21 | 1014.38 | 3.16 |
| Kenya | 3/18/2020 | 2.73 | 22.21 | 14.24 | 5.21 | 1015.38 | 2.93 |
| Kenya | 3/19/2020 | 2.88 | 22.08 | 13.98 | 5.41 | 1016.38 | 2.91 |
| Kenya | 3/20/2020 | 3.18 | 21.89 | 14.27 | 5.21 | 1016.18 | 3.21 |
| Kenya | 3/21/2020 | 2.42 | 22.59 | 13.09 | 5.21 | 1016.38 | 3.55 |
| Kenya | 3/22/2020 | 2.83 | 22.18 | 13.00 | 6.00 | 1016.38 | 2.96 |
| Kenya | 3/23/2020 | 3.39 | 21.87 | 13.48 | 6.21 | 1016.18 | 3.02 |
| Kenya | 3/24/2020 | 3.27 | 22.09 | 13.97 | 6.21 | 1015.18 | 2.98 |
| Kenya | 3/25/2020 | 3.64 | 21.68 | 14.57 | 5.21 | 1013.38 | 2.36 |
| Kenya | 3/26/2020 | 4.10 | 21.79 | 15.05 | 5.21 | 1012.38 | 1.65 |
| Kuwait | 2/25/2020 | 2.89 | 20.23 | 10.83 | 6.00 | 1011.00 | 5.04 |
| Kuwait | 2/26/2020 | 1.09 | 16.77 | 5.83 | 4.00 | 1018.00 | 5.51 |
| Kuwait | 2/27/2020 | 1.16 | 15.99 | 6.63 | 5.00 | 1021.00 | 4.31 |
| Kuwait | 2/28/2020 | 0.85 | 17.19 | 6.74 | 5.00 | 1019.00 | 3.23 |
| Kuwait | 2/29/2020 | 1.27 | 17.37 | 7.18 | 6.00 | 1015.00 | 2.29 |
| Kuwait | 3/1/2020 | 1.40 | 19.25 | 8.54 | 6.00 | 1014.00 | 2.87 |
| Kuwait | 3/2/2020 | 0.92 | 17.38 | 6.30 | 5.00 | 1019.00 | 4.18 |
| Kuwait | 3/3/2020 | 0.82 | 16.86 | 5.99 | 5.00 | 1021.00 | 5.01 |
| Kuwait | 3/4/2020 | 0.82 | 16.99 | 5.99 | 5.00 | 1019.00 | 5.34 |
| Kuwait | 3/5/2020 | 0.71 | 18.25 | 6.24 | 6.00 | 1018.00 | 3.19 |
| Kuwait | 3/6/2020 | 0.85 | 18.85 | 7.05 | 6.00 | 1016.00 | 2.81 |
| Kuwait | 3/7/2020 | 1.37 | 19.22 | 8.14 | 6.00 | 1013.00 | 3.14 |
| Kuwait | 3/8/2020 | 1.78 | 18.77 | 7.41 | 6.00 | 1014.00 | 3.96 |
| Kuwait | 3/9/2020 | 1.43 | 19.83 | 7.55 | 6.00 | 1017.00 | 3.83 |
| Kuwait | 3/10/2020 | 0.97 | 20.31 | 7.03 | 6.00 | 1018.00 | 1.81 |
| Kuwait | 3/11/2020 | 0.86 | 20.97 | 6.24 | 6.00 | 1019.00 | 1.27 |
| Kuwait | 3/12/2020 | 1.10 | 20.33 | 6.11 | 6.00 | 1018.00 | 1.86 |
| Kuwait | 3/13/2020 | 1.31 | 20.20 | 9.46 | 6.00 | 1016.00 | 4.55 |
| Kuwait | 3/14/2020 | 2.47 | 20.03 | 10.49 | 6.00 | 1011.00 | 4.55 |
| Kuwait | 3/15/2020 | 2.62 | 19.94 | 11.64 | 6.00 | 1013.00 | 2.22 |
| Kuwait | 3/16/2020 | 1.86 | 20.86 | 9.23 | 6.00 | 1015.00 | 1.55 |
| Kuwait | 3/17/2020 | 1.76 | 20.77 | 9.65 | 6.00 | 1015.00 | 3.05 |
| Kuwait | 3/18/2020 | 2.84 | 20.64 | 12.29 | 6.00 | 1012.00 | 5.34 |
| Kuwait | 3/19/2020 | 1.17 | 20.52 | 7.65 | 6.00 | 1015.00 | 4.40 |
| Kuwait | 3/20/2020 | 1.49 | 19.71 | 7.92 | 6.00 | 1014.00 | 2.83 |
| Kuwait | 3/21/2020 | 1.81 | 21.25 | 9.18 | 7.00 | 1007.00 | 2.18 |
| Kuwait | 3/22/2020 | 1.68 | 18.14 | 6.68 | 5.00 | 1011.00 | 7.10 |
| Kuwait | 3/23/2020 | 1.44 | 17.69 | 6.60 | 5.00 | 1015.00 | 4.75 |
| Kuwait | 3/24/2020 | 1.16 | 19.16 | 7.17 | 6.00 | 1015.00 | 2.40 |
| Kuwait | 3/25/2020 | 1.65 | 19.98 | 8.50 | 6.00 | 1013.00 | 3.16 |
| Kuwait | 3/26/2020 | 1.85 | 20.09 | 10.54 | 6.00 | 1012.00 | 3.83 |
| Kuwait | 3/27/2020 | 2.26 | 20.66 | 12.03 | 6.00 | 1013.00 | 5.07 |
| Kuwait | 3/28/2020 | 2.89 | 22.84 | 11.74 | 7.00 | 1011.00 | 5.89 |
| Kuwait | 3/29/2020 | 1.79 | 21.64 | 9.35 | 6.00 | 1010.00 | 5.60 |
| Kuwait | 3/30/2020 | 1.27 | 21.04 | 7.86 | 6.00 | 1014.00 | 3.43 |
| Kuwait | 3/31/2020 | 1.81 | 21.63 | 8.10 | 6.00 | 1017.00 | 2.25 |
| Kuwait | 4/1/2020 | 2.20 | 22.23 | 9.56 | 7.00 | 1015.00 | 5.01 |
| Kuwait | 4/2/2020 | 1.83 | 22.55 | 11.14 | 6.00 | 1011.00 | 3.26 |
| Kuwait | 4/3/2020 | 1.59 | 23.00 | 8.68 | 7.00 | 1014.00 | 3.87 |
| Kuwait | 4/4/2020 | 1.40 | 23.97 | 8.35 | 6.00 | 1017.00 | 3.22 |
| Kuwait | 4/5/2020 | 1.33 | 24.12 | 8.11 | 7.00 | 1019.00 | 3.95 |
| Kuwait | 4/6/2020 | 1.65 | 24.45 | 7.53 | 7.00 | 1017.00 | 2.08 |
| Kuwait | 4/7/2020 | 2.65 | 25.50 | 9.78 | 8.00 | 1012.00 | 3.20 |
| Kuwait | 4/8/2020 | 2.58 | 25.05 | 11.76 | 6.00 | 1008.00 | 1.81 |
| Kuwait | 4/9/2020 | 3.06 | 26.20 | 11.76 | 7.00 | 1007.00 | 4.78 |
| Kuwait | 4/10/2020 | 2.86 | 26.39 | 11.14 | 8.00 | 1006.00 | 3.23 |
| Kuwait | 4/11/2020 | 1.89 | 25.06 | 8.51 | 7.00 | 1010.00 | 5.74 |
| Kuwait | 4/12/2020 | 1.71 | 24.54 | 8.30 | 7.00 | 1014.00 | 4.93 |
| Kuwait | 4/13/2020 | 1.80 | 24.13 | 8.16 | 7.00 | 1014.00 | 4.90 |
| Kuwait | 4/14/2020 | 1.05 | 23.80 | 7.48 | 7.00 | 1016.00 | 2.42 |
| Kuwait | 4/15/2020 | 1.83 | 23.47 | 8.42 | 7.00 | 1014.00 | 3.32 |
| Kuwait | 4/16/2020 | 1.49 | 24.39 | 7.78 | 7.00 | 1013.00 | 1.88 |
| Kuwait | 4/17/2020 | 1.07 | 24.91 | 7.01 | 7.00 | 1011.00 | 2.04 |
| Kuwait | 4/18/2020 | 2.32 | 23.65 | 11.75 | 7.00 | 1010.00 | 4.94 |
| Kuwait | 4/19/2020 | 2.00 | 24.17 | 10.84 | 7.00 | 1011.00 | 3.24 |
| Kuwait | 4/20/2020 | 2.45 | 25.11 | 9.66 | 7.00 | 1012.00 | 2.65 |
| Kuwait | 4/21/2020 | 2.97 | 25.35 | 10.43 | 7.00 | 1013.00 | 2.02 |
| Kuwait | 4/22/2020 | 2.94 | 24.95 | 12.28 | 7.00 | 1013.00 | 1.99 |
| Kuwait | 4/23/2020 | 2.66 | 25.08 | 12.75 | 7.00 | 1010.00 | 4.10 |
| Kuwait | 4/24/2020 | 2.07 | 25.91 | 13.89 | 7.00 | 1008.00 | 2.95 |
| Kuwait | 4/25/2020 | 1.66 | 27.17 | 13.46 | 8.00 | 999.00 | 3.22 |
| Kuwait | 4/26/2020 | 1.97 | 27.86 | 11.57 | 8.00 | 999.00 | 4.13 |
| Kuwait | 4/27/2020 | 1.19 | 25.35 | 7.80 | 7.00 | 1013.00 | 3.53 |
| Kuwait | 4/28/2020 | 1.98 | 25.84 | 8.93 | 7.00 | 1012.00 | 2.48 |
| Kuwait | 4/29/2020 | 2.80 | 26.81 | 11.14 | 7.00 | 1010.00 | 4.56 |
| Kuwait | 4/30/2020 | 2.83 | 26.79 | 13.65 | 6.00 | 1009.00 | 3.20 |
| Kuwait | 5/1/2020 | 1.74 | 27.78 | 9.79 | 8.00 | 1010.00 | 1.98 |
| Kuwait | 5/2/2020 | 2.18 | 28.61 | 9.45 | 7.00 | 1009.00 | 1.57 |
| Kuwait | 5/3/2020 | 1.10 | 29.08 | 8.25 | 8.00 | 1011.00 | 1.48 |
| Kuwait | 5/4/2020 | 1.37 | 28.67 | 8.43 | 8.00 | 1012.00 | 1.91 |
| Kuwait | 5/5/2020 | 2.66 | 30.05 | 10.36 | 8.00 | 1010.00 | 3.66 |
| Kuwait | 5/6/2020 | 2.66 | 29.47 | 11.05 | 7.00 | 1011.00 | 4.21 |
| Kuwait | 5/7/2020 | 1.25 | 27.17 | 9.10 | 8.00 | 1013.00 | 3.80 |
| Kuwait | 5/8/2020 | 1.57 | 28.17 | 9.40 | 8.00 | 1013.00 | 2.82 |
| Kyrgyzstan | 3/16/2020 | 0.57 | 7.62 | 3.21 | 4.00 | 1022.00 | 1.59 |
| Kyrgyzstan | 3/17/2020 | 1.11 | 8.62 | 4.59 | 5.00 | 1025.00 | 1.43 |
| Kyrgyzstan | 3/18/2020 | 1.17 | 5.46 | 5.22 | 3.00 | 1027.00 | 1.08 |
| Kyrgyzstan | 3/19/2020 | 1.01 | 5.79 | 4.60 | 3.00 | 1025.00 | 1.43 |
| Kyrgyzstan | 3/20/2020 | 0.85 | 6.70 | 4.38 | 3.00 | 1020.00 | 1.46 |
| Kyrgyzstan | 3/21/2020 | 1.17 | 5.87 | 5.09 | 4.00 | 1020.00 | 1.91 |
| Kyrgyzstan | 3/22/2020 | 0.84 | 4.21 | 4.35 | 3.00 | 1025.00 | 1.48 |
| Kyrgyzstan | 3/23/2020 | 0.76 | 2.06 | 3.68 | 3.00 | 1024.00 | 2.12 |
| Kyrgyzstan | 3/24/2020 | 0.54 | 0.34 | 2.65 | 3.00 | 1023.00 | 2.63 |
| Kyrgyzstan | 3/25/2020 | 0.50 | -1.72 | 2.27 | 2.00 | 1025.00 | 2.72 |
| Kyrgyzstan | 3/26/2020 | 0.68 | 0.94 | 3.20 | 1.00 | 1025.00 | 1.46 |
| Kyrgyzstan | 3/27/2020 | 0.82 | 3.89 | 3.78 | 3.00 | 1025.00 | 1.65 |
| Kyrgyzstan | 3/28/2020 | 0.73 | 2.57 | 3.45 | 2.00 | 1023.00 | 2.04 |
| Kyrgyzstan | 3/29/2020 | 0.37 | 2.08 | 2.59 | 2.00 | 1025.00 | 1.77 |
| Kyrgyzstan | 3/30/2020 | 0.48 | 4.99 | 2.91 | 4.00 | 1022.00 | 1.80 |
| Kyrgyzstan | 3/31/2020 | 0.99 | 7.79 | 4.17 | 4.00 | 1015.00 | 1.39 |
| Latvia | 2/27/2020 | 0.86 | 1.18 | 3.83 | 1.00 | 992.00 | 2.64 |
| Latvia | 2/28/2020 | 0.73 | -0.40 | 3.26 | 1.00 | 999.00 | 3.31 |
| Latvia | 2/29/2020 | 0.68 | 0.38 | 3.36 | 1.00 | 1006.00 | 3.48 |
| Latvia | 3/1/2020 | 1.12 | 3.01 | 4.03 | 2.00 | 999.00 | 5.20 |
| Latvia | 3/2/2020 | 1.06 | 2.93 | 4.53 | 2.00 | 999.00 | 3.17 |
| Latvia | 3/3/2020 | 1.35 | 4.08 | 4.65 | 2.00 | 999.00 | 2.67 |
| Latvia | 3/4/2020 | 1.56 | 4.01 | 4.73 | 2.00 | 1007.00 | 4.14 |
| Latvia | 3/5/2020 | 1.00 | 3.04 | 4.22 | 1.00 | 1007.00 | 2.88 |
| Latvia | 3/6/2020 | 0.84 | 2.66 | 3.71 | 2.00 | 1008.00 | 2.15 |
| Latvia | 3/7/2020 | 1.21 | 2.46 | 4.00 | 1.00 | 1010.00 | 2.25 |
| Latvia | 3/8/2020 | 0.97 | 2.57 | 3.86 | 2.00 | 1016.00 | 1.33 |
| Latvia | 3/9/2020 | 0.84 | 2.84 | 4.02 | 3.00 | 1016.00 | 3.31 |
| Lebanon | 2/22/2020 | 1.24 | 9.01 | 6.81 | 3.00 | 1018.00 | 3.97 |
| Lebanon | 2/23/2020 | 1.30 | 8.94 | 6.31 | 3.00 | 1019.00 | 1.75 |
| Lebanon | 2/24/2020 | 1.28 | 9.67 | 6.27 | 5.00 | 1017.00 | 1.64 |
| Lebanon | 2/25/2020 | 1.37 | 9.48 | 7.16 | 5.00 | 1015.00 | 2.14 |
| Lebanon | 2/26/2020 | 0.93 | 9.97 | 7.02 | 4.00 | 1021.00 | 2.05 |
| Lebanon | 2/27/2020 | 0.97 | 10.87 | 6.34 | 5.00 | 1022.00 | 1.36 |
| Lebanon | 2/28/2020 | 1.47 | 12.59 | 6.05 | 5.00 | 1018.00 | 0.83 |
| Lebanon | 2/29/2020 | 1.23 | 9.38 | 6.52 | 4.00 | 1013.00 | 4.32 |
| Lebanon | 3/1/2020 | 0.98 | 7.93 | 6.01 | 3.00 | 1018.00 | 3.60 |
| Lebanon | 3/2/2020 | 1.08 | 9.42 | 6.64 | 4.00 | 1022.00 | 1.96 |
| Lebanon | 3/3/2020 | 0.93 | 10.39 | 6.07 | 5.00 | 1023.00 | 1.29 |
| Lebanon | 3/4/2020 | 0.82 | 13.59 | 6.13 | 6.00 | 1019.00 | 2.49 |
| Lebanon | 3/5/2020 | 0.87 | 14.87 | 6.04 | 6.00 | 1013.00 | 2.18 |
| Lithuania | 3/4/2020 | 1.80 | 5.06 | 5.16 | 2.00 | 1008.00 | 1.66 |
| Lithuania | 3/5/2020 | 1.03 | 2.51 | 4.05 | 2.00 | 1008.00 | 1.35 |
| Lithuania | 3/6/2020 | 0.79 | 1.57 | 3.69 | 2.00 | 1008.00 | 1.21 |
| Lithuania | 3/7/2020 | 1.36 | 2.81 | 4.48 | 1.00 | 1010.00 | 1.27 |
| Lithuania | 3/8/2020 | 1.18 | 3.15 | 4.71 | 2.00 | 1017.00 | 0.88 |
| Lithuania | 3/9/2020 | 1.05 | 3.04 | 4.37 | 2.00 | 1017.00 | 1.04 |
| Lithuania | 3/10/2020 | 1.10 | 3.44 | 4.45 | 2.00 | 1014.00 | 1.62 |
| Lithuania | 3/11/2020 | 1.12 | 4.38 | 4.70 | 3.00 | 999.00 | 3.49 |
| Lithuania | 3/12/2020 | 1.09 | 4.65 | 4.84 | 3.00 | 998.00 | 3.95 |
| Lithuania | 3/13/2020 | 0.84 | 3.62 | 4.05 | 1.00 | 999.00 | 3.45 |
| Macedonia | 2/26/2020 | 1.00 | 7.04 | 5.49 | 4.00 | 1011.00 | 2.74 |
| Macedonia | 2/27/2020 | 0.77 | 1.64 | 3.90 | 3.00 | 1015.00 | 2.28 |
| Macedonia | 2/28/2020 | 0.87 | 0.79 | 3.75 | 2.00 | 1019.00 | 3.01 |
| Macedonia | 2/29/2020 | 0.66 | 1.14 | 3.11 | 3.00 | 1022.00 | 1.88 |
| Macedonia | 3/1/2020 | 1.16 | 4.51 | 4.14 | 3.00 | 1020.00 | 1.73 |
| Macedonia | 3/2/2020 | 1.10 | 8.16 | 5.87 | 4.00 | 1015.00 | 2.71 |
| Macedonia | 3/3/2020 | 1.42 | 9.29 | 6.24 | 4.00 | 1010.00 | 2.83 |
| Macedonia | 3/4/2020 | 1.50 | 5.74 | 5.79 | 2.00 | 1009.00 | 2.03 |
| Macedonia | 3/5/2020 | 0.87 | 3.61 | 4.01 | 2.00 | 1011.00 | 1.92 |
| Macedonia | 3/6/2020 | 1.24 | 4.93 | 4.73 | 3.00 | 1010.00 | 2.10 |
| Macedonia | 3/7/2020 | 1.43 | 6.10 | 5.65 | 2.00 | 1012.00 | 1.79 |
| Macedonia | 3/8/2020 | 1.22 | 1.91 | 4.34 | 2.00 | 1017.00 | 2.05 |
| Macedonia | 3/9/2020 | 1.17 | 1.69 | 4.23 | 2.00 | 1016.00 | 1.71 |
| Macedonia | 3/10/2020 | 1.09 | 2.04 | 4.26 | 2.00 | 1019.00 | 2.26 |
| Macedonia | 3/11/2020 | 1.18 | 5.47 | 4.44 | 3.00 | 1023.00 | 1.19 |
| Macedonia | 3/12/2020 | 1.22 | 10.58 | 5.71 | 4.00 | 1023.00 | 1.06 |
| Madagascar | 3/14/2020 | 2.18 | 19.23 | 14.08 | 6.00 | 1015.00 | 2.09 |
| Madagascar | 3/15/2020 | 3.41 | 20.70 | 15.33 | 5.00 | 1013.00 | 1.01 |
| Madagascar | 3/16/2020 | 3.72 | 19.80 | 14.49 | 5.00 | 1012.00 | 1.41 |
| Madagascar | 3/17/2020 | 3.88 | 20.78 | 15.23 | 5.00 | 1013.00 | 1.22 |
| Madagascar | 3/18/2020 | 4.29 | 20.73 | 16.14 | 5.00 | 1015.00 | 1.13 |
| Madagascar | 3/19/2020 | 3.68 | 20.51 | 15.38 | 5.00 | 1016.00 | 1.12 |
| Madagascar | 3/20/2020 | 3.72 | 20.82 | 15.54 | 5.00 | 1017.00 | 0.91 |
| Madagascar | 3/21/2020 | 3.98 | 20.30 | 15.34 | 5.00 | 1018.00 | 1.76 |
| Madagascar | 3/22/2020 | 2.79 | 18.85 | 13.73 | 5.00 | 1020.00 | 2.20 |
| Madagascar | 3/23/2020 | 2.39 | 18.67 | 13.76 | 5.00 | 1020.00 | 2.30 |
| Madagascar | 3/24/2020 | 2.14 | 18.37 | 13.37 | 4.00 | 1019.00 | 1.86 |
| Malaysia | 2/19/2020 | 4.62 | 25.14 | 17.11 | 8.00 | 1014.00 | 1.74 |
| Malaysia | 2/20/2020 | 4.20 | 25.17 | 17.10 | 8.00 | 1013.00 | 1.92 |
| Malaysia | 2/21/2020 | 3.95 | 25.20 | 17.17 | 8.00 | 1014.00 | 1.95 |
| Malaysia | 2/22/2020 | 3.84 | 24.84 | 16.88 | 7.00 | 1015.00 | 2.37 |
| Malaysia | 2/23/2020 | 4.11 | 25.19 | 17.03 | 7.00 | 1014.00 | 1.99 |
| Malaysia | 2/24/2020 | 4.75 | 25.44 | 17.37 | 8.00 | 1013.00 | 1.81 |
| Malaysia | 2/25/2020 | 4.46 | 25.64 | 17.19 | 8.00 | 1012.00 | 1.52 |
| Malaysia | 2/26/2020 | 4.32 | 25.74 | 17.35 | 8.00 | 1011.00 | 1.61 |
| Malaysia | 2/27/2020 | 4.43 | 25.85 | 17.59 | 8.00 | 1011.00 | 1.65 |
| Malaysia | 2/28/2020 | 4.17 | 25.72 | 17.18 | 8.00 | 1011.00 | 1.83 |
| Malaysia | 2/29/2020 | 4.12 | 25.79 | 16.84 | 8.00 | 1010.00 | 1.43 |
| Malaysia | 3/1/2020 | 4.37 | 26.26 | 17.33 | 8.00 | 1011.00 | 1.10 |
| Malaysia | 3/2/2020 | 3.93 | 26.08 | 17.07 | 8.00 | 1011.00 | 1.11 |
| Malaysia | 3/3/2020 | 4.18 | 26.45 | 18.03 | 8.00 | 1011.00 | 1.09 |
| Maldives | 4/5/2020 | 4.94 | 28.63 | 19.52 | 7.00 | 1013.00 | 2.84 |
| Maldives | 4/6/2020 | 4.04 | 28.33 | 19.15 | 6.00 | 1013.00 | 2.40 |
| Maldives | 4/7/2020 | 4.25 | 28.41 | 18.17 | 7.00 | 1013.00 | 1.51 |
| Maldives | 4/8/2020 | 3.96 | 28.95 | 18.55 | 7.00 | 1013.00 | 2.87 |
| Maldives | 4/9/2020 | 4.85 | 29.03 | 18.97 | 7.00 | 1013.00 | 2.15 |
| Maldives | 4/10/2020 | 4.80 | 29.10 | 18.61 | 7.00 | 1012.00 | 1.72 |
| Maldives | 4/11/2020 | 4.65 | 28.66 | 18.21 | 6.00 | 1013.00 | 1.24 |
| Maldives | 4/12/2020 | 4.99 | 29.20 | 18.50 | 6.00 | 1013.00 | 2.02 |
| Maldives | 4/13/2020 | 5.09 | 28.83 | 19.59 | 6.00 | 1011.00 | 2.52 |
| Maldives | 4/14/2020 | 5.03 | 29.09 | 19.70 | 8.00 | 1011.00 | 2.25 |
| Maldives | 4/15/2020 | 5.67 | 28.85 | 19.30 | 6.00 | 1013.00 | 1.77 |
| Maldives | 4/16/2020 | 5.44 | 29.39 | 18.15 | 6.00 | 1011.00 | 1.15 |
| Maldives | 4/17/2020 | 5.18 | 29.39 | 19.63 | 6.00 | 1011.00 | 2.77 |
| Maldives | 4/18/2020 | 5.42 | 29.23 | 20.59 | 7.00 | 1011.00 | 3.77 |
| Maldives | 4/19/2020 | 5.51 | 29.06 | 20.86 | 6.00 | 1010.00 | 3.72 |
| Mali | 3/18/2020 | 2.76 | 32.21 | 8.32 | 9.00 | 1012.00 | 2.15 |
| Mali | 3/19/2020 | 2.26 | 31.40 | 6.02 | 8.00 | 1011.00 | 2.70 |
| Mali | 3/20/2020 | 2.08 | 31.27 | 5.15 | 8.00 | 1011.00 | 2.81 |
| Mali | 3/21/2020 | 2.98 | 31.86 | 7.47 | 8.00 | 1011.00 | 3.16 |
| Mali | 3/22/2020 | 3.50 | 32.56 | 10.03 | 8.00 | 1011.00 | 1.46 |
| Mali | 3/23/2020 | 3.87 | 31.70 | 10.85 | 8.00 | 1012.00 | 2.66 |
| Mali | 3/24/2020 | 3.24 | 30.70 | 8.15 | 8.00 | 1013.00 | 3.84 |
| Mali | 3/25/2020 | 2.28 | 29.15 | 6.01 | 8.00 | 1014.00 | 2.21 |
| Mali | 3/26/2020 | 2.52 | 29.49 | 5.89 | 8.00 | 1012.00 | 2.06 |
| Mali | 3/27/2020 | 2.89 | 30.75 | 6.94 | 8.00 | 1011.00 | 1.66 |
| Mali | 3/28/2020 | 3.80 | 32.44 | 11.37 | 8.00 | 1011.00 | 1.63 |
| Mali | 3/29/2020 | 3.98 | 28.86 | 14.12 | 8.00 | 1012.00 | 1.70 |
| Mali | 3/30/2020 | 2.97 | 29.30 | 10.53 | 9.00 | 1010.00 | 2.64 |
| Mali | 3/31/2020 | 2.54 | 29.89 | 9.49 | 9.00 | 1010.00 | 3.13 |
| Mali | 4/1/2020 | 2.52 | 30.08 | 9.41 | 9.00 | 1012.00 | 3.37 |
| Mali | 4/2/2020 | 2.67 | 31.13 | 8.66 | 9.00 | 1012.00 | 2.91 |
| Mali | 4/3/2020 | 2.39 | 31.09 | 7.41 | 9.00 | 1010.00 | 2.23 |
| Mali | 4/4/2020 | 2.14 | 31.46 | 7.22 | 9.00 | 1010.00 | 2.26 |
| Mali | 4/5/2020 | 2.12 | 32.15 | 8.74 | 9.00 | 1011.00 | 1.30 |
| Mali | 4/6/2020 | 1.73 | 30.97 | 6.71 | 9.00 | 1012.00 | 2.41 |
| Mali | 4/7/2020 | 1.73 | 30.53 | 6.14 | 9.00 | 1012.00 | 1.69 |
| Malta | 3/2/2020 | 1.60 | 16.30 | 10.04 | 5.00 | 1017.00 | 3.83 |
| Malta | 3/3/2020 | 1.79 | 15.62 | 8.43 | 5.00 | 1012.00 | 7.01 |
| Malta | 3/4/2020 | 1.81 | 14.68 | 7.84 | 4.00 | 1012.00 | 8.08 |
| Malta | 3/5/2020 | 1.59 | 14.74 | 7.59 | 3.00 | 1015.00 | 5.94 |
| Malta | 3/6/2020 | 1.92 | 15.73 | 9.94 | 5.00 | 1013.00 | 4.69 |
| Malta | 3/7/2020 | 1.57 | 14.54 | 7.61 | 4.00 | 1014.00 | 7.28 |
| Malta | 3/8/2020 | 1.52 | 13.86 | 7.33 | 3.00 | 1017.00 | 6.74 |
| Malta | 3/9/2020 | 1.46 | 14.23 | 7.21 | 3.00 | 1018.00 | 6.08 |
| Malta | 3/10/2020 | 1.89 | 13.87 | 7.57 | 3.00 | 1018.00 | 4.74 |
| Malta | 3/11/2020 | 1.40 | 14.59 | 6.96 | 4.00 | 1024.00 | 3.62 |
| Mauritania | 5/10/2020 | 2.42 | 29.06 | 7.49 | 7.00 | 1012.00 | 3.82 |
| Mauritania | 5/11/2020 | 2.52 | 29.40 | 7.27 | 7.00 | 1012.00 | 3.97 |
| Mauritania | 5/12/2020 | 2.17 | 27.74 | 8.02 | 7.00 | 1014.00 | 5.17 |
| Mauritania | 5/13/2020 | 1.87 | 25.71 | 8.62 | 6.00 | 1013.00 | 4.73 |
| Mauritania | 5/14/2020 | 1.97 | 26.48 | 8.11 | 7.00 | 1013.00 | 4.56 |
| Mauritania | 5/15/2020 | 2.33 | 26.70 | 8.27 | 6.00 | 1014.00 | 5.07 |
| Mauritania | 5/16/2020 | 1.98 | 27.10 | 6.81 | 7.00 | 1012.00 | 3.91 |
| Mauritania | 5/17/2020 | 1.74 | 27.17 | 6.64 | 7.00 | 1012.00 | 3.51 |
| Mauritania | 5/18/2020 | 1.89 | 26.30 | 8.14 | 6.00 | 1014.00 | 3.34 |
| Mauritania | 5/19/2020 | 2.64 | 24.82 | 10.63 | 6.00 | 1015.00 | 4.14 |
| Mauritania | 5/20/2020 | 2.63 | 24.78 | 10.68 | 6.00 | 1014.00 | 4.53 |
| Mauritania | 5/21/2020 | 2.44 | 26.74 | 10.23 | 6.00 | 1015.00 | 4.44 |
| Mauritania | 5/22/2020 | 2.24 | 28.57 | 8.60 | 7.00 | 1013.00 | 4.51 |
| Mauritania | 5/23/2020 | 2.35 | 27.84 | 10.28 | 7.00 | 1012.00 | 4.40 |
| Mauritania | 5/24/2020 | 1.96 | 27.68 | 10.18 | 7.00 | 1013.00 | 2.95 |
| Mauritania | 5/25/2020 | 1.79 | 27.74 | 9.91 | 7.00 | 1013.00 | 2.72 |
| Mauritania | 5/26/2020 | 1.91 | 28.41 | 9.58 | 7.00 | 1012.00 | 2.65 |
| Mauritania | 5/27/2020 | 2.01 | 29.08 | 9.49 | 7.00 | 1011.00 | 2.75 |
| Mauritania | 5/28/2020 | 2.42 | 28.95 | 9.58 | 7.00 | 1011.00 | 2.50 |
| Mauritania | 5/29/2020 | 2.40 | 28.79 | 9.59 | 7.00 | 1011.00 | 2.78 |
| Mauritania | 5/30/2020 | 2.61 | 28.63 | 9.05 | 7.00 | 1011.00 | 5.50 |
| Mauritius | 3/15/2020 | 3.58 | 26.78 | 17.76 | 6.00 | 1010.00 | 4.30 |
| Mauritius | 3/16/2020 | 4.95 | 26.59 | 17.23 | 7.00 | 1008.00 | 3.05 |
| Mauritius | 3/17/2020 | 3.92 | 26.77 | 17.32 | 7.00 | 1008.00 | 4.84 |
| Mauritius | 3/18/2020 | 3.38 | 26.78 | 18.15 | 7.00 | 1011.00 | 4.11 |
| Mauritius | 3/19/2020 | 3.38 | 26.60 | 16.87 | 6.00 | 1014.00 | 3.71 |
| Mauritius | 3/20/2020 | 3.17 | 26.26 | 15.59 | 7.00 | 1016.00 | 3.65 |
| Mauritius | 3/21/2020 | 2.53 | 26.30 | 14.99 | 6.00 | 1017.00 | 4.49 |
| Mauritius | 3/22/2020 | 3.49 | 26.85 | 17.26 | 6.00 | 1017.00 | 5.96 |
| Mauritius | 3/23/2020 | 5.80 | 27.49 | 19.48 | 6.00 | 1016.00 | 6.90 |
| Mauritius | 3/24/2020 | 6.02 | 27.08 | 18.98 | 6.00 | 1015.00 | 6.53 |
| Mauritius | 3/25/2020 | 4.96 | 27.06 | 18.22 | 6.00 | 1012.00 | 6.31 |
| Mauritius | 3/26/2020 | 3.98 | 26.80 | 17.58 | 6.00 | 1013.00 | 6.85 |
| Mauritius | 3/27/2020 | 3.61 | 26.87 | 17.45 | 6.00 | 1014.00 | 5.45 |
| Mauritius | 3/28/2020 | 4.20 | 27.07 | 18.49 | 7.00 | 1014.00 | 3.96 |
| Mauritius | 3/29/2020 | 5.20 | 27.56 | 19.50 | 6.00 | 1011.00 | 6.33 |
| Mexico | 2/29/2020 | 1.11 | 16.22 | 5.33 | 5.00 | 1024.00 | 0.44 |
| Mexico | 3/1/2020 | 1.36 | 17.39 | 6.50 | 6.00 | 1020.00 | 1.10 |
| Mexico | 3/2/2020 | 1.47 | 18.41 | 7.34 | 6.00 | 1017.00 | 1.39 |
| Mexico | 3/3/2020 | 1.02 | 17.04 | 6.25 | 6.00 | 1015.00 | 1.91 |
| Mexico | 3/4/2020 | 0.80 | 16.82 | 5.38 | 6.00 | 1017.00 | 1.34 |
| Mexico | 3/5/2020 | 0.88 | 15.96 | 5.02 | 6.00 | 1021.00 | 1.42 |
| Mexico | 3/6/2020 | 1.05 | 13.96 | 5.43 | 5.00 | 1025.00 | 1.38 |
| Mexico | 3/7/2020 | 1.22 | 14.72 | 5.43 | 5.00 | 1025.00 | 0.78 |
| Mexico | 3/8/2020 | 1.25 | 15.57 | 5.69 | 5.00 | 1021.00 | 0.88 |
| Mexico | 3/9/2020 | 1.20 | 16.89 | 5.72 | 6.00 | 1022.00 | 0.75 |
| Moldova | 3/3/2020 | 1.56 | 9.36 | 5.54 | 4.00 | 1010.00 | 2.81 |
| Moldova | 3/4/2020 | 1.51 | 11.24 | 5.86 | 5.00 | 1008.00 | 4.55 |
| Moldova | 3/5/2020 | 1.97 | 8.69 | 6.09 | 4.00 | 1008.00 | 3.11 |
| Moldova | 3/6/2020 | 1.44 | 8.22 | 4.88 | 3.00 | 1007.00 | 2.48 |
| Moldova | 3/7/2020 | 1.53 | 10.12 | 5.23 | 4.00 | 1012.00 | 2.42 |
| Moldova | 3/8/2020 | 2.01 | 8.41 | 5.91 | 4.00 | 1018.00 | 2.08 |
| Moldova | 3/9/2020 | 2.13 | 10.32 | 6.56 | 3.00 | 1018.00 | 1.48 |
| Moldova | 3/10/2020 | 1.96 | 9.03 | 6.09 | 4.00 | 1011.00 | 4.90 |
| Moldova | 3/11/2020 | 1.55 | 7.98 | 4.78 | 4.00 | 1015.00 | 3.68 |
| Moldova | 3/12/2020 | 1.22 | 9.34 | 5.40 | 3.00 | 1019.00 | 2.80 |
| Moldova | 3/13/2020 | 1.92 | 12.24 | 6.34 | 5.00 | 1015.00 | 2.89 |
| Moldova | 3/14/2020 | 0.80 | 6.49 | 3.72 | 3.00 | 1021.00 | 4.48 |
| Moldova | 3/15/2020 | 0.36 | 0.37 | 1.99 | 2.00 | 1032.00 | 3.88 |
| Moldova | 3/16/2020 | 0.30 | 1.55 | 1.79 | 3.00 | 1034.00 | 1.27 |
| Moldova | 3/17/2020 | 0.80 | 4.48 | 2.96 | 3.00 | 1031.00 | 1.98 |
| Moldova | 3/18/2020 | 0.88 | 5.81 | 3.61 | 2.00 | 1032.00 | 1.31 |
| Moldova | 3/19/2020 | 1.34 | 8.20 | 4.20 | 4.00 | 1029.00 | 1.84 |
| Moldova | 3/20/2020 | 1.56 | 10.78 | 5.47 | 3.00 | 1022.00 | 2.72 |
| Moldova | 3/21/2020 | 1.74 | 10.21 | 5.61 | 3.00 | 1019.00 | 2.66 |
| Moldova | 3/22/2020 | 1.05 | 3.78 | 3.53 | 2.00 | 1023.00 | 4.59 |
| Moldova | 3/23/2020 | 0.69 | 1.35 | 2.30 | 2.00 | 1030.00 | 4.90 |
| Moldova | 3/24/2020 | 0.56 | 0.28 | 1.71 | 1.00 | 1033.00 | 5.36 |
| Moldova | 3/25/2020 | 0.75 | 2.29 | 2.12 | 1.00 | 1034.00 | 4.55 |
| Monaco | 3/16/2020 | 1.05 | 8.41 | 5.32 | 4.00 | 1027.00 | 4.04 |
| Monaco | 3/17/2020 | 1.38 | 9.93 | 5.83 | 4.00 | 1029.00 | 1.85 |
| Monaco | 3/18/2020 | 1.14 | 10.90 | 5.87 | 5.00 | 1029.00 | 1.43 |
| Monaco | 3/19/2020 | 0.89 | 11.12 | 5.94 | 5.00 | 1026.00 | 2.23 |
| Monaco | 3/20/2020 | 1.18 | 9.94 | 6.23 | 3.00 | 1024.00 | 1.69 |
| Monaco | 3/21/2020 | 1.21 | 9.93 | 6.43 | 4.00 | 1019.00 | 1.41 |
| Monaco | 3/22/2020 | 1.43 | 9.39 | 6.25 | 4.00 | 1019.00 | 3.08 |
| Monaco | 3/23/2020 | 0.93 | 6.30 | 4.45 | 4.00 | 1025.00 | 4.82 |
| Monaco | 3/24/2020 | 0.62 | 4.10 | 3.58 | 2.00 | 1025.00 | 3.64 |
| Monaco | 3/25/2020 | 0.65 | 2.86 | 3.51 | 3.00 | 1023.00 | 5.10 |
| Monaco | 3/26/2020 | 0.92 | 3.08 | 3.67 | 3.00 | 1016.00 | 6.13 |
| Monaco | 3/27/2020 | 1.53 | 6.44 | 4.95 | 3.00 | 1016.00 | 5.47 |
| Monaco | 3/28/2020 | 1.43 | 9.36 | 5.83 | 4.00 | 1017.00 | 2.00 |
| Monaco | 3/29/2020 | 1.38 | 10.35 | 6.67 | 4.00 | 1014.00 | 1.03 |
| Monaco | 3/30/2020 | 1.46 | 9.21 | 6.37 | 4.00 | 1014.00 | 1.39 |
| Montenegro | 3/13/2020 | 1.62 | 13.22 | 7.02 | 5.00 | 1018.00 | 1.18 |
| Montenegro | 3/14/2020 | 1.91 | 10.10 | 6.45 | 5.00 | 1016.00 | 3.64 |
| Montenegro | 3/15/2020 | 1.19 | 6.47 | 4.71 | 3.00 | 1023.00 | 5.52 |
| Montenegro | 3/16/2020 | 0.63 | 7.11 | 4.37 | 4.00 | 1030.00 | 3.47 |
| Montenegro | 3/17/2020 | 0.70 | 8.89 | 4.54 | 4.00 | 1030.00 | 1.41 |
| Montenegro | 3/18/2020 | 0.89 | 9.27 | 5.18 | 4.00 | 1028.00 | 2.47 |
| Montenegro | 3/19/2020 | 0.62 | 9.40 | 4.58 | 4.00 | 1028.00 | 1.81 |
| Montenegro | 3/20/2020 | 0.92 | 11.22 | 5.27 | 4.00 | 1025.00 | 1.10 |
| Montenegro | 3/21/2020 | 0.85 | 11.04 | 5.47 | 4.00 | 1020.00 | 1.32 |
| Montenegro | 3/22/2020 | 1.32 | 9.44 | 5.59 | 3.00 | 1018.00 | 3.51 |
| Montenegro | 3/23/2020 | 0.88 | 2.93 | 3.46 | 3.00 | 1020.00 | 5.02 |
| Montenegro | 3/24/2020 | 0.62 | 1.40 | 3.29 | 1.00 | 1019.00 | 2.32 |
| Montenegro | 3/25/2020 | 1.03 | 4.91 | 4.24 | 2.00 | 1020.00 | 3.95 |
| Montenegro | 3/26/2020 | 1.51 | 7.05 | 5.00 | 3.00 | 1015.00 | 7.03 |
| Morocco | 3/4/2020 | 1.33 | 16.33 | 9.52 | 5.00 | 1026.00 | 4.10 |
| Morocco | 3/5/2020 | 1.76 | 16.24 | 9.53 | 5.00 | 1025.00 | 3.65 |
| Morocco | 3/6/2020 | 1.19 | 15.41 | 7.83 | 4.00 | 1027.00 | 4.79 |
| Morocco | 3/7/2020 | 1.02 | 15.68 | 7.99 | 4.00 | 1025.00 | 4.98 |
| Morocco | 3/8/2020 | 0.88 | 16.38 | 8.55 | 5.00 | 1026.00 | 4.69 |
| Morocco | 3/9/2020 | 1.05 | 16.71 | 9.14 | 5.00 | 1027.00 | 3.57 |
| Morocco | 3/10/2020 | 1.59 | 16.89 | 9.31 | 5.00 | 1026.00 | 3.05 |
| Morocco | 3/11/2020 | 1.55 | 18.42 | 9.47 | 6.00 | 1020.00 | 2.22 |
| Morocco | 3/12/2020 | 1.72 | 17.82 | 9.15 | 6.00 | 1021.00 | 1.86 |
| Morocco | 3/13/2020 | 2.11 | 16.43 | 9.70 | 5.00 | 1020.00 | 2.44 |
| Morocco | 3/14/2020 | 2.24 | 15.47 | 9.18 | 5.00 | 1019.00 | 1.81 |
| Morocco | 3/15/2020 | 2.33 | 15.71 | 9.35 | 4.00 | 1018.00 | 2.52 |
| Morocco | 3/16/2020 | 1.36 | 14.63 | 7.51 | 3.00 | 1022.00 | 4.90 |
| Morocco | 3/17/2020 | 1.53 | 13.66 | 6.90 | 4.00 | 1021.00 | 2.09 |
| Morocco | 3/18/2020 | 2.15 | 14.83 | 8.23 | 3.00 | 1014.00 | 2.44 |
| Morocco | 3/19/2020 | 2.11 | 16.74 | 10.05 | 4.00 | 1011.00 | 2.30 |
| Mozambique | 4/28/2020 | 4.17 | 23.22 | 13.17 | 6.00 | 1021.00 | 2.43 |
| Mozambique | 4/29/2020 | 3.42 | 22.25 | 12.38 | 5.00 | 1020.00 | 2.22 |
| Mozambique | 4/30/2020 | 3.00 | 23.64 | 12.43 | 5.00 | 1021.00 | 2.17 |
| Mozambique | 5/1/2020 | 2.35 | 23.16 | 12.53 | 6.00 | 1023.00 | 3.29 |
| Mozambique | 5/2/2020 | 1.90 | 22.97 | 11.93 | 7.00 | 1023.00 | 2.07 |
| Mozambique | 5/3/2020 | 2.00 | 23.38 | 12.26 | 7.00 | 1019.00 | 2.23 |
| Mozambique | 5/4/2020 | 2.89 | 23.20 | 13.74 | 5.00 | 1024.00 | 3.49 |
| Mozambique | 5/5/2020 | 2.77 | 23.47 | 12.92 | 7.00 | 1024.00 | 1.75 |
| Mozambique | 5/6/2020 | 3.03 | 22.89 | 12.61 | 5.00 | 1024.00 | 3.41 |
| Mozambique | 5/7/2020 | 2.81 | 21.99 | 12.09 | 6.00 | 1025.00 | 3.59 |
| Mozambique | 5/8/2020 | 2.72 | 22.49 | 11.89 | 6.00 | 1024.00 | 1.80 |
| Mozambique | 5/9/2020 | 2.77 | 22.23 | 12.38 | 6.00 | 1026.00 | 2.23 |
| Mozambique | 5/10/2020 | 2.35 | 22.25 | 12.06 | 6.00 | 1026.00 | 2.10 |
| Mozambique | 5/11/2020 | 1.81 | 22.11 | 11.37 | 6.00 | 1023.00 | 2.45 |
| Mozambique | 5/12/2020 | 1.82 | 21.98 | 11.19 | 6.00 | 1018.00 | 1.76 |
| Mozambique | 5/13/2020 | 2.28 | 23.23 | 11.59 | 7.00 | 1019.00 | 3.54 |
| Mozambique | 5/14/2020 | 2.60 | 21.81 | 11.78 | 5.00 | 1023.00 | 4.55 |
| Mozambique | 5/15/2020 | 2.18 | 22.19 | 10.99 | 6.00 | 1022.00 | 2.76 |
| Mozambique | 5/16/2020 | 2.15 | 21.15 | 10.71 | 6.00 | 1022.00 | 5.58 |
| Mozambique | 5/17/2020 | 2.06 | 21.91 | 10.53 | 5.00 | 1023.00 | 2.60 |
| Mozambique | 5/18/2020 | 1.78 | 22.06 | 10.43 | 7.00 | 1024.00 | 2.76 |
| Mozambique | 5/19/2020 | 2.13 | 21.61 | 12.01 | 6.00 | 1025.00 | 1.79 |
| Mozambique | 5/20/2020 | 2.44 | 21.72 | 12.49 | 6.00 | 1028.00 | 2.13 |
| Mozambique | 5/21/2020 | 2.83 | 21.47 | 12.41 | 6.00 | 1028.00 | 2.16 |
| Mozambique | 5/22/2020 | 2.67 | 21.98 | 12.12 | 6.00 | 1026.00 | 1.47 |
| Mozambique | 5/23/2020 | 2.09 | 21.93 | 11.72 | 6.00 | 1025.00 | 1.36 |
| Mozambique | 5/24/2020 | 1.93 | 21.49 | 11.51 | 6.00 | 1023.00 | 1.69 |
| Mozambique | 5/25/2020 | 2.11 | 22.63 | 10.25 | 7.00 | 1018.00 | 3.87 |
| Mozambique | 5/26/2020 | 2.06 | 22.76 | 9.38 | 7.00 | 1015.00 | 4.04 |
| Mozambique | 5/27/2020 | 1.63 | 19.33 | 7.29 | 5.00 | 1023.00 | 4.99 |
| Mozambique | 5/28/2020 | 1.48 | 19.74 | 7.04 | 6.00 | 1023.00 | 3.23 |
| Mozambique | 5/29/2020 | 1.25 | 20.57 | 6.65 | 6.00 | 1019.00 | 2.55 |
| Mozambique | 5/30/2020 | 2.55 | 21.62 | 11.49 | 6.00 | 1024.00 | 3.35 |
| Nepal | 4/19/2020 | 1.77 | 19.35 | 8.40 | 5.00 | 1014.00 | 1.86 |
| Nepal | 4/20/2020 | 1.89 | 19.76 | 8.45 | 6.00 | 1015.00 | 1.74 |
| Nepal | 4/21/2020 | 2.09 | 16.02 | 8.42 | 5.00 | 1016.00 | 1.35 |
| Nepal | 4/22/2020 | 2.05 | 14.96 | 8.93 | 4.00 | 1014.00 | 1.78 |
| Nepal | 4/23/2020 | 2.11 | 15.37 | 8.52 | 5.00 | 1014.00 | 1.42 |
| Nepal | 4/24/2020 | 2.21 | 15.93 | 8.97 | 5.00 | 1013.00 | 1.50 |
| Nepal | 4/25/2020 | 2.20 | 16.49 | 9.35 | 5.00 | 1013.00 | 1.49 |
| Nepal | 4/26/2020 | 2.60 | 15.86 | 10.14 | 4.00 | 1016.00 | 1.51 |
| Nepal | 4/27/2020 | 2.69 | 16.31 | 10.43 | 6.00 | 1018.00 | 1.52 |
| Nepal | 4/28/2020 | 2.69 | 15.53 | 10.74 | 4.00 | 1018.00 | 1.63 |
| Nepal | 4/29/2020 | 2.74 | 16.23 | 10.52 | 6.00 | 1015.00 | 1.58 |
| Nepal | 4/30/2020 | 2.86 | 17.15 | 10.82 | 5.00 | 1015.00 | 1.34 |
| Nepal | 5/1/2020 | 2.90 | 16.44 | 11.98 | 5.00 | 1014.00 | 1.54 |
| Nepal | 5/2/2020 | 2.90 | 16.19 | 11.12 | 6.00 | 1014.00 | 1.24 |
| Nepal | 5/3/2020 | 2.82 | 17.97 | 11.20 | 5.00 | 1014.00 | 1.47 |
| Nepal | 5/4/2020 | 2.53 | 18.46 | 11.42 | 5.00 | 1015.00 | 1.49 |
| Nepal | 5/5/2020 | 2.37 | 18.32 | 10.93 | 6.00 | 1014.00 | 1.40 |
| Nepal | 5/6/2020 | 2.24 | 17.55 | 9.99 | 5.00 | 1013.00 | 1.38 |
| Nepal | 5/7/2020 | 2.02 | 17.05 | 9.71 | 6.00 | 1013.00 | 1.50 |
| Nepal | 5/8/2020 | 2.07 | 17.14 | 9.28 | 6.00 | 1013.00 | 1.63 |
| Nepal | 5/9/2020 | 2.25 | 19.02 | 10.16 | 6.00 | 1014.00 | 1.63 |
| Nepal | 5/10/2020 | 2.65 | 18.06 | 11.09 | 6.00 | 1015.00 | 1.46 |
| Nepal | 5/11/2020 | 2.46 | 18.84 | 10.86 | 6.00 | 1016.00 | 1.43 |
| Nepal | 5/12/2020 | 2.87 | 18.17 | 11.49 | 6.00 | 1014.00 | 1.65 |
| Nepal | 5/13/2020 | 2.70 | 19.57 | 11.18 | 6.00 | 1012.00 | 1.76 |
| Nepal | 5/14/2020 | 2.68 | 20.65 | 11.30 | 6.00 | 1012.00 | 1.62 |
| Nepal | 5/15/2020 | 2.81 | 20.43 | 12.40 | 7.00 | 1010.00 | 1.55 |
| Nepal | 5/16/2020 | 2.22 | 21.14 | 11.00 | 7.00 | 1010.00 | 1.76 |
| Nepal | 5/17/2020 | 1.49 | 21.61 | 8.51 | 7.00 | 1008.00 | 1.70 |
| Nepal | 5/18/2020 | 2.12 | 21.20 | 10.26 | 7.00 | 1008.00 | 1.61 |
| Nepal | 5/19/2020 | 2.06 | 21.29 | 11.42 | 7.00 | 1009.00 | 1.41 |
| Nepal | 5/20/2020 | 2.11 | 20.74 | 10.98 | 6.00 | 1008.00 | 1.49 |
| Nepal | 5/21/2020 | 2.89 | 20.95 | 11.61 | 5.00 | 1007.00 | 1.73 |
| Nepal | 5/22/2020 | 2.37 | 22.05 | 10.68 | 6.00 | 1009.00 | 1.59 |
| Nepal | 5/23/2020 | 2.56 | 21.65 | 11.94 | 6.00 | 1010.00 | 1.63 |
| Nepal | 5/24/2020 | 2.41 | 22.12 | 11.36 | 6.00 | 1010.00 | 1.90 |
| Nepal | 5/25/2020 | 1.97 | 23.57 | 9.74 | 7.00 | 1008.00 | 1.70 |
| Nepal | 5/26/2020 | 1.92 | 22.32 | 9.77 | 7.00 | 1009.00 | 1.42 |
| Nepal | 5/27/2020 | 2.39 | 21.71 | 11.29 | 7.00 | 1008.00 | 1.46 |
| Nepal | 5/28/2020 | 2.81 | 19.28 | 11.92 | 5.00 | 1010.00 | 1.40 |
| Netherlands | 2/25/2020 | 0.97 | 6.44 | 5.00 | 2.00 | 999.00 | 5.87 |
| Netherlands | 2/26/2020 | 0.96 | 4.25 | 4.47 | 2.00 | 999.00 | 5.35 |
| Netherlands | 2/27/2020 | 0.96 | 3.23 | 4.29 | 1.00 | 999.00 | 4.47 |
| Netherlands | 2/28/2020 | 1.24 | 4.70 | 4.73 | 2.00 | 1015.00 | 6.30 |
| Netherlands | 2/29/2020 | 1.27 | 7.84 | 5.66 | 2.00 | 999.00 | 8.80 |
| Netherlands | 3/1/2020 | 1.00 | 6.34 | 5.13 | 2.00 | 992.00 | 7.16 |
| Netherlands | 3/2/2020 | 1.19 | 5.11 | 5.12 | 2.00 | 994.00 | 4.03 |
| Netherlands | 3/3/2020 | 1.00 | 5.23 | 4.78 | 2.00 | 999.00 | 3.89 |
| Netherlands | 3/4/2020 | 1.08 | 6.13 | 4.96 | 2.00 | 1010.00 | 3.16 |
| Netherlands | 3/5/2020 | 1.57 | 4.93 | 5.20 | 2.00 | 999.00 | 5.20 |
| Netherlands | 3/6/2020 | 1.24 | 6.17 | 5.26 | 2.00 | 999.00 | 4.80 |
| Netherlands | 3/7/2020 | 1.29 | 6.83 | 5.05 | 2.00 | 1017.00 | 4.48 |
| Netherlands | 3/8/2020 | 1.36 | 7.94 | 5.99 | 2.00 | 1015.00 | 6.57 |
| Netherlands | 3/9/2020 | 1.27 | 7.30 | 5.53 | 2.00 | 1013.00 | 5.10 |
| Netherlands | 3/10/2020 | 2.63 | 9.68 | 7.06 | 3.00 | 1010.00 | 8.23 |
| Netherlands | 3/11/2020 | 1.71 | 10.64 | 7.11 | 3.00 | 1011.00 | 5.92 |
| New Zealand | 3/2/2020 | 3.07 | 21.32 | 11.76 | 5.00 | 1023.00 | 6.18 |
| New Zealand | 3/3/2020 | 4.10 | 21.93 | 13.32 | 6.00 | 1019.00 | 4.21 |
| New Zealand | 3/4/2020 | 3.81 | 21.26 | 13.28 | 6.00 | 1016.00 | 4.47 |
| New Zealand | 3/5/2020 | 2.68 | 19.95 | 10.84 | 5.00 | 1024.00 | 5.85 |
| New Zealand | 3/6/2020 | 1.69 | 19.28 | 9.00 | 5.00 | 1025.00 | 6.24 |
| New Zealand | 3/7/2020 | 1.66 | 19.97 | 9.19 | 5.00 | 1024.00 | 4.26 |
| New Zealand | 3/8/2020 | 1.81 | 20.22 | 9.70 | 6.00 | 1018.00 | 3.37 |
| New Zealand | 3/9/2020 | 3.36 | 20.69 | 12.17 | 5.00 | 1015.00 | 5.03 |
| New Zealand | 3/10/2020 | 2.17 | 19.95 | 10.38 | 4.00 | 1011.00 | 6.92 |
| New Zealand | 3/11/2020 | 2.34 | 19.54 | 10.24 | 4.00 | 1015.00 | 4.66 |
| New Zealand | 3/12/2020 | 2.31 | 19.41 | 10.41 | 5.00 | 1022.00 | 3.40 |
| New Zealand | 3/13/2020 | 2.06 | 19.11 | 9.28 | 5.00 | 1023.00 | 4.38 |
| New Zealand | 3/14/2020 | 2.02 | 19.31 | 9.42 | 5.00 | 1020.00 | 3.31 |
| New Zealand | 3/15/2020 | 1.89 | 19.59 | 10.13 | 5.00 | 1015.00 | 2.71 |
| New Zealand | 3/16/2020 | 1.91 | 18.64 | 10.24 | 5.00 | 1014.00 | 2.73 |
| Nigeria | 4/3/2020 | 5.01 | 27.91 | 18.70 | 8.26 | 1009.74 | 1.94 |
| Nigeria | 4/4/2020 | 4.79 | 27.82 | 18.73 | 7.53 | 1011.21 | 2.09 |
| Nigeria | 4/5/2020 | 4.83 | 28.33 | 18.31 | 7.26 | 1012.21 | 1.35 |
| Nigeria | 4/6/2020 | 5.02 | 28.02 | 18.96 | 8.26 | 1011.47 | 2.23 |
| Nigeria | 4/7/2020 | 4.72 | 27.23 | 19.07 | 8.26 | 1011.00 | 1.64 |
| Nigeria | 4/8/2020 | 5.29 | 27.19 | 18.74 | 8.26 | 1010.47 | 1.85 |
| Nigeria | 4/9/2020 | 4.99 | 27.32 | 18.41 | 8.26 | 1010.74 | 1.63 |
| Nigeria | 4/10/2020 | 4.80 | 27.32 | 16.45 | 8.00 | 1012.26 | 1.40 |
| Nigeria | 4/11/2020 | 4.62 | 26.83 | 16.09 | 8.00 | 1012.26 | 1.25 |
| Nigeria | 4/12/2020 | 4.13 | 27.48 | 15.89 | 8.00 | 1012.53 | 1.43 |
| Nigeria | 4/13/2020 | 4.51 | 28.00 | 16.35 | 8.00 | 1012.26 | 1.80 |
| Nigeria | 4/14/2020 | 4.41 | 28.08 | 16.25 | 8.00 | 1011.53 | 1.75 |
| Nigeria | 4/15/2020 | 4.40 | 28.32 | 16.63 | 7.26 | 1011.26 | 1.73 |
| Nigeria | 4/16/2020 | 4.46 | 28.64 | 16.65 | 7.26 | 1010.53 | 1.66 |
| Nigeria | 4/17/2020 | 4.01 | 28.47 | 16.27 | 7.26 | 1010.00 | 2.03 |
| Nigeria | 4/18/2020 | 4.22 | 28.12 | 16.18 | 7.26 | 1011.74 | 1.96 |
| Nigeria | 4/19/2020 | 4.21 | 27.66 | 16.58 | 7.26 | 1012.74 | 2.14 |
| Nigeria | 4/20/2020 | 4.27 | 27.70 | 17.20 | 7.53 | 1013.47 | 1.87 |
| Nigeria | 4/21/2020 | 4.54 | 28.50 | 17.79 | 7.53 | 1012.00 | 1.57 |
| Nigeria | 4/22/2020 | 5.10 | 28.05 | 18.36 | 8.26 | 1010.74 | 1.69 |
| Nigeria | 4/23/2020 | 4.85 | 28.06 | 18.02 | 7.53 | 1010.74 | 1.57 |
| Nigeria | 4/24/2020 | 4.72 | 27.94 | 18.25 | 7.53 | 1010.74 | 1.60 |
| Nigeria | 4/25/2020 | 5.23 | 27.70 | 18.51 | 8.26 | 1010.74 | 2.43 |
| Nigeria | 4/26/2020 | 5.35 | 27.30 | 18.05 | 6.79 | 1011.47 | 1.99 |
| Nigeria | 4/27/2020 | 5.16 | 27.41 | 18.65 | 8.26 | 1012.47 | 1.71 |
| Norway | 2/19/2020 | 0.67 | 0.65 | 3.66 | 1.00 | 999.00 | 0.40 |
| Norway | 2/20/2020 | 1.21 | 1.54 | 4.35 | 2.00 | 999.00 | 0.69 |
| Norway | 2/21/2020 | 0.90 | 2.38 | 4.37 | 1.00 | 997.00 | 0.67 |
| Norway | 2/22/2020 | 1.02 | 3.94 | 4.58 | 2.00 | 990.00 | 0.88 |
| Norway | 2/23/2020 | 0.60 | 1.97 | 3.74 | 1.00 | 999.00 | 0.51 |
| Norway | 2/24/2020 | 0.52 | -0.97 | 3.08 | 2.00 | 1007.00 | 0.46 |
| Norway | 2/25/2020 | 0.70 | -1.19 | 3.01 | 2.00 | 999.00 | 0.73 |
| Norway | 2/26/2020 | 0.29 | -4.35 | 2.18 | 1.00 | 999.00 | 0.57 |
| Norway | 2/27/2020 | 0.45 | -4.92 | 1.98 | 2.00 | 999.00 | 0.27 |
| Norway | 2/28/2020 | 0.65 | -2.94 | 2.95 | 2.00 | 1002.00 | 0.38 |
| Norway | 2/29/2020 | 0.99 | -1.67 | 3.37 | 2.00 | 999.00 | 0.38 |
| Norway | 3/1/2020 | 1.04 | 0.06 | 3.97 | 1.00 | 981.00 | 0.47 |
| Norway | 3/2/2020 | 0.91 | 1.38 | 4.34 | 1.00 | 996.00 | 0.49 |
| Norway | 3/3/2020 | 0.99 | 0.58 | 4.01 | 1.00 | 999.00 | 0.43 |
| Norway | 3/4/2020 | 1.01 | -0.96 | 3.57 | 1.00 | 1006.00 | 0.39 |
| Oman | 3/8/2020 | 1.58 | 22.58 | 10.61 | 7.00 | 1016.00 | 2.58 |
| Oman | 3/9/2020 | 2.21 | 22.62 | 11.66 | 7.00 | 1014.00 | 2.87 |
| Oman | 3/10/2020 | 1.53 | 20.76 | 9.43 | 6.00 | 1017.00 | 3.85 |
| Oman | 3/11/2020 | 0.67 | 20.84 | 6.86 | 6.00 | 1019.00 | 2.68 |
| Oman | 3/12/2020 | 0.71 | 22.11 | 7.65 | 6.00 | 1018.00 | 1.93 |
| Oman | 3/13/2020 | 0.95 | 22.30 | 8.94 | 7.00 | 1019.00 | 1.99 |
| Oman | 3/14/2020 | 1.45 | 22.80 | 10.13 | 7.00 | 1018.00 | 2.25 |
| Oman | 3/15/2020 | 2.24 | 23.00 | 11.33 | 7.00 | 1016.00 | 2.36 |
| Oman | 3/16/2020 | 2.05 | 22.55 | 11.72 | 7.00 | 1016.00 | 3.66 |
| Oman | 3/17/2020 | 1.99 | 22.83 | 11.96 | 6.00 | 1017.00 | 2.88 |
| Oman | 3/18/2020 | 2.11 | 23.10 | 11.92 | 7.00 | 1017.00 | 2.54 |
| Oman | 3/19/2020 | 2.91 | 23.81 | 12.25 | 7.00 | 1015.00 | 2.05 |
| Oman | 3/20/2020 | 3.13 | 25.37 | 13.11 | 7.00 | 1012.00 | 1.68 |
| Oman | 3/21/2020 | 3.78 | 24.25 | 14.50 | 7.00 | 1012.00 | 2.72 |
| Oman | 3/22/2020 | 3.86 | 24.27 | 15.76 | 6.00 | 1012.00 | 3.15 |
| Oman | 3/23/2020 | 2.47 | 23.07 | 12.59 | 5.00 | 1012.00 | 4.27 |
| Oman | 3/24/2020 | 1.29 | 20.76 | 9.45 | 6.00 | 1013.00 | 4.09 |
| Oman | 3/25/2020 | 1.94 | 21.16 | 10.29 | 6.00 | 1013.00 | 2.49 |
| Oman | 3/26/2020 | 2.08 | 22.01 | 10.72 | 6.00 | 1014.00 | 1.87 |
| Oman | 3/27/2020 | 2.35 | 22.78 | 12.64 | 6.00 | 1017.00 | 2.05 |
| Oman | 3/28/2020 | 2.57 | 24.02 | 13.31 | 7.00 | 1017.00 | 3.19 |
| Oman | 3/29/2020 | 2.85 | 24.51 | 13.06 | 7.00 | 1013.00 | 4.10 |
| Oman | 3/30/2020 | 3.20 | 24.85 | 14.47 | 7.00 | 1011.00 | 4.42 |
| Oman | 3/31/2020 | 1.32 | 23.32 | 10.80 | 6.00 | 1014.00 | 3.00 |
| Oman | 4/1/2020 | 1.30 | 23.80 | 10.39 | 7.00 | 1015.00 | 2.52 |
| Oman | 4/2/2020 | 2.19 | 24.37 | 13.46 | 7.00 | 1014.00 | 2.12 |
| Oman | 4/3/2020 | 2.17 | 25.08 | 14.25 | 7.00 | 1014.00 | 1.93 |
| Oman | 4/4/2020 | 2.01 | 26.47 | 13.82 | 8.00 | 1015.00 | 1.31 |
| Oman | 4/5/2020 | 2.63 | 28.53 | 12.61 | 8.00 | 1015.00 | 1.08 |
| Oman | 4/6/2020 | 2.21 | 27.67 | 13.49 | 8.00 | 1014.00 | 2.24 |
| Oman | 4/7/2020 | 1.11 | 26.02 | 11.78 | 7.00 | 1013.00 | 3.38 |
| Oman | 4/8/2020 | 1.66 | 27.25 | 12.03 | 7.00 | 1013.00 | 1.87 |
| Oman | 4/9/2020 | 2.02 | 27.22 | 13.54 | 8.00 | 1013.00 | 1.90 |
| Pakistan | 3/10/2020 | 1.58 | 20.21 | 8.50 | 6.57 | 1011.43 | 2.91 |
| Pakistan | 3/11/2020 | 1.39 | 19.08 | 7.42 | 5.57 | 1010.57 | 3.88 |
| Pakistan | 3/12/2020 | 1.11 | 18.73 | 6.86 | 5.57 | 1014.00 | 3.94 |
| Pakistan | 3/13/2020 | 1.30 | 19.02 | 6.53 | 5.70 | 1016.43 | 1.86 |
| Pakistan | 3/14/2020 | 1.39 | 19.28 | 6.85 | 5.70 | 1018.00 | 1.59 |
| Pakistan | 3/15/2020 | 1.70 | 20.30 | 7.64 | 6.00 | 1017.87 | 1.34 |
| Pakistan | 3/16/2020 | 1.72 | 21.14 | 8.25 | 6.57 | 1016.57 | 2.06 |
| Pakistan | 3/17/2020 | 1.83 | 22.92 | 8.57 | 7.57 | 1016.00 | 1.92 |
| Pakistan | 3/18/2020 | 2.06 | 23.55 | 9.07 | 7.57 | 1015.00 | 2.03 |
| Pakistan | 3/19/2020 | 1.92 | 23.42 | 9.97 | 7.00 | 1011.57 | 2.25 |
| Pakistan | 3/20/2020 | 2.46 | 22.53 | 11.90 | 7.00 | 1011.00 | 2.93 |
| Pakistan | 3/21/2020 | 3.01 | 23.34 | 12.93 | 7.00 | 1011.43 | 2.67 |
| Pakistan | 3/22/2020 | 2.94 | 23.45 | 12.41 | 7.00 | 1013.43 | 2.19 |
| Pakistan | 3/23/2020 | 3.09 | 22.85 | 12.14 | 7.00 | 1012.57 | 1.91 |
| Pakistan | 3/24/2020 | 3.13 | 22.93 | 12.90 | 6.13 | 1011.57 | 2.44 |
| Pakistan | 3/25/2020 | 2.48 | 23.47 | 10.69 | 6.57 | 1012.43 | 2.34 |
| Pakistan | 3/26/2020 | 2.82 | 22.70 | 10.58 | 6.00 | 1013.57 | 2.10 |
| Pakistan | 3/27/2020 | 2.59 | 20.99 | 10.99 | 5.70 | 1015.13 | 2.49 |
| Pakistan | 3/28/2020 | 2.25 | 22.29 | 10.96 | 6.13 | 1015.87 | 2.81 |
| Palestina | 2/29/2020 | 1.49 | 14.79 | 7.90 | 5.00 | 1015.00 | 4.03 |
| Palestina | 3/1/2020 | 1.23 | 14.06 | 7.01 | 4.00 | 1021.00 | 3.83 |
| Palestina | 3/2/2020 | 1.18 | 13.38 | 6.90 | 4.00 | 1022.00 | 2.02 |
| Palestina | 3/3/2020 | 1.09 | 13.89 | 7.20 | 5.00 | 1022.00 | 2.20 |
| Palestina | 3/4/2020 | 0.86 | 16.81 | 6.83 | 6.00 | 1017.00 | 2.09 |
| Palestina | 3/5/2020 | 1.04 | 17.83 | 6.38 | 6.00 | 1012.00 | 2.49 |
| Palestina | 3/6/2020 | 1.83 | 14.52 | 7.88 | 4.00 | 1016.00 | 2.76 |
| Palestina | 3/7/2020 | 1.43 | 14.64 | 8.15 | 5.00 | 1021.00 | 2.16 |
| Palestina | 3/8/2020 | 1.36 | 16.24 | 8.40 | 5.00 | 1021.00 | 2.30 |
| Palestina | 3/9/2020 | 1.22 | 18.82 | 7.41 | 6.00 | 1017.00 | 3.34 |
| Palestina | 3/10/2020 | 1.56 | 15.10 | 8.64 | 5.00 | 1017.00 | 2.37 |
| Palestina | 3/11/2020 | 2.26 | 17.27 | 8.68 | 5.00 | 1016.00 | 2.13 |
| Palestina | 3/12/2020 | 3.17 | 19.99 | 10.57 | 5.00 | 999.00 | 4.45 |
| Palestina | 3/13/2020 | 2.23 | 15.25 | 8.48 | 4.00 | 999.00 | 6.25 |
| Palestina | 3/14/2020 | 2.38 | 15.72 | 9.45 | 4.00 | 1016.00 | 4.03 |
| Palestina | 3/15/2020 | 1.95 | 16.80 | 9.20 | 5.00 | 1018.00 | 1.29 |
| Palestina | 3/16/2020 | 1.61 | 17.28 | 9.04 | 6.00 | 1015.00 | 2.09 |
| Palestina | 3/17/2020 | 1.59 | 14.60 | 8.21 | 5.00 | 1014.00 | 4.31 |
| Palestina | 3/18/2020 | 1.16 | 13.40 | 6.69 | 3.00 | 1015.00 | 4.59 |
| Palestina | 3/19/2020 | 1.24 | 12.99 | 6.60 | 3.00 | 1016.00 | 4.86 |
| Palestina | 3/20/2020 | 1.41 | 11.56 | 6.41 | 3.00 | 1016.00 | 4.15 |
| Palestina | 3/21/2020 | 1.44 | 11.64 | 6.47 | 3.00 | 1016.00 | 3.05 |
| Palestina | 3/22/2020 | 1.04 | 13.23 | 6.50 | 4.00 | 1018.00 | 1.48 |
| Palestina | 3/23/2020 | 1.75 | 17.19 | 7.01 | 5.00 | 1017.00 | 2.61 |
| Panama | 3/5/2020 | 3.64 | 27.12 | 16.94 | 7.00 | 1013.00 | 6.19 |
| Panama | 3/6/2020 | 3.04 | 26.99 | 16.61 | 7.00 | 1014.00 | 6.83 |
| Panama | 3/7/2020 | 2.98 | 27.06 | 16.28 | 7.00 | 1014.00 | 7.17 |
| Panama | 3/8/2020 | 2.97 | 26.89 | 15.67 | 7.00 | 1014.00 | 7.95 |
| Panama | 3/9/2020 | 2.87 | 26.97 | 15.96 | 7.00 | 1014.00 | 8.21 |
| Panama | 3/10/2020 | 3.00 | 26.97 | 16.10 | 7.00 | 1015.00 | 7.18 |
| Panama | 3/11/2020 | 2.83 | 26.81 | 16.08 | 7.00 | 1014.00 | 6.47 |
| Panama | 3/12/2020 | 2.55 | 26.73 | 15.38 | 7.00 | 1012.00 | 5.77 |
| Panama | 3/13/2020 | 2.91 | 26.84 | 15.49 | 7.00 | 1014.00 | 5.73 |
| Panama | 3/14/2020 | 2.97 | 27.05 | 15.80 | 7.00 | 1016.00 | 6.23 |
| Panama | 3/15/2020 | 3.03 | 27.05 | 15.81 | 7.00 | 1015.00 | 7.00 |
| Paraguay | 2/29/2020 | 2.35 | 26.54 | 10.45 | 8.00 | 1017.00 | 2.31 |
| Paraguay | 3/1/2020 | 2.39 | 27.85 | 10.72 | 8.00 | 1017.00 | 2.20 |
| Paraguay | 3/2/2020 | 2.66 | 28.21 | 11.21 | 8.00 | 1016.00 | 2.06 |
| Paraguay | 3/3/2020 | 2.71 | 28.33 | 11.69 | 8.00 | 1015.00 | 1.94 |
| Paraguay | 3/4/2020 | 2.34 | 28.43 | 11.19 | 9.00 | 1014.00 | 1.89 |
| Paraguay | 3/5/2020 | 2.03 | 27.74 | 9.88 | 9.00 | 1015.00 | 1.89 |
| Paraguay | 3/6/2020 | 1.55 | 27.51 | 9.06 | 9.00 | 1014.00 | 1.41 |
| Paraguay | 3/7/2020 | 2.51 | 28.81 | 9.65 | 8.00 | 1010.00 | 1.76 |
| Paraguay | 3/8/2020 | 3.37 | 29.67 | 11.24 | 9.00 | 1009.00 | 1.44 |
| Paraguay | 3/9/2020 | 3.77 | 31.92 | 12.63 | 9.00 | 1013.00 | 1.73 |
| Paraguay | 3/10/2020 | 3.07 | 31.78 | 11.29 | 9.00 | 1014.00 | 2.27 |
| Paraguay | 3/11/2020 | 2.43 | 30.82 | 9.77 | 9.00 | 1012.00 | 2.61 |
| Paraguay | 3/12/2020 | 3.07 | 32.29 | 11.06 | 9.00 | 1011.00 | 2.36 |
| Paraguay | 3/13/2020 | 3.99 | 33.74 | 13.00 | 9.00 | 1012.00 | 2.61 |
| Paraguay | 3/14/2020 | 3.89 | 34.27 | 11.80 | 9.00 | 1013.00 | 2.32 |
| Paraguay | 3/15/2020 | 3.46 | 33.24 | 10.29 | 9.00 | 1011.00 | 2.15 |
| Peru | 2/26/2020 | 4.21 | 22.58 | 15.45 | 6.00 | 1017.00 | 2.88 |
| Peru | 2/27/2020 | 4.02 | 22.64 | 15.57 | 6.00 | 1018.00 | 2.61 |
| Peru | 2/28/2020 | 4.71 | 22.46 | 15.73 | 7.00 | 1017.00 | 2.54 |
| Peru | 2/29/2020 | 5.28 | 22.37 | 15.71 | 7.00 | 1018.00 | 1.99 |
| Peru | 3/1/2020 | 4.82 | 22.49 | 15.63 | 7.00 | 1017.00 | 2.21 |
| Peru | 3/2/2020 | 4.50 | 22.76 | 15.42 | 7.00 | 1017.00 | 2.92 |
| Peru | 3/3/2020 | 4.27 | 22.80 | 15.39 | 7.00 | 1016.00 | 3.57 |
| Peru | 3/4/2020 | 4.11 | 22.58 | 15.37 | 7.00 | 1015.00 | 3.64 |
| Peru | 3/5/2020 | 4.43 | 22.62 | 15.22 | 6.00 | 1016.00 | 2.91 |
| Peru | 3/6/2020 | 4.63 | 22.34 | 15.08 | 6.00 | 1017.00 | 2.66 |
| Peru | 3/7/2020 | 4.40 | 22.39 | 14.90 | 6.00 | 1016.00 | 2.20 |
| Peru | 3/8/2020 | 4.66 | 22.16 | 14.96 | 5.00 | 1016.00 | 2.07 |
| Philippines | 2/27/2020 | 3.08 | 23.41 | 15.66 | 7.00 | 1015.00 | 1.78 |
| Philippines | 2/28/2020 | 3.40 | 23.20 | 15.39 | 7.00 | 1014.00 | 1.90 |
| Philippines | 2/29/2020 | 3.89 | 23.66 | 15.70 | 7.00 | 1014.00 | 1.54 |
| Philippines | 3/1/2020 | 3.55 | 23.87 | 15.72 | 7.00 | 1014.00 | 1.25 |
| Philippines | 3/2/2020 | 2.93 | 23.65 | 15.62 | 7.00 | 1014.00 | 1.49 |
| Philippines | 3/3/2020 | 3.32 | 23.71 | 16.47 | 7.00 | 1015.00 | 1.53 |
| Philippines | 3/4/2020 | 2.86 | 23.58 | 15.88 | 7.00 | 1014.00 | 1.44 |
| Philippines | 3/5/2020 | 3.03 | 24.07 | 15.54 | 8.00 | 1014.00 | 1.19 |
| Philippines | 3/6/2020 | 4.28 | 24.52 | 16.68 | 7.00 | 1014.00 | 1.36 |
| Philippines | 3/7/2020 | 4.75 | 25.62 | 18.13 | 7.00 | 1013.00 | 0.96 |
| Philippines | 3/8/2020 | 5.07 | 26.50 | 18.61 | 8.00 | 1012.00 | 0.72 |
| Philippines | 3/9/2020 | 4.37 | 26.82 | 18.26 | 7.00 | 1011.00 | 0.81 |
| Philippines | 3/10/2020 | 4.35 | 27.63 | 17.81 | 8.00 | 1013.00 | 0.44 |
| Philippines | 3/11/2020 | 4.47 | 25.79 | 17.88 | 7.00 | 1015.00 | 1.21 |
| Philippines | 3/12/2020 | 3.64 | 24.90 | 16.90 | 8.00 | 1016.00 | 1.60 |
| Philippines | 3/13/2020 | 3.05 | 25.47 | 16.94 | 8.00 | 1017.00 | 1.09 |
| Philippines | 3/14/2020 | 3.35 | 25.85 | 17.31 | 8.00 | 1014.00 | 0.94 |
| Philippines | 3/15/2020 | 4.12 | 24.71 | 17.30 | 7.00 | 1015.00 | 1.04 |
| Philippines | 3/16/2020 | 4.13 | 24.04 | 16.88 | 6.00 | 1015.00 | 1.69 |
| Philippines | 3/17/2020 | 3.62 | 23.53 | 16.37 | 7.00 | 1017.00 | 1.75 |
| Philippines | 3/18/2020 | 3.12 | 24.16 | 16.34 | 7.00 | 1015.00 | 1.34 |
| Philippines | 3/19/2020 | 3.26 | 24.91 | 16.74 | 8.00 | 1015.00 | 1.08 |
| Poland | 2/24/2020 | 0.91 | 3.48 | 4.17 | 1.00 | 1009.56 | 5.74 |
| Poland | 2/25/2020 | 1.54 | 6.03 | 5.59 | 2.17 | 999.00 | 4.67 |
| Poland | 2/26/2020 | 0.96 | 3.72 | 4.63 | 2.00 | 997.20 | 3.56 |
| Poland | 2/27/2020 | 0.74 | 1.01 | 3.59 | 1.55 | 999.00 | 3.58 |
| Poland | 2/28/2020 | 0.77 | 0.49 | 3.47 | 1.45 | 1016.20 | 3.12 |
| Poland | 2/29/2020 | 1.11 | 2.30 | 3.62 | 1.62 | 1016.20 | 4.99 |
| Poland | 3/1/2020 | 1.30 | 5.39 | 4.78 | 3.00 | 1004.82 | 4.40 |
| Poland | 3/2/2020 | 1.25 | 3.72 | 4.67 | 3.00 | 999.00 | 2.90 |
| Poland | 3/3/2020 | 1.58 | 5.71 | 5.47 | 3.17 | 999.00 | 3.49 |
| Poland | 3/4/2020 | 1.06 | 3.38 | 4.16 | 2.00 | 1011.10 | 2.51 |
| Poland | 3/5/2020 | 0.71 | 2.27 | 3.56 | 2.38 | 1011.10 | 2.19 |
| Poland | 3/6/2020 | 1.31 | 2.56 | 4.11 | 3.00 | 1002.53 | 4.12 |
| Portugal | 2/23/2020 | 0.92 | 14.27 | 6.95 | 5.00 | 1031.00 | 2.50 |
| Portugal | 2/24/2020 | 0.96 | 13.99 | 7.90 | 5.00 | 1030.00 | 2.88 |
| Portugal | 2/25/2020 | 1.79 | 13.13 | 8.16 | 3.00 | 1028.00 | 4.36 |
| Portugal | 2/26/2020 | 1.47 | 12.93 | 7.16 | 3.00 | 1027.00 | 4.95 |
| Portugal | 2/27/2020 | 1.05 | 13.30 | 7.62 | 4.00 | 1027.00 | 2.98 |
| Portugal | 2/28/2020 | 1.08 | 12.33 | 7.19 | 4.00 | 1024.00 | 2.84 |
| Portugal | 2/29/2020 | 2.16 | 14.45 | 8.77 | 3.00 | 1019.00 | 4.86 |
| Portugal | 3/1/2020 | 2.53 | 15.21 | 9.75 | 5.00 | 1020.00 | 6.12 |
| Portugal | 3/2/2020 | 1.71 | 14.09 | 7.67 | 3.00 | 1024.00 | 5.90 |
| Portugal | 3/3/2020 | 1.95 | 14.72 | 9.29 | 5.00 | 1025.00 | 4.44 |
| Portugal | 3/4/2020 | 1.77 | 15.46 | 9.36 | 4.00 | 1025.00 | 2.98 |
| Portugal | 3/5/2020 | 1.64 | 14.35 | 7.89 | 3.00 | 1025.00 | 6.25 |
| Portugal | 3/6/2020 | 1.24 | 13.41 | 6.57 | 3.00 | 1029.00 | 6.43 |
| Portugal | 3/7/2020 | 0.79 | 12.81 | 6.68 | 5.00 | 1029.00 | 3.89 |
| Portugal | 3/8/2020 | 1.34 | 12.78 | 7.33 | 4.00 | 1029.00 | 3.29 |
| Portugal | 3/9/2020 | 1.40 | 13.62 | 7.35 | 3.00 | 1030.00 | 4.45 |
| Romania | 3/2/2020 | 1.50 | 10.02 | 5.38 | 3.15 | 1013.56 | 2.94 |
| Romania | 3/3/2020 | 1.61 | 13.92 | 6.50 | 4.88 | 1008.56 | 3.44 |
| Romania | 3/4/2020 | 1.78 | 12.25 | 7.06 | 4.64 | 1007.66 | 3.94 |
| Romania | 3/5/2020 | 1.76 | 7.70 | 5.45 | 2.88 | 1008.47 | 1.57 |
| Romania | 3/6/2020 | 1.30 | 8.92 | 4.81 | 3.73 | 1008.88 | 1.59 |
| Romania | 3/7/2020 | 1.75 | 11.90 | 5.70 | 4.64 | 1012.36 | 1.49 |
| Romania | 3/8/2020 | 2.19 | 11.33 | 7.22 | 4.03 | 1016.24 | 1.74 |
| Romania | 3/9/2020 | 1.93 | 8.49 | 6.14 | 2.88 | 1016.24 | 1.24 |
| Romania | 3/10/2020 | 1.97 | 8.85 | 6.00 | 3.61 | 1014.20 | 3.08 |
| Romania | 3/11/2020 | 1.88 | 9.43 | 5.43 | 3.00 | 1018.20 | 3.72 |
| Romania | 3/12/2020 | 1.51 | 11.54 | 5.92 | 5.00 | 1021.00 | 1.47 |
| Romania | 3/13/2020 | 1.50 | 13.51 | 6.03 | 4.88 | 1017.51 | 1.99 |
| Romania | 3/14/2020 | 1.59 | 9.66 | 5.17 | 3.00 | 1019.09 | 2.13 |
| Romania | 3/15/2020 | 0.90 | 3.83 | 3.07 | 2.12 | 1031.76 | 4.39 |
| Romania | 3/16/2020 | 0.30 | 2.07 | 2.02 | 3.00 | 1034.64 | 1.59 |
| Russia | 2/29/2020 | 0.56 | -3.12 | 2.83 | 1.00 | 1010.00 | 4.08 |
| Russia | 3/1/2020 | 0.89 | -2.49 | 3.06 | 2.00 | 1012.00 | 5.15 |
| Russia | 3/2/2020 | 1.25 | 0.54 | 3.83 | 1.00 | 1013.00 | 5.14 |
| Russia | 3/3/2020 | 1.30 | 0.34 | 3.81 | 1.00 | 1013.00 | 3.21 |
| Russia | 3/4/2020 | 1.15 | 0.59 | 3.84 | 1.00 | 1020.00 | 1.72 |
| Russia | 3/5/2020 | 1.56 | 0.55 | 3.80 | 1.00 | 1020.00 | 6.14 |
| Russia | 3/6/2020 | 1.51 | 1.02 | 3.93 | 1.00 | 1013.00 | 5.41 |
| Russia | 3/7/2020 | 1.66 | 1.86 | 4.21 | 2.00 | 1015.00 | 2.86 |
| Russia | 3/8/2020 | 1.42 | 1.77 | 4.19 | 2.00 | 1020.00 | 2.81 |
| Russia | 3/9/2020 | 1.19 | 3.02 | 4.31 | 3.00 | 1020.00 | 1.88 |
| Russia | 3/10/2020 | 1.47 | 2.52 | 4.45 | 2.00 | 1018.00 | 1.95 |
| Russia | 3/11/2020 | 1.66 | 3.01 | 4.81 | 2.00 | 999.00 | 2.05 |
| Russia | 3/12/2020 | 0.98 | 3.71 | 4.31 | 2.00 | 998.00 | 6.04 |
| Russia | 3/13/2020 | 0.98 | 3.38 | 4.12 | 1.00 | 999.00 | 9.12 |
| Russia | 3/14/2020 | 0.82 | 1.27 | 3.67 | 2.00 | 999.00 | 4.48 |
| Russia | 3/15/2020 | 0.42 | -4.39 | 2.07 | 1.00 | 1022.00 | 5.91 |
| Russia | 3/16/2020 | 0.36 | -1.38 | 2.63 | 2.00 | 1025.00 | 4.60 |
| Russia | 3/17/2020 | 0.81 | 1.90 | 3.52 | 1.00 | 1025.00 | 3.13 |
| Russia | 3/18/2020 | 1.09 | 2.90 | 3.87 | 2.00 | 1025.00 | 5.47 |
| Russia | 3/19/2020 | 1.12 | 4.48 | 4.26 | 2.00 | 1014.00 | 5.72 |
| Russia | 3/20/2020 | 0.78 | 1.35 | 3.73 | 2.00 | 1016.00 | 4.03 |
| Russia | 3/21/2020 | 0.45 | -1.69 | 2.56 | 1.00 | 1027.00 | 3.70 |
| Russia | 3/22/2020 | 0.33 | -3.58 | 2.02 | 2.00 | 1036.00 | 4.72 |
| Saudi Arabia | 2/29/2020 | 1.23 | 20.13 | 2.94 | 6.00 | 1016.00 | 3.08 |
| Saudi Arabia | 3/1/2020 | 1.46 | 22.61 | 3.62 | 7.00 | 1014.00 | 3.82 |
| Saudi Arabia | 3/2/2020 | 0.70 | 16.93 | 3.24 | 5.00 | 1020.00 | 3.68 |
| Saudi Arabia | 3/3/2020 | 0.48 | 15.32 | 2.51 | 5.00 | 1023.00 | 1.84 |
| Saudi Arabia | 3/4/2020 | 0.65 | 16.99 | 2.72 | 6.00 | 1021.00 | 1.79 |
| Saudi Arabia | 3/5/2020 | 0.75 | 17.98 | 2.82 | 6.00 | 1020.00 | 1.44 |
| Saudi Arabia | 3/6/2020 | 0.99 | 20.00 | 2.91 | 6.00 | 1017.00 | 2.57 |
| Saudi Arabia | 3/7/2020 | 1.06 | 21.49 | 3.21 | 7.00 | 1015.00 | 3.82 |
| Saudi Arabia | 3/8/2020 | 1.00 | 18.62 | 3.62 | 6.00 | 1018.00 | 3.65 |
| Saudi Arabia | 3/9/2020 | 0.83 | 17.32 | 4.56 | 5.00 | 1019.00 | 3.06 |
| Saudi Arabia | 3/10/2020 | 0.74 | 18.91 | 3.69 | 6.00 | 1019.00 | 1.53 |
| Saudi Arabia | 3/11/2020 | 0.77 | 19.36 | 3.13 | 6.00 | 1020.00 | 2.14 |
| Saudi Arabia | 3/12/2020 | 0.93 | 20.14 | 3.02 | 5.00 | 1018.00 | 3.28 |
| Senegal | 3/2/2020 | 1.49 | 25.00 | 9.14 | 7.00 | 1014.00 | 4.74 |
| Senegal | 3/3/2020 | 1.36 | 25.12 | 8.37 | 7.00 | 1014.00 | 4.05 |
| Senegal | 3/4/2020 | 1.13 | 24.56 | 10.22 | 7.00 | 1014.00 | 4.37 |
| Senegal | 3/5/2020 | 0.97 | 24.02 | 11.32 | 7.00 | 1013.00 | 4.17 |
| Senegal | 3/6/2020 | 0.83 | 24.10 | 11.70 | 7.00 | 1012.00 | 4.03 |
| Senegal | 3/7/2020 | 0.98 | 23.90 | 12.39 | 7.00 | 1011.00 | 3.51 |
| Senegal | 3/8/2020 | 1.31 | 23.99 | 12.95 | 7.00 | 1011.00 | 3.33 |
| Senegal | 3/9/2020 | 1.41 | 23.89 | 13.20 | 7.00 | 1012.00 | 3.35 |
| Senegal | 3/10/2020 | 1.25 | 24.53 | 11.62 | 7.00 | 1012.00 | 2.14 |
| Senegal | 3/11/2020 | 1.41 | 25.34 | 11.52 | 7.00 | 1011.00 | 2.33 |
| Senegal | 3/12/2020 | 1.25 | 25.55 | 9.76 | 7.00 | 1011.00 | 4.61 |
| Senegal | 3/13/2020 | 1.54 | 24.86 | 10.45 | 7.00 | 1012.00 | 4.58 |
| Senegal | 3/14/2020 | 1.85 | 24.23 | 12.04 | 6.00 | 1013.00 | 4.27 |
| Senegal | 3/15/2020 | 1.82 | 23.25 | 12.42 | 6.00 | 1013.00 | 5.02 |
| Senegal | 3/16/2020 | 1.58 | 22.79 | 11.94 | 6.00 | 1013.00 | 6.10 |
| Senegal | 3/17/2020 | 1.55 | 22.41 | 12.52 | 6.00 | 1013.00 | 5.29 |
| Senegal | 3/18/2020 | 1.95 | 21.92 | 12.43 | 6.00 | 1013.00 | 4.82 |
| Senegal | 3/19/2020 | 2.38 | 21.97 | 11.93 | 6.00 | 1014.00 | 4.67 |
| Serbia | 3/2/2020 | 1.48 | 12.95 | 6.34 | 4.00 | 1007.00 | 3.73 |
| Serbia | 3/3/2020 | 2.12 | 12.91 | 7.52 | 4.00 | 1003.00 | 3.21 |
| Serbia | 3/4/2020 | 1.52 | 7.17 | 5.54 | 2.00 | 1013.00 | 3.19 |
| Serbia | 3/5/2020 | 0.82 | 5.39 | 3.80 | 2.00 | 1013.00 | 2.25 |
| Serbia | 3/6/2020 | 1.77 | 7.54 | 5.57 | 3.00 | 1007.00 | 2.70 |
| Serbia | 3/7/2020 | 1.48 | 7.38 | 5.62 | 3.00 | 1017.00 | 2.37 |
| Serbia | 3/8/2020 | 1.36 | 4.33 | 4.71 | 2.00 | 1020.00 | 2.15 |
| Serbia | 3/9/2020 | 1.25 | 6.15 | 4.73 | 2.00 | 1019.00 | 1.18 |
| Serbia | 3/10/2020 | 1.36 | 5.90 | 4.47 | 2.00 | 1019.00 | 2.44 |
| Serbia | 3/11/2020 | 1.84 | 8.64 | 5.31 | 3.00 | 1022.00 | 1.83 |
| Serbia | 3/12/2020 | 1.61 | 14.42 | 6.59 | 4.00 | 1022.00 | 1.72 |
| Serbia | 3/13/2020 | 1.92 | 11.45 | 6.31 | 5.00 | 1021.00 | 3.23 |
| Serbia | 3/14/2020 | 1.13 | 7.52 | 4.56 | 3.00 | 1022.00 | 1.76 |
| Serbia | 3/15/2020 | 0.87 | 5.47 | 3.93 | 3.00 | 1030.00 | 3.09 |
| Serbia | 3/16/2020 | 0.39 | 4.80 | 3.00 | 3.00 | 1031.00 | 4.54 |
| Serbia | 3/17/2020 | 0.90 | 7.63 | 4.20 | 4.00 | 1030.00 | 2.20 |
| Serbia | 3/18/2020 | 1.09 | 9.25 | 4.41 | 4.00 | 1030.00 | 1.70 |
| South Africa | 2/27/2020 | 2.08 | 21.75 | 9.52 | 7.00 | 1015.00 | 2.10 |
| South Africa | 2/28/2020 | 2.91 | 20.16 | 11.42 | 6.00 | 1020.00 | 3.25 |
| South Africa | 2/29/2020 | 2.75 | 15.81 | 9.33 | 4.00 | 1022.00 | 2.43 |
| South Africa | 3/1/2020 | 2.83 | 16.78 | 10.65 | 4.00 | 1021.00 | 1.34 |
| South Africa | 3/2/2020 | 2.97 | 16.54 | 11.41 | 4.00 | 1018.00 | 1.13 |
| South Africa | 3/3/2020 | 2.27 | 18.13 | 10.32 | 5.00 | 1020.00 | 1.40 |
| South Africa | 3/4/2020 | 1.75 | 19.53 | 9.93 | 7.00 | 1022.00 | 1.62 |
| South Africa | 3/5/2020 | 1.66 | 20.27 | 9.12 | 7.00 | 1023.00 | 1.99 |
| South Africa | 3/6/2020 | 1.19 | 18.96 | 7.54 | 7.00 | 1022.00 | 2.24 |
| South Africa | 3/7/2020 | 1.32 | 18.24 | 8.12 | 7.00 | 1023.00 | 1.84 |
| South Africa | 3/8/2020 | 1.61 | 20.63 | 8.37 | 7.00 | 1021.00 | 1.06 |
| South Africa | 3/9/2020 | 1.86 | 21.35 | 8.56 | 7.00 | 1018.00 | 1.15 |
| South Africa | 3/10/2020 | 2.18 | 21.01 | 9.65 | 7.00 | 1016.00 | 2.01 |
| South Africa | 3/11/2020 | 2.56 | 21.10 | 10.87 | 7.00 | 1016.00 | 1.52 |
| South Africa | 3/12/2020 | 2.90 | 19.96 | 12.30 | 7.00 | 1017.00 | 1.59 |
| South Africa | 3/13/2020 | 2.41 | 20.15 | 11.01 | 5.00 | 1018.00 | 1.63 |
| South Africa | 3/14/2020 | 1.91 | 20.62 | 9.48 | 6.00 | 1018.00 | 1.93 |
| South Africa | 3/15/2020 | 1.35 | 20.80 | 7.90 | 7.00 | 1016.00 | 2.13 |
| Spain | 2/17/2020 | 1.45 | 7.94 | 6.44 | 4.00 | 1031.00 | 1.83 |
| Spain | 2/18/2020 | 0.97 | 7.55 | 5.28 | 4.00 | 1032.00 | 2.00 |
| Spain | 2/19/2020 | 0.76 | 7.60 | 4.33 | 4.00 | 1033.00 | 1.37 |
| Spain | 2/20/2020 | 0.75 | 8.43 | 4.53 | 4.00 | 1027.00 | 0.92 |
| Spain | 2/21/2020 | 0.66 | 9.20 | 4.27 | 4.00 | 1029.00 | 1.51 |
| Spain | 2/22/2020 | 0.55 | 10.20 | 4.26 | 5.00 | 1033.00 | 1.16 |
| Spain | 2/23/2020 | 0.67 | 10.95 | 4.42 | 5.00 | 1034.00 | 1.10 |
| Spain | 2/24/2020 | 0.67 | 11.03 | 4.40 | 5.00 | 1030.00 | 1.12 |
| Spain | 2/25/2020 | 0.90 | 9.13 | 5.30 | 4.00 | 1024.00 | 2.32 |
| Spain | 2/26/2020 | 1.09 | 7.81 | 4.92 | 4.00 | 1024.00 | 2.06 |
| Spain | 2/27/2020 | 0.86 | 8.00 | 4.79 | 4.00 | 1024.00 | 2.18 |
| Spain | 2/28/2020 | 0.95 | 9.29 | 5.27 | 5.00 | 1025.00 | 1.52 |
| Spain | 2/29/2020 | 1.52 | 8.27 | 6.54 | 3.00 | 1016.00 | 4.28 |
| Spain | 3/1/2020 | 1.52 | 8.87 | 6.87 | 3.00 | 1016.00 | 4.99 |
| Sweden | 2/24/2020 | 0.51 | 0.87 | 3.39 | 2.00 | 999.00 | 2.54 |
| Sweden | 2/25/2020 | 0.72 | 1.18 | 3.58 | 2.00 | 999.00 | 2.89 |
| Sweden | 2/26/2020 | 0.77 | 0.19 | 3.12 | 1.00 | 994.00 | 4.93 |
| Sweden | 2/27/2020 | 0.76 | -0.29 | 3.18 | 1.00 | 998.00 | 3.08 |
| Sweden | 2/28/2020 | 0.62 | -1.26 | 2.79 | 1.00 | 999.00 | 1.59 |
| Sweden | 2/29/2020 | 0.89 | 1.08 | 3.92 | 2.00 | 999.00 | 2.04 |
| Sweden | 3/1/2020 | 1.17 | 2.79 | 4.55 | 1.00 | 989.00 | 2.93 |
| Sweden | 3/2/2020 | 1.02 | 3.33 | 4.63 | 1.00 | 999.00 | 1.79 |
| Sweden | 3/3/2020 | 1.21 | 3.08 | 4.64 | 1.00 | 999.00 | 2.03 |
| Switzerland | 2/16/2020 | 1.19 | 5.47 | 4.82 | 3.00 | 1025.39 | 2.57 |
| Switzerland | 2/17/2020 | 1.40 | 4.89 | 5.23 | 3.00 | 1026.79 | 2.12 |
| Switzerland | 2/18/2020 | 0.68 | 1.87 | 4.06 | 2.00 | 1031.39 | 1.99 |
| Switzerland | 2/19/2020 | 0.84 | 0.48 | 3.87 | 2.00 | 1029.39 | 2.60 |
| Switzerland | 2/20/2020 | 0.77 | 2.13 | 3.86 | 2.00 | 1028.00 | 1.62 |
| Switzerland | 2/21/2020 | 0.80 | 2.42 | 3.69 | 3.00 | 1031.61 | 1.49 |
| Switzerland | 2/22/2020 | 0.94 | 4.84 | 3.87 | 3.00 | 1031.79 | 1.40 |
| Switzerland | 2/23/2020 | 1.44 | 6.97 | 4.78 | 3.39 | 1032.18 | 2.65 |
| Switzerland | 2/24/2020 | 1.13 | 8.11 | 6.22 | 3.39 | 1028.18 | 2.22 |
| Switzerland | 2/25/2020 | 1.04 | 4.68 | 5.11 | 3.00 | 1016.18 | 3.02 |
| Switzerland | 2/26/2020 | 0.67 | -0.27 | 3.78 | 1.00 | 1015.58 | 4.22 |
| Switzerland | 2/27/2020 | 0.82 | -0.85 | 3.63 | 1.00 | 1016.58 | 4.08 |
| Switzerland | 2/28/2020 | 0.74 | 0.06 | 3.18 | 1.00 | 1024.00 | 1.90 |
| Switzerland | 2/29/2020 | 1.28 | 3.83 | 4.50 | 2.39 | 1016.00 | 3.84 |
| Switzerland | 3/1/2020 | 0.94 | 3.79 | 4.69 | 2.00 | 1002.94 | 2.82 |
| Switzerland | 3/2/2020 | 0.99 | 1.55 | 4.29 | 1.61 | 999.00 | 2.32 |
| Switzerland | 3/3/2020 | 0.69 | 0.07 | 3.69 | 1.00 | 1003.91 | 1.89 |
| Switzerland | 3/4/2020 | 0.74 | 0.72 | 3.53 | 2.00 | 1017.39 | 1.18 |
| Togo | 3/10/2020 | 5.70 | 29.31 | 18.77 | 7.00 | 1011.00 | 3.62 |
| Togo | 3/11/2020 | 5.70 | 28.70 | 19.77 | 7.00 | 1011.00 | 3.41 |
| Togo | 3/12/2020 | 5.30 | 28.65 | 19.99 | 7.00 | 1010.00 | 2.85 |
| Togo | 3/13/2020 | 5.11 | 29.17 | 20.06 | 8.00 | 1011.00 | 2.93 |
| Togo | 3/14/2020 | 5.44 | 28.79 | 19.13 | 8.00 | 1011.00 | 2.24 |
| Togo | 3/15/2020 | 4.68 | 29.25 | 19.10 | 8.00 | 1011.00 | 3.14 |
| Togo | 3/16/2020 | 4.22 | 28.92 | 19.43 | 8.00 | 1011.00 | 3.13 |
| Togo | 3/17/2020 | 4.96 | 28.81 | 19.80 | 8.00 | 1010.00 | 3.22 |
| Togo | 3/18/2020 | 4.92 | 29.42 | 19.38 | 8.00 | 1011.00 | 3.39 |
| Togo | 3/19/2020 | 5.10 | 28.92 | 19.32 | 8.00 | 1011.00 | 4.45 |
| Togo | 3/20/2020 | 4.20 | 28.85 | 17.77 | 6.00 | 1011.00 | 3.63 |
| Togo | 3/21/2020 | 4.72 | 28.14 | 19.18 | 7.00 | 1011.00 | 2.75 |
| Togo | 3/22/2020 | 5.49 | 27.99 | 19.15 | 6.00 | 1011.00 | 3.20 |
| Togo | 3/23/2020 | 5.60 | 27.61 | 18.73 | 7.00 | 1012.00 | 3.77 |
| Togo | 3/24/2020 | 5.66 | 26.83 | 18.02 | 6.00 | 1013.00 | 2.82 |
| Togo | 3/25/2020 | 5.57 | 27.06 | 17.72 | 7.00 | 1012.00 | 3.19 |
| Togo | 3/26/2020 | 5.44 | 27.93 | 18.07 | 8.00 | 1011.00 | 2.86 |
| Togo | 3/27/2020 | 5.29 | 27.50 | 18.27 | 7.00 | 1011.00 | 2.32 |
| Togo | 3/28/2020 | 4.92 | 27.79 | 18.60 | 7.00 | 1012.00 | 2.88 |
| Togo | 3/29/2020 | 4.87 | 28.06 | 18.83 | 7.00 | 1012.00 | 3.10 |
| Trinidad and Tobago | 3/10/2020 | 3.08 | 26.33 | 15.37 | 7.00 | 1017.00 | 6.09 |
| Trinidad and Tobago | 3/11/2020 | 3.10 | 26.20 | 15.02 | 7.00 | 1014.00 | 5.15 |
| Trinidad and Tobago | 3/12/2020 | 3.28 | 26.15 | 15.61 | 7.00 | 1013.00 | 5.72 |
| Trinidad and Tobago | 3/13/2020 | 3.59 | 26.61 | 16.10 | 7.00 | 1014.00 | 5.41 |
| Trinidad and Tobago | 3/14/2020 | 4.23 | 26.53 | 16.67 | 7.00 | 1015.00 | 3.82 |
| Trinidad and Tobago | 3/15/2020 | 4.45 | 26.31 | 17.48 | 7.00 | 1014.00 | 4.01 |
| Trinidad and Tobago | 3/16/2020 | 3.90 | 26.23 | 16.60 | 7.00 | 1015.00 | 3.87 |
| Trinidad and Tobago | 3/17/2020 | 3.20 | 26.67 | 16.29 | 7.00 | 1015.00 | 4.73 |
| Trinidad and Tobago | 3/18/2020 | 3.22 | 26.74 | 16.10 | 7.00 | 1016.00 | 4.87 |
| Trinidad and Tobago | 3/19/2020 | 3.31 | 26.86 | 16.14 | 7.00 | 1016.00 | 5.63 |
| Trinidad and Tobago | 3/20/2020 | 3.56 | 26.70 | 16.00 | 7.00 | 1016.00 | 5.55 |
| Trinidad and Tobago | 3/21/2020 | 4.08 | 26.23 | 16.18 | 7.00 | 1015.00 | 5.20 |
| Trinidad and Tobago | 3/22/2020 | 4.36 | 26.78 | 16.23 | 6.00 | 1016.00 | 6.37 |
| Tunisia | 3/5/2020 | 2.01 | 14.59 | 7.82 | 5.00 | 1017.00 | 2.63 |
| Tunisia | 3/6/2020 | 1.82 | 14.98 | 7.54 | 5.00 | 1010.00 | 4.19 |
| Tunisia | 3/7/2020 | 1.43 | 11.52 | 6.55 | 4.00 | 1016.00 | 5.96 |
| Tunisia | 3/8/2020 | 1.27 | 11.94 | 6.34 | 3.00 | 1022.00 | 5.43 |
| Tunisia | 3/9/2020 | 1.48 | 11.59 | 6.76 | 4.00 | 1022.00 | 4.06 |
| Tunisia | 3/10/2020 | 1.55 | 12.14 | 6.65 | 3.00 | 1022.00 | 5.24 |
| Tunisia | 3/11/2020 | 1.94 | 13.05 | 7.53 | 4.00 | 1025.00 | 3.57 |
| Tunisia | 3/12/2020 | 1.36 | 14.15 | 7.03 | 5.00 | 1025.00 | 1.69 |
| Tunisia | 3/13/2020 | 1.52 | 17.13 | 7.44 | 6.00 | 1020.00 | 1.82 |
| Tunisia | 3/14/2020 | 1.94 | 15.54 | 7.92 | 5.00 | 1018.00 | 3.69 |
| Tunisia | 3/15/2020 | 1.56 | 14.10 | 7.79 | 5.00 | 1019.00 | 3.34 |
| Tunisia | 3/16/2020 | 2.05 | 14.36 | 8.18 | 5.00 | 1023.00 | 4.16 |
| Turkey | 3/9/2020 | 1.62 | 13.99 | 6.28 | 5.00 | 1016.00 | 1.50 |
| Turkey | 3/10/2020 | 1.98 | 10.59 | 7.02 | 3.00 | 1015.00 | 2.14 |
| Turkey | 3/11/2020 | 1.52 | 11.46 | 6.39 | 3.00 | 1020.00 | 1.45 |
| Turkey | 3/12/2020 | 1.23 | 11.01 | 6.11 | 4.00 | 1021.00 | 1.87 |
| Turkey | 3/13/2020 | 0.97 | 9.76 | 6.33 | 4.00 | 1019.00 | 2.25 |
| Turkey | 3/14/2020 | 1.53 | 10.72 | 6.34 | 4.00 | 1016.00 | 1.95 |
| Turkey | 3/15/2020 | 1.82 | 7.25 | 5.64 | 3.00 | 1025.00 | 5.08 |
| Turkey | 3/16/2020 | 0.74 | 5.33 | 3.79 | 2.00 | 1032.00 | 5.79 |
| Turkey | 3/17/2020 | 0.52 | 4.32 | 3.34 | 3.00 | 1032.00 | 3.08 |
| Ukraine | 3/6/2020 | 1.73 | 8.36 | 5.89 | 3.48 | 1008.56 | 2.59 |
| Ukraine | 3/7/2020 | 1.66 | 8.25 | 5.80 | 3.52 | 1012.99 | 3.22 |
| Ukraine | 3/8/2020 | 1.80 | 7.97 | 6.01 | 3.48 | 1018.54 | 2.22 |
| Ukraine | 3/9/2020 | 2.01 | 8.92 | 6.41 | 3.23 | 1018.27 | 2.00 |
| Ukraine | 3/10/2020 | 1.71 | 8.64 | 5.88 | 2.96 | 1012.91 | 3.64 |
| Ukraine | 3/11/2020 | 1.60 | 7.50 | 5.36 | 3.27 | 1010.52 | 4.60 |
| Ukraine | 3/12/2020 | 1.06 | 8.54 | 5.01 | 3.46 | 1015.27 | 5.30 |
| Ukraine | 3/13/2020 | 1.42 | 8.94 | 4.87 | 3.48 | 1011.75 | 4.61 |
| Ukraine | 3/14/2020 | 0.82 | 4.80 | 3.83 | 2.94 | 1016.59 | 5.50 |
| Ukraine | 3/15/2020 | 0.42 | -0.33 | 2.21 | 1.21 | 1028.11 | 5.43 |
| Ukraine | 3/16/2020 | 0.38 | 1.26 | 2.31 | 2.73 | 1031.94 | 2.53 |
| Ukraine | 3/17/2020 | 0.73 | 2.39 | 3.07 | 1.42 | 1030.18 | 2.89 |
| Ukraine | 3/18/2020 | 0.70 | 4.52 | 3.41 | 3.00 | 1031.54 | 1.81 |
| Ukraine | 3/19/2020 | 1.37 | 7.56 | 4.43 | 3.79 | 1027.18 | 2.92 |
| Ukraine | 3/20/2020 | 1.02 | 7.78 | 4.30 | 3.46 | 1021.25 | 3.00 |
| Ukraine | 3/21/2020 | 1.20 | 5.99 | 4.10 | 2.94 | 1020.00 | 2.42 |
| United Arab Emirates | 3/4/2020 | 1.06 | 20.20 | 9.54 | 6.00 | 1015.00 | 6.67 |
| United Arab Emirates | 3/5/2020 | 1.52 | 20.87 | 9.76 | 6.00 | 1015.00 | 2.00 |
| United Arab Emirates | 3/6/2020 | 1.21 | 21.80 | 9.20 | 6.00 | 1016.00 | 2.12 |
| United Arab Emirates | 3/7/2020 | 1.25 | 22.97 | 10.02 | 6.00 | 1015.00 | 1.72 |
| United Arab Emirates | 3/8/2020 | 1.61 | 22.68 | 10.36 | 7.00 | 1014.00 | 2.52 |
| United Arab Emirates | 3/9/2020 | 1.72 | 21.67 | 11.17 | 6.00 | 1013.00 | 2.98 |
| United Arab Emirates | 3/10/2020 | 1.46 | 21.52 | 9.49 | 6.00 | 1016.00 | 3.19 |
| United Arab Emirates | 3/11/2020 | 0.79 | 21.04 | 7.95 | 6.00 | 1018.00 | 2.11 |
| United Arab Emirates | 3/12/2020 | 0.83 | 21.15 | 9.57 | 6.00 | 1018.00 | 2.40 |
| United Arab Emirates | 3/13/2020 | 1.00 | 22.40 | 8.12 | 6.00 | 1018.00 | 1.86 |
| United Arab Emirates | 3/14/2020 | 1.73 | 23.47 | 8.90 | 7.00 | 1017.00 | 2.23 |
| United Arab Emirates | 3/15/2020 | 1.88 | 24.45 | 9.46 | 7.00 | 1014.00 | 2.51 |
| United Arab Emirates | 3/16/2020 | 2.35 | 23.77 | 11.81 | 7.00 | 1015.00 | 3.41 |
| United Arab Emirates | 3/17/2020 | 2.35 | 24.06 | 12.69 | 6.00 | 1016.00 | 2.84 |
| United Arab Emirates | 3/18/2020 | 2.46 | 25.11 | 12.38 | 7.00 | 1015.00 | 2.91 |
| United Arab Emirates | 3/19/2020 | 3.28 | 25.95 | 12.09 | 7.00 | 1014.00 | 2.77 |
| United Arab Emirates | 3/20/2020 | 2.90 | 25.99 | 12.77 | 7.00 | 1011.00 | 1.42 |
| United Arab Emirates | 3/21/2020 | 3.77 | 25.38 | 13.69 | 6.00 | 1010.00 | 2.97 |
| United Arab Emirates | 3/22/2020 | 3.45 | 25.23 | 14.51 | 6.00 | 1010.00 | 5.33 |
| United Arab Emirates | 3/23/2020 | 1.27 | 21.05 | 8.79 | 6.00 | 1013.00 | 7.17 |
| United Kingdom | 2/17/2020 | 1.03 | 6.68 | 5.31 | 3.00 | 1018.00 | 6.99 |
| United Kingdom | 2/18/2020 | 1.13 | 6.12 | 5.30 | 3.00 | 1020.00 | 6.37 |
| United Kingdom | 2/19/2020 | 1.33 | 5.48 | 5.29 | 2.00 | 1022.00 | 5.21 |
| United Kingdom | 2/20/2020 | 1.22 | 7.02 | 5.75 | 2.00 | 1022.00 | 7.22 |
| United Kingdom | 2/21/2020 | 1.25 | 6.73 | 5.25 | 3.00 | 1023.00 | 6.71 |
| United Kingdom | 2/22/2020 | 1.98 | 9.17 | 6.46 | 3.00 | 1022.00 | 7.67 |
| United Kingdom | 2/23/2020 | 2.24 | 9.48 | 6.57 | 3.00 | 1025.00 | 6.41 |
| United Kingdom | 2/24/2020 | 1.72 | 8.58 | 6.54 | 3.00 | 1025.00 | 6.53 |
| United Kingdom | 2/25/2020 | 0.93 | 4.89 | 4.93 | 3.00 | 999.00 | 5.50 |
| United Kingdom | 2/26/2020 | 0.81 | 3.61 | 4.35 | 2.00 | 999.00 | 5.17 |
| United Kingdom | 2/27/2020 | 0.94 | 3.32 | 4.37 | 2.00 | 999.00 | 4.13 |
| United Kingdom | 2/28/2020 | 1.54 | 5.82 | 5.67 | 3.00 | 999.00 | 5.83 |
| United Kingdom | 2/29/2020 | 1.25 | 6.85 | 5.45 | 2.00 | 994.00 | 8.49 |
| United Kingdom | 3/1/2020 | 0.94 | 5.31 | 4.89 | 2.00 | 990.00 | 5.88 |
| United Kingdom | 3/2/2020 | 0.99 | 5.09 | 4.94 | 2.00 | 998.00 | 3.95 |
| United Kingdom | 3/3/2020 | 0.91 | 4.27 | 4.71 | 3.00 | 999.00 | 3.75 |
| United Kingdom | 3/4/2020 | 1.24 | 4.12 | 4.92 | 3.00 | 1011.00 | 2.72 |
| United Kingdom | 3/5/2020 | 1.50 | 4.87 | 5.25 | 2.00 | 999.00 | 5.53 |
| United Kingdom | 3/6/2020 | 0.91 | 4.17 | 4.25 | 2.00 | 999.00 | 2.81 |
| United Kingdom | 3/7/2020 | 1.41 | 6.51 | 5.78 | 2.00 | 1017.00 | 4.46 |
| United Kingdom | 3/8/2020 | 1.37 | 8.49 | 6.21 | 3.00 | 1010.00 | 6.67 |
| United Kingdom | 3/9/2020 | 1.56 | 6.85 | 5.58 | 3.00 | 1015.00 | 5.58 |
| United Kingdom | 3/10/2020 | 2.57 | 10.91 | 7.67 | 3.00 | 1008.00 | 6.93 |
| Uruguay | 3/7/2020 | 2.50 | 24.46 | 11.31 | 7.00 | 1014.00 | 3.25 |
| Uruguay | 3/8/2020 | 3.59 | 26.00 | 11.92 | 7.00 | 1012.00 | 3.47 |
| Uruguay | 3/9/2020 | 3.75 | 22.47 | 11.54 | 6.00 | 1019.00 | 4.57 |
| Uruguay | 3/10/2020 | 3.04 | 21.79 | 10.18 | 5.00 | 1021.00 | 3.96 |
| Uruguay | 3/11/2020 | 4.21 | 21.51 | 12.63 | 7.00 | 1017.00 | 3.44 |
| Uruguay | 3/12/2020 | 2.93 | 22.96 | 13.95 | 7.00 | 1013.00 | 2.35 |
| Uruguay | 3/13/2020 | 2.50 | 25.22 | 12.95 | 7.00 | 1016.00 | 3.31 |
| Uruguay | 3/14/2020 | 4.10 | 22.76 | 14.17 | 7.00 | 1016.00 | 4.04 |
| Uruguay | 3/15/2020 | 2.25 | 19.36 | 10.37 | 5.00 | 1019.00 | 3.52 |
| Uzbekistan | 3/6/2020 | 0.79 | 2.88 | 3.18 | 3.00 | 1024.00 | 1.95 |
| Uzbekistan | 3/7/2020 | 0.77 | 1.91 | 3.05 | 3.00 | 1030.00 | 1.50 |
| Uzbekistan | 3/8/2020 | 0.74 | 1.79 | 3.00 | 3.00 | 1030.00 | 1.41 |
| Uzbekistan | 3/9/2020 | 0.78 | 4.54 | 3.01 | 4.00 | 1029.00 | 1.93 |
| Uzbekistan | 3/10/2020 | 0.94 | 4.81 | 3.18 | 4.00 | 1026.00 | 1.48 |
| Uzbekistan | 3/11/2020 | 0.82 | 5.65 | 3.36 | 4.00 | 1025.00 | 1.18 |
| Uzbekistan | 3/12/2020 | 0.80 | 7.42 | 3.34 | 4.00 | 1024.00 | 1.53 |
| Uzbekistan | 3/13/2020 | 0.60 | 8.84 | 3.19 | 5.00 | 1024.00 | 1.56 |
| Uzbekistan | 3/14/2020 | 0.78 | 11.60 | 3.78 | 5.00 | 1022.00 | 1.72 |
| Uzbekistan | 3/15/2020 | 0.57 | 13.22 | 3.75 | 5.00 | 1020.00 | 1.60 |
| Uzbekistan | 3/16/2020 | 1.92 | 16.31 | 6.09 | 6.00 | 1017.00 | 1.61 |
| Uzbekistan | 3/17/2020 | 2.14 | 15.27 | 7.40 | 6.00 | 1022.00 | 1.79 |
| Uzbekistan | 3/18/2020 | 1.42 | 15.15 | 6.21 | 5.00 | 1023.00 | 1.75 |
| Uzbekistan | 3/19/2020 | 1.27 | 14.64 | 5.81 | 6.00 | 1020.00 | 1.82 |
| Uzbekistan | 3/20/2020 | 1.49 | 16.14 | 6.19 | 6.00 | 1013.00 | 1.49 |
| Uzbekistan | 3/21/2020 | 1.64 | 11.28 | 7.08 | 3.00 | 1020.00 | 2.13 |
| Uzbekistan | 3/22/2020 | 1.22 | 11.72 | 5.21 | 5.00 | 1022.00 | 1.83 |
| Uzbekistan | 3/23/2020 | 1.90 | 14.44 | 6.23 | 4.00 | 1016.00 | 0.81 |
| Uzbekistan | 3/24/2020 | 2.14 | 13.61 | 7.29 | 4.00 | 1014.00 | 1.44 |
| Uzbekistan | 3/25/2020 | 1.23 | 9.18 | 4.95 | 4.00 | 1019.00 | 2.29 |
| Uzbekistan | 3/26/2020 | 1.17 | 8.89 | 4.71 | 5.00 | 1022.00 | 1.49 |
| Uzbekistan | 3/27/2020 | 1.42 | 10.93 | 4.83 | 5.00 | 1021.00 | 1.30 |
| Uzbekistan | 3/28/2020 | 0.86 | 10.22 | 3.99 | 5.00 | 1021.00 | 1.87 |
| Uzbekistan | 3/29/2020 | 0.64 | 11.22 | 3.63 | 5.00 | 1020.00 | 1.42 |
| Uzbekistan | 3/30/2020 | 1.47 | 14.41 | 4.92 | 6.00 | 1014.00 | 1.42 |
| Uzbekistan | 3/31/2020 | 2.06 | 14.45 | 7.00 | 5.00 | 1010.00 | 1.34 |
| Uzbekistan | 4/1/2020 | 1.90 | 11.95 | 7.26 | 4.00 | 1019.00 | 2.30 |
| Uzbekistan | 4/2/2020 | 1.30 | 11.77 | 5.95 | 4.00 | 1021.00 | 1.21 |
| Uzbekistan | 4/3/2020 | 1.27 | 15.50 | 5.40 | 6.00 | 1018.00 | 1.47 |
| Uzbekistan | 4/4/2020 | 1.80 | 17.60 | 6.71 | 6.00 | 1018.00 | 1.86 |
| Uzbekistan | 4/5/2020 | 2.15 | 13.81 | 7.74 | 6.00 | 1018.00 | 2.37 |
| Venezuela | 3/12/2020 | 3.44 | 25.00 | 12.97 | 6.00 | 1012.00 | 1.77 |
| Venezuela | 3/13/2020 | 3.72 | 24.81 | 13.77 | 6.00 | 1014.00 | 2.02 |
| Venezuela | 3/14/2020 | 2.63 | 24.55 | 13.65 | 7.00 | 1016.00 | 2.02 |
| Venezuela | 3/15/2020 | 2.19 | 24.35 | 13.21 | 7.00 | 1016.00 | 1.89 |
| Venezuela | 3/16/2020 | 2.01 | 24.35 | 13.09 | 7.00 | 1015.00 | 1.87 |
| Venezuela | 3/17/2020 | 2.25 | 24.38 | 13.51 | 6.00 | 1016.00 | 2.03 |
| Venezuela | 3/18/2020 | 2.50 | 24.65 | 13.37 | 5.00 | 1016.00 | 2.01 |
| Venezuela | 3/19/2020 | 2.37 | 24.60 | 13.29 | 5.00 | 1017.00 | 2.18 |
| Venezuela | 3/20/2020 | 2.23 | 24.28 | 13.14 | 6.00 | 1017.00 | 2.24 |
| Venezuela | 3/21/2020 | 2.50 | 24.34 | 13.02 | 6.00 | 1016.00 | 1.99 |
| Venezuela | 3/22/2020 | 3.31 | 24.86 | 13.88 | 6.00 | 1016.00 | 2.01 |
| Venezuela | 3/23/2020 | 3.90 | 25.23 | 14.23 | 5.00 | 1017.00 | 2.05 |
| Venezuela | 3/24/2020 | 3.92 | 25.26 | 14.19 | 5.00 | 1016.00 | 2.05 |
| Venezuela | 3/25/2020 | 3.32 | 25.16 | 13.16 | 7.00 | 1014.00 | 2.05 |
| Venezuela | 3/26/2020 | 2.78 | 25.13 | 13.53 | 7.00 | 1013.00 | 2.02 |
| Venezuela | 3/27/2020 | 2.08 | 24.55 | 12.84 | 7.00 | 1013.00 | 2.08 |
| Venezuela | 3/28/2020 | 2.23 | 24.72 | 13.59 | 6.00 | 1013.00 | 2.13 |
| Venezuela | 3/29/2020 | 3.34 | 24.97 | 14.25 | 5.00 | 1015.00 | 2.42 |
| Venezuela | 3/30/2020 | 3.01 | 24.81 | 13.27 | 5.00 | 1014.00 | 2.29 |
| Venezuela | 3/31/2020 | 3.05 | 25.00 | 13.13 | 6.00 | 1013.00 | 1.83 |
| Venezuela | 4/1/2020 | 3.48 | 26.00 | 13.09 | 7.00 | 1012.00 | 1.37 |
| Venezuela | 4/2/2020 | 3.68 | 26.33 | 13.28 | 7.00 | 1013.00 | 1.29 |
| Venezuela | 4/3/2020 | 4.78 | 26.89 | 15.14 | 7.00 | 1013.00 | 1.24 |
| Venezuela | 4/4/2020 | 4.87 | 27.60 | 15.57 | 5.00 | 1012.00 | 1.45 |
| Venezuela | 4/5/2020 | 4.02 | 27.31 | 14.30 | 7.00 | 1012.00 | 1.80 |
| Venezuela | 4/6/2020 | 4.18 | 27.71 | 15.78 | 7.00 | 1014.00 | 1.75 |
| Venezuela | 4/7/2020 | 3.90 | 27.89 | 15.67 | 7.00 | 1014.00 | 1.64 |
| Venezuela | 4/8/2020 | 3.50 | 27.74 | 14.93 | 7.00 | 1013.00 | 1.82 |
| Vietnam | 2/26/2020 | 3.68 | 29.64 | 14.18 | 8.00 | 1013.00 | 2.75 |
| Vietnam | 2/27/2020 | 3.22 | 30.08 | 13.00 | 8.00 | 1013.00 | 3.14 |
| Vietnam | 2/28/2020 | 3.09 | 29.29 | 12.09 | 8.00 | 1013.00 | 2.52 |
| Vietnam | 2/29/2020 | 3.61 | 29.40 | 14.50 | 8.00 | 1012.00 | 3.39 |
| Vietnam | 3/1/2020 | 3.84 | 29.83 | 15.40 | 8.00 | 1012.00 | 3.28 |
| Vietnam | 3/2/2020 | 4.07 | 29.76 | 15.10 | 8.00 | 1012.00 | 3.89 |
| Vietnam | 3/3/2020 | 4.34 | 29.97 | 14.17 | 8.00 | 1013.00 | 4.12 |
| Vietnam | 3/4/2020 | 4.32 | 29.60 | 15.18 | 8.00 | 1013.00 | 3.72 |
| Vietnam | 3/5/2020 | 4.67 | 29.53 | 15.33 | 8.00 | 1013.00 | 4.12 |
| Vietnam | 3/6/2020 | 4.86 | 29.16 | 14.97 | 8.00 | 1014.00 | 3.60 |
| Vietnam | 3/7/2020 | 5.02 | 30.19 | 15.52 | 8.00 | 1014.00 | 2.47 |
| Vietnam | 3/8/2020 | 4.67 | 30.73 | 15.55 | 8.00 | 1013.00 | 2.00 |
| Vietnam | 3/9/2020 | 4.63 | 30.88 | 16.73 | 8.00 | 1011.00 | 2.68 |
| Vietnam | 3/10/2020 | 4.59 | 30.63 | 16.80 | 8.00 | 1011.00 | 2.54 |
| Vietnam | 3/11/2020 | 4.88 | 29.55 | 16.40 | 7.00 | 1013.00 | 3.74 |
| Vietnam | 3/12/2020 | 4.42 | 30.38 | 15.49 | 8.00 | 1014.00 | 3.85 |
| Vietnam | 3/13/2020 | 4.20 | 30.18 | 16.17 | 8.00 | 1015.00 | 3.53 |
| Vietnam | 3/14/2020 | 3.50 | 30.10 | 14.99 | 8.00 | 1014.00 | 4.04 |
| Vietnam | 3/15/2020 | 3.31 | 30.17 | 15.10 | 8.00 | 1014.00 | 4.91 |
| Vietnam | 3/16/2020 | 3.41 | 30.35 | 14.97 | 8.00 | 1013.00 | 4.61 |
| Vietnam | 3/17/2020 | 3.33 | 30.32 | 14.91 | 8.00 | 1015.00 | 4.25 |
| Vietnam | 3/18/2020 | 3.43 | 29.95 | 14.74 | 8.00 | 1013.00 | 3.55 |